\documentclass[a4paper,12pt, DIV13, smallheadings, headsepline, headinclude, footexclude, bibtotoc, BCOR14mm, pointlessnumbers ]{scrbook}

\usepackage{uqthesis}

 \usepackage{makeidx}  

\usepackage[square,comma,numbers,sort&compress]{natbib}

\usepackage[nottoc]{tocbibind}  

\usepackage{amsmath,amsfonts} 

\usepackage{graphicx} 

\usepackage{epsfig} 

\copyrightpagetrue 

\examinerscopytrue   

\titlesmallcapstrue

\begin{document}
\setkomafont{caption}{\itshape}
\setkomafont{pagehead}{\itshape}

\def\bea{\begin{eqnarray}}
\def\eea{\end{eqnarray}}
\def\beq{\begin{equation}}
\def\eeq{\end{equation}}
\def\f{\frac}
\def\k{\kappa}
\def\sx{\sigma_{xx}}
\def\sy{\sigma_{yy}}
\def\sxy{\sigma_{xy}}
\def\e{\epsilon}
\def\ve{\varepsilon}
\def\ex{\epsilon_{xx}}
\def\ey{\epsilon_{yy}}
\def\exy{\epsilon_{xy}}
\def\be{\beta}
\def\D{\Delta}
\def\h{\theta}
\def\r{\rho}
\def\a{\alpha}
\def\s{\sigma}
\def\kb{k_B}
\def\la{\langle}
\def\ra{\rangle}
\def\nn{\nonumber}
\def\bu{{\bf u}}
\def\bn{\bar{n}}
\def\br{{\bf r}}
\def\up{\uparrow}
\def\dn{\downarrow}
\def\S{\Sigma}
\def\dg{\dagger}
\def\d{\delta}
\def\p{\partial}
\def\l{\lambda}
\def\G{\Gamma}
\def\o{\omega}
\def\g{\gamma}
\def\kv{\bar{k}}
\def\ha{\hat{A}}
\def\hv{\hat{V}}
\def\hg{\hat{g}}
\def\hG{\hat{G}}
\def\hTT{\hat{T}}
\def\noi{\noindent}
\def\a{\alpha}
\def\d{\delta}
\def\p{\partial} 
\def\nn{\nonumber}
\def\r{\rho}
\def\xv{\vec{x}}
\def\rv{\vec{r}}
\def\fv{\vec{f}}
\def\ov{\vec{0}}
\def\vv{\vec{v}}
\def\la{\langle}
\def\ra{\rangle}
\def\e{\epsilon}
\def\o{\omega}
\def\n{\eta}
\def\g{\gamma}
\def\th{\hat{t}}
\def\uh{\hat{u}}
\def\break#1{\pagebreak \vspace*{#1}}
\def\f{\frac}
\def\hf{\frac{1}{2}}
\def\uu{\vec{u}}

\frontmatter
\large\sffamily
\title{\sf{\LARGE\bfseries  Equilibrium and transport properties of constrained systems}} 
\author{\sf{\large\bfseries Debasish Chaudhuri}} 
\dept{\sf{\LARGE\bfseries S. N. B. N. C. B. S., Kolkata}} 

\titlepage 

\signaturepage  

\chapter{Acknowlegements}

This thesis is a result of almost five years of work. During this
period I have been accompanied and supported by many people. It is
my pleasure that I have now the opportunity to express my gratitude
to all of them.

First and foremost, I would like to thank Dr. Surajit Sengupta, my
thesis supervisor, for being a continuous source of support, enthusiasm,
many new ideas, constructive criticisms, for giving me  
independence to pursue research on the few ideas that ever came across
my mind. His simplicity, knowledge, open mind, simple and critical outlook
towards science always remain an inspiration. It helped me a lot, that I
could always discuss with him on any topic without any hesitation. I 
admire his ability of concentrating in his work even in very disturbing
situations.

It is a pleasure to thank Dr. Abhishek Dhar, Raman Research Institute,
Bangalore. I found his support at a time I needed it most. His interests
and clear understanding in a wide spectrum of physics, his simple 
and clean reasonings and his dedication in research has always 
been an inspiration. I gained a lot from the opportunity of working 
with him, his teaching, his criticism, his friendship and many many 
discussions that I had with him.

I thank  Dr. Madan Rao and Dr. Yashodhan Hatwalne of Raman Research Institute,
Bangalore, for introducing me to the fascinating field of soft condensed 
matter physics. Madan remains a source of inspiration with his enthusiasm, 
non- stop flow of new ideas, knowledge, critical outlook, his views towards 
science, many useful discussions and all his encouragements.

Two of my teachers in Kalyani University, Prof. Padmanabha Dasgupta and
Dr. Somenath Chakraborty inspired with their teaching and encouragements.
I thank Dr. Srikanth Sastry (~Jawaharlal Nehru Centre for Advanced
Scientific Research, Bangalore~) from whom I have learnt
computational techniques, that remains very useful in my research.
A very special thanks to Dr. P. A. Sreeram for all the
time that he gave to maintain a very good computational facility
at S. N. Bose National Centre for Basic Sciences (SNBNCBS). 
I thank Sreeram also for a lot of useful discussions.

I would like to acknowledge helpful discussions with V. B. Shenoy 
(Indian Institute of Science, Bangalore), 
Y. Kantor (Tel Aviv University, Israel), 
M. Barma (Tata Institute of Fundamental Research, Mumbai),
E. Frey (Ludwig- Maximilians- Universi{\"t}at M{\"u}nchen, Germany) and
A. M. Jayannavar (Institute of Physics, Bhubaneswar). 
I also thank Raman Research Institute, Bangalore, where 
a significant part of my work on semi- flexible polymers and heat conduction
was carried out, for hospitality. I would also like to acknowledge the 
Council of Scientific and Industrial Research, Government of India for 
the CSIR (NET) research fellowship. The computational facility from
DST Grant No. SP/S2/M-20/2001 is gratefully acknowledged.

It is a pleasure to acknowledge all the supports of non- academic staffs 
of SNBNCBS. I would like to specially mention the efficient handling of
CSIR related matters by Mr. Sukanta Mukherjee of accounts section and 
prompt helps by Mr. Swapan Ghosh and Ms. Ruma Mazumdar of library. 

I have found good support from a lot of friends and colleagues. Abhishek 
Chaudhuri and Sudipto Muhuri are the two with whom I shared a lot of good
times, common concerns, many worries and a good lot of
discussions, scientific and otherwise that helped me to grow. 
The regular discussions with Abhishek on our work was helpful impetus 
and often helped to understand things better. Many discussions with
Gautam Mukhopadhyay, Tamoghna K. Das, Arya Paul and Monodeep Chakraborty
was useful. At various times 
I remained fortunate enough to get accompaniements of good friends like,
Anuj Nandi, Ain-ul Huda, Dipanjan Bhattacharya, Dipanjan Chakraborty, 
Indranil Chattopadhyay, Kamal K Saha,  Ram Narayan Deb, Samir Mandal, 
Santabrata Das and Utpal Chatterjee.

Last, but not the least, I would like to thank my family members. 
It is my pleasure to thank my wife Swarnali Bandopadhyay who has always
been supportive through many ups and downs, in times of anxiety, frustrations
and few small successes, who has allowed all my unwillingness to do anything 
outside the academics and always remained a source of encouragement. I have also
benefitted a lot from continuous scientific discussions with her.
I thank my father for being so supportive that he is. I thank
Swapan and Kakima for being there, always, whenever the going got tough.

\chapter{List of Publications}

\noindent {\bf Publications: in refereed journals}  
\newcounter{refno10}
\begin{list}{[\arabic{refno10}]}{\usecounter{refno10}}
\item ``Heat conduction in confined solid strip: response to external strain",
{\bf Debasish Chaudhuri} and Abhishek Dhar, communicated to {\em Physical
Review E} (cond-mat/0601310).
\item ``Electrical transport in deformed nanostrips: electrical signature of reversible mechanical failure",  Soumendu Datta, {\bf Debasish Chaudhuri}, Tanusri Saha-Dasgupta, Surajit Sengupta, {\em EuroPhysics Letters} {\bf 73}, 765 (2006);
(cond-mat/0511457); [~not included in this thesis~].
\item ``Direct test of defect mediated laser induced melting theory for two dimensional solids", {\bf Debasish Chaudhuri} and Surajit Sengupta, {\em Physical 
Review E} {\bf 73}, 011507 (2006); (cond-mat/0508514).
\item ``Constrained Deformation of a Confined Solid: Anomalous Failure by Nucleation of Smectic Bands", {\bf Debasish Chaudhuri} and Surajit Sengupta, {\em Physical Review Letters}, {\bf 93}, 115702 (2004); (cond-mat/0401121).
\item ``A numerical renormalization group study of laser-induced freezing", {\bf Debasish Chaudhuri} and Surajit Sengupta, {\em EuroPhysics Letters}, {\bf 67}, 814- 819 (2004); also see ``{\em Erratum} A numerical renormalization group study of laser-induced freezing", {\bf Debasish Chaudhuri} and Surajit Sengupta, {\em EuroPhysics Letters}, {\bf 68}, 160 (2004); (cond-mat/0403319).
\item ``Triple Minima in the Free Energy of Semiflexible Polymers", Abhishek Dhar and {\bf Debasish Chaudhuri}, {\em Physical Review Letters}, {\bf 89}, 065502 (2002); (cond-mat/0203482).
\end{list}

\vskip .2cm
\noindent{\bf Publications: in proceedings}
\newcounter{refno11}
\begin{list}{[\arabic{refno11}]}{\usecounter{refno11}}

\item ``Mechanical Failure of a Small and Confined Solid", {\bf Debasish Chaudhuri} and Surajit Sengupta, {\em Indian Journal of Physics}, {\bf 79}, 941-945 (2005); (cond-mat/0508513).
\item ``Induced interfaces at nano-scales: structure and dynamics", Abhishek Chaudhuri, {\bf Debasish Chaudhuri} and Surajit Sengupta, {\em International Journal of Nanoscience}, {\bf 4}, 995 (2005).
\item ``Elastic properties, structures and phase transitions in model colloids", P. Nielaba, K. Binder, {\bf D. Chaudhuri}, K. Franzrahe, P. Henseler, M. Lohrer, A. Ricci, S. Sengupta, and W.Strepp, {\em J. Phys.: Condens. Matter}, {\bf 16}, S4115- S4136 (2004).
\end{list}

\chapter{Abstract}

Systems under external confinement and constraints 
often show interesting properties. In this thesis,
we study some systems under external confinement. We begin by finding out
the probability distribution of end-to-end separation of a semiflexible polymer
within the Worm Like Chain model using Monte- Carlo simulations (MC). In the 
constant extension ensemble, where
the two ends of the polymer are placed in stiff potential traps created by 
laser tweezers, the probability distribution shows triple maxima indicating a
non-monotonic force- extension in some intermediate regime of polymer stiffness.
Whereas this feature is absent in constant force ensemble. Our study on this 
system revealed the ensemble dependence of physical properties for finite sized
systems.Then, we fix the orientations at the ends of a polymer and find that the
orientation, as well as the ensemble, control the statistics and mechanical
properties of a semi- flexible polymer. We present an exact theory to
calculate the partition functions in both the Helmholtz and Gibbs ensembles 
and this theory takes care of the orientations at the ends of the polymer. 
We find multimodality in Helmholtz ensemble as a generic signature of 
semi- flexibility.

Secondly, we study Laser Induced Freezing (LIF\index{LIF}) where a colloidal 
liquid is constrained by an external laser field periodic in one direction. 
Using 
a Kosterlitz- Thouless type renormalization group calculation and a restricted
MC simulation we  calculate the phase diagrams for model 
colloids interacting via Hard Disk, Soft Disk and DLVO potentials. The phase
diagrams match exactly with the corresponding phase diagrams simulated by other
groups, thereby proving that LIF is indeed a dislocation mediated transition.

Lastly, we study the phase behaviors and failure mechanism of a two-
dimensional solid confined within a hard wall channel using MC and 
Molecular dynamics simulations. This system fails by 
nucleation of smectic phase within the solid. We have shown that thinner 
strips are stronger! The failure is ductile showing reversible plasticity. 
Density functional arguments can capture some of these features qualitatively.
We have used a mean field calculation to find out the phase diagram of
this system in density- channel width plane.
We show that fluctuations in quasi one dimension lead to very strange
behavior, namely a system that looks solid considering its structure factor 
shows vanishingly small shear modulus like a liquid!
We study the impact of this reversible failure on transport properties.
We find that the heat current in response to tensile strain varies differently 
depending on whether the strain is imposed in the confining direction or
the other. We propose a simple free volume calculation that captures the strain
response of heat current, exactly, within small strains.

\mainmatter

\tableofcontents

\chapter{Introduction}
\begin{verse}
\begin{flushright}
{
\it {Every year during the month of March a family of ragged gypsies
would set up their tents near the village, and with a great uproar
of pipes and kettledrums they would display new innovations.}
\hfill -- G. G. M{\'a}rquez}
\end{flushright}
\end{verse}
\vskip 1cm

\noindent
Traditionally, the science of thermodynamics and statistical mechanics were 
concerned with determining the properties of materials in the so-called 
{\it thermodynamic limit} \cite{griffiths}. 
In this limit, relevant for most materials and experimental 
situations, the number of particles $N\to \infty$ and
all statistical mechanical ensembles e.g. 
Helmholtz (constant number, volume, strain, etc.) and Gibbs (constant 
chemical potential, pressure, stress, etc.) are rigorously equivalent.
Also, in this limit, for most systems with short ranged interactions 
and compact shape, the nature of the boundaries and boundary fields have 
no impact on bulk properties. 

\noindent
Historically, the need for examining systems far removed from the 
thermodynamic limit arose first with the advent of computers and computer 
simulation techniques in the early 1950s and 60s. Early computers were 
not very powerful and the largest system sizes that could be handled 
were small. The need for extrapolating results obtained from small 
system sizes used in computer simulations for predicting phase boundaries, 
susceptibilities, critical exponents etc. as measured in the laboratory 
spawned the discipline of finite size scaling 
\cite{fisher-rmp1,fisher-rmp2,stanley}. The early 
works of Binder \cite{bincu,bin1st} and 
Fisher\cite{fisher-1,fisher-2,fisher-3,fisher-4} are significant
in this context. The emerging ideas of the renormalization group 
\cite{wil-kogu,golden}
had a direct impact on this endeavor 
and aided immensely our understanding of the thermodynamics of small 
systems. 

\noindent
Simultaneously, technological breakthroughs in semiconductors, magnetic 
recording devices and experiments on surfaces, thin films and adsorbtion 
pointed out the importance of boundaries and surfaces in determining 
materials properties \cite{sutton,safran}.
Surface phases, wetting, the physics of interfaces 
and surface phase transformations became topics of intense study using 
computer simulations, renormalization group theory and experiments. In 
the later part of the last century rapid advances in primarily two areas 
of science and technology were responsible for a further spurt of activity 
in trying to understand the role of finite size in determining the properties 
of materials. Firstly, new techniques like the use of lasers in manipulating 
objects and novel microscopic techniques meant that one could measure 
properties of systems down to the size of a large molecule e.g. biologically
important polymers like proteins and DNA \cite{ott, gittes, cell}. 
Rapid advances in biotechnology 
made it possible for us to ultimately measure forces involved, say, in 
the replication of DNA or in protein synthesis \cite{1dna,smol}. 
Nanotechnology on the other hand \cite{nanobook2}, 
using techniques like laser trapping, atomic force and 
scanning tunneling microscopy could finally make the study of finite 
systems useful and imperative for its own sake, rather than a precursor 
to taking the thermodynamic limit. 

\noindent
It is this general context which provides the backdrop of this thesis. 
In this thesis we have focussed on a number of such systems which are 
either ``small'' in the sense of being far from the thermodynamic limit
and/or are acted upon by external fields which produce severe constraints
that leads to significant change in their behaviour which may be strongly 
ensemble dependent. We study their statistical and
mechanical properties, phase behaviors etc. We also study transport properties
of one such system as it undergoes structural transformations that are
controlled by external confinement and strains. The structure of this thesis
is as follows.

In the next two chapters we study the properties of a semiflexible polymer.
We work within a coarse grained model -- the worm like chain model --
of such a polymer. This model has been successful in predicting mechanical
properties obtained from single molecule experiments on real biological 
polymers like, DNA, microtubules, actin filaments etc. In chapter-2, 
we show that in the Helmholtz ensemble these polymers can show a non- monotonic
force versus extension which is an impossibility if the experiment on this 
finite chain is made in the Gibbs ensemble or in the thermodynamic limit. 
This behavior is obtained in a certain range of stiffness of a semiflexible
polymer and gives a qualitative signature of semiflexibility vis-a-vis a
flexible polymer. In this chapter the non- monotonic force- extension curve 
was obtained for a polymer whose boundaries were free to rotate. In chapter-3,
we present an exact theory to calculate the properties of a semi- flexible 
polymer for all possible bending rigidities taking into account the
orientations of ends of a polymer. Using this theory and simulations we
establish that imposing constraints in boundary orientations vary the
statistics and mechanical properties of the polymer. Thus in these
two chapters we establish that for semi- flexible polymers, both ensembles
and boundary orientations leave important impact on the system properties.

In chapter-3, we study the phenomenon of laser induced freezing and reentrant
melting. This comes about due to constraining a two dimensional system by
imposing an external potential that is constant in one direction and 
periodically modulated
in the other. We take the simplest case where the periodicity of the external
potential is commensurate with spacings between the lattice planes of the 
system. Thus the external constraint permeates the whole system. With
change in the strength of this potential phase transitions occur
between a modulated
liquid and a locked floating solid phase. We use a constrained Monte- Carlo
simulation to obtain input parameters which are used in a numerical 
renormalization group scheme to test a recently proposed defect mediated 
melting theory for this system. 

In the following chapter, we study the impact of a different kind of 
external potential on a two dimensional solid. We confine the 
solid in a quasi one dimensional hard channel such that, in the direction
of confinement the solid is only a few atomic layers thick and study its 
mechanical and phase behavior. Study of this system is important due to 
the recent interest in nano- technology. This confinement is a boundary
effect and drastically changes the properties of the solid and introduces
many layering transitions, which are absent in the thermodynamic limit. 
 We also study the anomalous and reversible 
failure of this system under tensile strain. 
We present mean field calculations to substantiate
our simulation data. We end this chapter with a discussion on enhanced
flcutuations in this system due to the reduced dimensionality.

In chapter-6, we find out the impact of reversible failure on the 
heat transport properties of this system. We use molecular dynamics simulation
to calculate the change in heat current as the system undergoes tensile strain. 
The change in heat current in response to strain imposed in the confining
direction is very different from the case when the strain is imposed in the
perpendicular direction.
We introduce a free volume like theory to calculate the heat current and 
obtain exact match with simulation results up to small strains. 

In summary, in this thesis we have taken up a set of independent problems 
which are however connected by the theme of finite size effects and the effect 
of boundaries and constraints on the determination of overall behavior. Each 
of the chapters in the thesis are fairly self-contained and more or less 
independent of each other.

\chapter{Nonmonotonic Force-extension in Semi-flexible Polymer}
\begin{verse}
\begin{flushright}
\it {``Things have a life of there own", the gypsy proclaimed with a
harsh accent. ``It's simply a matter of waking up their souls".}
\hfill -- G. G. M{\'a}rquez
\end{flushright}
\end{verse}
\vskip 1cm

\vskip 1.5cm
The simplest model for describing semiflexible polymers without
self-avoidance is the so called Worm-Like-Chain (WLC) model \cite{doi,saito,frey-wlc}. In this
model the polymer is modeled as a continuous curve that can be specified by a
$d-$dimensional ($d > 1$) vector $\bar{x}(s)$, $s$ being the distance,
measured along the length of the curve, from one fixed end. The energy of
the WLC model is just the bending energy due to curvature and is given by 
\index{bending energy}
\index{WLC}
\bea
\f{H}{k_BT}=\f{\kappa}{2} \int_0^L (\f{\p \uh (s)}{\p s})^2 ds,
\eea
where $\uh (s)=\p{\bar{x}}/\p{s} $ is the tangent vector and satisfies
$\uh^2=1$. The parameter $\kappa$ specifies the stiffness of the
\index{stiffness}
chain and is related to the persistence length $\lambda$ defined
\index{persistence length}
through $\la \uh(s). \uh(s') \ra = e^{-|s-s'|/ \lambda}$. It can
be shown that $\kappa= (d-1) \lambda/2$ .

The thermodynamic properties of such a chain can be obtained from the
free energy which can be either the Helmholtz $(F)$ free energy or
\index{Helmholtz}
the Gibbs $(G)$ 
\index{Gibbs}
energy. In the former case one considers a polymer whose ends are kept
at a fixed distance $r$ [one end fixed at the origin and the other end
at $\rv =(0,...0,r)$] 
 by an average force $ \la f \ra = \p F(r,L)/\p r$, while in the latter
case one fixes the force and the average extension is given by $\la r \ra= -\p G(f,L)/
\p f$. It can be shown that in the thermodynamic limit $L \to \infty$
\index{thermodynamic limit}
the two ensembles are equivalent and related by the usual Legendre 
\index{Legendre transform}
transform $G=F-fr$.        
For a system with finite $L/\lambda$, the equivalence of the two
ensembles is not guaranteed, especially when fluctuations become
large. We note that real polymers come with a wide range of 
values of the parameter $ t=L/\lambda$ [e.g. $\lambda \approx 0.1 \mu m$ for
DNA while $\lambda \approx 1 \mu m$ for 
\index{DNA}
Actin and their lengths can be varied] and fluctuations in $r$ (or $f$)
can be very large. Then the choice of the ensemble depends on 
\index{ensemble}
the experimental conditions. Experiments on stretching polymers are usually
performed by fixing one end of the polymer and attaching the other end
to a bead which is then pulled by various means (magnetic, optical,
mechanical, etc.). In such experiments one can either
 fix the force on
the bead and measure the average  polymer extension, or, one could constrain
the bead's position and look at the average force on the polymer.
\index{bead}
In the former case, the Gibbs free energy is relevant while
\index{Gibbs}
it is the Helmholtz   in the second case. This point has been carefully
\index{Helmholtz}
analyzed by Kreuzer and Payne in the context of atomic force microscope
\index{atomic force microscope}
experiments \cite{kreu}.

Theoretically, the constant-force
ensemble is easier to treat, and infact an exact numerical solution
\index{ensemble}
has been obtained \cite{marko} (though only for $ t >> 1$).  
In two extreme limits of small force and large force WLC model can be
\index{WLC}
solved to obtain force extension relations. In 3D this relation in the
small force limit is $\la z \ra/L = 2\l f/3\kb T$ and in the high force
limit is $\la z \ra/L = 1 - \sqrt{\kb T/4\l f}$.
Data on force-extension experiments on DNA \cite{smith}
\index{force-extension}
\index{DNA}
have been explained using this ensemble  \cite{marko}.    
The case of constant-extension ensemble turns out to be much harder
\index{constant-extension}
\index{ensemble}
and no exact solution is available. The $t \to 
0$ and $t \to \infty $ cases correspond to the solvable limits  of the
hard rod and the Gaussian chain. The small and large $t$ 
cases have been treated analytically  by perturbation theory about
\index{perturbation theory}
these two limits \cite{dan,gob,nori}. Numerical simulations for
different values of $t$ have been reported by Wilhelm and Frey
\cite{wilh}, who have also obtained series expansions valid in the
small $t$ limit. A mean-field treatment has also recently been
\index{mean-field}
reported \cite{thiru}.

In this chapter we probe the nature of the transition from the Gaussian
to the rigid rod with change of stiffness 
\index{rigid rod}
\index{stiffness}
as shown by the form of the Helmholtz free energy of the  WLC model
\index{Helmholtz}
(or equivalently the distribution of end-to-end distance). Extensive
simulations are performed in two and three dimensions using the
equivalence of the WLC 
\index{WLC}
model to a random walk with one-step memory. We find the surprising
\index{one-step memory}
\index{random walk}
result that, over a range of values of $t$, the free energy has three 
minima. This is verified in a one-dimensional version of the model
which is exactly solvable. 

We first note that the WLC model describes a particle in
\index{WLC}
$d-$dimensions moving with a constant speed (set to unity) and with a random
acceleration. It is thus described by the propagator
\index{propagator}
\bea
&& Z(\xv,\uh,L|\xv',\uh',0)
 = \f{ \int^{(\xv,\uh)}_{(\xv',\uh')} {\cal{D}}[\xv(s)] 
e^{-H/k_BT}}{  \int {\cal{D}}[\xv{(s)}]
e^{-H/k_BT} }  
\eea
where in the numerator only paths $\xv(s)$, satisfying
$\xv(0)=\xv',\xv(L)=\xv, \uh(0)=\uh' $ and $ \uh(L)=\uh$ are
considered. It can be shown that the corresponding probability
\index{probability}
distribution $W(\xv,\uh,L)$ 
satisfies the following Fokker-Planck equation \cite{dan,gob}:
\index{Fokker-Planck}
\bea
\f{\p W}{\p L}+\uh.\nabla_{\xv} W - \f{1}{2 \kappa} \nabla_{\uh}^2 W=0
\label{fkpl}
\eea
where $\nabla_{\uh}^2$ is the diffusion operator on the surface of the
\index{diffusion operator}
unit sphere in $d-$dimensions. 
The discretized version of this model is the freely
\index{freely rotating chain}
rotating chain model (FRC) of semiflexible polymers \cite{doi}. In the FRC one
\index{FRC}
considers a polymer with $N$ segments, each of length
$b=L/N$. Successive segments are constrained to be at a fixed angle,
$\theta$, with each other. The WLC model is obtained, in the limit
$\theta , b \to 0$, $ N \to \infty$ keeping $\lambda =2 b/ \theta^2$
and $L=Nb$ finite. In this chapter we report the simulation results of this
\index{simulation}
FRC model. In the next chapter we shall see
\index{FRC}
how WLC can also be discretized to a Heisenberg spin model with nearest 
\index{WLC}
neighbour coupling.

Here we will consider the  situation where the 
ends are kept at a fixed separation $r$ [with $\xv'$ at the origin and
$\xv=\rv=(0,...0,r)$]
but there is no constraint on $\uh$ and 
$\uh'$ and they are taken as uniformly distributed. Thus we will be
interested in the distribution 
$P(r,L)=\la \d (\xv-\rv) \ra= \int d\uh W(\rv,\uh,L) $: this gives
the Helmholtz free energy $F(r,L)=-Log[P(r,L)]$. 
\index{Helmholtz}
For the spherically symmetric situation we are considering, $P(r,L)$
\index{spherically symmetric}
is simply related to the radial probability distribution $S(r,L)$
\index{probability}
through $S(r,L)=C_d r^{d-1}P(r,L)$, $C_d$ being a constant equal to the area of
the $d-$dimensional unit sphere. 
It may be
noted that the WLC Hamiltonian is  equivalent to spin $O(d)$ 
\index{WLC}
models in one dimension in the limit of the exchange constant $J \to
\index{dimension}
\infty$ (with $Jb =\kappa$ finite) and 
all results can be translated into spin language.
However, for spin systems, the present free energy is not very relevant since it
corresponds to putting unnatural constraints on the magnetization
\index{constraints}
vector.

\begin{figure}[t]
\begin{center}
\includegraphics[width=8.0cm]{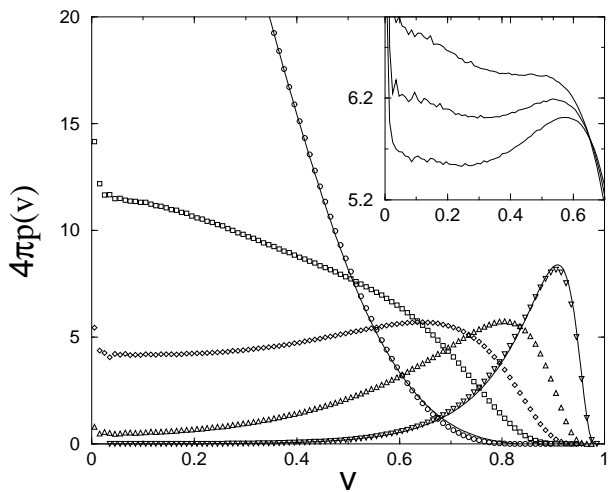}
\end{center}
\caption{ Monte-Carlo data for $p(v,t)$ for the $3-$dimensional WLC for
\index{WLC}
values of $t=10 (\circ),5,3.33,2$ and $1 (\nabla)$. The inset is a
blowup of curves in the transition region ($t=4,3.85,3.7$) and shows the
presence of the two maxima. 
}
\label{wlc3d}
\end{figure}

\section{Numerical Simulations}
\label{numerics}
The simulations were performed by generating
random configurations of the FRC model and computing the distribution
\index{FRC}
of end-to-end distances. 
To obtain equivalence with the WLC model the appropriate limits were
\index{WLC}
taken. We note that 
because these simulations do not require equilibration, they are much
faster than simulations on equivalent spin models and give
better statistics. The number of configurations generated was 
around $10^8$ for  chains of size $N=10^3$. We verified that
increasing $N$ did not change the data significantly. As a
check on our numerics we evaluated $\la r^2 \ra $ and $\la r^4
\ra$. Using Eq.~(\ref{fkpl}) and following \cite{herm} we can compute these (in all dimensions):
\bea
&&\la r^2 \ra=\f{4 \kappa L}{d-1}-\f{8 \kappa^2(1-e^{-\f{(d-1)L}{2
\kappa}})}{(d-1)^2}   \nn \\ 
&& \la r^4 \ra =\f{64 \kappa^4(d-1)}{d^3 (d+1)^2}e^{-\f{dL}{\kappa}}-\f{128 \kappa^4
(d+5)^2}{(d-1)^4 (d+1)^2}e^{-\f{(d-1)L}{2\kappa}} +\f{64 \kappa^3 L
(d^2-8 d+7)}{(d-1)^4(d+1)} e^{-\f{(d-1)L}{2 \kappa}}\nn \\ 
&& +\f{64 \kappa^4 (d^3+23 d^2-7d+1)}{(d-1)^4 d^3}  -\f{64 \kappa^3 L
(d^3+5 d^2-7d+1)}{(d-1)^4 d^2}+\f{16 \kappa^2 L^2 (d^3-3d+2)}{d(d-1)^4}.
\eea
Infact it is straightforward to compute all even moments, though it
becomes increasingly tedious to get the higher moments. 
Our numerics agrees with the exact results to around
$0.1 \% $ for $\la r^2 \ra$ and $0.5\%$ for $\la r^4 \ra$.

\begin{figure}[t]
\begin{center}
\includegraphics[width=8.0cm]{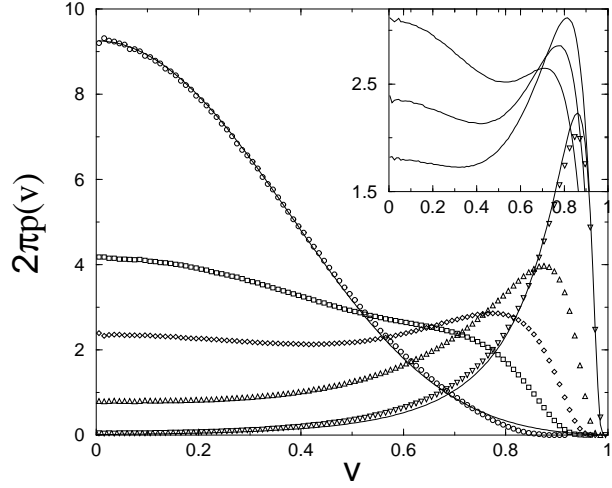}
\end{center}
\caption{ Monte-Carlo data for $p(v,t)$ for the $2-$dimensional WLC for
\index{WLC}
values of $t=10 (\circ),5,3.33,2$ and $1 (\nabla)$. The inset is a
blowup of curves in the transition region ($t=4,3.33,2.86$) and clearly shows the
presence of the two maxima. Note that because of $\pm v$ symmetry, we
have plotted data for positive $u$ values only. For the fits at large
and small $t$ see text.}     
\label{wlc2d}
\end{figure}

The function $P$ has the scaling form $P(r,L)=\f{1}{L^d} p(r/L,L/\lambda )$ and we will
\index{scaling}
focus on determining the function $p(v,t)$ 
{\footnote {Wilhelm and Frey \cite{wilh} have looked at the radial 
distribution $S(r,L)$ which however misses the interesting details of the
transition. We note that the relevant distribution here is indeed
$p(v,t)$ since this gives the Helmholtz free energy}}
\index{Helmholtz}
In Fig.~(\ref{wlc2d}) and Fig.~(\ref{wlc3d}), we show the results of
our simulations in two and three dimensions. 
At large values of $t$ there is a
single maximum at $v=r/L=0$ corresponding to a Gaussian distribution while at
small $t$, the maximum is close to the fully extended value of
$v=\pm 1$. The transition is {\it first-order-like}: as we decrease $t$,
\index{first-order}
at some critical value, $p$ develops two additional
maxima at non-zero values of $v$. Further decreasing $t$ weakens the
maximum at $v=0$ until it finally disappears and there are
just two maxima which correspond to the rigid chain.

For the limiting cases of small and large values of $t$ there are
analytic results for the distribution function and as can be seen in
Fig.~(\ref{wlc2d},\ref{wlc3d}) our data agrees with them.
 For large $t$ we find that Daniels approximation \cite{dan},
\index{Daniels approximation}
which is a perturbation about the Gaussian, fits the data quite
well. In the other limit of small $t$ the series solutions provided in
\cite{wilh} fits our data. 
For intermediate values of $t$ neither of the two forms are able to
capture, even qualitatively, the features of the free energy. 
Specifically, we note that all the analytic theories(perturbative, series
expansions and mean-field)  
\index{mean-field}
predict a second-order-like
transition and do not give triple minima of the free energy for any
\index{triple minima}
parameter value.

\section{Exact Calculation in 1D}
\label{1d-wlc}
It is instructive to study the following one-dimensional version of the WLC
\index{WLC}
which shows the same qualitative features (the equivalent spin problem
is the Ising model). We consider a $N$ step random walk,
\index{random walk}
with step-size $b$ which, with probability 
\index{probability}
$\epsilon$, reverses its direction of motion and with $1-\epsilon$,
continues to move in the same direction. 

\begin{figure}[t]
\begin{center}
\includegraphics[width=8.0cm]{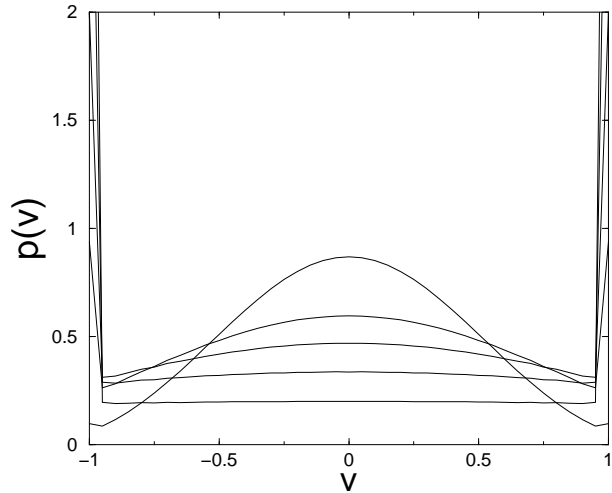}
\end{center}
\caption{ The exact distribution $p(v,t)$ of the $1-$dimensional WLC
\index{WLC}
[ Eq.~(\ref{p1d})] for different
values of $t$ ($10,5,3.33,2,1$). Even for the most stiff chain
considered here ($t=1$), the distribution has a peak at the centre (in
addition to the $\delta-$function peaks at ends) though it looks flat.
}
\label{wlc1d}
\end{figure}

The appropriate scaling limit is: $b \to 0$, $\epsilon
\index{scaling}
\to 0$, $N \to \infty$ keeping $L=Nb$,  $t=L/\lambda =2N\epsilon $ finite.
Defining $Z_{\pm}(x,L)$ as the probability of the walker to be at $x$
\index{probability}
with either positive or negative velocity, we have the following
Fokker-Planck equation:
\index{Fokker-Planck}
\bea
\f{\p Z_{\pm}}{\p L}= \mp \f{\p Z_{\pm}}{\p x} \mp \f{1}{2 \lambda}
(Z_+-Z_-)
\eea    
This can be solved for $P(x,L)=Z_++Z_-=\f{1}{L} p(x/L,L/\lambda)$. We get
\bea
&& p(v,t)=  \f{te^{-t/2}}{4}
[\f{I_1(\f{t}{2}\sqrt{1-v^2})}{\sqrt{1-v^2}}
+I_0(\f{t}{2}\sqrt{1-v^2})] \nn \\
&&~~~~~~~~~  + \f{e^{-t/2}}{2} [\d (v-1) + \d (v+1) ],
\label{p1d}
\eea
where $I_0$ and $I_1$ are modified Bessel functions. 
In Fig.~\ref{wlc1d}
we  have plotted $p(v,t)$ for different 
values of stiffness. We find that the probability distribution has three
\index{probability}
peaks for all values of $t$. Unlike in $2$ and $3$ dimensions, the
$\delta-$function peaks at $v=\pm1$ (which corresponds to fully
extended chains) persist at all values of stiffness though their
\index{stiffness}
weight decays exponentially. Similarly the peak at $v=0$ is always present.   

\section{Discussion} 
\label{discuss}
\begin{figure}[t]
\begin{center}
\includegraphics[width=8.0cm]{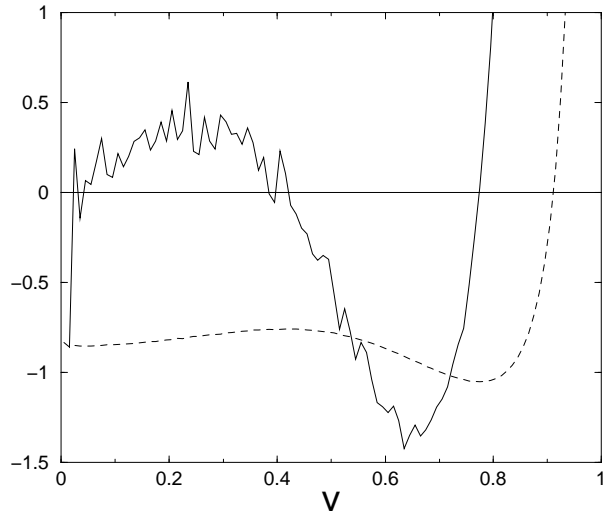}
\index{WLC}
\end{center}
\caption{ The free energy (dotted line) and the corresponding
force-extension curve (solid line) for a $2-$dimensional chain with $t=3.33$.
\index{force-extension}
\index{solid}
}
\label{fext}
\end{figure}

The most interesting result of this chapter is the triple
\index{triple minima}
minima seen in the Helmholtz free energy of the WLC. 
\index{WLC}
\index{Helmholtz}
Physically, this results from the competing effects of
entropy, which tries to pull in the polymer and 
the bending energy,
\index{bending energy}
which tries to extend it. 
This form of the free energy 
leads to a highly {\it counterintuitive force-extension curve},  
very different from what one obtains from the
constant force ensemble or from approximate theories. In
Fig.~(\ref{fext}) we show the 
force-extension curve for a two dimensional chain with $t=3.33$. We
see that there are two stable positions for which the force is zero.   
In the constant-force ensemble, it is easy
to show that $\p \la r \ra /\p f=\la r^2 \ra-\la r \ra^2$ and so the
force-extension is always monotonic. However, in the
\index{force-extension}
constant-extension ensemble, there 
\index{constant-extension}
\index{ensemble}
is no analogous result ({\it{for finite systems}}), and consequently monotonicity is not guaranteed.  

\begin{figure}[t]
\begin{center}
\includegraphics[width=8.0cm]{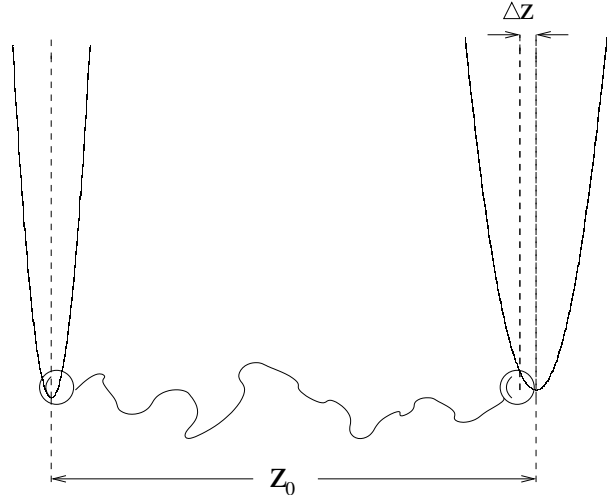}
\index{WLC}
\end{center}
\caption{ 
A schematic of the experimental set-up required to realize the
constant-extension ensemble discussed in this chapter. 
\index{constant-extension}
\index{ensemble}
For a stiff trap the average 
displacement of the bead $\la \Delta z \ra $ from the trap center is
\index{bead}
\index{trap}
\index{displacement}
small and the average force on the polymer is: $\la f \ra =-k \la \Delta z
\ra $. 
}
\label{schm}
\end{figure}

Most of the recent experiments on stretching DNA have  $t
\index{DNA}
\geq 100 $. The distribution is 
then sharply peaked at
zero and one expects the equivalence of different ensembles.    
Experimentally the value of $t$ can be tuned by
various means, for example, 
by changing the length of the polymer or
the temperature. 
Polymer-stretching experiments can thus be performed
for intermediate  $t$ values.
Since we consider the tangent vectors at the polymer-ends to
be unconstrained an accurate experimental realization of our set-up
would be one in which both ends are attached to beads [see
Fig.~(\ref{schm})]. The beads are
put in optical traps and so are free to rotate (this setup is
identical to the one used in refn.~\cite{mein}). Making the traps stiff
corresponds to working in the constant-extension ensemble 
\index{constant-extension}
\index{ensemble}
\cite{kreu} and one can measure the average force.  
Our predictions can then be experimentally
verified.

\vskip .1cm
We make some estimates on the experimental
requirements (for a $3$-d polymer with stiffness $t=3.85$). 
\index{stiffness}
Assume that at one end, the origin, the trap is so stiff that 
the bead can only rotate. We make measurements at the other end. The
trap-center is placed at $\rv_0=(0,0,z_0)$ and the mean bead displacement
\index{bead}
\index{displacement}
$\Delta z= \la (z-z_0) \ra $ gives the mean force $\la f \ra$ on the polymer. 
 We then consider the problem of the polymer in the presence of a  trap
\index{trap}
\index{trap potential}
potential $V=[k_t(x^2+y^2)+k (z-z_0)^2]/2$. Assume $k_t >> k$ so we
can neglect fluctuations in the transverse directions. The 
distribution of the bead's position in the presence of 
the potential is given by $Q(\rv)=e^{-\beta [F(\rv)+V(\rv)]}/\int d^3
\rv e^{-\beta [F(\rv)+V(\rv)]}$. For a stiff trap, we can expand $F$
\index{trap}
about $\rv=\rv_0$ and find the average displacement
\index{displacement}
of the bead. 
\index{bead}
\bea
F(\rv) = F(\rv_0) + \f{\p F}{\p x_i}|_{\rv_0} (x_i-x_{0_i})
       + \f{1}{2}\f{\p^2 F}{\p x_i \p x_j}|_{\rv_0} (x_i-x_{0_i})(x_j-x_{0_j})
\nn.
\eea
Due to spherical symmetry $F(\rv)=F(r)$. The operator identity $\f{\p}{\p x_i} =
\f{\p}{\p r} \f{\p r}{\p x_i} = \f{x_i}{r} \f{\p}{\p r}$ leads to 
$\f{\p F}{\p x_i}=\f{x_i}{r} \f{\p F}{\p r}$ and
\bea
\f{\p^2 F}{\p x_i \p x_j} = \f{1}{r}\f{\p F}{\p r}\d_{ij}+\f{x_i x_j}{r^2}
                  \left(\f{\p^2 F}{\p r^2} - \f{1}{r}\f{\p F}{\p r}\right)\nn
\eea
At the point of measurement, $\rv_0=(0,0,z_0)$, $\f{\p F}{\p x}= \f{\p F}{\p y}
=0$, $\f{\p F}{\p z}= F'_0$, $\f{\p^2 F}{\p x_i \p x_j}=0$ for $i\neq j$, 
$\f{\p^2 F}{\p x^2}=\f{\p^2 F}{\p y^2}=F'_0/z_0\equiv G_0$ and 
$\f{\p^2 F}{\p x^2}=F''_0$. Therefore,
\bea
F(\rv)+V(\rv) = F_0 + F'_0 (z-z_0) + \hf (G_0+k_t) (x^2+y^2) + \hf (F''_0+k) (z-z_0)^2\nn
\eea
writing $F(r_0)=F_0$.
Then the average displacement can be written as,
\index{displacement}
\bea
\Delta z &=& \f{\int d^3\rv (z-z_0) e^{-\be(F+V)}}{\int d^3\rv e^{-\be(F+V)}}\nn\\
&=& \f{\int dz (z-z_0) e^{-\be\hf k'(z-z_0+F'_0/k')^2}}{\int dz e^{-\be\hf k'(z-z_0+F'_0/k')^2}} = - \f{F'_0}{k'}\nn
\eea
where $k'=k+F''(z_0)$.
In short, we have shown that $\Delta z= \int d^3 \rv
(z-z_0) Q(\rv)= -\la f \ra/k'$, where $k'=k+F''(z_0) \approx k$ 
(valid except when $z_0 \approx L$) and $\la f \ra=F'_0=F'(z_0)$. 
The rms fluctuation of the
\index{fluctuation}
bead about the trap center is given by 
\index{bead}
\index{trap}
$z^2_{rms}=k_B T /k$. Hence we get $  \Delta z = -\la f\ra z_{rms}^2/(k_B T)=
-(\la f\ra L/k_B T) (z_{rms}^2/L)$. The scaled force $\la f \ra L/(k_B T)$ is of order
$0.1$. The different minima are separated by  distances $\approx 0.2 L$, hence to
see the effect we need to have $z_{rms}/L \leq 0.1$. Thus finally
we find that the typical displacement of the bead $\Delta z$ is about  $0.01
\index{displacement}
z_{rms}$. This is quite small and means that 
it is necessary to collect data on the bead position over long periods of
\index{bead}
time.  

As suggested in \cite{wilh}, a more direct way of measuring the
Helmholtz free energy would be to attach marker molecules at the ends of the
\index{Helmholtz}
\index{marker molecules}
polymer and determine the distribution of end-to-end
distances. Fluorescence microscopy as in \cite{ott} could be another
possible method. 
It is to be remembered of course that real polymers are well-modeled
by the WLC model provided we can neglect monomer-monomer interactions
(steric, electrolytic etc.).
\index{steric}
Thus the experiments would really test the relevance of the WLC model in
\index{WLC}
describing real semiflexible polymers in different stiffness regimes.
\index{stiffness}

\section{Conclusion}
\label{conclusion}
In conclusion we have presented some new and interesting properties of
the WLC model and have pointed out that polymer
\index{WLC}
properties are ensemble-dependent. This is a finite size effect.
\index{ensemble}
In this chapter we have given one
example of qualitative differences in force-extension measurements in
\index{force-extension}
different ensembles. Other quantitative differences will occur even in
more flexible chains and should be easier to observe
experimentally. If the ends of the polymer are free to rotate but the
end to end separation is constrained to be at a constant distance, such that 
the system is in constant extension ensemble, then the free energy of 
\index{ensemble}
the polymer is shown to have a triple
minima structure for some intermediate values of stiffness. In the next chapter
\index{stiffness}
we shall investigate the robustness of this triple minima structure as 
\index{triple minima}
some fixed orientations.
\index{}

\chapter{Semiflexibile Polymer: Ensemble and End- Orientation}
\begin{verse}
\begin{flushright}
\it {The children would remember for the rest of their lives the august
solemnity with which their father, devastated by his prolonged vigil
and by the wrath of his imagination, revealed his discovery to them:
``The earth is round like an orange".}
\hfill -- G. G. M{\'a}rquez
\end{flushright}
\end{verse}
\vskip 1cm

In the last chapter we have studied the statistical and mechanical
properties of a semiflexible finite polymer with its ends left free
to rotate. The bending rigidity coupled with the finite size of the
polymer gave rise to inequivalence of ensemble and
a very interesting triple minima in free energy
\index{triple minima}
in the Helmholtz ensemble and consequently non- monotonic
\index{Helmholtz}
force extension which is absent in Gibbs ensemble. In this chapter,
\index{Gibbs}
\index{ensemble}
we shall discuss an exact theory that gives results which match 
exactly with the simulations. Further, we shall go over to different
boundary conditions by fixing the orientations at the ends of a polymer
\index{boundary conditions}
and study its impact, which is very non- trivial, on the end to end
probability distributions and mechanical properties. 
\index{probability}

Microtubules and actin polymers constitute the structure of 
cytoskeleton that gives shape, strength and motility to most of the 
\index{motility}
\index{cytoskeleton}
living cells. They are semiflexible polymers in the sense that 
\index{cells}
their persistence lengths $\l$ are of the order of their chain 
lengths $L$ such that $t=L/\l$ is small and finite. For example,
Actin has $\l=16.7~\mu m$, $L\sim30~\mu m$\cite{ott,gittes}, Microtubule
\index{microtubule}
has $\l=5.2~mm$ and statistical contour length $L\sim 10~\mu m$\cite{gittes},
double stranded DNA has $\l=50~nm$ and contour length 
\index{double stranded}
\index{contour length}
\index{DNA}
$L\sim 300~nm$\cite{1dna}.
While it is obvious that in the thermodynamic limit of $t\to\infty$, 
\index{thermodynamic limit}
Gibbs (constant force) and Helmholtz (constant extension) ensemble
\index{Gibbs}
\index{Helmholtz}
\index{ensemble}
predict identical properties, the same is not true for real semi- flexible
polymers which are far away from this limit. 
In biological cells actin filaments remain dispersed throughout 
the cytoplasm with higher concentration in the cortex region, 
just beneath the plasma membrane. Microtubules, on the other hand,
have one end attached to a microtubule- organizing centre,  centrosome
\index{centrosome}
in animal cells. Another polymer, microtubule- associated proteins (MAP)
\index{cells}
\index{microtubule- associated proteins}
attach one or both their ends to microtubules to arrange them in
microtubule bundles \cite{cell}. Thus, end point orientation of polymers 
\index{microtubule bundles}
\index{microtubule}
play a crucial role in many important phenomena. For instance, gene- 
\index{gene- regulation}
regulation in the cell is controlled by DNA- binding proteins, many
of which loop DNA with fixed end orientations
\cite{dna-loop-1,dna-loop-2,dna-loop-3}.
\index{DNA}
Thus it becomes important to understand the statistics and the mechanical
properties of semi- flexible polymers with different possibilities
of end orientations and ensembles.

\begin{figure}[t]
\begin{center}
\includegraphics[width=15cm]{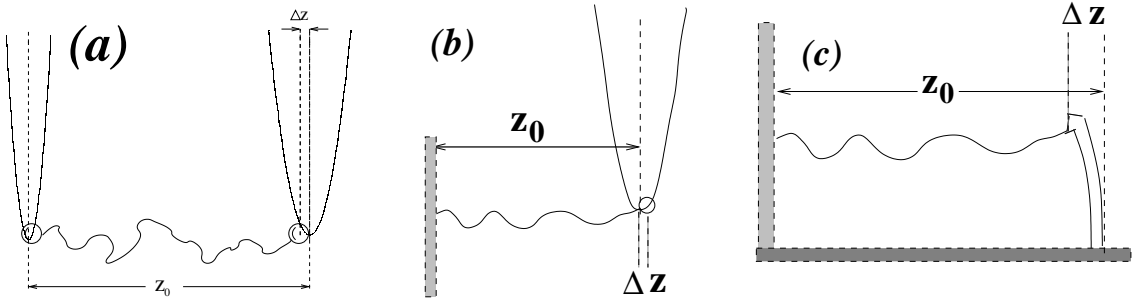}
\index{WLC}
\end{center}
\caption{Three possible experimental set- up for force- extension measurements.
\index{force- extension}
(a)~left and right ends are held by optical traps, (b)~left end is anchored to
a surface and the right end is held by optical trap, (c)~left end is anchored to
a surface and the right end is held by the functionalized tip of an AFM 
\index{AFM}
cantilever. $z_0$ denote the position of the free dielectric bead (for optical 
\index{bead}
\index{dielectric bead}
trap) or the free cantilever tip in absence of the polymer, $\D z$ denote the
\index{trap}
displacement associated with the force exerted by the polymer.
\index{displacement}
}
\label{expt-wlc2}
\end{figure}

During the last decade many single molecule experiments have been 
\index{single molecule experiments}
performed on  semi- flexible polymers\cite{smith,busta,smol,1dna}. 
This has been done by using optical tweezers\cite{busta}, 
magnetic tweezers\cite{magtwz}, AFMs\cite{afm} etc.
\index{AFM}
In optical tweezer experiments one end of a polymer is attached to
a dielectric bead which is, in turn, trapped by the light intensity profile 
of a laser tweezer. In this case the dielectric bead is free to rotate 
\index{laser tweezer}
\index{dielectric bead}
\index{laser}
within the optical trap. On the other hand,
attaching an end of a polymer to a super-paramagnetic bead,
one can use magnetic field gradients to trap the polymer using a magnetic 
\index{magnetic tweezer}
\index{trap}
tweezer setup. In this case one can rotate the bead while holding it fixed
\index{bead}
in position by changing the direction of the external magnetic field. 
\index{magnetic field}
In AFM experiments on the other hand one end of polymer is trapped
by a functionalized tip of an AFM cantilever.
\index{AFM}
In Fig.\ref{expt-wlc2} we have shown cartoons of three possible
experimental setups.
The two distinct procedures which can be followed to measure
force- extension are: 
\index{force- extension}
(a) Both the ends of the polymer are held via laser or magnetic tweezers. 
(b) One end of the polymer is attached to a substrate such that the position
\index{substrate}
and orientation of this end is fixed while the other end is trapped via
laser tweezer or magnetic tweezer or AFM cantilever. 
\index{magnetic tweezer}
\index{laser tweezer}
\index{AFM}
\index{laser}

\index{AFM}
\index{substrate}

While optical tweezers allow free rotation of dielectric beads 
within the trap thereby allowing free orientations of the polymer end,
\index{trap}
magnetic tweezers fix the orientation of the ends and one can study the 
dependence of polymer properties on orientation of its ends by controlled 
change of the direction of external magnetic field . 
\index{magnetic field}
In this chapter, we call this fixing of orientation of an end of 
a polymer as grafting. By changing the trapping potential 
from stiff to soft trap one can go from Helmholtz to Gibbs 
\index{Gibbs}
\index{Helmholtz}
\index{trap}
ensemble\cite{sam-ensemble}. 
Before we proceed, let us first elaborate on how to fix the
ensemble of a mechanical measurement. In a simplest case we can assume 
\index{ensemble}
that one end of the polymer is trapped in a harmonic well,
\bea
\be V(z) = C \f{(z-z_0)^2}{2}
\eea
with ($0,0,z_0$) being the position of potential- minimum. 
The polymer end will undergo continuous thermal motion. One can
use a feedback circuit to shift $z_0$
to force back the fluctuating polymer end to its original position.
This will ensure a Helmholtz ensemble.
\index{Helmholtz}
This can also be achieved by taking $C\to\infty$.
On the other hand one can use a feedback circuit to fix the force 
\index{feedback circuit}
$-C(z-z_0)$ by varying $C$ depending on the position $z$ of the
polymer end. This will ensure a Gibbs ensemble.
\index{Gibbs}
\index{ensemble}
This can also be achieved by taking a vanishingly soft
($C\to 0$) trap to infinitely large distance ($z_0\to\infty$)
\index{trap}
such that within the length scale of fluctuation the polymer end 
\index{fluctuation}
feels a constant slope of the parabolic potential. Surely, in experiments,
using a feedback circuit is easier to implement a particular ensemble.
\index{feedback circuit}
\index{ensemble}
However, the other procedure is mathematically well defined and one
can seek recourse of it to show that the partition function
\index{partition function}
of two ensembles are related by a Laplace transform~\cite{sam}.
\index{Laplace transform}
This does not depend on the choice of Hamiltonian for the polymer.
An exact relation between the two ensembles for worm like
\index{worm like chain}
chain (WLC) model is shown in Sec.\ref{pathint}.
\index{WLC}

From the above discussion on possible experiments, it is clear 
that there can be three possibilities of boundary conditions 
\index{boundary conditions}
in terms of orientation.
In an experiment the possibilities are, 
\index{experiment}
(a) free end: both the ends of a polymer can remain free to 
rotate\cite{mywlc,sam},
(b) one end grafted: one end may be grafted and the other can take all possible 
orientations\cite{munk-frey} and 
(c) both ends grafted: the orientations of both the ends are kept fixed.
\index{grafted}
Thus, in experiments, we can have two possible ensembles and 
three possible boundary conditions. 
\index{boundary conditions}
We investigate the probabilitiy distribution, free 
energy profile and force extension relation for each of 
these cases in this chapter. We shall see that the properties
of a semiflexible polymer depend both on the choice of the ensemble and
\index{ensemble}
\index{semiflexible polymer}
the boundary condition. 

WLC model is a simple coarse grained way to capture
\index{WLC}
\index{coarse grained}
bending rigidity of an unstretchable polymer~\cite{kratky,doi}
embedded in a thermal environment. 
Recent single molecule experiments in biological
\index{single molecule experiments}
physics~\cite{smith,busta,smol,1dna} renewed interest in this old model
of polymer physics. It was successfully employed by 
Bustamante, Marko, Siggia and Smith~\cite{bmss,marko} 
to model data from force- extension
\index{force- extension}
experiment~\cite{smith} on double stranded DNA molecules.
\index{double stranded}
\index{experiment}
\index{DNA}
Mechanical properties of giant muscle protein titin~\cite{titin,sac-tit},
\index{protein}
polysaccharide dextrane~\cite{afm,sac-tit} and single molecule of 
\index{AFM}
xanthane~\cite{xan} were also explained using WLC model. 
Due to the inextensibility constraint WLC model is hard to tract
\index{WLC}
analytically except for in the two limits of flexible chain ($t\to\infty$) 
\index{flexible chain}
and rigid rod ($t\to 0$), about which perturbative calculations have been 
\index{rigid rod}
done~\cite{dan,gob,nori}. A key quantity that describes statistical
property of such polymers is the end-to-end distance distribution.
Numerical simulations for
different values of $t$ have been reported by Wilhelm and Frey
\cite{wilh}, who have also obtained series expansions valid in the
small $t$ limit.  Mean-field treatments by Thirumalai and
\index{mean-field}
his collaborators has also been reported~\cite{thiru,jkb}. 
In an earlier study\cite{mywlc} we have investigated the free energy profile of 
a semiflexible polymer whose ends were free to rotate in the 
\index{semiflexible polymer}
constant extension ensemble and in the stiffness regime of 
\index{stiffness}
$1\le t\le 10$. This has been predicted that a clear qualitative signature of
semi- flexibility would be a non- monotonic force extension 
for stiffnesses around $t=4$ in Helmholtz (constant end to 
\index{Helmholtz}
end separation) ensemble. This comes from bimodality of probability
\index{bimodality}
distribution of end to end separation. This non- monotonicity must be absent 
in Gibbs (constant force) ensemble\cite{mywlc}. 
\index{Gibbs}
\index{ensemble}
Multiple maxima in the probability distribution of end to end separation
was due to a competition between entropy, that prefers a maximum near 
zero separation, and energy, that likes an extended polymer.
A lot of theoretical works followed
to reproduce and understand the probability density at all stiffnesses 
\index{probability}
including the very interesting bimodality using analytic techniques
\index{bimodality}
\cite{sam,kleinert-1,kleinert-2,stepanow}. 
Recently, Frey and his collaborators studied the interesting multimodality
in transverse fluctuations of a grafted polymer using simulations
\index{grafted polymer}
\index{grafted}
\cite{munk-frey} and approximate theory \cite{kroy,munk-frey-th}.
In a separate study Spakowitz and Wang used Greens function technique 
that takes into account the orientations of the polymer ends \cite{wang}.
WLC model has been extended to study double stranded to single stranded DNA 
\index{WLC}
\index{double stranded}
\index{single stranded}
\index{DNA}
transition~\cite{sunil} and  to incorporate twist degree of 
freedom~\cite{twist-1,twist-2,twist-3}. 

The construction of this chapter is as follows. 
In Sec.\ref{pathint} we present a path integral technique
for exact calculation of WLC model for all the three boundary 
\index{boundary conditions}
conditions and two ensembles. Then in Sec.\ref{simu} we discuss 
the different discretized versions of WLC model and the Monte- 
\index{WLC}
\index{Monte- Carlo}
Carlo (MC) simulation procedures followed in this work.
\index{simulation}
\index{MC}
In Sec.\ref{result} we present all the results obtained from
theory and simulations. In this section we present statisticl
and mechanical properties for all the possible situations.
Then we summarise our results and
conclude with some discussions in Sec.\ref{conclusion}.

\section{Theory}
\label{pathint}

In WLC model the polymer is taken as a continuous curve denoted
\index{WLC}
by a $d$- dimensional vector $\rv(s)$ where $s$ is a distance measured over the 
contour of the curve from any end of it. This curve has a bending 
rigidity and thus the Hamiltonian is given by
\bea
\be{\cal H} &=& \int_0^L ds~ \f{\k}{2} \left(\f{\p\th(s)}{\p s}\right)^2,
\label{eqwlc}
\eea
where $\th(s)=\p\rv(s)/\p s$ is the tangent vector and the polymer is
inextendible i.e. $\th^2=1$, $\be$ is the inverse temperature. 
The bending rigidity $\k$ is  related to
persistence length $\l$ via $\k = (d-1)\l/2$. Persistence length is a
\index{persistence length}
measure of the distance up to which the consecutive tangent vectors on
the contour do not bend appreciably and is defined by
$\la \th(s).\th(0)\ra = \exp(-s/\l)$.

In this section we present a theoretical method to solve WLC model to
\index{WLC}
any desired accuracy for both the Helmholtz and Gibbs ensemble and all three 
\index{Gibbs}
\index{Helmholtz}
\index{ensemble}
possible boundary orientations over the entire range of stiffness parameter $t$.
\index{stiffness}
We first develop the method for a free polymer. Then we extend
\index{free polymer}
it to calculate properties of grafted [~one/both end(s)~] polymers.
\index{grafted}

If the tangent vectors of two ends of a polymer are held fixed at $\th_i$ and
$\th_f$, the probability distribution of end to end vector 
\index{probability}
in constant extension ensemble can be written in path integral notation as
\index{ensemble}
\bea
P(\rv) = {\cal N} \int_{\th_i}^{\th_f} {\cal D}[\th(s)] \exp\left(-\be{\cal H}\right) 
\times \d^d\left( \rv - \int_0^L \th ds\right)
\eea
where $\be{\cal H}$ is given by Eq.\ref{eqwlc} 
and ${\cal D}[\th(s)]$ denotes integration over all possible paths in
tangent vector space from tangent at one end $\th_i$ to tangent vector 
at the other end $\th_f$. In $d$- dimensions $\rv=(r_1,r_2,\dots,r_d)$.
There is no eaxct analytic calculation of this distribution because of 
the difficulty presented by the inextensibility constraint 
introduced via the Dirac- delta function, though some
mean field way of enforcing this constraint exist\cite{thiru,jkb}.  
Unfortunately that do not capture the very interesting triple maxima 
feature of the radial distribution at intermediate stiffness values as obtained
\index{stiffness}
in Ref.\cite{mywlc}. Recently, following our earlier work\cite{mywlc},
J. Samuel {\em et. al.} \cite{sam} developed a path integral 
Greens function formulation to evaluate the distribution for a free
\index{free polymer}
polymer in 3D. We closely follow that method to generalize that to obtain
results for various orientation constraints on polymer ends.
\index{constraints}

The integrated (projected) probability distribution is given by,
\index{probability}
\bea
P_x(x) = \int d\rv P(\rv)\d (r_1 -x).
\eea
We define the generating function of $P_x(x)$ via Laplace transform as,
\index{Laplace transform}
\bea
\tilde P(f) &=& \int_{-L}^L dx \exp(F x/\kb T) P_x(x) \nn\\
&=& \int_{-L}^L dx \exp(fx/\l) P_x(x)
\label{zf}
\eea
where $f$ is the force in units of $\kb T/\l$ i.e. $f=F\l/\kb T$ applied
along the $x$- axis.
This gives,
\bea
\tilde P(f) &=& {\cal N} \int_{\th_i}^{\th_f} {\cal D}[\th(s)] e^{\left(
-\f{(d-1)\l}{4}\int_0^L ds \left( \f{\p \th (s)}{\p s}\right)^2 
+ \f{f}{\l}\int_0^L \th_x ds\right)} \nn\\
&=& {\cal N} \int_{\th_i}^{\th_f} {\cal D}[\th(\tau')] e^{\left[
-\int_0^{t} \left\{ \f{(d-1)}{4} \left(\f{\p \th (\tau')}{\p \tau'}\right)^2 
- f \th_x \right\} d\tau' \right]}\nn\\
\eea
The last step is obtained by replacing $\tau' = s/\l$, $t=L/\l $ and
using the identity $\k = (d-1)\l/2$.
Note that, $\tilde P(f)$,  
is the partition function  in constant force ensemble where $t$ behaves 
\index{partition function}
\index{ensemble}
like an inverse temperature such that the Gibb's free energy can be written 
as $G(f) = - 1/t~ \ln \tilde P(f)$.
Now considering $\tau'$ as imaginary time and replacing $\tau = - i \tau'$ 
we get,
\bea
\tilde P(f) &=& {\cal N} \int_{\th_i}^{\th_f} {\cal D}[\th(\tau)] e^{\left[
i \int_0^{-it} \left\{ \f{(d-1)}{4} \left(\f{\p \th (\tau)}{\p \tau}\right)^2 
+ f \th_x \right\} d\tau \right]} \nn\\
&=& {\cal N} \int_{\th_i}^{\th_f} {\cal D}[\th(\tau)] e^{ \left[i \int_0^{-it} L d\tau \right]}
\eea
With the identification of 
$L= \f{(d-1)}{4} \left(\f{\p \th (\tau)}{\p \tau}\right)^2 + f \th_x$ 
as the Lagrangian, $\tilde P(f)$
[~= $\tilde Z(f)/\tilde Z(0)$~] in the above expression 
is the path integral representation for the propagator
\index{propagator}
of a {\em quantum} particle, on the surface of a $d$- dimensional sphere,
that takes a state $|\th_i\rangle$ to $|\th_f\rangle$. 
In Schr\"odinger picture this can be written
as the inner product of a state $|\th_i\rangle$ and another state 
$|\th_f\rangle$ evolved by imaginary time $-it$, 
\bea
\tilde Z(f) = \langle \th_i| \exp(-i \hat H (-it))|\th_f\rangle = 
\langle \th_i| \exp(-t \hat H )|\th_f\rangle .
\label{prop}
\eea

Once $\tilde P(f)=\tilde Z(f)/\tilde Z(0)$ is calculated, 
performing an inverse Laplace transform we can 
\index{Laplace transform}
obtain the projected probability density $P_x(x)$. 
\index{probability}
\index{projected probability}
We now describe how to do that.
Eq.\ref{zf} can be written as,   
\bea
\tilde P(f) &=& \int_{-L}^L dx \exp(fx/\l) P_x(x) \nn\\
&=& \int_{-1}^1 dv_x \exp(tf~v_x) LP_x(x)\nn\\ 
&=& \int_{-1}^1 dv_x \exp(tf~v_x) p_x(v_x)
\eea
where $v_x = x/L$ and $p_x(v_x)=LP_x(x)$ is a scaling relation.  
\index{scaling}
Note that the Helmholtz free energy is given by 
\index{Helmholtz}
${\cal F}_x(v_x)=-(1/t)\ln p_x(v_x)$. Therefore, we can write the above
expression as, 
\bea
\exp[-tG(f)]=\int_{-1}^1 dv_x \exp(tf~v_x) \exp[-t{\cal F}_x(v_x)].\nn
\eea
This is the relation between free energies of Helmholtz and Gibbs ensemble
\index{Gibbs}
\index{Helmholtz}
\index{ensemble}
for a finite chain (finite $t$). In thermodynamic limit of $t\to\infty$
\index{thermodynamic limit}
a steepest descent approximation of the above integral relation gives 
$G(f)={\cal F}_x(v_x)-fv_x$, the normal Legendre transformation.
Let us use the identity,
\bea
p_x(v_x) = \f{1}{2\pi} \int_{-\infty}^\infty e^{iuv_x} du \int_{-1}^1 p_x(w)
e^{-iuw} dw 
\eea
In it we can define $\tilde p_x(u) = \int_{-1}^1 p_x(w) \exp(-iuw) dw$ as the
Fourier transform and 
\bea
p_x(v_x) = \f{1}{2\pi} \int_{-\infty}^\infty du  \tilde p_x(u) \exp(iuv_x) 
\label{nlap}
\eea
as the inverse Fourier transform. Now, with $-iu = tf$ we
get $\tilde P(f) = \tilde p_x(u=ift) $ and the inverse Fourier transform can be
written as an inverse Laplace transform,
\index{Laplace transform}
\bea
p_x(v_x) = t \f{1}{2\pi i}\int_{-i\infty}^{i\infty} df
\tilde P(f) \exp(-tfv_x)
\eea
This gives the relation between the partition function $\tilde P(f)$ in the
\index{partition function}
constant force ensemble and the projected probability density of end to end
\index{probability}
\index{projected probability}
separation  $p_x(v_x)$ along any given direction $x$ in the constant extension
ensemble.  
\index{ensemble}
In numerical evaluation, the simplest way to obtain $p_x(v_x)$ is 
to replace $f=-iu/t$ in the
expression for $\tilde Z(f)$ to obtain $\tilde p_x(u)$ 
and evaluate the inverse Fourier transform (Eq.\ref{nlap}).

\begin{figure}[t]
\begin{center}
\includegraphics[width=9cm]{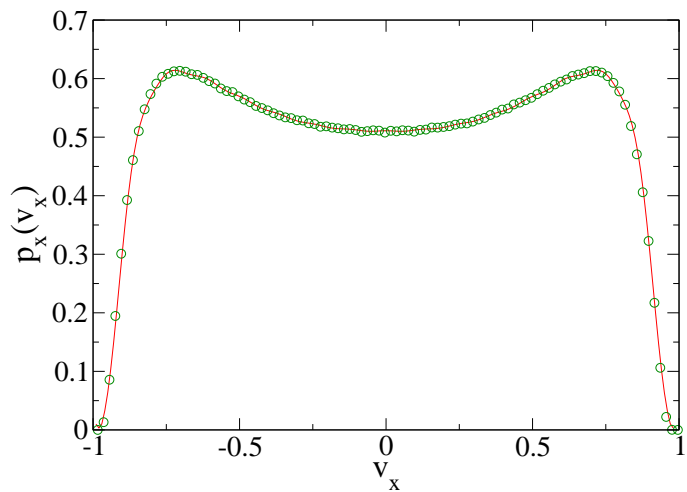}
\index{WLC}
\end{center}
\caption{ 
For a semiflexible polymer having its ends free to rotate 
\index{semiflexible polymer}
$p_x(v_x)~(~=p_y(v_y)~)$ is plotted at stiffness parameter $t=2$.  
\index{stiffness}
The points are collected from Monte- Carlo simulation
\index{Monte- Carlo}
in freely rotating chain model (see Sec.\ref{simu}). The line is calculated
\index{freely rotating chain}
from theory (see Sec.\ref{pathint}). The theory matches exactly with 
simulation.  It clearly shows bistability via two maxima in 
\index{simulation}
integrated probability density at the two near complete extensions.
\index{probability}
}
\label{t2-fx-free}
\end{figure}

Up to this point everything has been treated in $d$- dimensions. 
Experiments on single polymer can be performed in 3D as well as in 2D.
In 3D polymers are left in a solution whereas one can float the polymer
on a liquid film to measure its properties in 2D \cite{gittes}.
Moreover, polymers embedded in 2D are more interesting because of the 
following reason.
It was shown that in Helmholtz ensemble in three dimensions\cite{sam},
\index{Helmholtz}
\index{ensemble}
\bea
p(v_x) = -\f{1}{2\pi v_x}~\f{dp_x}{dv_x}
\eea
where $p(v_x)$ is the scaled radial distribution function $L^2 P(r)=p(v)$
with $v$ replaced by $v_x$. Since $p(v)$ is a probability density
\index{probability}
$p(v)\geq 0$ and therefore $dp_x/dv_x\leq 0$ for $v_x>0$ thus ruling
out multiple peaks in $p_x(v_x)$ \cite{sam-ensemble} and showing that 
\index{ensemble}
$p_x(v_x)$ will have single maximum at $v_x=0$ for all stiffness parameter $t$.
\index{stiffness}
No such simple relation exists between $p(v_x)$ and $p_x(v_x)$ in two 
dimensions. Therefore, the two dimensional
WLC polymer having its ends free to rotate may show more than one maximum
\index{WLC}
in $p_x(v_x)$ and therefore non-monotonicity. Indeed our calculation
and simulation (see Sec.\ref{simu}) does show non-monotonicity 
\index{simulation}
(Fig.\ref{t2-fx-free}). This is a curious difference between semiflexible 
polymers in 2D and 3D. Because of this and the fact that experiments in
2D are possible, in this work we focus on the 2D WLC model.
\index{WLC}

We have already given a general form of $\tilde Z(f)$ (Eq.\ref{prop})
which depends on the dimensionality $d$ of the embedding space of the
polymer.
$d=3$ formulation of this theory for a polymer with its ends free to rotate
has been carried out in detail in Ref.\cite{sam}. For $d=2$, we can
assume $\th = (\cos\h,\sin\h)$. Therefore, $L = \{1/4~\dot\h^2 + f cos\h\}$ and 
angular momentum $p_\h = \f{\p L}{\p \dot\h} = \dot\h/2$. 
Then the corresponding Hamiltonian is 
$H = \dot\h p_\h - L = p_\h^2 - f cos\h$ and in planar polar coordinates,
replacing $p_\h \to -i\f{\p}{\p \h}$ we obtain the quantum hamiltonian
operator, $\hat H = - \f{\p^2}{\p \h^2} - f cos\h$. In this 
representation of $\th$ we can write,
\bea
\tilde Z(f) &=& \langle \h_i| \exp(-t \hat H )|\h_f\rangle \nn\\
&=& \sum_{n,n'} \langle \h_i| n\rangle \langle n| \exp(-t \hat H )| n'\rangle
\langle n'| \h_f \rangle \nn\\
&=& \sum_{n,n'} \phi_n^\ast(\h_i) \phi_{n'}(\h_f) \langle n| \exp(-t \hat H )| n'\rangle
\label{zsym0}
\eea
This propagator takes a definite tangent vector state $|\h_i\rangle$ 
\index{propagator}
at one end of the polymer to a definite final tangent vector state 
$|\h_f\rangle$ at the other end of it. 

If external force is applied along $x$- direction as in Eq.\ref{zf},
$\hat H = \hat H_0 + \hat H_I = 
- \f{\p^2}{\p \h^2} - f cos\h$ where 
$\hat H_0= -\f{\p^2}{\p \h^2}$ 
is the Hamiltonian for a free rigid rotor in 2D and 
$\hat H_I= -f \cos\h$ 
is the part of Hamiltonian introduced by an external field. 
Thus the total Hamiltonian 
$\hat H$ denotes a rigid rotor in presence of a constant external field. The
eigenvalues of $ \hat H_0 $, the hamiltonian for a 2D rigid rotor, are $E_n = 
\index{rigid rotor}
n^2$ and the complete set of ortho- normalized eigen- functions are given by 
$\phi_n(\h) = \exp(i n \h)/\sqrt{2\pi}$ where $n=0,\pm 1,\pm 2,\dots$. 
The orthonormality condition is 
$\la n|n'\ra = (1/2\pi)\int_0^{2\pi}d\h \exp[i(n-n')\h]=\d_{n,n'}$.
In this basis 
$\langle n|\hat H_I| n'\rangle = -(f/2)(\d_{n',n+1} + \d_{n',n-1})$. 
Therefore, 
$\langle n|\hat H| n'\rangle = n^2 \d_{n',n} - (f/2)(\d_{n',n+1} + \d_{n',n-1})$.
If the external force were applied in $y$- direction $\hat H_I= -f \sin\h$ and
$\langle n|\hat H| n'\rangle = n^2 \d_{n',n}-(f/2i)(\d_{n',n+1} - \d_{n',n-1})$.
$\langle n|\exp(-t\hat H)| n'\rangle$ can be calculated by exponentiating the
matrix $\langle n|\hat H| n'\rangle$. Thus one can find $\tilde Z(f)$.

\subsection{ Free Polymer}  
\index{free polymer}
For a polymer which has both its ends free to rotate,
integrating Eq.\ref{zsym0} over all possible initial and final tangent vectors
in rigid rotor basis we get
\index{rigid rotor}
\index{rigid rotor}
\bea
\tilde Z(f) = 2\pi~\langle 0| \exp(-t \hat H )| 0\rangle
\eea
This means that $\tilde Z(f)$ is given by the $(0,0)$ element of the matrix
$\langle n| \exp(-t \hat H )| n'\rangle$.  
In this case $\tilde Z(0) = 2\pi~\la 0|\exp(-t H_0)|0\ra = 2\pi~\exp(-t~0^2) 
=2\pi$ and therefore 
$\tilde P(f)=\tilde Z(f)/\tilde Z(0)= \langle 0| \exp(-t \hat H )| 0\rangle$.
To evaluate the matrix element $\la 0| \exp(-t \hat H )| 0\ra$
we exponentiate the matrix 
$- t \langle n|\hat H| n'\rangle$ 
and pick up the $(0,0)$-th element. 
Thus, if the external force is applied in $x$- direction, 
remembering $\tilde p_x(u)=\tilde P(f=-iu/t)$ we calculate 
the inverse Fourier transform (Eq.\ref{nlap}) to obtain $p_x(v_x)$.
Here it is useful to note that due to spherical symmetry of a polymer whose 
ends are free to rotate $p_x(v_x)=p_y(v_y)$. 

\subsection{One End Grafted}
\index{grafted}
This symmetry breaks down immediately if we hold one end of the polymer to a
specific direction, namely along the $x$- direction {\em i.e.} $\h_i = 0$. 
Then on Eq.(\ref{zsym0}) integrating over all possible $\h_f$  and 
leaving $\h_i=0$ we obtain
\bea
\tilde Z(f) = \sum_{n}  \langle n| \exp(-t \hat H )| 0\rangle
\eea
in the rigid- rotor basis. 
Note for this case $\tilde Z(0) = \sum_n \la n|\exp(-t H_0)|0\ra 
= \sum_n \exp(-t n^2) \d_{n,0} =1$ and therefore we have 
$\tilde P(f)=\tilde Z(f)$.
If the external force acts in  $x$- direction, 
the Laplace transform of $\tilde Z(f)$, defined in the way described above, 
gives the projected probability distribution in $x$- direction, $p_x(v_x)$. 
On the other hand, if the external force acts in  $y$- direction, 
the Laplace transform of $\tilde Z(f)$ gives the projected probability 
\index{probability}
\index{projected probability}
\index{Laplace transform}
distribution in $y$- direction $p_y(v_y)$, the distribution of transverse 
\index{transverse fluctuation}
fluctuation while one end of the polymer is grafted in $x$- direction.
\index{fluctuation}

\subsection{Both Ends Grafted}
Two ends of a polymer can be grafted in infinitely different ways.
\index{grafted}
Let us fix the orientation of one end along $x$- direction ($\h_i=0$)
and the other end is orientated along any direction $\h_f$, 
then Eq.(\ref{zsym0}) gives
\bea
\tilde Z(f) = \f{1}{2\pi}~\sum_{n,n'} e^{in'\h_f} \langle n| \exp(-t \hat H )| n'\rangle.
\eea
As above, to obtain $p_x(v_x)$ we use 
$\langle n|\hat H_I| n'\rangle = -(f/2)(\d_{n',n+1} + \d_{n',n-1})$,
whereas to obtain $p_y(v_y)$ we use 
$\langle n|\hat H_I| n'\rangle = -(f/2i)(\d_{n',n+1} - \d_{n',n-1})$
to calculate $\tilde P(f)$ and perform inverse Laplace transform.
\index{Laplace transform}
For this case of grafting 
$2\pi\tilde Z(0) = \sum_{n,n'} e^{in'\h_f}\la n|e^{-t H_0}|n'\ra
= \sum_{n} e^{in\h_f-t n^2}$ and therefore
$\tilde P(f)=\sum_{n,n'}e^{in'\h_f}\langle n| e^{-t \hat H }| n'\rangle/[\sum_{n} e^{in\h_f-t n^2}]$.

\vskip .5cm
Up to this point all the relations are exact. Since an infinite dimensional
calculation of $\la n|\exp(-t\hat H)|n' \ra$ is not feasible, we calculate
it numerically up to a dimension $N_d$, that controls the accuracy, limited
\index{dimension}
only by computational power
\footnote{We use the MatrixExp function of Mathematica\cite{mathematica}.
}. 
We use $N_d=10$ which already gives 
numbers for probability distribution which differs within $1\%$
\index{probability}
from that obtained from $N_d=20$ and gives a fair comparison
with simulated data (see Sec.\ref{result}). The inverse Laplace transform 
\index{Laplace transform}
to obtain $p_x(v_x)$ etc. from $\tilde P(f)$ is also done numerically.

\section{Simulation}
\index{simulation}
\label{simu}
In this section, we introduce two discretized models
used to simulate semi- flexible polymers. Both of these are
derived from the WLC model which is used for our analytical
\index{WLC}
treatment in Sec.\ref{pathint}. After introducing the discretized 
models we have listed the various boundary conditions used. We 
\index{boundary conditions}
perform Monte- Carlo (MC) simulations to obtain probability 
\index{probability}
\index{Monte- Carlo}
\index{MC}
distributions in Helmholtz ensemble. 
\index{Helmholtz}
\index{ensemble}

One discretized version of this model is the freely rotating chain (FRC)
\index{freely rotating chain}
model\cite{doi}. In the FRC model one considers a polymer as a random walk
\index{FRC}
\index{random walk}
of $N$ steps each of length $b=L/N$ with one step memory, such that, successive
steps are constrained to be at an fixed angle $\h$ with $\l=2b/\h^2$. 
The continuum WLC model is obtained in the limit $\h,~b\to 0$, $N\to \infty$
\index{WLC}
keeping $\l$ and $L$ finite. To simulate a polymer with ends free to rotate
a large number of configurations are generated with first step taken in any
random direction. Whereas if one choses the first step to be in some specific
direction, this will simulate a polymer with one end grafted in that 
\index{grafted}
direction.

A straight discretization of Eq.\ref{eqwlc} in 3D (2D) 
is an 1d Heisenberg (classical $xy$) model:
\bea
\be{\cal H} &=& \f{\k}{2} \sum_{i=1}^N \f{(\th_i-\th_{i-1})^2}{b} 
= \sum_{i=1}^N (-J~\th_i.\th_{i-1})
\eea
with a nearest neighbour coupling $J=\k /b$ between `spins' $\th_i$.
We have ignored a constant term in energy. 
The appropriate continuum limit is recovered for $b\to 0$, $J\to \infty$ with
$Jb=\k$ finite. In this model grafting is simulated by fixing end spins on the
1D chain. If an end is free then the end spin takes up any orientation
that are allowed by the energy and entropy. In this model, by fixing the
two end- spins, one can easily simulate a polymer with both its ends
grafted in some fixed orientations. We follow the normal Metropolis 
\index{grafted}
algorithm to perform MC simulation in this model.
\index{simulation}
\index{MC}

We restrict ourselves to two dimensions.
The numerics were checked via exact calculation of $\la r^2\ra$ and
$\la r^4\ra$ which match within $0.5\%$. In the FRC model 
\index{FRC}
simulations we have used a chain length of $N=10^3$ and generated around $10^8$
configurations. This simulation does not require equilibration run. 
\index{simulation}
Therefore all the $10^8$ configurations were used for data collection.
In $xy$ model we have simulated $N=50$ spins and equilibrated over $10^6$
MC steps. A further $10^6$ configurations were generated 
\index{MC}
to collect data. We have averaged over $10^3$ initial
configurations, each of which were randomly chosen from 
nearly minimum energy configurations that conforms with the boundary
\index{boundary conditions}
conditions. Increasing the number of spins do not change the averaged data.

\section{Results}
\label{result}
Once all these theoretical and simulation tools are available, 
\index{simulation}
we apply them to bring out statistical and mechanical properties 
of a semiflexible polymer. We have three different boundary conditions
\index{boundary conditions}
\index{semiflexible polymer}
depending on the orientations of polymer ends and two different ensembles.
For each case we look at the various probability densities, ensemble 
\index{probability}
\index{ensemble}
dependence of force- extension etc. For the case of a polymer with both 
\index{force- extension}
ends grafted we find that the properties depend on the relative 
\index{grafted}
orientation of the two ends.

\subsection{Free Polymer:}
\index{free polymer}

\index{WLC}
\index{FRC}
\index{simulation}

\begin{figure}[t]
\begin{center}
\includegraphics[width=9.cm]{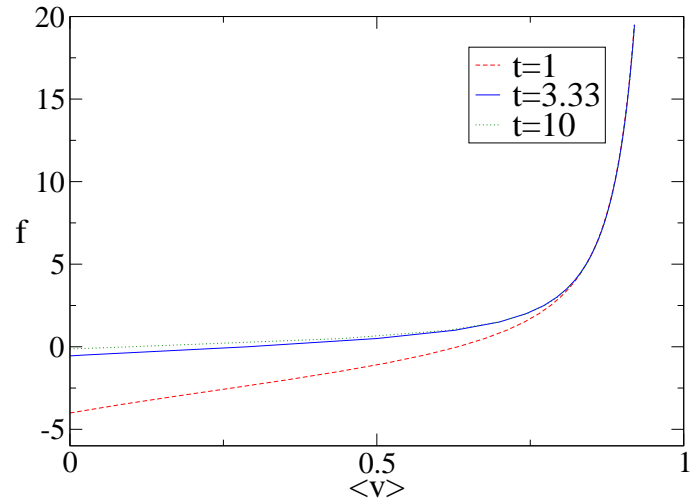}
\index{WLC}
\end{center}
\caption{Compare this force- extension relation with that in Fig.2.4 of 
previous chapter. None of the curves including that at $t=3.33$ show non-
monotonic behaviour. This shows the force- extension behaviour in constant 
\index{force- extension}
force ensemble. Forces are expressed in units of $\kb T/\l$.
\index{ensemble}
}
\label{fx-free}
\end{figure}

{\em Helmholtz ensemble:}
\index{Helmholtz}
\index{ensemble}
We employ the theory as described in Sec.\ref{pathint} to calculate
$p_x(v_x)$ and $p_y(v_y)$ for a polymer with both its ends free to rotate.
We compare the probability distribution obtained at stiffness parameter
\index{probability}
\index{stiffness}
$t=2$ with that obtained from MC simulation (Sec.\ref{simu}) using the 
\index{MC}
FRC model (see Fig.\ref{t2-fx-free}). This shows exact 
\index{FRC}
agreement between theory and simulation. For a free polymer $p_x(v_x)$
\index{free polymer}
\index{simulation}
and $p_y(v_y)$ are same due to the spherical symmetry.
Note that, ${\cal F}(v_x) = -1/t \ln p_x(v_x)$ would give a 
non- monotonic force- extension via 
$\la f_x\ra = (\p {\cal F}/\p v_x)$. 
The force- extension obtained from the projected probability density $p_x(v_x)$
\index{probability}
\index{projected probability}
\index{force- extension}
will describe the experimental scenario in which the external potential
traps the polymer end only in $x$- direction and constant in $y$.
In general, if the external potential traps the polymer in $d_r$ dimensions
($d_r\leq d$) then a $d_r$ dimensional projection
(~($d-d_r$) dimensional integration~) of the probability distribution
\index{probability}
of end to end vector $p(\vv)$ gives the free energy and decides the
force- extension relation.
\index{force- extension}
This understanding is general and do not depend on the orientational
boundary conditions at the polymer ends or the dimensionality of
\index{boundary conditions}
embedding space. This is important to keep in mind while analyzing
experimental data.

{\em Gibbs ensemble:}
We have already mentioned that the non- 
monotonic nature of free energy, a strong qulitative signature of semi- 
flexibility, is observable only in Helmholtz ensemble not in Gibbs 
\index{Helmholtz}
ensemble \cite{mywlc}. 
The partition function in Gibbs ensemble with force $f$ applied in 
\index{Gibbs}
\index{partition function}
\index{ensemble}
$x$- direction gives the Gibb's free energy 
$G(f) = -(1/t)\ln[\tilde P(f)]$. From this the averaged
extension comes out to be $\la v_x \ra= -(\p G/\p f)$. 
For a polymer with its ends free to rotate, this force extension 
relations, that have been calculated from theory,  at various $t$ 
are shown in Fig.\ref{fx-free}.
Notice that $\p \la v_x \ra/\p f = t[\la v_x^2\ra - \la v_x\ra^2] \geq 0$.
Similar relation for response function does not exist in Helmholtz ensemble.
Therefore, the force- extension in Gibbs ensemble has to be
\index{Gibbs}
\index{force- extension}
monotonic (Fig.\ref{fx-free}) in contrast to Helmholtz ensemble.
\index{Helmholtz}
\index{ensemble}
Note that, for any non-zero stiffness, negative force is required to bring 
the end to end separation to zero. The amount of this force is larger for 
larger stiffness (smaller $t$).
\index{stiffness}
At large and positive force polymer goes to fully extended limit whichafter
inextensibility constraint stops the polymer to extend any more.

\begin{figure}[t]
\begin{center}
\includegraphics[width=9.cm]{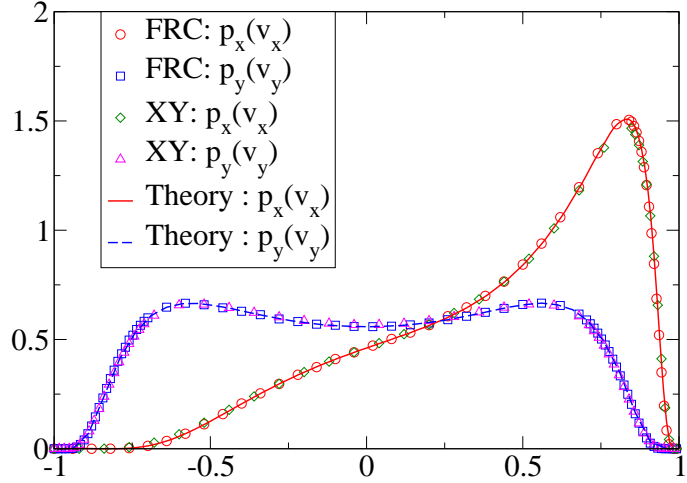}
\index{WLC}
\end{center}
\caption{The simulation data for $p_x(v_x)$ and $p_y(v_y)$ from FRC model and
\index{FRC}
\index{simulation}
XY- model simulations are compared with their theoretical estimates. Simulations
and calculations were done at $t=2$ for a polymer with one end grafted in $x$-
\index{grafted}
direction. The data labelled LMF is taken from Ref.\cite{munk-frey}.
}
\label{pxpy}
\end{figure}

\begin{figure}[t]
\begin{center}
\includegraphics[width=10cm]{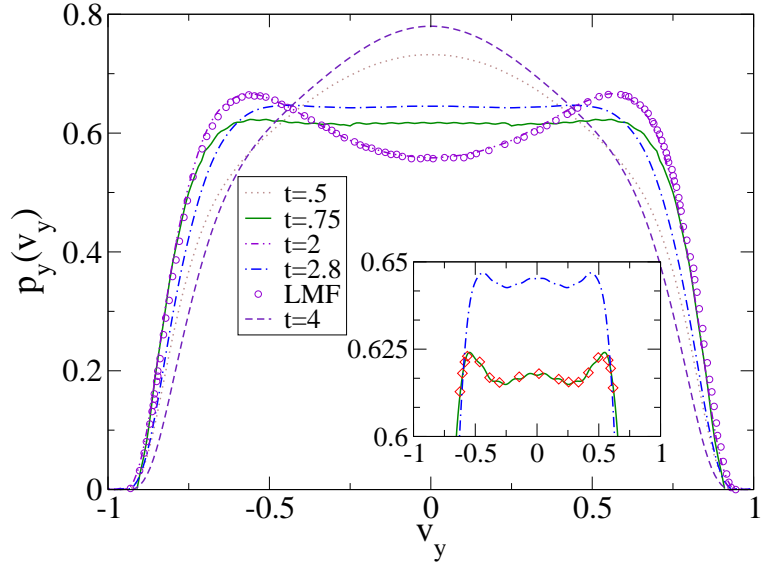}
\index{WLC}
\end{center}
\caption{The integrated probability density $p_y(v_y)$ is plotted at various
\index{probability}
stiffnesses $t$. At $t=4$ there is a single maximum at $v_y=0$.
Decreasing $t$  we see at $t=2.8$ emergence of two more 
peaks at nonzero $v_y$ except for the one at $v_y=0$ (See inset). 
At $t=2$ central peak vanishes, the
trimodal distribution becomes bimodal. 
The circles labeled LMF are data taken from Ref.\cite{munk-frey} at $t=2$
to show exact agreement with our theory.
At $t=0.75$ we see re-emergence of the central peak and tri-modality in
$p_y(v_y)$ (See inset, $\Diamond$s are from our MC simulation in the FRC 
\index{FRC}
\index{simulation}
\index{MC}
model at $t=0.75$, the lines are calculated from theory). 
}
\label{manyt-py}
\end{figure}

\subsection{Grafted Polymer: One End} 
\index{grafted polymer}
\index{probability}
\index{grafted}

\begin{figure}[t]
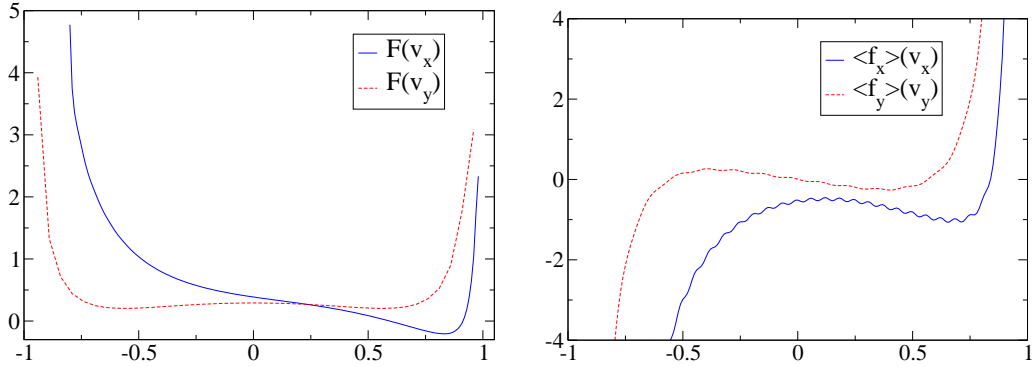

\begin{center}
\includegraphics[width=6.5cm]{t2-Feng-teth.eps}
\hskip .5cm
\includegraphics[width=6.5cm]{t2-f-constext-teth.eps}
\index{WLC}
\end{center}
\caption{The left panel shows the Helmholtz free energies ${\cal F}(v_x)$ and
${\cal F}(v_y)$ of a polymer with $t=2$ and one end grafted in $x$- direction.
\index{grafted}
The right panel shows the corresponding force- extensions in Helmholtz 
\index{Helmholtz}
ensemble. Both $\la f_x\ra$- $v_x$ and $\la f_y\ra$- $v_y$ show regions of 
\index{ensemble}
negative slope. Free energies are expressed in units of $\kb T$ and forces
are expressed in units of $\kb T/\l$.
}
\label{cext}
\end{figure}

\begin{figure}[t]
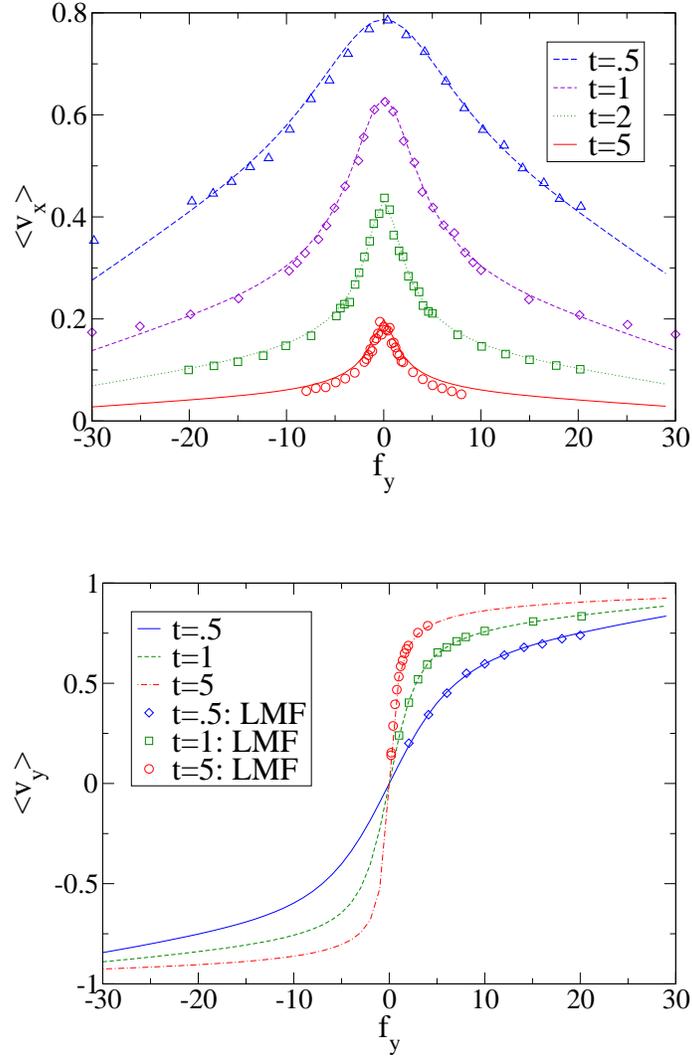

\begin{center}
\includegraphics[width=9.cm]{f-xavg.eps}
\vskip 1cm
\includegraphics[width=9.cm]{f-yavg.eps}
\index{WLC}
\end{center}
\caption{Average displacements along $x$- direction $\la v_x\ra$ and $y$-
direction $\la v_y\ra$ as a function of transverse force (transverse to 
grafting direction) in constant force ensemble. Lines denote our theoretical
\index{ensemble}
calculation while points denote the MC simulation data taken from
\index{simulation}
\index{MC}
Ref.\cite{munk-frey}. Forces ($F_y$) are expressed in units of $\kb T/\l$, i.e.
$f_y=F_y\l/\kb T$.
}
\label{fx-teth}
\end{figure}

{\em Helmholtz ensemble:}
\index{Helmholtz}
\index{ensemble}
Next, we use our thery to plot $p_x(v_x)$ and $p_y(v_y)$ at $t=2$ 
(Fig.\ref{pxpy}) for a semiflexible polymer with one end grafted 
\index{grafted}
\index{semiflexible polymer}
in $x$- direction. In Fig.\ref{pxpy} we have also plotted MC
\index{MC}
data from the FRC model and $xy$ model simulation. The exact match
\index{FRC}
validates our theory and both the simulation techniques. In $p_x(v_x)$
\index{simulation}
the peak in near complete extension along positive $x$ is due to
the large bending energy and orientation of the other end towards
\index{bending energy}
this direction (also see $p(v_x,v_y)$ in Fig.\ref{comp3}).
We then explore the transverse fluctuation $p_y(v_y)$ of this system,
\index{transverse fluctuation}
\index{fluctuation}
in detail, for different $t$ (Fig.\ref{manyt-py}).
At large $t(=10)$, $p_y(v_y)$ has single maximum at 
$v_y=0$. At such low stiffnesses entropy takes over energy contributions.
Number of possible configurations and thus entropy gains if end to
end separation remains close to zero. This gives rise to this single
central maximum.  The emergence of multiple maxima at nonzero $v_y$
in polymers, the multi- stability, with larger stiffness ($t=2.8$) is 
\index{stiffness}
due to the entropy- energy competition. The central peak is due to the
entropy driven Gaussian behavior. The other two peaks emerge as 
entropy tries the polymer to bend around $v_y=0$ and energy restricts the amount
of bending. Since bending in positive and negative $y$- directions are 
equally likely, the transverse fluctuation shows two more maxima at non- zero 
\index{transverse fluctuation}
\index{fluctuation}
$v_y$. With further increase in stiffness ($t=2$), 
the central Gaussian peak vanishes (see Fig.\ref{comp3}) and therefore
$p_y(v_y)$ becomes bistable with two maxima (Fig.\ref{manyt-py}). 
At even higher stiffness ($t=0.75$) 
\index{stiffness}
central peak reappears, this time due to higher bending energy.
\index{bending energy}
At $t=0.5$ the distribution again becomes
single peaked at $v_y=0$ as entropy almost completely loses out.
This is the rigid rod limit.
\index{rigid rod}
Notice that we have plotted MC data as obtained in Ref.\cite{munk-frey}
\index{MC}
for $xy$ model simulation at $t=2$. This agrees exactly with our theory.
In the inset of Fig.\ref{manyt-py} we have blown up the multistability
at $t=2.8$ and $t=0.75$. We have also plotted our FRC model simulation 
\index{FRC}
\index{simulation}
data at $t=0.75$ for comparison.  

\index{grafted polymer}
\index{grafted}
At this point it is instructive to look at the force extension behavior  
in Helmholtz ensemble, the ensemble in which 
\index{Helmholtz}
\index{ensemble}
$p_y(v_y)$ and $p_x(v_x)$ have been calculated above. In it the extension $v_x$
[~or $v_y$~] is held constant and the corresponding average force in $x$- 
[~or $y$-~] direction is found from the 
relation $\la f_x\ra = \p {\cal F}(v_x)/\p v_x$
(~or $\la f_y\ra = \p {\cal F}(v_y)/\p v_y$~). In Fig.\ref{cext} we
show the Helmholtz free energies ${\cal F}(v_x) = -(1/t)\ln~p_x(v_x)$ and
\index{Helmholtz}
${\cal F}(v_y) = -(1/t)\ln~p_y(v_y)$ and the corresponding force extension 
curves in constant extension ensemble. 
Note that unlike the monotonicity beared by
$\la v_y\ra$-$f_y$ curve (Fig.\ref{fx-teth}) obtained in Gibbs ensemble 
the $\la f_y\ra$-$v_y$ curve in Fig.\ref{cext} clearly shows  non-monotonicity,
a signature of the Helmholtz ensemble.
\index{Helmholtz}

\index{grafted polymer}
{\em Gibbs ensemble:}
\index{Gibbs}
\index{ensemble}
From our theory we can also explore the transverse response 
of a polymer which has one of its ends grafted to a substrate and a constant 
\index{substrate}
\index{grafted}
force is applied to the other end in a direction transverse to the grafting. 
Assume that the
grafting direction is $x$ and we apply a force $f_y$ in $y$- direction to
study the response. A linear response theory was proposd earlier\cite{kroy} 
to tackle this qustion. Our theory can predict the effect of externally applied
force $f_y$ of arbitrary magnitude on the average positions $\la v_x\ra$ and 
$\la v_y\ra$.
As the force is applied in $y$-direction i.e. $\vec f=\hat y f_y$, 
we have $H_I = -f_y \sin\h$. Because one end of the polymer is 
grafted in $x$- direction we use
\index{grafted}
$\langle n|\hat H_I| n'\rangle = -(f_y/2i)(\d_{n',n+1} - \d_{n',n-1})$
to evaluate $\tilde Z(f_y)$, whereas to calculate 
$\la v_x\ra = -(\p G/\p f_x)$ 
[~or, $\la v_y\ra = -(\p G/\p f_y)$~] we introduce a small 
perturbing force $\d f_x$ (~or, $\d f_y$~) in the Hamiltonian matrix
to obtain the partial derivatives.  Thus we obtain
the corresponding force- extensions shown in Fig.\ref{fx-teth}.
As the grafted end is oriented in $x$- direction we expect 
\index{grafted}
in absence of any external force $\la v_x\ra$ will be maximum 
and will keep on reducing due to the
bending of the other end generated by the external force 
imposed in $y$- direction. 
Thus $\la v_x\ra$ is expected to be independent of the sign of $f_y$. Similarly
$\la v_y\ra$ should follow the direction of external force and therefore
is expected to carry the same sign as $f_y$.
Fig.\ref{fx-teth} verifies these expectations and
shows the comparison of our theory with simulated data taken
from Ref.\cite{munk-frey}. 

\begin{figure}[t]
\begin{center}
\includegraphics[width=9.cm]{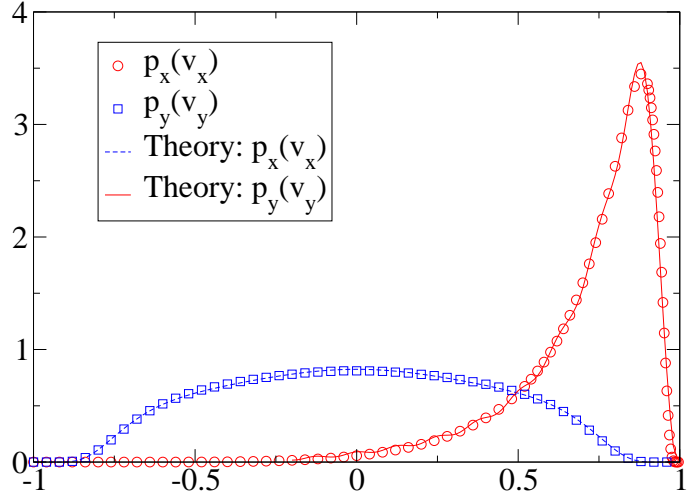}
\end{center}
\caption{The simulation data for $p_x(v_x)$ and $p_y(v_y)$ from 
\index{simulation}
XY- model simulations of a WLC polymer are compared with their theoretical 
estimates. Simulations and calculations were done at $t=2$ for a WLC polymer 
\index{WLC}
with both its ends grafted in $x$- direction. 
\index{grafted}
}
\label{both-pxpy}
\end{figure}

\index{Helmholtz}
\index{ensemble}
\begin{figure}[t]
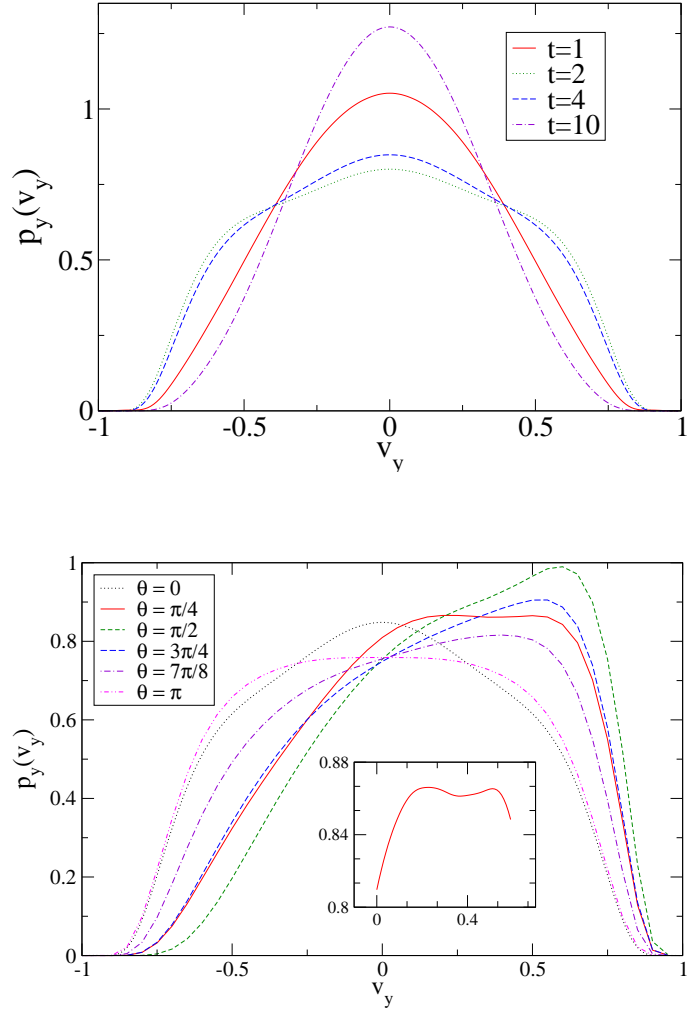

\begin{center}
\includegraphics[width=9cm]{2end-pyvy-manyt.eps}
\vskip 1cm
\includegraphics[width=9cm]{manytheta-pyvy-t4.eps}
\index{WLC}
\end{center}
\caption{The upper panel shows $p_y(v_y)$ response of a polymer with both 
ends grafted along the same direction at various stiffness parameters $t$.
\index{grafted}
\index{stiffness}
They always remain single moded. In lower panel, $p_y(v_y)$ plotted for various relative angle between the orientations of two ends $\h$ at $t=4$. 
The inset shows imergence of bistability at $\h=\pi/4$. 
}
\label{bothend1}
\end{figure}
\subsection{Grafted Polymer: Both Ends}
\index{grafted polymer}
\index{grafted}
{\em Helmholtz ensemble:}
\index{Helmholtz}
\index{ensemble}
Let us first fix the orientations of the polymer at 
both its ends along $x$- axis and 
compare $p_x(v_x)$ and $p_y(v_y)$ obtained from our $xy$ model simulation
and our theory (Fig.\ref{both-pxpy}).
This validates both our theory and simulation.
\index{simulation}
Next, we go on to explore the properties of this system using our theory.
We take a polymer with both its ends grafted, orientation of first end is 
\index{grafted}
fixed in $x$- dirction ($\h_i=0$) and that in the other end ($\h_f$) 
is varied to study the change in transverse fluctuation $p_y(v_y)$.
\index{transverse fluctuation}
\index{fluctuation}
To begin with let us find $p_y(v_y)$ for different stiffnesses
$t$ with $\h_f=0$ (Fig.\ref{bothend1}). The height of central peak shows
non- monotonicity; with increase in $t$ from $t=1$ the
height of central peak first decreases up to $t=2$ and then eventually again
increases. The initial decrease in the height of the maximum is due to the 
fact that with increase in $t$ the other end of the polymer (relative to the
first end) starts to sweep larger angles about $x$- axis. With further
increase in $t$ entropic contributions that favor $\vv=0$ 
win over to increase the height
of the maximum (see Fig.\ref{comp3}). 
Note that, $p_y(v_y)$ does not show multimodality as has been seen in
the transverse fluctuation of a polymer with one end grafted.
\index{transverse fluctuation}
\index{grafted}
\index{fluctuation}

\begin{figure}[t]
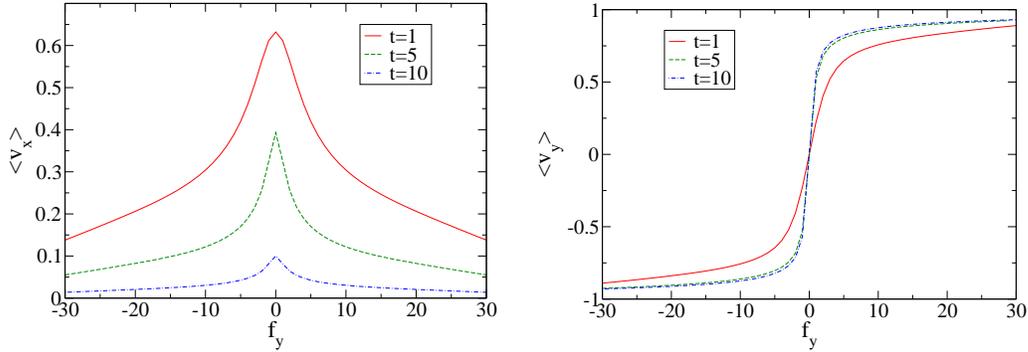

\begin{center}
\includegraphics[width=6.5cm]{bothend-f-xavg.eps}
\hskip .4cm
\includegraphics[width=6.5cm]{bothend-f-yavg.eps}
\index{WLC}
\end{center}
\caption{The left panel shows the transverse response and the right 
panel shows the longitudinal response of a polymer with both ends grafted
\index{grafted}
along the same direction. 
}
\label{bothend2}
\end{figure}

\begin{figure}[t]
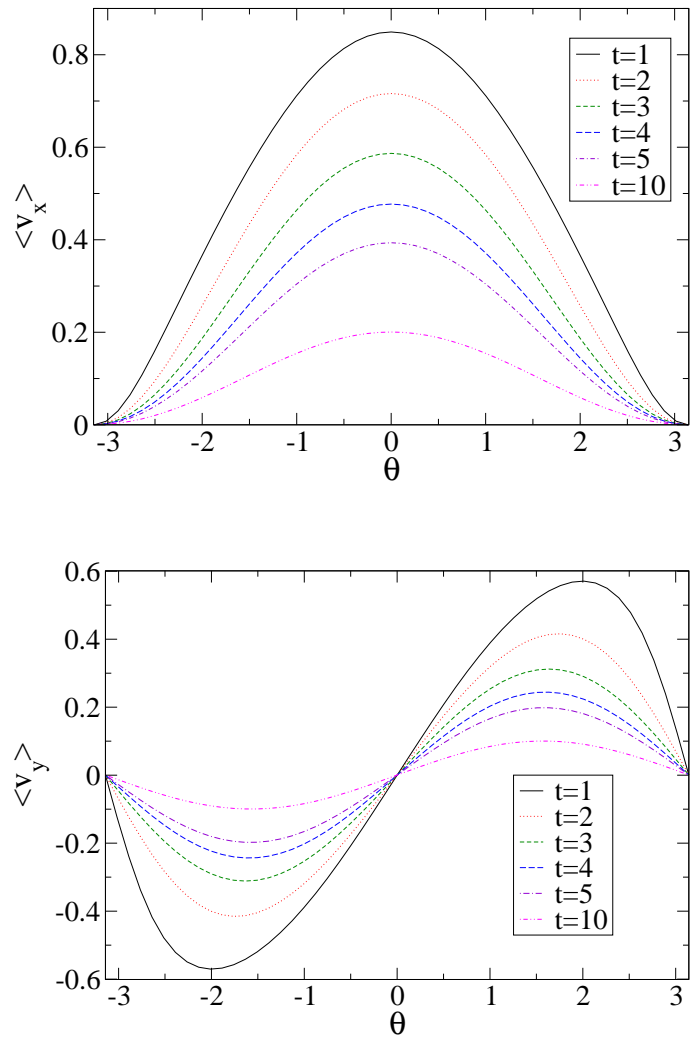

\begin{center}
\includegraphics[width=9cm]{bothend-theta-xavg.eps}
\vskip 1cm
\includegraphics[width=9cm]{bothend-theta-yavg.eps}
\index{WLC}
\end{center}
\caption{The upper panel shows the variation of $\la v_x\ra$ as a function
of $\h$ and the lower  panel shows the variation of $\la v_y\ra$ as a function
of $\h$. $\la v_x\ra$ and $\la v_y\ra$ are calculated for stiffness
\index{stiffness}
parameters $t=1,2,3,4,5,10$.
}
\label{bothend3}
\end{figure}

We then fix the stiffness at $t=4$ and rotate the orientation of the other
\index{stiffness}
end and find out the transverse fluctuation $p_y(v_y)$
\index{transverse fluctuation}
(transverse to the orientation of first end, see Fig.\ref{bothend1}). 
At $\h=\h_f=0$ the fluctuation is unimodal with the maximum at $v_y=0$.
\index{fluctuation}
With increase in $\h$ the orientation of the other end rotates from
positive $x$- axis towards positive $y$- axis. 
\index{bending energy}
\index{probability}
Energetically the polymer gains the most if it bends on the perimeter
of a circle. 
Therefore,  energetically at any $\h$, the peak of
$p_y(v_y)$ should be at $v_y = (~1.-\cos\h~)/\h$.
Thus at $\h=0,\pi/4,\pi/2,3\pi/4,7\pi/8,\pi$ the peak of
$p_y(v_y)$ should be, respectively, at $v_y = 0,0.37,0.64,0.72,0.69,0.64$.
Fig.\ref{bothend1} shows that the peak positions almost follow these values
up to $\h=\pi/2$, above which entropic contribution dominates. However,
entropy always play a crucial role, e.g. at $\h=\pi/4$ showing a double peak
around $v_y = 0.37$.
At $\h=\pi$ the two ends of the polymer are anti- parallel.  
Notice that, as $\h=\pi$ and $\h=-\pi$
are physically the same thing, $v_y=\pm 0.64$ are equally likely. 
Entropy would like the two ends to bend to $v_y=0$. Competition 
between energy and entropy leads to almost a constant 
distribution up to $|v_y|\sim 0.5$. 
Notice that the bending energy couples to end orientations. 
\index{bending energy}
For end orientations perpendicular to each- other, this coupling is 
strongest leading to a peak position that is closest to energetic 
expectation.  The behavior of $p_y(v_y)$
for $-\pi\le \h\le 0$ is mirror symmetric about $v_y=0$ with respect
to the behaviour of $p_y(v_y)$ in the region $0\le \h\le \pi$.

{\em Gibbs ensemble:}
\index{Gibbs}
We then work in the constant force ensemble by applying a force 
\index{ensemble}
$\vec f=\hat y f_y$. Let us fix $\h=0$, i.e. both ends are oriented along
$x$- axis. We find out the corresponding transverse
and longitudinal response to this force in the similar manner as in
the case of a polymer with one end grafted (see Fig.\ref{bothend2}).
The force extension carries the same qualitative features as for a single
end grafted polymer. 
\index{grafted polymer}
\index{grafted}
For other end orientations ($\h\neq0$) $\la v_x\ra$ and $\la v_y\ra$
does not remain at zero at $f_y=0$. Otherwise the nature of responses 
remain the same.
To see the impact of the change in relative angle $\h$ between the 
orientations at the two ends we calculate the average extension of the 
polymer ends in $x$- and $y$- directions as we vary $\h$ in constant 
force ensemble at $f=0$. Due to the bending energy, 
\index{bending energy}
\index{ensemble}
the $\la v_x\ra$ is expected 
to by highest at $\h=0$ and lowest for $\h=\pm \pi$. Similarly, $\la v_y\ra$
is expected to be highest at $\h=\pm~\pi/2$ and lowest for $\h=0$ and 
$\h=\pm \pi$. Fig.\ref{bothend3} shows, while $\la v_x\ra-\h$ bears out
the above expectations for all $t$, the expectation of getting highest 
magnitude for $\la v_y\ra$ at $\h=\pm~\pi/2$ is fulfilled only for lower
stiffnesses ($t\ge 5$). At lower $t$ the highest $\la v_y\ra$ shifts 
towards higher relative angles $\h$ as bending energy takes over entropy.
\index{bending energy}
Note that, the smaller $t$ gets, larger gets $\la v_y\ra$, an impact of 
larger stiffness.
\index{stiffness}

\subsection{End to End Vector Distribution}
\begin{figure}[t]
\begin{center}
\includegraphics[width=9.0cm]{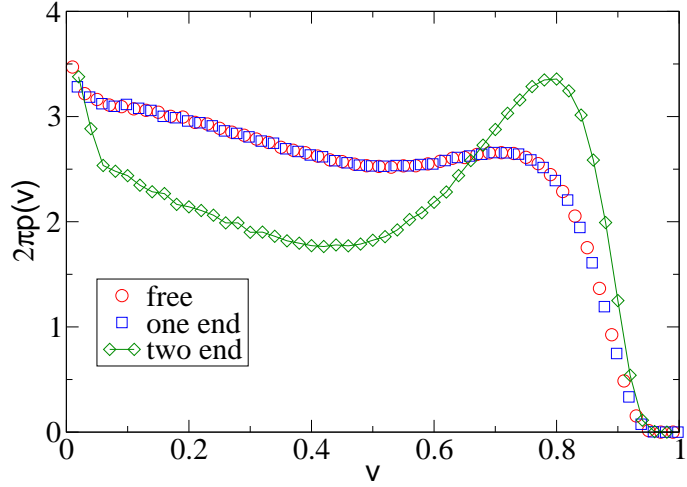}
\index{WLC}
\end{center}
\caption{The radial distribution function $p(v)$ at stiffness $t=4$ 
\index{stiffness}
is plotted for the three different boundary conditions -- 
\index{boundary conditions}
(a) both ends free, 
(b) one end oriented in $x$- direction and the other is free, 
(c) both ends oriented in $x$- direction. Radial distribution of 
first two cases are equal, whereas for the third case it is different.
However, all three show double maxima.
\index{substrate}
\index{grafted}
\index{stiffness}
}
\label{comp1}
\end{figure}

\begin{figure}[t]
\begin{center}
\includegraphics[width=10cm]{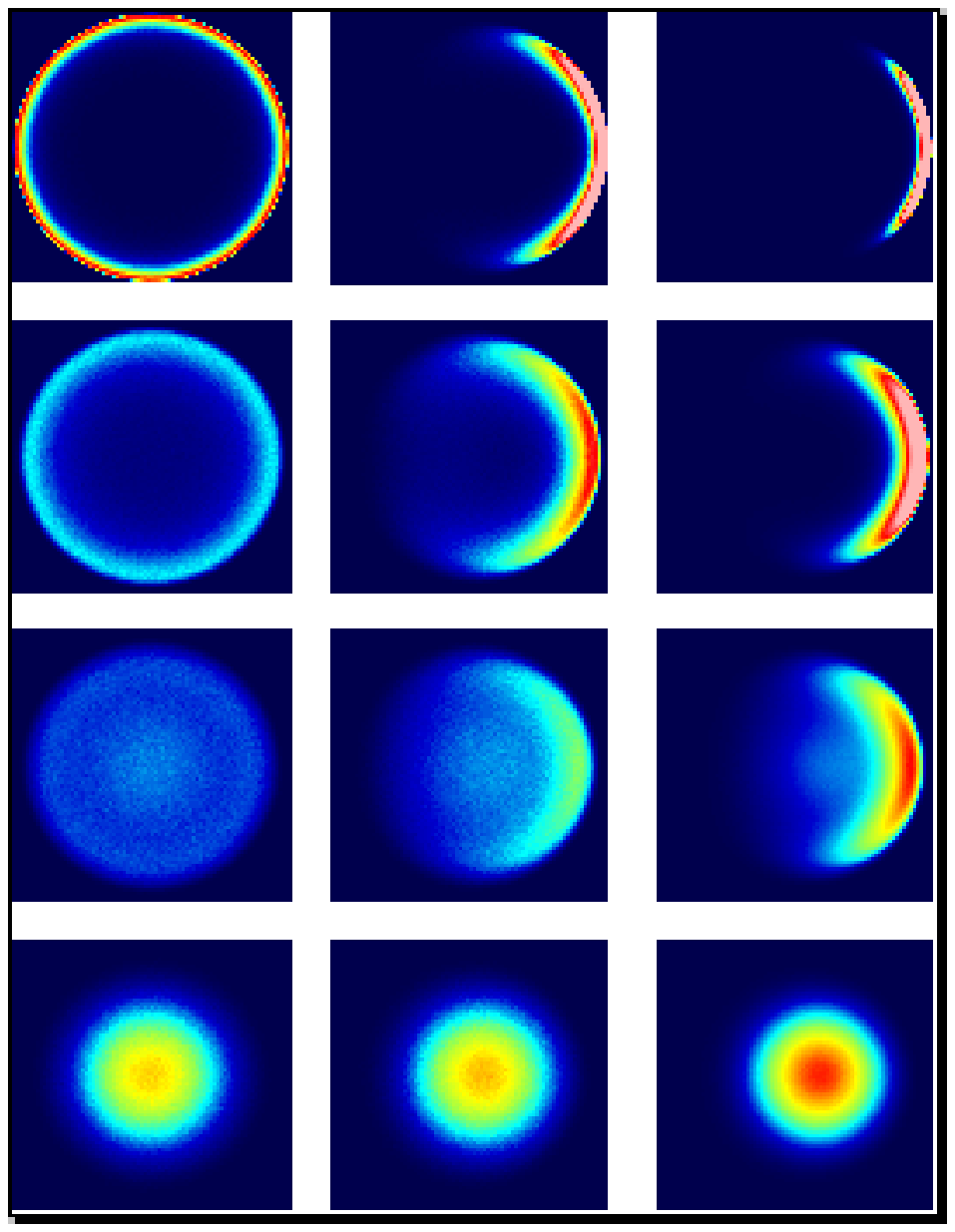}
\index{WLC}
\end{center}
\caption{Density plot of $p(v_x,v_y)$. Color code : red - high density,
\index{Density plot}
blue - low density. Left panel is for free polymer, middle
\index{free polymer}
panel is for polymer having one end grafted in $x$- direction and the 
right panel is for polymer having both ends grafted in $x$- direction.
\index{grafted}
From top to bottom four panels denote stiffness values $t=.5,2,4,10$.
\index{stiffness}
Note that the double maxima feature in $p(v_x,v_y)$ 
(one maximum near the centre and another near the rim) 
at $t=4$ persists for all three boundary consitions.
}
\label{comp3}
\end{figure}

We now employ MC simulations to study some other aspects of
\index{MC}
probability distribution in Helmholtz ensemble. 
\index{Helmholtz}
\index{probability}
\index{ensemble}
We first examine the radial distribution 
function $p(v)$ ($v=r/L$, $p(v)=L^2 P(r)$). It
is clear from Fig.\ref{comp1} that grafting one end does not change the
double maxima feature in $p(v)$ at intermediate values of stiffness 
\index{stiffness}
($4\le t\le2$). Though grafting both the ends change the distribution
function, the double maxima feature persists.
Here we note that once one end of a polymer is grafted immediately the system
\index{grafted}
loses its spherical symmetry, more so, since we restrict ourselves to 
semi-flexible regime. 
\index{probability}
\index{projected probability}
\index{scaling}
We have already seen that the projected probability distribution
\index{probability}
\index{projected probability}
$p_x(v_x)$ and $p_y(v_y)$ are very different
depending on the orientations of polymer ends. However, for grafting
of one end, this does not show up in radial distribution function $p(v)$
(Fig.\ref{comp1}).  This is because, fixing one end only shifts
the probability weight distributed over all possible angles at a given
\index{probability}
radial distance $v$ towards the direction of orientation but on the 
same radial distance.

The full statistics is encoded in the vector distribution function 
$p(v_x,v_y)$. To see the complete structure of it we next 
obtain $p(v_x,v_y)$ from MC simulations in FRC (for one or both ends
\index{FRC}
\index{MC}
of the polymer free to rotate) and $xy$- model (for polymer with both its
ends fixed in some given orientation) and plot
the total distribution $p(v_x,v_y)$ as a two dimensional density plot. 
\index{Density plot}
We compare $p(v_x,v_y)$ of free polymer, polymer with one end grafted 
\index{free polymer}
and polymer with both ends grafted (Fig.\ref{comp3}). For definiteness we
\index{grafted}
chose all the graftings to be in the $x$- direction.
We plot $p(v_x,v_y)$ over a range of stiffnesses ($t=0.5,2,4,10$). 
The distribution has finite values for $v\leq 1$ and zero for 
$v>1$. This is due to the inextensibility constraint in WLC model.
\index{WLC}
In these density plots high probability is shown in red and low 
\index{probability}
in blue (Fig.\ref{comp3}). At small stiffness (large $t$) $p(v_x,v_y)$
\index{stiffness}
shows a single entropic peak, at $\vv=0$ for free polymer and slightly
\index{free polymer}
shifted towards the direction of end- orientations in grafted polymers.
\index{grafted}
This shifted entropic peak slowly moves to $\vv=0$ in the 
limit $t\to\infty$. 
\index{probability}
\index{spherically symmetric}

With increase in stiffness ($t=4$) an energy dominated probability peak starts
\index{stiffness}
to appear near the full extension limit $v=1$ of the polymer. This peak
forms a circular ring for free polymers. For a grafted polymer this peak
\index{grafted polymer}
\index{grafted}
is aligned in the direction of grafting. The probability distribution
$p(v_x,v_y)$ at $t=4$ clearly shows two regions of probability maxima,
one near zero extension another near full extension, that gives rise to
the double maxima in radial distribution function \cite{mywlc}.

\index{stiffness}
\index{free polymer}
\index{probability}

At larger stiffness ($t=2$) the entropic maximum near the centre ($\vv=0$)
\index{stiffness}
disappears. For the free polymer, one energy dominated maximum gets equally
\index{free polymer}
distributed over all angles. This way the system uses its spherical symmetry 
to gain in entropy. For grafted polymers, probability maximum near the
full extension fans a finite solid angle around the direction of grafting.
\index{solid}
The distribution around the grafting direction is narrower for the polymer
with both its ends grafted along the same direction. End orientation and
\index{grafted}
bending energy couples to decide the probability distribution. This 
\index{bending energy}
\index{probability}
coupling is higher at smaller $t$ \cite{wang}. This fact is more pronounced
in plot of $p(v_x,v_y)$ at $t=0.5$ (Fig.\ref{comp3}).
Spakowitz {\em et. al.} have investigated $p(v_x,v_y)$ using their
Greens function calculations for a polymer with one of its ends 
grafted \cite{wang}. In this section we have used MC simulations 
\index{grafted}
\index{MC}
to study $p(v_x,v_y)$ for all the possible boundary conditions. 
We have shown that multiple maxima in $p(v_x,v_y)$ persists near 
$t=4$ for all these boundary conditions. 
\index{boundary conditions}

\begin{figure}[t]
\begin{center}
\includegraphics[width=9cm]{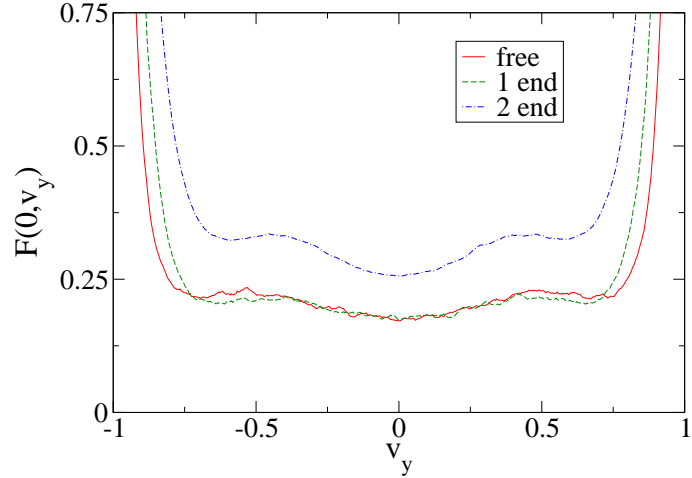}
\index{WLC}
\end{center}
\caption{At $t=4$ free energy profile ${\cal F}(v_y)$ in $y$- direction at 
$v_x=0$ corresponding to the probability distributions shown in Fig.\ref{comp3}
\index{probability}
are plotted.
This clearly shows that the tripple minima feature in free energy for
a polymer with both ends free persists
even after grafting one or both ends of the polymer. 
}
\label{comp4}
\end{figure}

As mentioned earlier, in Helmholtz ensemble the free energy is given by 
\index{Helmholtz}
\index{ensemble}
${\cal F}(v_x,v_y) = -(1/t) \ln [p(v_x,v_y)]$.
In Fig.\ref{comp4} we plot this free energy profile on constant $v_x=0$
line at $t=4$ and compare the three different boundary conditions. 
\index{boundary conditions}
This is obtaind from $v_x=0$ cut of 2D probability density $p(v_x,v_y)$.
\index{probability}
This plot clearly shows that triple minima in free energy \cite{mywlc} 
\index{triple minima}
prevails even after grafting one or both ends of a semi- flexible polymer. 
This means that in constant extension ensemble if one fixes the position
\index{ensemble}
of one end and traps the other end in 2D (both $x$- and $y$- directions) 
to extend the end
to end separation of a semiflexible polymer from the first free energy 
\index{semiflexible polymer}
minimum at $v_y=0$ to full extension
of $v_y$ along $v_x=0$ line, for small extensions an average force will
try to pull back the separation to zero corresponding to the central minimum
of the free energy. However, once the separation is taken beyond the maxima 
in the free energy near $v_y=\pm 0.5$ the average force will push the separation
away from the center towards the non- zero $v_y$ minima in free energy.
Further extension will cause a huge force towards this second minimum in the 
free energy. This is due to the inextensibility constraint in WLC model.
\index{WLC}
Thus, in force- extension
\index{force- extension}
experiments on a polymer in constant extension ensemble, this multi- stability
\index{ensemble}
(non- monotonic force extension) at intermediate stiffness values will be 
\index{stiffness}
measurable for all kind of boundary conditions\cite{mywlc}. The coupling
\index{boundary conditions}
of end orientation and bending energy have raised the free energy of a polymer
\index{bending energy}
with both its ends grafted, with respect to that of a free polymer and a 
\index{free polymer}
polymer with one of its grafted (Fig.\ref{comp4}).
\index{grafted}

We again emphasize that whether the end to end vector distribution or
some projected probability distribution will capture the results of
\index{probability}
\index{projected probability}
a force- extension measurement will depend on the kind of trapping 
potential (what are the directions in which it traps a polymer end).
Apart from this, as we have shown, the orientational boundary conditions at
\index{boundary conditions}
the ends of a polymer and the ensemble of experiment will affect 
\index{experiment}
\index{ensemble}
the force- extension behavior non- trivially.
\index{force- extension}

\section{Conclusion}
\label{conclusion}
In this chapter, we found the exact solution of WLC model of semi- flexible
\index{WLC}
polymers for all possible end orientations and in both Gibbs and Helmholtz
\index{Gibbs}
\index{Helmholtz}
ensembles. In previous chapter, we had emphasized on the ensemble
\index{ensemble}
dependence of mechanical properties \cite{mywlc}. In this chapter
we have shown that polymer properties also crucially depend on end
orientations. Conditions of fixed end orientations are biologically 
important in several processes. Magnetic tweezer experiments can be
\index{magnetic tweezer}
used in laboratory to study orientation dependent properties of
semi- flexible polymers. We have presented an exact theory 
to obtain end to end distribution function of WLC polymers by calculating
\index{WLC}
the propagator for a quantum particle moving on a unit sphere. 
\index{propagator}
For a finite chain, the free energies in two ensembles were shown 
to be related via a Laplace transform relation which in thermodynamic 
\index{Laplace transform}
\index{thermodynamic limit}
limit goes over to the typical Legendre transform relation. Two discretized
\index{Legendre transform}
versions of WLC model have been used in MC simulations. In Helmholtz 
\index{WLC}
\index{Helmholtz}
\index{MC}
ensemble, for stiffness parameters around $t=4$, we found multimodality
\index{ensemble}
\index{stiffness}
(multiple minima in free energy) for all kinds of boundary orientations.
Thus we have generalized the finding of triple minima in free energy 
\index{triple minima}
that we obtained for free polymers in the previous chapter \cite{mywlc}. 
The radial distribution function of a free polymer and a polymer with 
\index{free polymer}
a fixed orientation in one end show the same behavior. 
This we trace back to the properties of full
distribution of end to end vector. We have calculated various projected
\index{projected probability}
probability distributions. The transverse fluctuation of a polymer with
\index{transverse fluctuation}
\index{probability}
a fixed orientation at one end had been studied using MC simulation
\index{MC}
\cite{munk-frey} and an effective medium theory \cite{munk-frey-th} which
approximately captures the qualitative features of the simulations.
We have calculated this fluctuation from our theory that shows exact
\index{fluctuation}
numerical agreement with our simulation and Lattanzi {\em et. al.}'s 
\cite{munk-frey} simulation results. For a polymer with orientations
\index{simulation}
at both its ends fixed, the end to end distribution shows dependence
on relative angle $\h$ between the two ends. The end orientations couple
to the bending energy to take an unimodal distribution through
\index{bending energy}
bimodality with change in $\h$. We have seen the general presence of
\index{bimodality}
non- monotoinic force extension in Helmholtz ensemble. We have also
studied the force- extension behavior in Gibbs ensemble. These never 
\index{force- extension}
show non- monotonicity. It is important to note that,
multiple maxima in end to end distribution means a non- monotonic
force extension in Helmholtz ensemble, {\em not} in Gibbs ensemble.
\index{Helmholtz}
As we have shown, the response function in Gibbs ensemble is always
\index{Gibbs}
positive, thus non- monotonicity in this ensemble is impossible.
\index{ensemble}
We have outlined the general theoretical framework in $d$- dimensions.
However, in this study we have looked at the properties of a 2D polymer.
At the onset we have shown that an important point of polymer statistics,
multimodality, is dependent on the dimensionality of embediing space.
In three dimensional free polymer,
\index{free polymer}
multimodality in projected probability distribution is impossible,
\index{probability}
\index{projected probability}
however presence of this is a reality in 2D. Similar studies in 3D remains
to be an interesting direction forward. The presence of multimodality in
intermediate range of stiffness (near $t=4$) gives rise to a 
\index{stiffness}
possibility of multi-stability in the dynamics of semi- flexible polymers. 
The simplest question remains, for a free polymer how does the 
\index{free polymer}
presence of a triple minima in free energy affects its dynamics. 
\index{triple minima}
\index{dynamics}
This might be important in the context of understanding the 
time- scale of a messanger- RNA finding out an active site on a DNA chain.
\index{}
\index{DNA}

\index{simulation}
\index{ensemble}
\index{force- extension}
\index{simulation}
\index{MC}
\index{ensemble}
\index{WLC}
\index{boundary conditions}
\index{MC}
\index{ensemble}
\index{stiffness}
\index{grafted}

\chapter{Laser Induced Freezing and Re-entrant Melting}
\begin{verse}
\begin{flushright}
\it {Reviewing Melquiades' notes, serene now, without the exaltation
of novelty, in prolonged and patient sessions they tried to separate
\'Ursula's gold from the debris that was stuck to the bottom of
the pot.}
\hfill -- G. G. M{\'a}rquez
\end{flushright}
\end{verse}
\vskip 1cm

In the previous two chapters we have studied the properties of a finite 
semiflexible polymer in different ensembles and under different conditions
\index{semiflexible polymer}
of its boundary orientations. These conditions can be implemented on the
polymer by imposing various kinds of external trapping and constraints.
\index{constraints}
The finite size and bending rigidity of such molecules coupled with the
boundary orientations rendered the different statistical and mechanical 
properties. In this chapter we consider another kind of trapping. 
A one dimensional modulating external trapping potential that couples 
with the local density is applied on a two dimensional system of particles 
interacting via various spherically symmetric short range repulsions.
\index{spherically symmetric}
This trapping potential constrains the system in one direction.
We find out the phase behavior of this system with change in the strength
\index{phase}
of the trapping potential. It is known that in two dimensions melting
occurs by the unbinding of dislocation pairs \cite{kt,kthny1,kthny2}.
\index{dislocation}
\index{dislocation}
How does this scenario change if there are periodic constraining 
potentials which may reduce the dimensionality of the system from two
to one?

Examples of phase transitions mediated by the unbinding of defect pairs abound
\index{defect}
\index{phase}
in two dimensions. The quasi- long- ranged order to disorder transition in the 
XY and planar rotor models\cite{kt,xy-simu1,xy-simu2,xy-simu3,rev1,surajitxy}, 
the melting transition of two dimensional solid\cite{morf-esol,kt,kthny1,
kthny2,rev2,sura-hdmelt},
the superconductor to normal phase transition in two dimensional 
\index{phase transition}
Josephson junction arrays\cite{joseph}, the 
commensurate- incommensurate transition of the striped  phase of 
\index{phase}
\index{commensurate}
smectic liquid crystals on anisotropic substrates\cite{stripe}, and the more 
\index{smectic}
recent discovery of a defect mediated re-entrant freezing transition in 
two dimensional colloids in an external periodic potential \cite{chowdhury,wei}
are all understood within such defect unbinding theories. While the very first  
defect mediated  transition theory for the phase 
\index{phase transition}
\index{phase}
transition in the  XY-model by Kosterlitz and Thouless (KT)\cite{kt} enjoyed 
almost immediate acceptance and was verified in simulations\cite{xy-simu1,xy-simu2,xy-simu3,surajitxy}
as well as experiments\cite{xy-expt-1,xy-expt-2}, defect mediated theories of 
\index{defect}
two dimensional melting took a long time to gain general acceptance in the 
community\cite{fight}. There were several valid reasons for this reticence 
however. 

Firstly, as was recognized even in the earliest papers\cite{kthny1,kthny2} 
on this subject, 
the dislocation unbinding transition, which represents an instability 
\index{dislocation}
\index{dislocation}
of the solid phase, may always be pre-empted by a 
\index{phase}
first order\cite{strand,nelson-dg} transition 
from a metastable solid to a stable liquid. Whether such a first order 
\index{solid}
melting transition actually occurs or not depends on the temperature 
of instability $T_{KT}$; so that if the transition temperature 
$T_c < T_{KT}$ the unbinding of dislocations does not occur. Clearly,
neither this condition nor its converse can hold for all 2d systems
in general. This is because 
$T_{KT}$ is a non-universal number which depends on the ``distance''
in coupling parameter space between the bare and the fixed point 
Hamiltonian and hence on the details of the interaction. 
Secondly, the renormalization group flow equations derived in all defect 
\index{renormalization}
mediated theories to date are perturbative expansions in the defect density 
\index{defect density}
\index{defect}
(fugacity) in the ordered phase. How fast does this perturbation series 
\index{phase}
converge? The answer lies again in the position of the bare Hamiltonian 
in the coupling parameter space. For the planar rotor model\cite{kt,surajitxy},
past calculations show that next to leading order terms in the flow equations 
are essential to reproduce the value of the transition
temperature obtained in simulations\cite{surajitxy}.   
Thirdly, defect mediated transitions predict an essential singularity\cite{kt} 
\index{defect}
of the correlation length at the transition temperature. This means that 
effects of finite size\cite{clock} would be substantial and may thoroughly 
mask the true thermodynamic
result. A rapid increase of the correlation length also implies that the 
\index{correlation}
relaxation time diverges as the transition temperature is approached -- 
critical slowing down. For the two dimensional solid, this last effect is 
\index{solid}
particularly crucial since, even far from the transition, the motion of 
defects is mainly thermally assisted and diffusional and therefore slow. 
The equilibration of defect configurations\cite{dyn-2dmelt}
\index{defect}
 is therefore often an issue 
even in solids of macroscopic dimensions. 

On the other hand, over the last few years it has been possible to test 
quantitatively some of the non-universal predictions of defect mediated 
\index{defect}
theories of phase transitions using simulations of restricted 
\index{phase}
systems\cite{surajitxy,sura-hdmelt,mylif}. A simulation of a system without 
\index{simulation}
defects is used to obtain the values
for the bare coupling constants which are then taken as inputs to the 
renormalization group equations for the appropriate defect unbinding theory 
\index{renormalization}
\index{defect}
to obtain quantities like the transition temperature. Needless to say, the 
simulated system does not undergo a phase transition and therefore problems
\index{phase transition}
\index{phase}
typically related to diverging correlation lengths and times do not occur.
\index{correlation}
Numerical agreement of the result of this calculation with that of 
unrestricted simulations or experiments is proof of the validity of the RG 
\index{RG}
flow equations\cite{kt,kthny1,kthny2,sura-hdmelt}. 
This idea has been repeatedly applied in the past to analyze 
defect mediated phase transitions in the planar rotor model\cite{surajitxy}, 
\index{defect}
\index{phase}
two dimensional melting of hard disks\cite{sura-hdmelt} and the re-entrant 
freezing of hard disks in an external periodic potential \cite{mylif,myerrlif}.
The last system is particularly interesting in view of its 
close relation with experiments on laser induced re-entrant freezing transition 
\index{laser}
in charge stabilized colloids \cite{chowdhury,wei} and this constitutes the 
subject of the present chapter as well. 

In this chapter we show in detail how restricted simulations of systems 
of particles interacting among themselves via a variety of interactions and
with a commensurate external periodic potential can be used to obtain phase 
\index{phase}
\index{commensurate}
diagrams showing the re-entrant freezing transition. The results obtained are 
compared to earlier unrestricted simulations for the same systems. Briefly
our results are as follows. Firstly, we observe that, as in an earlier 
study of the planar rotor model\cite{surajitxy}, next to leading order 
corrections to the
renormalization flow equations are {\em essential} to reproduce even the 
\index{renormalization}
gross features of the phase diagram. Specifically, the re-entrant portion of 
the phase diagram can be reproduced {\em only} if such correction terms 
are taken into account. Secondly while we find almost complete agreement with 
earlier results for the hard disk system which has been studied most 
\index{hard disk}
extensively, our phase diagram for the other forms of interaction is shifted 
\index{phase diagram}
\index{phase}
\index{phase diagram}
with respect to the results available in the literature. This may mean either
of two things --- inadequacy of the RG theory used by us or finite size 
\index{RG}
effects in the earlier results. 
Lastly, as a by product of our 
calculations, we have obtained the core energy for defects (dislocations) in 
these systems and studied its dependence on thermodynamic and potential 
parameters.          

The problem of re-entrant freezing transition of a system of interacting 
colloidal particles in a periodic potential has an interesting history 
involving experiments\cite{chowdhury,wei}, 
simulations\cite{jcdlvo,cdas1,cdas2,cdas3,lif-hd,lif-dlvo,lif-sd1,lif-sd2} 
\index{DLVO}
and theory\cite{jay,frey-lif-prl,frey-lif-pre}. 
\index{LIF}
In last couple of decades soft systems
like colloids have been studied extensively\cite{colbook} both for their
own sake and as typical toy models to study various important
condensed matter questions like structural and phase transitions through
\index{phase}
experiments that allow real space imaging. Charged colloids
confined within two glass plates form a model 2-d system as 
the electrostatic force from the plates almost completely suppresses the 
fluctuations of colloids perpendicular to the plates, practically confining 
them to a 2-d plane. In their pioneering 
experiment Chowdhury\cite{chowdhury} {\em et. al.} imposed a simple static 
\index{experiment}
background potential which is periodic in one direction and constant in the 
other (except for an overall Gaussian profile of intensity- variation) by 
interfering two laser beams. This potential immediately induces a density 
modulation in the colloidal system. The potential minima are spaced to 
overlap with the close packed lines of the ideal lattice of the colloidal 
system at a given density. With increase in potential strength 
such a colloidal liquid has been observed to solidify. This is known as laser 
induced freezing (LIF). In a recent experiment\cite{wei} it has been shown 
\index{LIF}
\index{experiment}
that with further increase in potential strength, surprisingly, the solid 
\index{solid}
phase re-melts into a modulated liquid. This phenomenon is known as re-entrant 
\index{modulated liquid}
laser induced freezing (RLIF).
\index{RLIF}
\index{laser}
\index{laser induced freezing}
Qualitatively, starting from a liquid phase, the external periodic potential
\index{phase}
immediately induces a density modulation, reducing fluctuations which 
eventually leads to solidification. Further increase in the amplitude of 
the potential reduces the system to a collection of decoupled 1-d strips. 
The dimensional reduction now {\em 
increases} fluctuations remelting the system.

The early mean field theories, namely, Landau theory\cite{chowdhury} 
and density functional theory\cite{jay} predicted a change from a first 
\index{density functional theory}
order to continuous transition with increase in
potential strength and failed to describe the re-entrant behavior, 
a conclusion seemingly confirmed by early experiments\cite{chowdhury} and 
some early simulations\cite{jcdlvo}.
Overall, the results from early simulations remained inconclusive
however, while one of them\cite{jcdlvo}
claimed to have found a tri-critical point at intermediate laser intensities 
\index{laser}
and re-entrance, later studies refuted these results
\cite{cdas1,cdas2,cdas3}. All of these studies used the change in order
parameter and the maximum in the specific heat to identify the phase 
\index{phase transition}
\index{phase}
transition points. While some later studies\cite{cdas1,cdas2,cdas3} found 
RLIF for hard disks they reported 
\index{RLIF}
laser induced freezing and absence of any re-entrant melting for the DLVO 
\index{DLVO}
\index{laser}
\index{laser induced freezing}
potential\cite{cdas3} in direct contradiction to experiments\cite{wei}.

Following the defect mediated disordering approach of Kosterlitz and Thouless
\index{defect}
\cite{kt}(KT), Frey, Nelson and Radzihovsky\cite{frey-lif-prl}(FNR) 
\index{LIF}
proposed a detailed theory for the re-entrant transition based 
on the unbinding of dislocations
with Burger's vector parallel to the line of potential minima. This theory 
\index{Burger's vector}
predicted RLIF and no tricritical point. The results of this work were in 
qualitative agreement with experiments\cite{wei} and provided a framework
for understanding RLIF in general. More accurate simulation studies on 
\index{RLIF}
\index{simulation}
systems of hard disks\cite{lif-hd}, soft disks\cite{lif-sd1,lif-sd2}, 
DLVO\cite{lif-dlvo} etc. confirmed the re-entrant freezing-melting transition 
\index{DLVO}
in agreement with experiments\cite{wei} and FNR 
theory\cite{frey-lif-prl,frey-lif-pre}. In these 
studies the phase transition point was found from the crossing of 
\index{phase transition}
\index{phase}
Binder-cumulants\cite{bincu,lan-bin} of order parameters corresponding
to translational and bond- orientational order, calculated for various 
sub- system sizes. A systematic finite size scaling analysis\cite{lif-hd} of 
\index{scaling}
\index{LIF}
\index{finite size scaling}
simulation results for the 2-d hard disk system in a 1-d modulating 
\index{hard disk}
\index{simulation}
potential showed, in fact, several universal features consistent with the 
predictions of FNR theory. It was shown in these studies that 
the resultant phase diagram remains system size dependent and the 
\index{phase diagram}
\index{phase}
\index{phase diagram}
cross- over to the zero field KTHNY melting\cite{kthny1,kthny2} 
\index{KTHNY}
plays a crucial role in understanding the results for small values of the 
external potential. 
While the data collapse and critical exponents were consistent with KT
\index{critical exponents}
theory for stronger potentials, for weaker potentials they match better 
with critical scaling\cite{lif-hd}. A common problem with all the simulation
\index{scaling}
\index{LIF}
\index{simulation}
studies might be equilibration with respect to dislocation movements along
\index{dislocation}
\index{dislocation}
climb (or even glide) directions. Also, non universal predictions, namely 
the phase diagram are difficult to compare because 
\index{phase diagram}
\index{phase diagram}
the FNR approach (like KT theory) is expressed in terms of the appropriate 
elastic moduli which are notoriously time-consuming to compute near a 
continuous phase transition. Diverging correlation lengths and times near
\index{correlation}
the phase transition further complicate an accurate evaluation of the 
\index{phase transition}
non universal predictions of the theory. 
 
We calculate the phase diagrams of three different 2-d systems 
with a 1-d modulating potential (see Fig.~\ref{cartoon}) following the
technique of restricted Monte Carlo simulations\cite{surajitxy, sura-hdmelt,
mylif}, to be discussed below. For the laser induced transition
\index{laser}
we use this method to generate whole phase diagrams. 
\index{phase}
We reject Monte Carlo moves which tend to distort an unit cell in a way which 
changes the local connectivity\cite{sura-hdmelt}. The percentage of moves thus 
rejected is a measure of the dislocation fugacity\cite{sura-hdmelt}. This, 
together with the elastic constants of the dislocation free lattice obtained 
\index{dislocation}
\index{dislocation}
separately, are our inputs (bare values) to the renormalization flow 
\index{renormalization}
equations\cite{frey-lif-prl,frey-lif-pre} 
\index{LIF}
to compute the melting points and hence the phase 
\index{phase diagram}
\index{phase}
\index{phase diagram}
diagram. Our results (Fig.~\ref{hd-phdia},\ref{dlvo-phdia},\ref{soft-phdia}) 
\index{DLVO}
clearly show a modulated liquid (ML) $\to$ locked floating solid (LFS) $\to$ 
\index{solid}
\index{modulated liquid}
ML re-entrant transition with increase in the amplitude ($V_0$) of the 
potential. In general, we find,  the predictions of FNR theory to be valid. 

Lastly, we must mention that our technique, as summarized above and used in
this as well as earlier\cite{mylif} work, corresponds closely with early 
studies of the melting of the electron solid by Morf\cite{morf-esol,morf}. 
\index{solid}
In the
latter case, the dislocation fugacity, which is one of the important inputs
to the KTHNY flow equations was obtained by a careful and direct calculation
\index{KTHNY}
of the dislocation core energy at $T=0$. Our approach is somewhat cruder but 
\index{dislocation}
\index{dislocation}
gives us numbers for nonzero $T$ which automatically contain the effects of
phonon fluctuations.

In section \ref{method} we first briefly discuss the FNR theory and then 
go on to show in detail the restricted simulation scheme used by us to 
obtain the various quantities required to calculate the phase diagram. In  
section \ref{result} we give the simulation results. We describe, 
\index{simulation}
in detail, the various quantities leading to the phase diagram for one of the
\index{phase diagram}
\index{phase diagram}
systems, viz. the hard disks\cite{jaster,sura-hdmelt}. Then 
we present the phase diagrams for the other two systems we study. We compare
\index{phase}
our results with earlier simulations. Lastly, in section \ref{conclusion} we 
summarize our main results and conclude.

\vskip .2cm

\begin{figure}[t]
\begin{center}
\includegraphics[width=7cm]{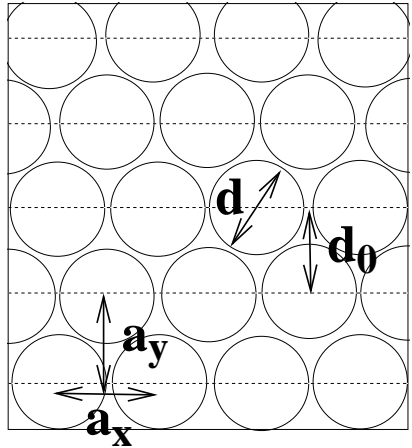}
\index{LIF}
\end{center}
\caption{This cartoon shows a typical 2-d system under consideration. 
$d$ is the length scale over which repulsive two body potentials are
operative. The dashed lines indicate minima of external modulating potential 
$\beta V(y)= -\beta V_0 \cos(2\pi y/d_0)$. $a_x=a_0$ is the lattice parameter 
fixed by the density $\rho$ and $a_y$ indicate the average separation 
between two layers along $y$-direction perpendicular to a set of close-packed 
\index{layers}
planes. For a perfect triangular lattice $a_y=\sqrt{3}a_0/2$.
The modulating potential is commensurate with the lattice such that
\index{commensurate}
$d_0=a_y$.}
\label{cartoon}
\end{figure}
\vskip .2cm

\section{Method}
\label{method}
A cartoon corresponding to the systems considered for our study is given in 
Fig.~\ref{cartoon}. The elastic free energy of the solid is given in terms of
the spatial derivatives of the displacement field 
$\vec u(\vec r) = \vec r - \vec r_0$ with $\vec r_0$ being the lattice vectors 
of the undistorted reference triangular lattice. For a solid in presence of a 
\index{solid}
modulating potential $\beta V(y)$ (Fig.~\ref{cartoon}) the
displacement mode $u_y$ becomes massive, leaving massless $u_x$ modes. After
\index{displacement}
integrating out the $u_y$ modes the free energy of the LFS 
may be expressed in terms of gradients of $u_x$ and 
elastic moduli\cite{frey-lif-prl,frey-lif-pre}, namely, the Young's
\index{LIF}
modulus $K(\beta V_0,\rho)$ and shear modulus $\mu(\beta V_0,\rho)$,
\index{shear modulus}
\begin{equation}
 {\cal H}_{el} = \int dx dy \left[ \frac{1}{2} K\left(\frac{\partial u_x}{\partial x}\right)^2+ \frac{1}{2} \mu\left(\frac{\partial u_x}{\partial y}\right)^2\right]
\label{hamiltonian} 
\end{equation}

Similar arguments\cite{frey-lif-prl,frey-lif-pre} 
\index{LIF}
show that among the three sets of low 
energy dislocations available in the 2-d triangular lattice, only those 
(type I) with Burger's vector parallel to the line of potential minima survive 
at large $\beta V_0$. Dislocations with Burger's vector pointing along the 
\index{Burger's vector}
other two possible close-packed directions (type II) in the 2-d triangular 
lattice have larger energies because the surrounding atoms are forced to ride 
the crests of the periodic potential\cite{frey-lif-prl,frey-lif-pre}. 
\index{LIF}
Within this set of 
assumptions, the system therefore shares the same symmetries as 
the XY model. Indeed, 
a simple rescaling of $x\to\sqrt{\mu}x$ and $y\to\sqrt{K}y$ leads this free
energy to the free energy of the XY-model with spin-wave stiffness 
\index{stiffness}
$K_{xy}=\sqrt{K\mu}a_0^2/4\pi^2$ and spin angle $\theta=2\pi u_x/a_0$:
$$
{\cal H}_{el} = \int dx dy ~ \left[\frac{1}{2}K_{xy}(\nabla\theta)^2\right]
$$ 
This immediately leads to the identification of a vortex in XY model 
($\oint d\theta = 2\pi$) with a dislocation of Burger's vector $ \vec b = 
\index{Burger's vector}
\hat i a_0$ ($\oint du_x = a_0$, $\hat i =$ unit vector along $x$- 
direction) parallel to the potential minima {\em i.e.} the dislocation of 
\index{dislocation}
\index{dislocation}
type I. The corresponding theory for phase transitions can then be recast as a  
\index{phase}
KT theory\cite{kt} and is described in the framework of a two parameter 
renormalization flow for the spin-wave stiffness 
\index{renormalization}
\index{stiffness}
$\beta K_{xy}(l)$ and the fugacity of type I dislocations $y'(l)$,
where $l$ is a measure of length scale as $l=\ln(r/a_0)$, $r$ being the 
size of the system. The flow equations are expressed in
terms of $x'=(\pi \tilde K_{xy}-2)$ where $\tilde K_{xy}=\beta K_{xy}$
and $y'=4\pi~exp(-\beta E_c)$ where $E_c$ is the
core energy of type I dislocations which is obtained from the dislocation 
\index{dislocation probability}
\index{dislocation}
\index{dislocation}
probability\cite{sura-hdmelt,morf}.
\index{probability}
Keeping  up to next to leading order terms in $y'$ 
the renormalization group flow equations\cite{amit,surajitxy} 
\index{renormalization}
are,
\begin{eqnarray}
\frac{dx'}{dl}&=& -y'^2 - y'^2x' \nonumber \\ 
\frac{dy'}{dl}&=& -x'y' + \frac{5}{4}y'^3.
\label{floweq}
\end{eqnarray}
Flows in $l$ generated by the above equations starting from initial or ``bare''
values of $x'$ and $y'$ fall in two categories. If, as  $l\to\infty$,  $y'$ 
diverges, the thermodynamic phase is disordered (i.e. ML), while on the other 
hand if $y'$ vanishes, it is an ordered phase 
(a LFS)\cite{frey-lif-prl,frey-lif-pre}. 
\index{LIF}
The two kinds of flows are demarcated by the {\em separatrix} 
which marks the phase transition point. For the linearized equations, that
\index{phase transition}
\index{phase}
keeps i only the leading order terms in $y'$, the 
separatrix is simply the straight line $y' = x'$, whereas for the full 
\index{separatrix}
non-linear equations one needs to calculate this 
numerically\cite{amit,surajitxy,sura-hdmelt}.   

\begin{figure}[t]
\begin{center}
\includegraphics[width=13cm]{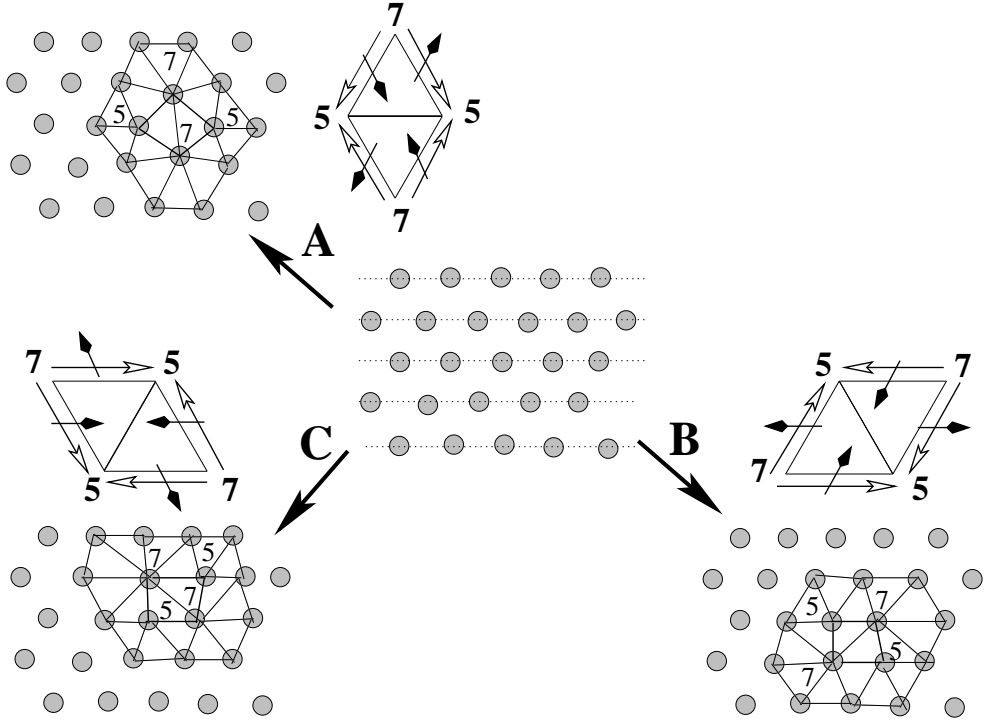}
\index{LIF}
\end{center}
\caption{
This diagram depicts all the possible dislocation generating moves that we 
\index{dislocation generating moves}
reject. Starting from the triangular lattice shown in the centre (the dotted 
lines show the potential minima), in all, there can be three types of  
dislocation- pair generating moves shown as A, B \& C. The numbers $7$ 
\index{dislocation}
\index{dislocation}
and $5$ denote the positions of two types of disclinations having seven 
nearest neighbours and five nearest neighbours respectively. Only those bonds, 
which are necessary to show distortions due to the generation of disclination 
quartets, have been drawn. The rhombi near each of the distorted lattice 
denote the unit cells and open arrows from $7 \to 5$ show the direction of 
\index{cells}
dislocation generating moves. The probabilities of these moves are $P_{m1}$(A),
\index{dislocation generating moves}
\index{dislocation}
\index{dislocation}
$P_{m2}$(B) and $P_{m3}$(C). Corresponding Burger's vectors (filled arrows) 
are bisectors pointing towards a direction rotated counter-clockwise starting 
from $7 \to 5$ directions and are parallel to one of the lattice planes. 
Notice, the separation between Burger's vectors of a pair along the glide 
direction (parallel to the Burger's vectors) is a single lattice separation 
($a_0$) and within this construction it is impossible to draw Burger's loop 
that can generate non-zero Burger's vector. Depending on which of the two 
\index{Burger's vector}
possible disclination pairs separate out any one dislocation- antidislocation 
\index{dislocation}
\index{dislocation}
pair will be formed.   
}
\label{dislo-cartoon}
\end{figure}

The bare numbers for $x'$ and $y'$ are 
relatively insensitive to system size since our Monte Carlo simulation does not
\index{simulation}
involve a  diverging correlation length associated with a phase transition. 
\index{correlation}
\index{phase transition}
\index{phase}
This is achieved as follows\cite{surajitxy,sura-hdmelt}. 
We monitor individual random moves of the particles in a system
and look for distortions of the neighboring unit cells. If in 
any of these unit cells the length of a next nearest neighbor bond 
\index{cells}
becomes smaller than the nearest neighbor bond, the move is rejected. 
All such  moves generate disclination quartets and are shown in the  
Fig.~\ref{dislo-cartoon}. Notice that each of these moves break a nearest 
neighbour bond to build a new next nearest neighbour bond, in the process 
generating two $7$-$5$ disclination pairs. These are
the moves rejected in the restricted simulation scheme we follow. 
\index{simulation}
The probabilities of such bond breaking moves are however computed by 
keeping track of the number of such rejected moves. One has to keep 
track of all the three possible distortions of the unit rhombus  with 
measured probabilities $P_{mi}, i = 1,2,3$ (see Fig.~\ref{dislo-cartoon}),
$$ 
P_{mi} = \frac{\mbox{number of rejected bond breaking moves of type i}}{\mbox{Total number of MC moves}}
\index{MC}
$$ 
Each of these distortions involves four $7-5$ disclinations {\em i.e.} 
two possible dislocation- antidislocation pairs which, we assume, occur 
mutually exclusively in a way that we explain shortly.
For a free ($V_0 = 0$) two dimensional system the dislocation 
\index{dislocation}
\index{dislocation}
core energy $E_c^t$ can be found through the relation\cite{morf}
\begin{equation}
\Pi = \exp(-\beta\,\, 2E_c^t)Z(\tilde K)
\label{ec0}
\end{equation}
where $\Pi=\sum_{i=1}^3 P_{mi}$ is the total number density of dislocation
\index{dislocation}
\index{dislocation}
pairs per particle
and $Z(\tilde K)$ is the ``internal partition function" incorporating all three types of degenerate orientations of 
\index{partition function}
dislocations, 
\begin{equation}
Z(\tilde K) = \frac{2\pi\sqrt{3}}{\tilde K/8\pi -1}
\left( \frac{r_{min}}{a_0}\right)^{2-\tilde K/4\pi}
I_0\left(\frac{\tilde K}{8\pi}\right)
\exp\left(\frac{\tilde K}{8\pi}\right)\nonumber
\end{equation}
where $I_0$ is a modified Bessel function, $\tilde K = \beta K a_0^2$ is a
dimensionless Young's modulus renormalized over phonon modes,
$a_0$ being the lattice parameter and $r_{min}$
is the separation between dislocation-antidislocation above which one counts 
the pairs. The above expression for $Z(\tilde K)$ and Eq.(\ref{ec0}) have been 
used previously in simulations\cite{sura-hdmelt,morf} of phase transitions of
\index{phase}
2-d systems in absence of any external potential to find the dislocation core 
\index{dislocation}
\index{dislocation}
energy $E_c^t$.

\vskip .2cm

We now show how the probabilities for generating pairs of specific types
of dislocations $P_{di}$ for $V_0\neq 0$ are related to $P_{mi}$. Consider 
Fig.\ref{dislo-cartoon} where each of the three varieties of bond breaking 
moves are depicted. It is clear that each given distortion can occur due to
the presence of two possible dislocation- antidislocation pairs acting
independently. For example a distortion of type- A can take place either due to
dislocation dipoles with Burgers vectors making an angle of $60^o$ with the
horizontal or an angle of $120^o$. Both of these dislocation dipoles are
\index{dislocation}
\index{dislocation}
of type II. If this bond breaking move were to be accepted, then at a 
subsequent time step the individual dislocations making up any one of the 
two possible pairs could separate, the two possible events being mutually
exclusive. This allows us to write down the following relations among the
various probabilities.
$$ P_{m1} = P_{d2}+P_{d3},~ P_{m2} = P_{d2}+P_{d1},~P_{m3} = P_{d3}+P_{d1}.$$
Solving for $P_{di}$s and remembering that $P_{d2}=P_{d3}$  by symmetry, 
we get
$P_{d1}=\frac{1}{2}(P_{m2}+P_{m3}-P_{m1})$ and  
$P_{d2} =  {P_{m1}}/{2}$.
The above expressions are motivated and illustrated in  
Fig.\ref{dislo-cartoon}. Once the probability of dislocation pairs are obtained
\index{probability}
in this fashion, we may proceed to calculate the dislocation core energy $E_c$
and the dislocation fugacity $y'$.

An argument following the lines of Fisher {\em et. al.}\cite{morf} 
shows that the dislocation probability (number density of dislocation pair 
\index{dislocation probability}
\index{dislocation}
\index{probability}
\index{dislocation}
per particle) for our system,
\begin{equation}
P_{d1} = \exp(-\beta\,\, 2 E_c) Z(\tilde K_{xy})
\label{fuga1}
\end{equation}
where $2E_c$ is the core energy and $Z(\tilde K_{xy})$ is the internal 
partition function of dislocation pair of type I (single orientation). 
\index{partition function}
\index{dislocation}
\index{dislocation}
\begin{eqnarray}
Z(\tilde K_{xy}) &=& \int_{r>r_{min}} \frac{d^2 r}{A_c} 
\exp\left[ -2\pi \tilde K_{xy} \log\left(\frac{r}{a_0}\right)\right] \nonumber \\ 
&=& \frac{2\pi}{\sqrt{3}}\,\, \frac{(r_{min}/a_0)^{2-2\pi \tilde K_{xy}}}{\pi \tilde K_{xy}-1}
\label{fuga2}
\end{eqnarray}
with $\tilde K_{xy} = \beta K_{xy}$ and $A_c = \sqrt3a_0^2/2$ being the area of an unit cell in the undistorted lattice. We choose $r_{min}=2a_0$. At this point
this choice is arbitrary. We give the detailed reasoning for this choice at the
end of the discussions on hard disks in the section \ref{result}.
Eq.\ref{fuga1} and Eq.\ref{fuga2} straightaway yield the required core energy 
$E_c$. The corresponding fugacity contribution to RG flow equations 
\index{RG}
(Eq.\ref{floweq}) is given via
\begin{equation}
y'=4\pi\sqrt{P_{d1}/Z(\tilde K_{xy})}
\label{y'}
\end{equation}

In the above, the following assumption is, however, implicit. Once a 
nearest neighbor bond breaks and a potential dislocation pair is formed,
\index{dislocation}
\index{dislocation}
they separate with probability one
\index{probability}
\footnote{This assumption is similar in spirit to assuming that a particle 
which reaches the saddle point in the Kramers barrier crossing problem 
would automatically cross the barrier \cite{kramers}.}.
This assumption goes into the identity Eq.\ref{fuga1} as well as in 
Eq.\ref{ec0}\cite{sura-hdmelt}.
Taking the rejection ratios due to bond- breaking as the dislocation 
probabilities, as well, require this assumption
\footnote{Note that the calculation of the bare fugacity from the 
dislocation probability is, we believe, more accurate that the 
\index{dislocation probability}
\index{dislocation}
\index{probability}
\index{dislocation}
procedure used in \cite{mylif}.}.
\index{LIF}

The same restricted Monte Carlo simulation can be used to find out the stress 
\index{simulation}
tensor, and the elastic moduli from the stress-strain curves.
\index{strain}
The dimensionless stress tensor for a free ($V_0 = 0$) system is given 
\index{stress}
by\cite{elast}, 
\begin{equation}
\beta \sigma_{\lambda\nu} {\rm d}^2 = -\frac{{\rm d}^2}{S}\left(-\sum_{<ij>}\left< \beta\frac{\partial \phi}{\partial r^{ij}}~\frac{r_\lambda^{ij}r_\nu^{ij}}{r^{ij}}\right> + N\delta_{\lambda\nu}\right)
\label{sts1}
\end{equation} 
where $i$, $j$ are particle indices and $\lambda$, $\nu$ denote directions $x$, $y$; $\phi(r^{ij})$ is the two- body interaction, $S/{\rm d}^2$ is the 
dimensionless area of the simulation box 
\index{simulation}
\footnote{In the presence of an external 1D modulating potential periodic 
  in the $y$-
  direction the stress has contribution from another virial- like additive
\index{stress}
  term, $-\frac{\beta{\rm d}^2}{S}\left<\sum_\lambda y^\lambda
  f_y^\lambda\right>$, where $y^\lambda$ is the $y$-component of position
  vector of particle $\lambda$. This contribution comes from the part of the
  free energy that involves higher energy (massive) excitations. For the
  elastic free energy which is lowest order in the displacement gradient
\index{displacement}
  (Eq.\ref{hamiltonian}) this part does not contribute towards the elastic
  constants, as the $x$- and $y$- component of gradient remain uncoupled. This
  extra term in stress remains a constant background without disturbing the
\index{stress}
  elastic constants connected to the Young and shear modulus that corresponds
\index{shear modulus}
  to distortions of the system in the low energy directions. We therefore
  neglect this background in calculating stresses where from we obtain the
  elastic moduli.}.

\section{Results and Discussion}
\label{result}
In this section we present the results from our simulations
for three different 2-d systems, namely hard disks, soft disks and 
a system of colloidal particles interacting via the DLVO 
\index{DLVO}
(Derjaguin-Landau-Verwey-Overbeek)\cite{dlvo1,dlvo2} potentials.
We discuss, first, our calculation for a two dimensional 
system of hard disks, in detail. The procedure followed in other
systems is almost identical.

{\it Hard disks:} 
The bulk system of hard disks where particles $i$ and $j$, in 2-d, 
interact via the potential $\phi(r^{ij}) = 0$ for $r^{ij} > {\rm d}$ 
and $\phi(r^{ij}) = \infty$ for $r^{ij} \leq {\rm d}$, where 
${\rm d}$ is the hard disk diameter and 
\index{hard disk}
$r^{ij} = |{\bf r}^j - {\bf r}^i|$ the 
relative separation of the particles, is known 
to melt\cite{al,zo,web,jaster,sura-hdmelt} from a high 
density triangular lattice to an isotropic liquid with a narrow 
intervening hexatic phase\cite{kthny1,kthny2,jaster,sura-hdmelt}. 
\index{phase}
The hard disk free energy is entirely entropic in 
\index{hard disk}
origin and the only thermodynamically relevant variable is the number density   
$\rho = N/V$ or the packing fraction $\eta = (\pi/4) \rho {\rm d}^2$.
Simulations\cite{jaster}, experimental\cite{colbook} and 
theoretical\cite{rhyzov} studies of hard 
disks show that for $\eta > .715$ the system exists as a triangular 
lattice which transforms to a liquid below $\eta = .706$. The small 
intervening region contains a hexatic phase predicted by 
\index{phase}
the KTHNY theory\cite{kthny1,kthny2} of 2-d melting. 
\index{KTHNY}
 Apart from being easily 
accessible to theoretical treatment\cite{hansen-macdonald}, experimental systems
with nearly ``hard'' interactions viz. sterically stabilized 
colloids\cite{colbook} are available.  
\begin{figure}[t]
\begin{center}
\includegraphics[width=9cm]{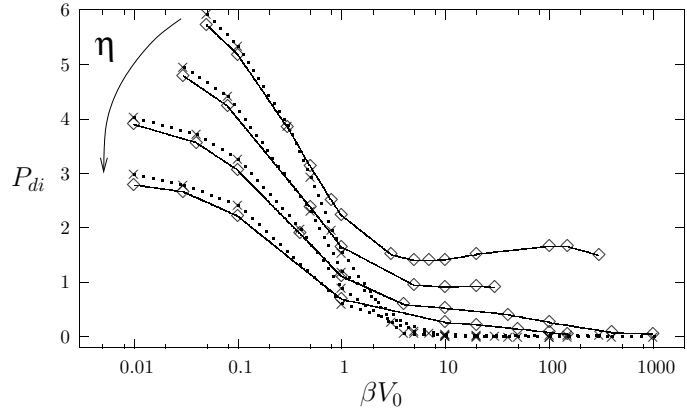}
\index{LIF}
\end{center}
\caption{
Number density of dislocation pairs of type I and II per particle as
\index{dislocation}
\index{dislocation}
a function of the amplitude of the laser potential $\beta V_0$.
\index{laser}
In this plot the {\Large $\diamond$} 
symbols correspond to $P_{d1}$, the probability for type I dislocations and 
the {\large $\times$} symbols to  $P_{d2}$ the probability for type II 
\index{probability}
dislocations obtained from the $P_{mi}$ (see text and Fig.\ref{dislo-cartoon})
for various $\eta$ values, arrow denoting the direction of increasing
$\eta$($=.69,.696,.7029,.71$). The $P_{di}$ for $i=1,2$ are multiplied by 
$10^4$.
These probabilities are plotted against the  potential strength 
$\beta V_0$. Note that for $\beta V_0 > 1$, the probability for type I
\index{probability}
dislocations is larger than that of type II. The dots and solid lines are
\index{solid}
only guides to eye.
}
\label{dislo}
\end{figure}

In presence of a periodic external potential, the only other energy scale 
present in the system is the relative potential
\footnote{
This interaction in colloids is due to polarization of the dielectric colloidal
  particles by the electric field of the laser. Though experiments of
\index{laser}
  Refs.\cite{chowdhury,wei} use charged colloids, the interaction of hard
  sphere colloids with lasers is similar.}
strength $\beta V_0$. 
If the modulating potential is commensurate with the spacing 
\index{commensurate}
between close- packed lines, the elastic free energy of this system in it's
solid phase follows Eq.\ref{hamiltonian} and the corresponding renormalization 
\index{renormalization}
\index{solid}
\index{phase}
flow equations are given by Eq.\ref{floweq}.

We obtain the bare $y'$ and $x'$ from Monte Carlo simulations of
$43\times 50=2150$ hard disks and use them as initial 
values for the numerical solution of Eqs.\,(\ref{floweq}). 
The Monte Carlo simulations for hard disks is done in the usual\cite{frenkel}
way viz. we perform individual
random moves of hard disks after checking for overlaps with neighbours.
When a move is about to be accepted, however, we  
look for the possibility of bond breaking as described in the previous section 
(Fig.\ref{dislo-cartoon}). We reject any such move and the rejection ratios
for specific types of bond breaking moves give us the
dislocation probabilities of type I and II, separately (Fig.\ref{dislo}).
\index{dislocation}
\index{dislocation}

From Fig.\ref{dislo} it is clear that the probability of type
\index{probability}
II  dislocations {\em i.e.} $P_{d2}$ drops down to zero for all packing 
fractions at higher 
potential strengths $\beta V_0$. The external potential suppresses formation
of this kind of dislocations. For small $\beta V_0$ on the other hand, the 
probabilities of type I and type II dislocations are roughly the same.
This should be a cause of concern since we neglect the contribution of type
II dislocations for {\em all} $\beta V_0$. We comment on this issue later
in this section.
\begin{figure}[t]
\begin{center}
\includegraphics[width=13cm]{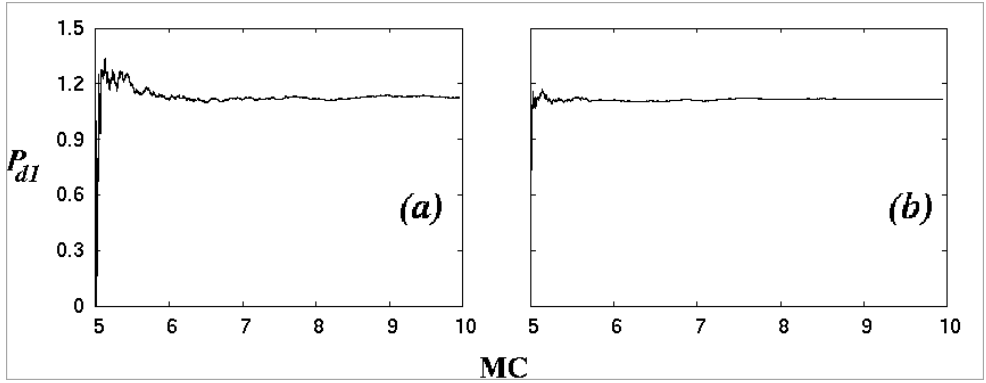}
\index{LIF}
\end{center}
\caption{ 
$P_{d1}$ as a function of MC steps.
$P_{d1}$ has been multiplied by $10^4$ and MC steps has been multiplied 
by $10^{-4}$ for clarity. The data have been collected for $\eta=.7029$
and $\beta V_0=1$.
Panel (a) is for system size of
$2150$ particles whereas (b) for $21488$ particles. Within $10^5$ MC steps
\index{MC}
all fluctuations die out. 
Clearly, the dependence of the dislocation number density
on the system sizes and the Monte Carlo errors are negligible. To calculate
dislocation fugacity we use averaging of data between $5\times10^5$ to $10^6$ 
\index{dislocation}
\index{dislocation}
MC steps.
\index{MC}
}
\label{Pd_err}
\end{figure}
Using Eq.\ref{y'} and Eq.\ref{fuga2} along
with the identity $r_{min}=2 a_0$ gives us the initial value $y'_0$ to be
used in renormalization flow Eq.\ref{floweq}.
\index{renormalization}

Before we move on, we comment on the magnitude of the errors for $P_{mi}$
and hence for $y_0'$. There are two main sources of errors for these 
quantities. This may arise from (a)finite simulation times and (b)small size 
\index{simulation}
of the system. In order to check for this, we have plotted the accumulated 
values for the probability $P_{d1}$ as a 
\index{probability}
function of Monte Carlo step for $2150$ and $21488$ hard disks 
(Fig.\ref{Pd_err}). It is clear that our estimates for the probabilities are
virtually error free! This demonstrates clearly the usefulness of our 
restricted Monte Carlo scheme.

\begin{figure}[t]
\begin{center}
\includegraphics[width=8.6cm]{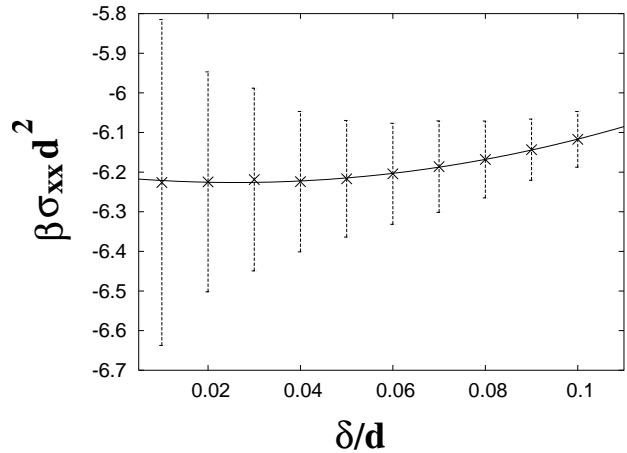}
\index{LIF}
\end{center}
\caption{Plot of $\beta d^2\sigma_{xx}$ vs. $\delta/{\rm d}$ at a strain value $\epsilon_{xx}=.02$ for packing fraction $\eta = .7029$ and potential strength $V_0=1$. A second order polynomial fit (solid line) utilizing the error bars to 
\index{solid}
\index{strain}
assign weights to each data points gives $\lim_{\delta\to 0}\beta d^2\sigma_{xx}=-6.21$ with an error within $.08\%$.
}
\label{ydel}
\end{figure}
\begin{figure}[t]
\begin{center}
\includegraphics[width=8.6cm]{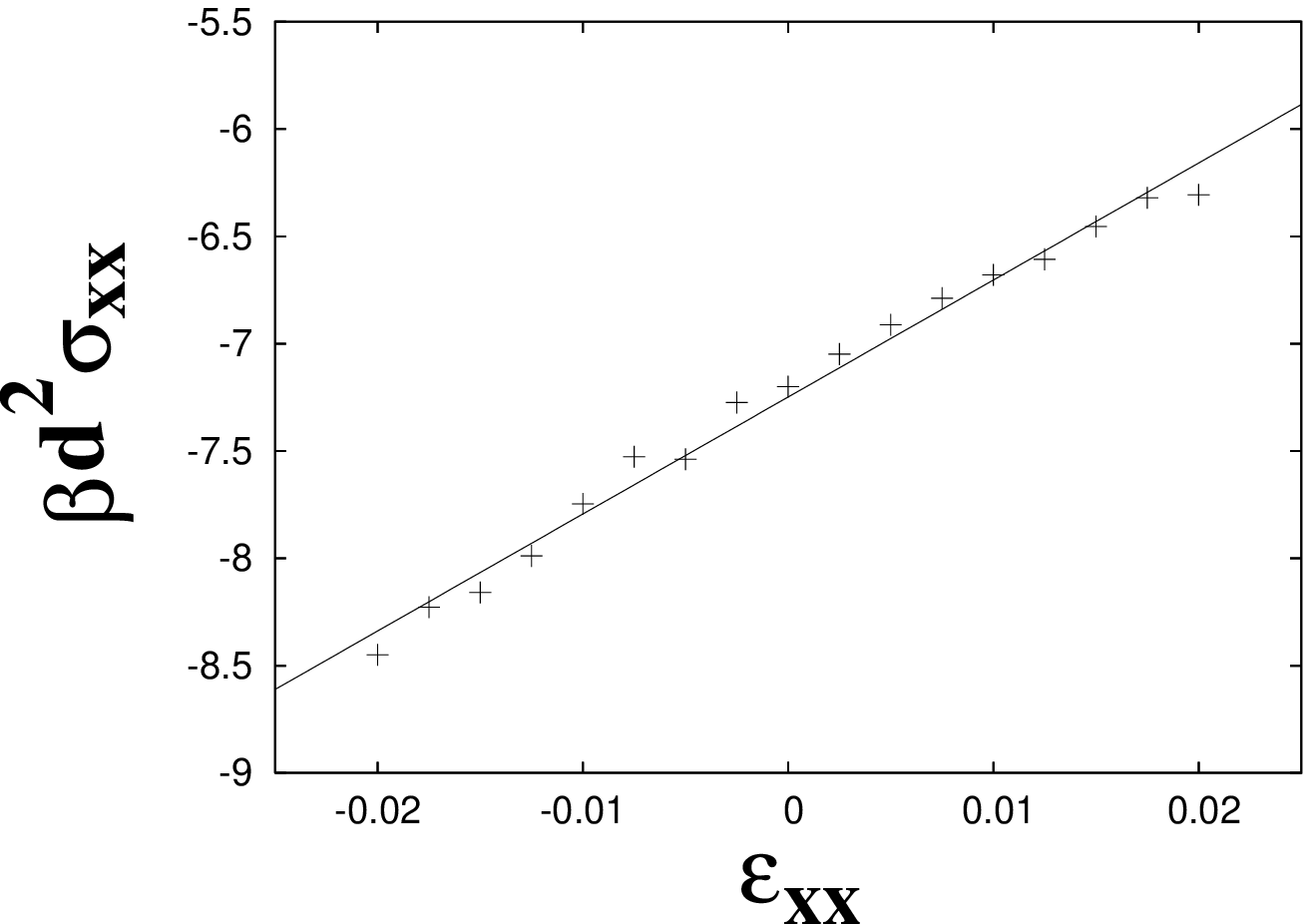}
\index{LIF}
\end{center}
\caption{A typical stress-strain curve used to obtain the Young's modulus from a linear fit (solid line). The graph is plotted at $\eta = .7029,~ V_0=1.0$. The fitted Young's modulus $\beta K {\rm d}^2=54.5$ with an error within $2.9\%$. 
\index{solid}
\index{strain}
The error bars in stress are 
\index{stress}
less than $.2\%$ and much smaller than the point sizes plotted in this graph.
}
\label{ymod}
\end{figure}
\begin{figure}[h]
\begin{center}
\includegraphics[width=8.6cm]{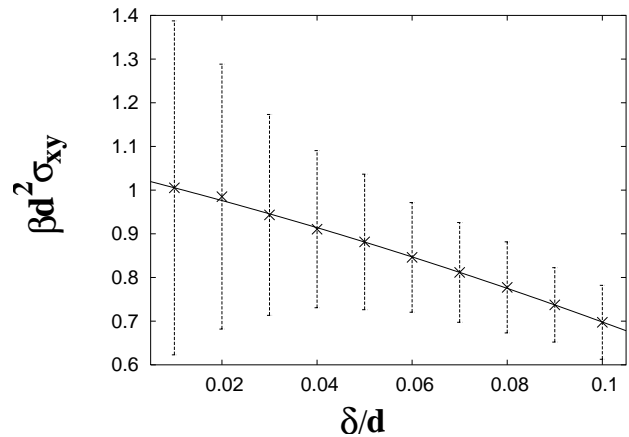}
\index{LIF}
\end{center}
\caption{ Plot of $\beta d^2\sigma_{xy}$ vs. $\delta/{\rm d}$ at strain value $\epsilon_{xy}=.079$ at the packing fraction $\eta = .7029$ and potential strength $V_0=1$. A second order polynomial fit (solid line) gives $\lim_{\delta\to 0}\beta d^2\sigma_{xy}=1.033$ with an error within $.5\%$.
\index{solid}
\index{strain}
}
\label{shdel}
\end{figure}
\begin{figure}[h]
\begin{center}
\includegraphics[width=9.0cm]{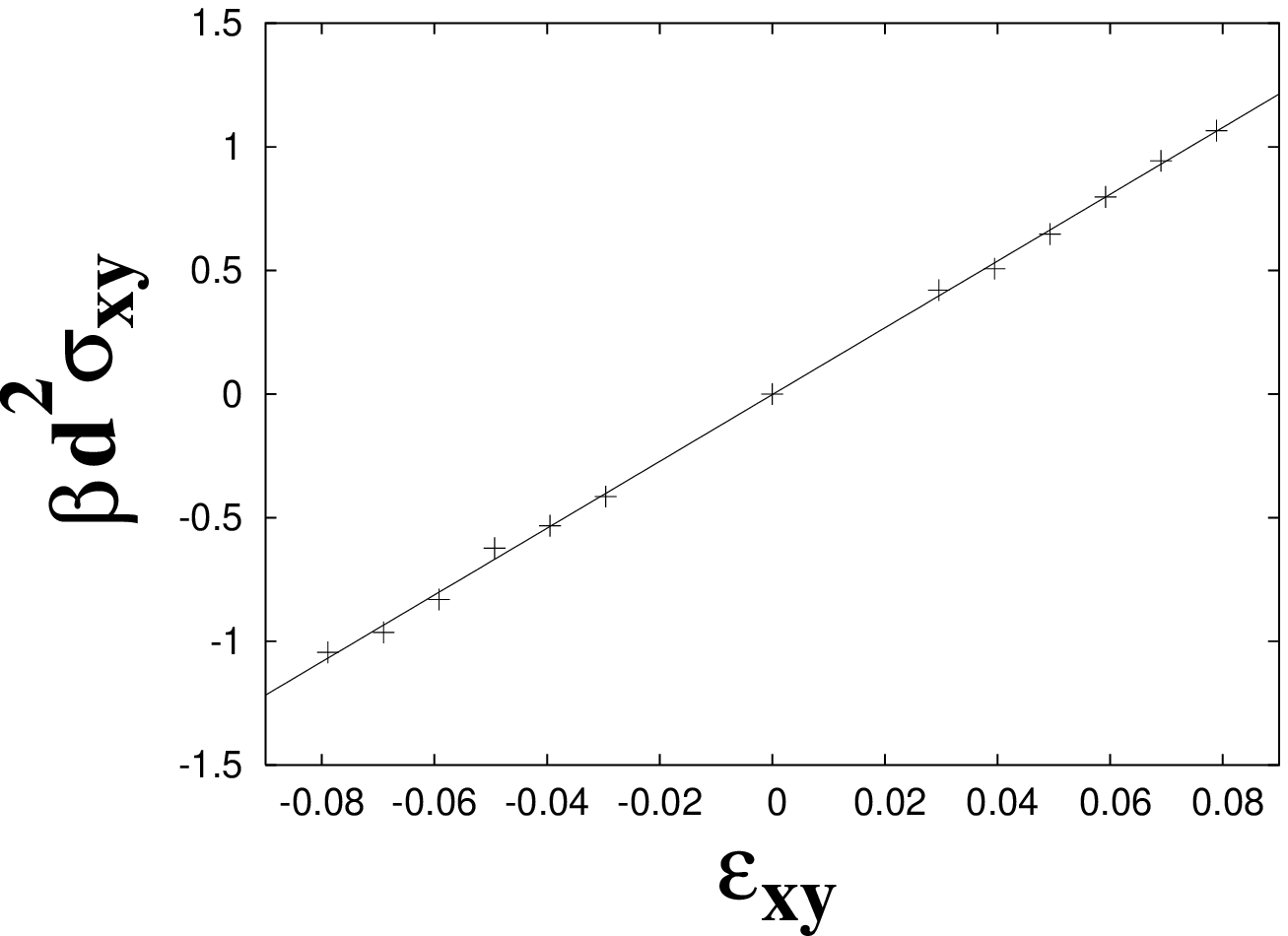}
\index{LIF}
\end{center}
\caption{
A typical stress-strain curve used to obtain shear modulus from a linear
\index{strain}
fit (solid line). The graph is plotted at $\eta = .7029,~ V_0=1.0$. The fitted shear modulus $\beta \mu {\rm d}^2=13.5$ with an error within $.9\%$.
\index{solid}
\index{shear modulus}
The error bars in stress are 
\index{stress}
less than $.5\%$ and much smaller than the point sizes plotted in this graph.
}
\label{shmod}
\end{figure}

To obtain $K_{xy}$ we
need to calculate the Young's modulus $K$ and shear modulus $\mu$.
\index{shear modulus}
In order to do that consider
Eq.\ref{sts1}, the expression for stress tensor. For hard disk 
\index{hard disk}
potentials the derivative $\partial\phi/\partial r^{ij}$ becomes a 
Dirac delta function and the expression for stress can be recast 
\index{stress}
into\cite{elast}
\begin{equation}
\beta \sigma_{\lambda\nu} {\rm d}^2 = -\frac{{\rm d}^2}{S}\left(\sum_{<i,j>}\left< \frac{r_\lambda^{ij}r_\nu^{ij}}{r^{ij}}\delta(r^{ij}-d)\right> + N\delta_{\lambda\nu}\right)
\label{sts2}
\end{equation}
The presence of the Dirac delta function $\delta (r^{ij}-d)$ in the above 
expression requires 
that the terms under the summation contribute, strictly, when two 
hard disks touch each other {\em i.e.} $r^{ij}\equiv r = \sigma$.
In practice, we implement this, by  adding the
terms under summation when each pair of hard disks come within a small 
separation $r = \sigma + \delta$. We then find 
$\beta \sigma_{\lambda\nu} {\rm d}^2$ as function 
of $\delta$ and fit the curve to a second order polynomial.
Extrapolating to the $\delta \to 0$ limit obtains the 
value of a given component of stress tensor at each strain value 
\index{strain}
\index{stress}
$\epsilon_{\lambda\nu}$\cite{elast}(Figs \ref{ydel},\ref{shdel}). 

For completeness, now we show how we calculate the two relevant elastic
moduli from the
stresses : $\sigma_{xx}$ at a given longitudinal strain $\epsilon_{xx}$
(Fig.~\ref{ydel}) and $\sigma_{xy}$ for a shear strain $\epsilon_{xy}$
\index{strain}
(Fig.~\ref{shdel}). All the data points are from our MC simulations averaged 
between $10,000$-$20,000$ MC steps. Increasing the number of configurations
does not change the values significantly.
The total errors arising from the MC simulations
\index{MC}
and the fit for a typical calculation of stresses is less than a percent.
We thus calculate the stress at each value of strain 
and from the slopes of stress-strain curves find out the bare Young's modulus 
\index{stress}
$\beta K {\rm d}^2$ (Fig.~\ref{ymod}) and shear modulus 
\index{shear modulus}
$\beta \mu {\rm d}^2$(Fig.~\ref{shmod}). 
We impose an elongational strain in $x$- direction which is parallel to the 
direction of potential minima to obtain $\beta K {\rm d}^2$.
Imposition of a shear strain in the same direction gives us
 $\beta \mu {\rm d}^2$. Any strain that forces the system to 
ride potential hills will give rise to massive displacement modes which do 
\index{displacement}
not contribute to elastic theory. Our results for the stress strain curves
for obtaining $\beta K{\rm d}^2$ and $\beta \mu{\rm d}^2$ are shown 
in Figs \ref{ymod},\ref{shmod} respectively. 
Note that the errors for the calculation of the elastic constants arise
solely from the fitting of the stress- strain curves. These can be made as
small as possible by increasing the number of strain values at which the 
\index{strain}
stresses are calculated. The values of the stress are also free from any 
\index{stress}
residual finite size effects which we checked by simulating systems of sizes
$10\times 10$ to $136\times 158$. 
From these elastic moduli we get the `bare' $\tilde K_{xy}$ 
(and hence $x'_0 = \pi \tilde K_{xy}-2$, see section~\ref{method}).
Our final estimate for the `bare' $\tilde K_{xy}$ are also correct 
to within a percent.

\begin{figure}[t]
\begin{center}
\includegraphics[width=8.6cm]{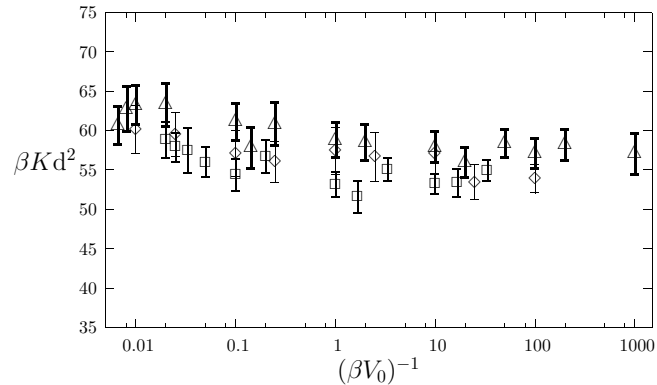}
\index{LIF}
\end{center}
\caption{
Young's modulus $\beta K$d$^2$ as a function of inverse laser potential
\index{laser}
$(\beta V_0)^{-1}$. Various symbols denote different densities --
$\diamond$ denotes $\eta = .7029$, $\triangle$ denotes $\eta=.705$ 
and $\Box$ denotes $\eta=.7$. The data for Figs \ref{K-V0} and 
\ref{mu-V0} were obtained from a separate run with a slightly higher 
error than that in Figs. \ref{ymod} and \ref{shmod}.
}
\label{K-V0}
\end{figure}

\begin{figure}[t]
\begin{center}
\includegraphics[width=8.6cm]{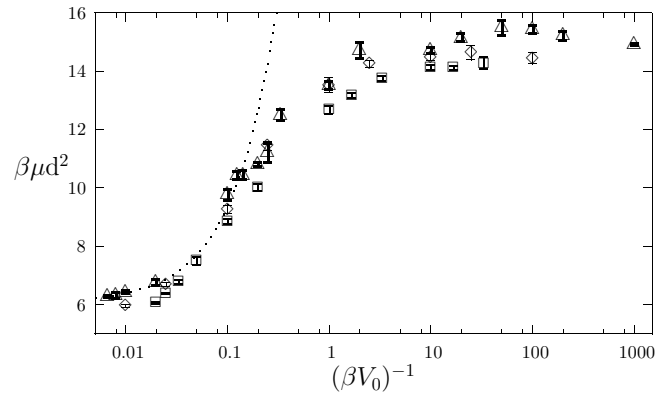}
\index{LIF}
\end{center}
\caption{
Shear modulus $\beta \mu$d$^2$ as a function of inverse laser potential
\index{shear modulus}
\index{laser}
$(\beta V_0)^{-1}$. Various symbols denote different densities --
$\diamond$ denotes $\eta = .7029$, $\triangle$ denotes $\eta=.705$ 
and $\Box$ denotes $\eta=.7$. 
The dotted line is a linear fit of the form $\beta \mu$d$^2 = 
a/\beta V_0 + b$ in the large $\beta V_0$ limit\cite{frey-lif-pre}.
\index{LIF}
}
\label{mu-V0}
\end{figure}

In Ref.\cite{frey-lif-pre} it is argued that the elastic constant 
\index{LIF}
$\beta K$d$^2$ 
remains more or less independent of amplitude of the laser potential 
\index{laser}
$\beta V_0$, while the shear modulus decreases linearly with increasing 
\index{shear modulus}
$\beta V_0$ for large $\beta V_0$. In figures \ref{K-V0} and \ref{mu-V0}
we have plotted the values of $\beta K$d$^2$ and $\beta \mu$d$^2$ 
respectively as a function of $\beta V_0$. It is apparent that the 
expectations of Ref.\cite{frey-lif-pre} are borne out by our data.
\index{LIF}
Incidentally, the behaviour of $\beta K$d$^2$ and $\beta \mu$d$^2$ with
increasing $\beta V_0$ offers an intuitive interpretation of the RLIF
\index{RLIF}
transition which we offer below.

\begin{figure}[t]
\begin{center}
\includegraphics[width=8.6cm]{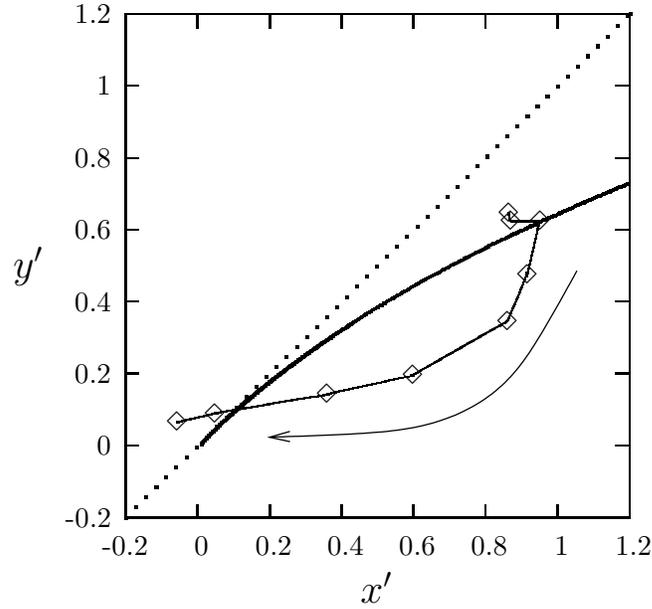}
\index{LIF}
\end{center}
\caption{
The initial values of $x'$ and $y'$ obtained from the
elastic moduli and dislocation probability
\index{dislocation probability}
\index{dislocation}
\index{probability}
\index{dislocation}
at $\eta=.7029$ plotted in $x'-y'$ plane. The line connecting the points
is a guide to eye. The arrow shows the direction of increase in
$\beta V_0$($=.01, .04, .1, .4, 1, 4, 10, 40, 100$).
The dotted line denotes the separatrix ($y'=x'$) incorporating only the
leading order term in KT flow equations. The solid curve is the separatrix
\index{solid}
when next to leading order terms are included. In
$l\to\infty$ limit any initial value below the separatrix flows to $y'=0$ line
whereas that above the separatrix flows to $y'\to\infty$. The intersection
points of the line of initial values with
the separatrix gives the phase transition points. The plot shows a
\index{phase transition}
\index{phase}
\index{separatrix}
freezing transition at $\beta V_0=.1$ followed by a melting at $\beta V_0=30$.
}
\label{flow}
\end{figure}

\begin{figure}[t]
\begin{center}
\includegraphics[width=8.6cm]{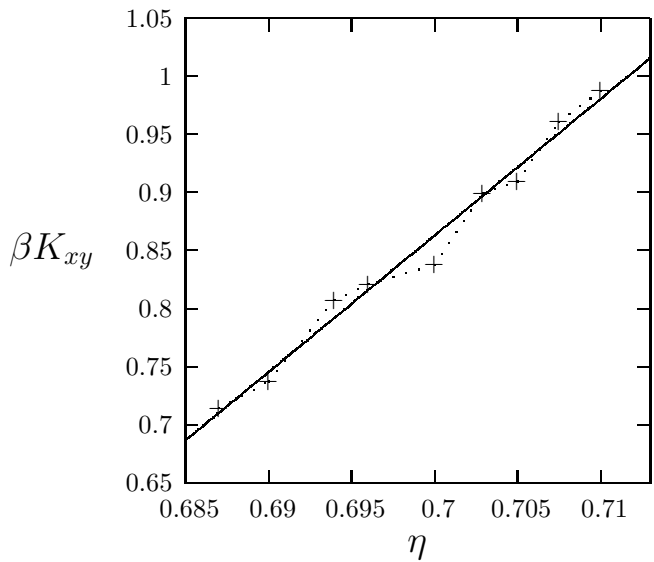}
\index{LIF}
\end{center}
\caption{For hard disk system $\tilde K_{xy}=\beta K_{xy}$ varies linearly with
\index{hard disk}
$\eta$. Data plotted at $V_0 = 1$. The solid line is a linear fit to the form $f(x)=a+b x$ with $a=-7.37$ and $b=11.76$. At each $V_0$ the error in 
\index{solid}
$\tilde K_{xy}$  determines the error in $\eta$: $\delta \eta/\eta = 
|1 + a/\eta b|  (\delta \tilde K_{xy}/\tilde K_{xy})$.
} 
\label{error1}
\end{figure}

\begin{figure}[h]
\begin{center}
\includegraphics[width=8.6cm]{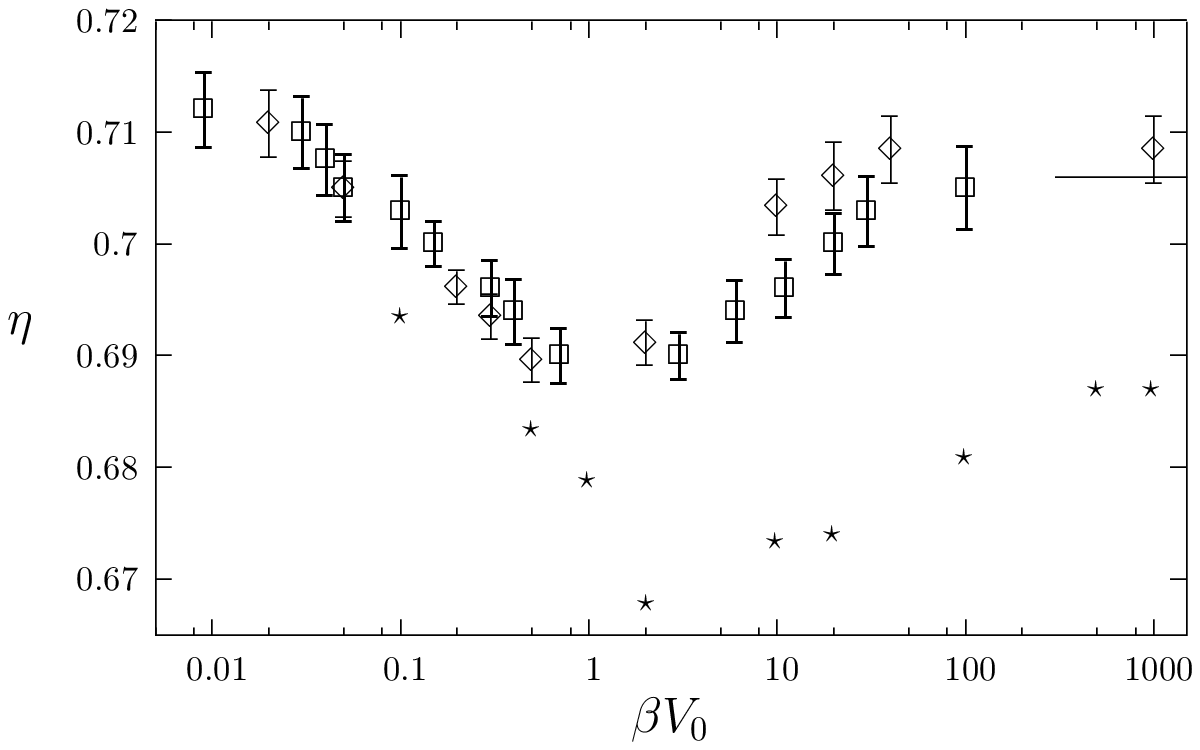}
\index{LIF}
\end{center}
\caption{The phase diagram of the hard disk system in the presence of a
\index{hard disk}
\index{phase diagram}
\index{phase}
\index{phase diagram}
1-d, commensurate, periodic potential in the packing fraction ($\eta$) -
\index{commensurate}
potential strength ($\beta V_0$) plane. The points denoted by $\Box$ 
correspond to our RG calculation using the techniques described in this chapter.
\index{RG}
The points denoted by $\Diamond$\cite{lif-hd} and $\ast$\cite{cdas3} are 
\index{LIF}
taken from earlier simulations. The vertical bars denote estimate of error. 
Our data clearly matches with Ref[7].The horizontal line at $\eta = .706$ 
denotes the calculated asymptotic phase transition point at $\beta V_0=\infty$.} 
\index{phase transition}
\index{phase}

\label{hd-phdia}
\end{figure}

In Fig.~\ref{flow} we have plotted $x'_0$ and $y'_0$ the bare values of 
$x'$ and $y'$ for various potential strengths $\beta V_0$ at
packing fraction $\eta=.7029$  
along with the separatrices for the linearized and the non-linear flow 
equations (Eq. \ref{floweq}). The line of initial conditions is seen to 
cross the non-linear separatrix twice (signifying re-entrant behaviour)  
while crossing  the  corresponding linearized separatrix only once at high 
\index{separatrix}
potential strengths. For small $\beta V_0$ the freezing transition is seen to
be driven mainly by the decrease of $y'$ (the dislocation density) since 
\index{dislocation}
\index{dislocation}
$\beta K$d$^2$ and $\beta \mu$d$^2$ are virtually constant. For large 
$\beta V_0$, the shear modulus $\beta \mu$d$^2$ vanishes and this results 
\index{shear modulus}
in the second point of intersection with the separatrix (remelting).
The phase diagram (Fig.~\ref{hd-phdia}) is obtained by computing the 
\index{phase diagram}
\index{phase}
\index{phase diagram}
points at which the line of 
initial conditions cut the non-linear separatrix using a simple 
\index{separatrix}
interpolation scheme.  
It is interesting to note that within a linear theory the KT flow 
equations {\em fail to predict a RLIF transition}. 
Performing the same calculation for different packing fractions $\eta$
we find out the whole phase diagram of RLIF in the $\eta$- $\beta V_0$
\index{RLIF}
\index{phase diagram}
\index{phase}
\index{phase diagram}
plane. 
\vskip .1cm

\begin{figure}[h]
\begin{center}
\includegraphics[width=8.6cm]{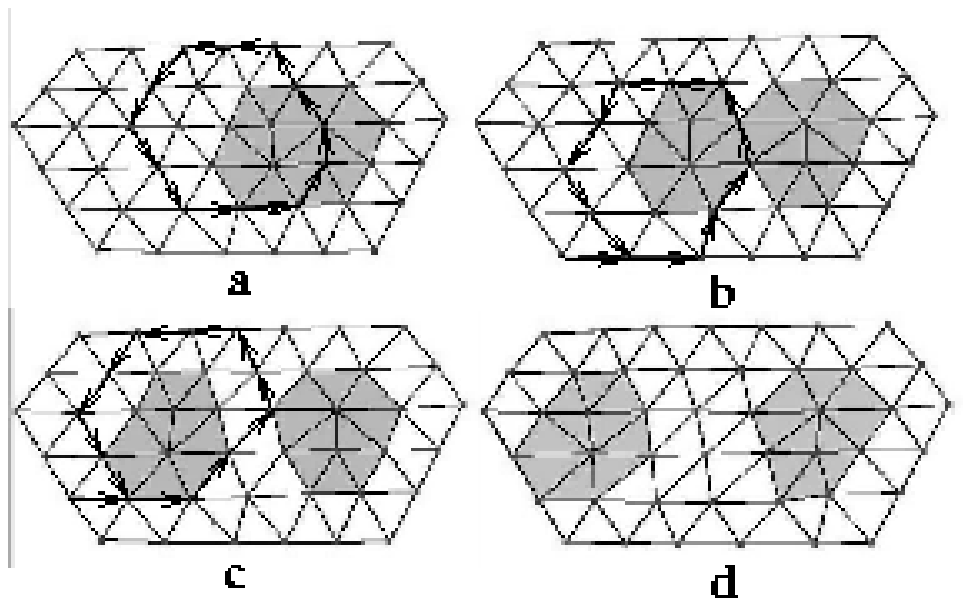}
\index{LIF}
\end{center}
\caption{The figures a -- d which we have drawn using the applet "voroglide"\cite{voro} show  four steps of separation of a type I dislocation pair, from a separation of $a_0$ to  $4a_0$. The shaded regions show the $5-7$ disclination pairs constituting the dislocations. Burger's circuits are shown in a -- c. Note that for separations $\geq 2 a_0$ separate Burger's circuits around  each disclination pair give rise to non-zero Burger's vectors, giving the dislocations their individual identity. This shows that the minimum meaningful separation 
\index{Burger's circuits}
between dislocation cores $r_{min} = 2 a_0$.}
\index{dislocation}
\index{dislocation}
\label{mv-dislo}
\end{figure}
\begin{figure}[h]
\begin{center}
\includegraphics[width=7cm]{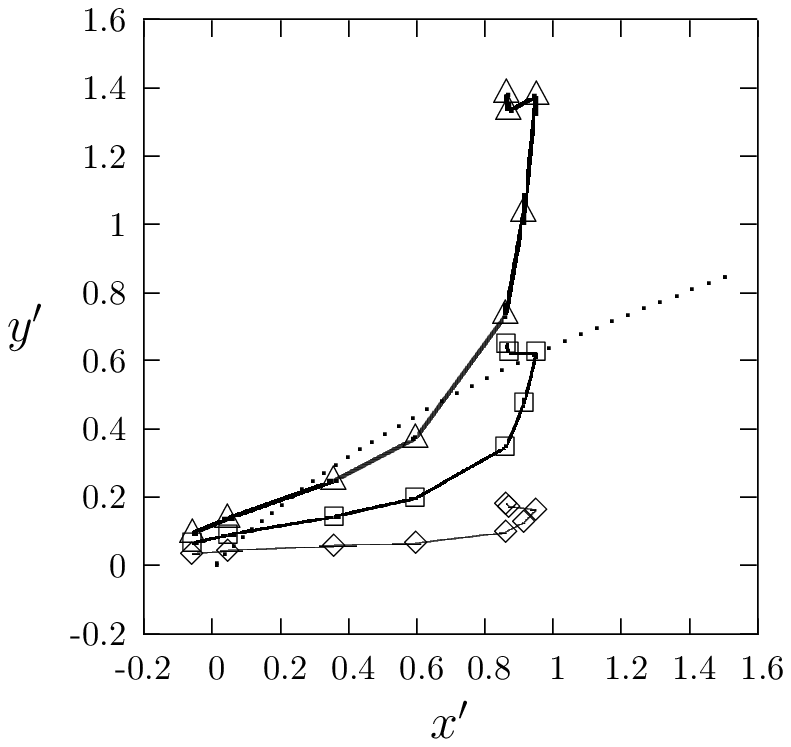}
\index{LIF}
\end{center}
\caption{Similar to Fig.\ref{flow}. The initial conditions $x'_0$ and
$y'_0$ are plotted as a function of $\beta V_0$. The different data sets
are created for different values of $r_{min}$. The symbols mean the
following :  $\Diamond$ denotes data for $r_{min} = a_0$, $\Box$ denotes that 
for $r_{min} = 2 a_0$ and $\triangle$ denote data for $r_{min} = 3 a_0$.
The dotted line denotes the non- linear separatrix. 
\index{separatrix}
}
\label{comp_rmin}
\end{figure}

Small, residual numerical errors in $x_0'$ and $y_0'$ translate into
errors in the location of the phase transition points. 
\index{phase transition}
\index{phase}
These are calculated as follows.
The quantity $\beta K_{xy}$ varies linearly with $\eta$ at all potential
strengths. Therefore the numerical error in $\eta$ is proportional to
the error in $\beta K_{xy}$ (see Fig. \ref{error1}). 
The error in $y_0'$ is neglected
\footnote{
The error in $y_0'$ has contributions from both $P_{di}$s and from $K_{xy}$
  while the former is practically zero, the contribution from the latter is
  neglected in this work.}.
\index{LIF}
The final error 
estimates are shown (as vertical error bars) in our results for the  
phase diagram of hard disks in an external potential in 
\index{phase diagram}
\index{phase}
\index{phase diagram}
Fig.\ref{hd-phdia}.

Comparing with previous computations\cite{lif-hd, cdas3} of the 
phase diagram for this system (also shown in Fig.~\ref{hd-phdia})
\index{phase diagram}
\index{phase}
\index{phase diagram}
we find that, within error- bars, our results agree at all values 
of $\eta$ and $\beta V_0$ with the results of  Strepp {\em et. al.}
\cite{lif-hd}. In numerical details, they, however, disagree with the results 
of C. Das {\em et. al.}\cite{cdas3}, though even these results show RLIF and 
\index{RLIF}
are in qualitative agreement with ours. 
This validates both our method and the quantitative predictions of 
Ref. \cite{frey-lif-prl,frey-lif-pre}. 
\index{LIF}

The effect of higher order terms in 
determining non-universal quantities has been pointed out 
earlier\cite{surajitxy} for the planar rotor model but in the present case 
their inclusion appears to be crucial.   
Nevertheless, we expect our procedure to break down 
in the $\beta V_0 \to 0$ limit where effects due to the cross-over 
from a KT to a KTHNY\cite{kthny1,kthny2} transition at $\beta V_0 = 0$ become 
\index{KTHNY}
significant. Indeed, as is evident from Fig.~\ref{dislo} for $\beta V_0 < 1$ 
the dislocation probabilities of both type I and type II 
\index{dislocation}
\index{dislocation}
dislocations are similar
\footnote
{
In analysing Fig.3 we must keep in mind that we can calculate from our
  simulations only the probability of formation of a disclination quartet.
  While we can, perhaps, safely assume that if type I dislocations are
  involved, they will seperate out with unit probability, the same can not be
  said of type II dislocations. This means that the probability of type II
\index{probability}
  dislocations could be much lower than what Fig.3 suggests.
}
\index{LIF}
and the assumptions of FNR theory and
our process (which involves only type I 
dislocations) need not be valid at small potential strengths. 
This fact is also supported by  
Ref.\cite{lif-hd} where it was shown that though at $\beta V_0 = 1000$ the 
\index{LIF}
scaling of susceptibility and order parameter cumulants gave good data 
collapse with values of critical exponents close to FNR predictions, at 
\index{critical exponents}
$\beta V_0 = .5$, on the other hand, ordinary critical scaling gave better data collapse than
the KT scaling form, perhaps due to the above mentioned crossover effects. 
\index{scaling}
In the asymptotic limit of $\beta V_0\to\infty$ the system
freezes above $\eta = .706$ which was determined from a separate simulation 
\index{simulation}
in that limit. This number is very close to the earlier 
value $\eta\sim.71$ quoted in Ref.\cite{lif-hd}. As expected, the 
\index{LIF}
freezing density in the $\beta V_0 \to \infty$ limit is lower than 
the value without the periodic potential {\em i.e.} $\eta \simeq .715$.

Before we go on to discuss other systems,
we discuss the reasons behind the particular choice of $r_{min}$ that we
made throughout this work. 
After a disclination quartet is formed, they get separated out 
and the easy direction of separation is the glide 
direction which is parallel to the Burger's vector. In Fig.\ref{mv-dislo} we 
show four steps of separation of such a dislocation pair of type I.
It is clear that, it is possible to  give individual identification to a 
dislocation only when the Burger's vector separation within a pair is
\index{dislocation}
\index{dislocation}
$\geq 2 a_0$ (Fig.\ref{mv-dislo}) {\em i.e.} $r_{min} = 2a_0$. 
For $r\geq 2 a_0$ Burger's loops can be drawn around each $5-7$ 
disclination pair (Fig.\ref{mv-dislo}) giving rise to a 
non-zero Burger's vector.  
\index{Burger's vector}
After motivating $r_{min} = 2a_0$ we show, in Fig.\ref{comp_rmin}, the 
three sets of initial values corresponding to $r_{min} = a_0,~2a_0,~3a_0 $
along with the non-linear separatrix at $\eta = .7029$ of hard disk system.
\index{hard disk}
\index{separatrix}
It is clear from the figure that $r_{min} = a_0$ 
predicts the system to be in the solid phase for any arbitrarily 
\index{solid}
\index{phase}
small amount of external potential and to melt at larger $\beta V_0$. This 
behaviour contradicts physical expectation that the melting density at 
$\beta V_0 = 0$ has to be larger than that at  $\beta V_0 = \infty$. 
On the other hand, while $r_{min} = 3a_0$ does not produce any unphysical 
prediction, it shrinks the region of re-entrance in the $\beta V_0$ direction. 
It is therefore satisfying to note that $r_{min}=2 a_0$, the minimum possible
value for the separation between members of a dislocation- antidislocation
\index{dislocation}
\index{dislocation}
pair which allows unambiguous identification also produces physically 
meaningful results for the phase diagram in closest agreement with earlier 
\index{phase diagram}
\index{phase}
\index{phase diagram}
simulation data.
\index{simulation}
\begin{figure}[t]
\begin{center}
\includegraphics[width=8.6cm]{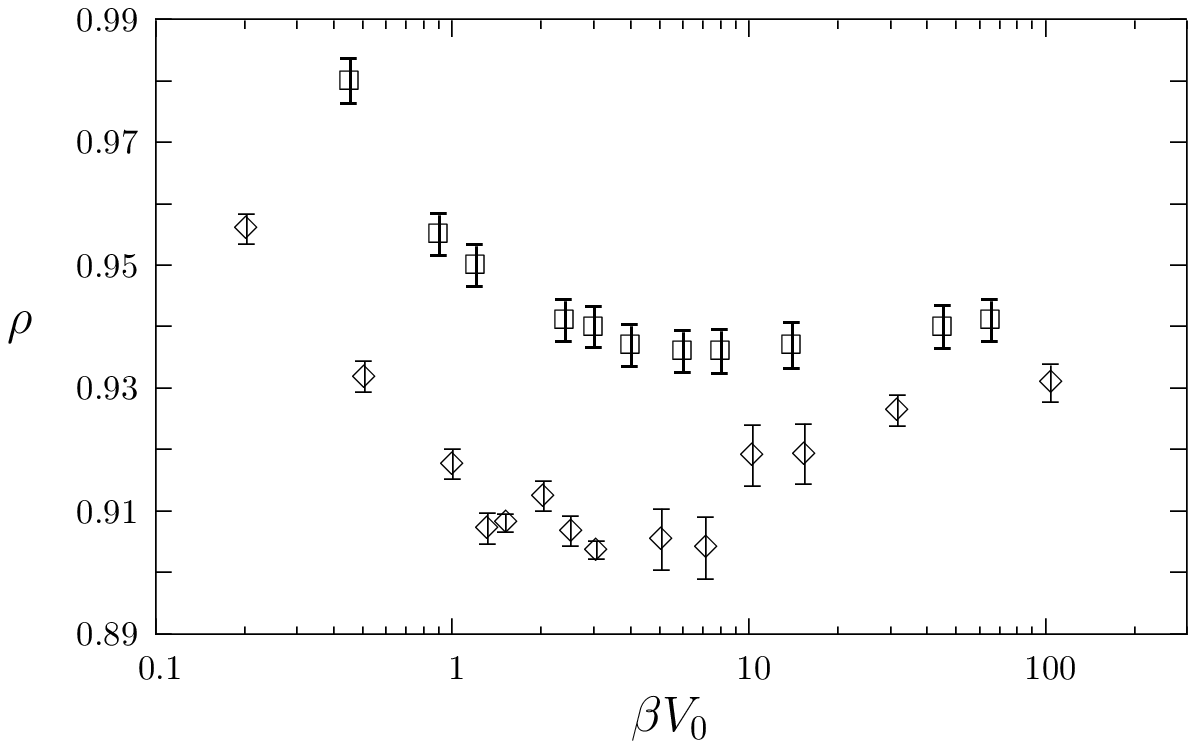}
\end{center}
\caption{ Phase diagram for soft disks:
\index{phase diagram}
\index{phase}
\index{phase diagram}
$\Box$   denote our calculation,
$\Diamond$ indicate earlier simulation data\cite{lif-sd1,lif-sd2}.
\index{LIF}
\index{simulation}
The vertical lines are the error- bars. 
}
\label{soft-phdia}
\end{figure}

It is possible to find out phase diagrams of any 2-d system in presence 
of external modulating potential commensurate with the density of the 
\index{commensurate}
system in a similar fashion. We illustrate this by calculating similar
phase diagrams for two other systems, viz. soft disks and the DLVO system.
\index{DLVO}
\index{phase}

{\it Soft disks:} Soft disks interact via the potential :
$$
\phi(r) = \frac{1}{r^{12}} 
$$
where $r$ denotes the separation between particles. In simulations,
the cutoff distance is chosen to be $r_c=2$ above which the particles are
assumed to be non- interacting. Apart from the external
potential strength $\beta V_0$ the relevant thermodynamic quantity is the
number density $\rho = N/L_xL_y$. To obtain `bare'
elastic moduli from restricted simulations the stress is calculated using
\index{stress}
Eq.\ref{sts1}. As this expression does not involve any Dirac delta 
functions (unlike hard disks), we do not require any fitting and extrapolation
to obtain the stresses and the errors are purely due to random statistical
fluctuations in our MC simulations. The elastic moduli are
\index{MC}
again found from stress- strain curves like Figs.\ref{ymod},\ref{shmod}. The
\index{strain}
\index{stress}
dislocation fugacity of type I is calculated from rejection ratio of 
dislocation generating moves. All these, at a given $\rho$ value
\index{dislocation generating moves}
\index{dislocation}
\index{dislocation}
generate the initial conditions $x'_0$ and $y'_0$ in RG flow diagrams.
\index{RG}
Again, the crossing of these initial conditions with the separatrix found
\index{separatrix}
from Eq.\ref{floweq} gives the phase transition points. The phase diagram 
\index{phase transition}
is plotted and compared with phase diagram from earlier 
\index{phase diagram}
\index{phase}
\index{phase diagram}
simulations\cite{lif-sd1,lif-sd2} in 
\index{LIF}
Fig.\ref{soft-phdia}. The error bar in $\rho$ is found from the
error in $\tilde K_{xy}$, as $\tilde K_{xy}$ varies linearly with 
$\rho$, through the 
relation $\delta \rho/\rho = |1 + a/\rho b|  
(\delta \tilde K_{xy}/\tilde K_{xy})$
where $a$ and $b$ are obtained from a linear $(a+bx)$ fit 
of the $\tilde K_{xy}$ vs. $\rho$ curve, at any given $\beta V_0$.
The phase diagram (Fig. \ref{soft-phdia}) again clearly shows 
\index{phase diagram}
\index{phase}
\index{phase diagram}
re-entrance (RLIF). This is in qualitative agreement with earlier 
\index{RLIF}
simulations\cite{lif-sd1,lif-sd2} (see Fig.\ref{soft-phdia}). 
\index{LIF}

 \begin{figure}[t]
\begin{center}
\includegraphics[width=8.6cm]{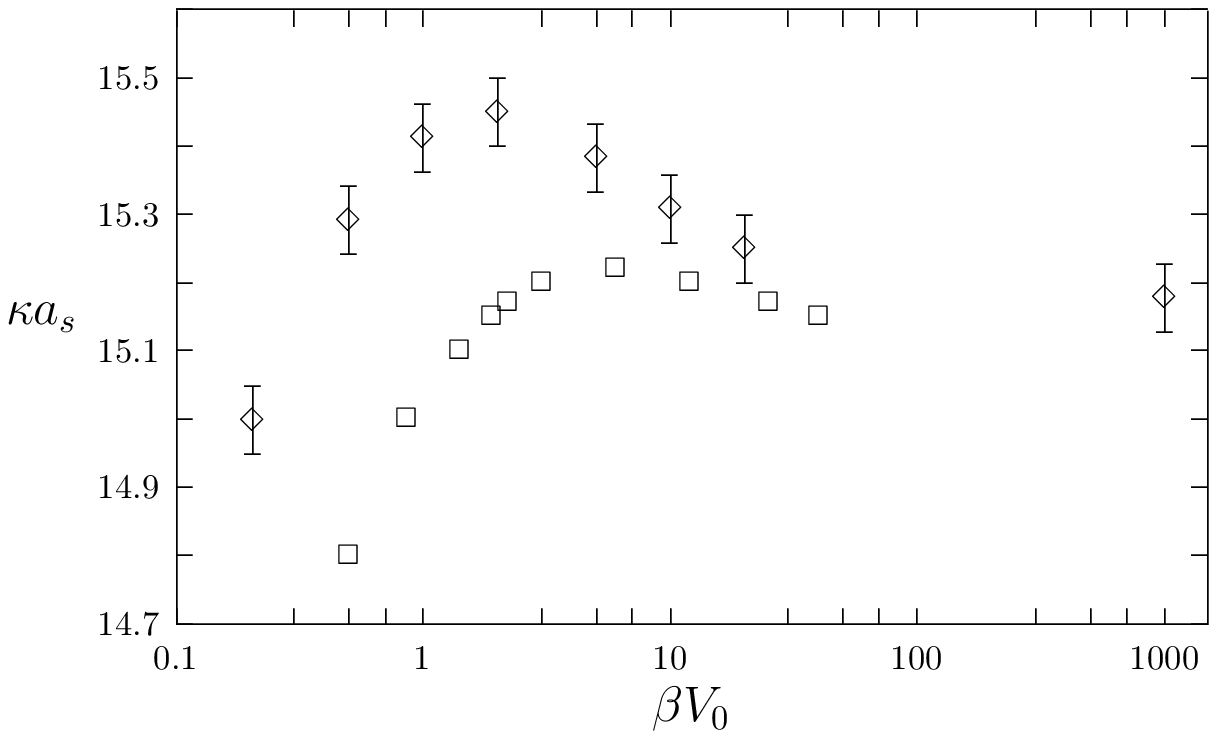}
\end{center}
\caption{ Phase diagram for particles interacting via the DLVO potential.
\index{phase diagram}
\index{phase}
\index{phase diagram}
$\Box$ denote our calculation, $\Diamond$ show the earlier simulation 
\index{simulation}
data\cite{lif-dlvo}. The vertical lines are the error- bars. Error bars
\index{LIF}
in our calculation being smaller than the symbol size are not shown.
}
\label{dlvo-phdia}
\end{figure}

{\it DLVO:}
For charge stabilized colloids the inter-particle potential that operates is
approximately given by the DLVO potential \cite{dlvo1,dlvo2}:
\index{DLVO}
$$
\phi(r) = \frac{(Z^\ast e)^2}{4\pi\epsilon_0\epsilon_r}\left( \frac{exp(.5\kappa {\rm d})}{1 + .5\kappa {\rm d}}\right)^2 \frac{exp(-\kappa r)}{r}
$$
where $r$ is the separation between two particles,
d is the diameter of the colloids, $\kappa$ is the inverse Debye 
screening length, $Z^\ast$ is the amount of effective surface charge and
$\epsilon_r$ is the dielectric constant of the water in which the colloids 
are floating. In order to remain close to experimental situations and to
be able to compare our phase diagram with the simulations of 
\index{phase diagram}
\index{phase}
\index{phase diagram}
Strepp {et. al.}\cite{lif-hd} we use $T=293.15 K$, 
\index{LIF}
d$=1.07\mu m$, $ Z^\ast = 7800$, $\epsilon_r=78$. 
In experiments, the  dimensionless inverse Debye screening length
$\kappa a_s$ can be varied either by changing $\kappa$ through the 
change in counter-ion concentration or by changing $a_s$ by varying 
density\cite{bch-fr}.  
In our restricted MC simulations we vary $\kappa$ keeping 
\index{MC}
the density fixed at $0.18 \mu m^{-2}$ by fixing the 
lattice parameter of the in ital configuration of ideal triangular lattice 
at $a_s=2.52578 \mu m$.
Further, we use a cut- off radius $r_c$ such that, $\phi(r>r_c) = 0$ where
$r_c$ is found from the condition $\beta\phi(r_c)=.001$. We find out phase 
\index{phase transition}
transition points (in $\kappa a_s$) at different external potential strengths 
$\beta V_0$ in the same fashion as described earlier.
The phase diagram in $\kappa a_s-\beta V_0$ plane is shown in 
\index{phase diagram}
\index{phase}
\index{phase diagram}
Fig. \ref{dlvo-phdia}. To obtain error bars in this case we note that
$\tilde K_{xy}$ varies linearly with $\kappa a_s$ and therefore the error in 
$\tilde K_{xy}$ is proportional to the error in $\kappa a_s$  
(Fig.\ref{dlvo-phdia}) through the relation 
\index{DLVO}
$\delta (\kappa a_s)/(\kappa a_s) = 
|1 + a/b \kappa a_s|(\delta \tilde K_{xy}/\tilde K_{xy})$.
The quantities $a$ and $b$ are found from  fitting 
$\tilde K_{xy}$ to a linear form of $\kappa a_s$, at any given $\beta V_0$.

Though there is a quantitative mismatch between our data and that of Strepp
{\em et. al.}\cite{lif-dlvo}, our data shows a clear region in 
\index{LIF}
\index{DLVO}
$\kappa a_s$ (between $15.1$ and $15.2$) where we obtain re-entrance (RLIF).
\index{RLIF}
This is in contrast to the simulated phase diagram of C. Das 
\index{phase diagram}
\index{phase}
\index{phase diagram}
{\em et. al.}\cite{cdas3}, where they observe absence of re-entrance at high
field strengths. We do not plot their data as the parameters these
authors used are not the same as the ones used in Fig.\ref{dlvo-phdia}.
\index{DLVO}
\vskip .2cm

It is interesting to note that, with increase in the range of the 
two- body interaction potentials the depths of re-entrance (in $\eta$,
$\rho$ or $\kappa a_s$) decreases. This is again in agreement with the 
understanding that, the re-entrant melting comes about due to decoupling of the
1-d trapped layers of particles that reduces the effective dimensionality 
\index{layers}
thereby increasing fluctuations. With an increase in range of the interacting 
potentials this decoupling gets more and more suppressed, thereby reducing
the region of re-entrance.

One aspect of our study which stands out is the exceptionally better
agreement of our results with previous simulations for hard disks as opposed
to systems with soft potentials like the soft disks and the DLVO. 
\index{DLVO}
This could, in principle, be due either (a) to the failure of the RG 
\index{failure}
equations used by us or some other assumptions in our 
calculations (b) or to unaccounted finite size effects in earlier simulations.
While it is difficult to estimate the effect of (a) since RG equations to 
\index{RG}
higher orders in $y$ are unknown at present, we may be able to motivate an 
estimation for (b). In order to explain the  
discrepancy in the positions of the phase boundaries, we need to go into 
some details of how the phase diagrams were
obtained in the earlier simulations. In these simulations 
\cite{lif-hd,lif-sd1,lif-sd2,lif-dlvo} the phase boundaries were obtained 
\index{LIF}
\index{DLVO}
\index{phase boundaries}
\index{phase}
from the crossing of the order parameter cumulants \cite{bincu,lan-bin} for 
various coarse graining sizes.
The system sizes simulated in these studies are the same ($N = 1024$). However,
the range of interaction differs. 
To obtain an objective measure we define the range of the potentials $\xi$  
as that at which the interaction potential $\phi$ is only 
$1 \%$ of its value at the lattice parameter. 
In units of lattice parameter, we obtain, for soft disks  $\xi = 1.47$ 
and for the DLVO potential $\xi= 1.29$ at typical screening
\index{DLVO}
of $\kappa a_s = 15$. By definition, for hard disks $\xi=1$.
The particles within the range of the potential 
are highly correlated and we calculate the number $N_{corr}$ of such 
independent bare {\em uncorrelated particles} within the full system size. 
$N_{corr}$ takes the values $N_{corr} =  1024,~473.88,~615.35$ for hard disks, 
soft disks and the DLVO potential respectively.
Since the effective system sizes are smaller for the soft potentials, finite 
size effects are expected to be larger. In this connection, it is of interest 
to note that in the same publications \cite{lif-hd,lif-sd1,lif-sd2,lif-dlvo}
\index{LIF}
a systematic finite size 
analysis showed that the phase diagrams shift towards higher (lower) 
\index{phase}
density (kappa) for hard and soft disks (DLVO).
A look at Fig.\ref{soft-phdia} and \ref{dlvo-phdia} should convince the 
\index{DLVO}
reader that such a shift would actually make the agreement with our results 
better. We emphasize here that our present restricted simulations are virtually 
free of finite size effects since the system does not undergo 
any phase transition.  
\index{phase transition}
\index{phase}

\section{Conclusion}
\label{conclusion}
We have presented a complete numerical renormalization group scheme to
\index{renormalization}
calculate phase diagrams for 2-d systems under a commensurate modulating 
\index{commensurate}
potential. We have used FNR theory along with this scheme to calculate
phase diagrams for three different systems, namely, the hard disks, the 
\index{phase}
DLVO and the soft disks. In all the cases we have found laser induced freezing
\index{DLVO}
\index{laser induced freezing}
followed by a re-entrant laser induced melting. 
\index{laser}
We show that the re-entrance behavior is built into the `bare' quantities 
themselves. We find extremely good agreement with earlier simulation results. 
\index{simulation}
In particular the phase diagram for hard disk comes out to be exactly the same 
\index{hard disk}
as found from one set of earlier simulations\cite{lif-hd}.To obtain the correct 
\index{LIF}
phase diagram, however, flow equations
\index{phase diagram}
\index{phase}
\index{phase diagram}
need to be correct at least upto next to leading order terms in the dislocation
\index{dislocation}
\index{dislocation}
fugacity. Our results, especially for small potential strengths, is 
particularly sensitive to these terms. Cross-over effects
from zero potential KTHNY melting transition are also substantial at small
\index{KTHNY}
values of the potential.

In this chapter we have studied the phenomena of RLIF, that comes about
\index{RLIF}
due to a confining potential which is constant in one direction and modulating 
in the other. In next chapter we shall study the effect of another kind of 
confinement. We shall confine a two dimensional solid in a narrow but long
channel and will find out its properties, phases, strain induced failure 
and phase diagram.

\chapter{Confined Solid: Phases and Failure}
\begin{verse}
\begin{flushright}
\it {One afternoon the boys grew enthusiastic over the flying carpet
that went swiftly by the laboratory at the window level \dots}
\hfill -- G. G. M{\'a}rquez
\end{flushright}
\end{verse}
\vskip 1cm

In the last chapter we studied phase transitions in a two dimensional solid
\index{solid}
\index{phase}
driven by a periodic confining potential which caused a dimensional crossover
from two to one dimension as the amplitude of the periodic potential was 
\index{dimension}
increased. In this chapter we study the effect of a different kind of 
external potential which forces a system of ``hard-disk'' atoms to remain 
in-between one and two dimensions. Specifically, we consider here the 
mechanical and thermodynamic properties of a narrow strip of crystalline 
solid with one dimension much longer than the other other i.e. a quasi 
one dimensional (Q1D) system. In the short dimension, the solid is confined
\index{dimension}
by hard, featureless walls.  

\noindent
We shall show in this chapter, that such a Q1D solid strip has rather anomalous
properties, which are quite different from bulk one, two or three dimensional 
systems. The Q1D solid is shown to have a non zero Young's modulus which 
\index{solid}
offers resistance to tensile deformations and approximate two dimensional 
hexagonal crystalline order. On the other hand, the shear modulus of the 
\index{shear modulus}
system is vanishingly small. Large wavelength displacement fluctuations
\index{displacement}
are seen to destabilize crystalline order beyond a certain length scale at 
low densities. At high densities these fluctuations appear to be kinetically 
suppressed. The failure properties of this quasi solid is also rather 
\index{failure}
interesting. In the constant extension (Helmholtz) ensemble, the initial 
\index{Helmholtz}
\index{ensemble}
rise in the tensile stress with tensile loading is interrupted at a limiting 
value of strain and on further extension the stress rapidly falls to zero 
\index{stress}
accompanied by a reduction in the number of solid layers parallel to the hard 
\index{solid}
\index{layers}
wall by one. However, this failure is reversible and the system completely 
recovers the initial structure once the strain is reduced. 
The critical strain for failure by this novel mechanism, for small channel
\index{strain}
widths, decreases with increase in channel width so that thinner strips 
are {\em more} resistant to failure. We have used an idealized model 
\index{failure}
solid to illustrate these phenomena. Our model solid has particles (disks)
\index{solid}
which interact among themselves only through excluded volume or ``hard''
repulsion. We have reasons to believe, however, that for the questions 
dealt with in this chapter, the detailed nature of the inter particle 
interactions are relatively irrelevant and system behaviour is largely
determined by the nature of confinement and the constraints. 
\index{constraints}
Our results may be directly verified in experiments on sterically stabilized
``hard sphere'' colloids\cite{colbook} confined in glass channels
and may also be relevant for similarly confined atomic systems interacting
with more complex potentials. We have also speculated
on applications of this reversible failure as accurate strain transducers or
strain induced electrical or thermal switching devices. In the next chapter 
\index{strain}
we shall study thermal transport across this Q1D solid, especially with 
\index{solid}
respect to the effects of reversible failure on the transport coefficients. 
\index{reversible failure}
\index{failure}
\index{transport}

\noindent 
Studies of small assemblages of molecules with one or 
more dimensions comparable to a few atomic spacings are 
significant in the context of nano-technology\cite{nanostuff-1,nanostuff-2}. 
Designing nano-sized machines requires a knowledge of the mechanical 
\index{nano}
behavior of systems up to atomic scales, where, a priori, there is no 
reason for our ideas, derived from macroscopic continuum elasticity 
theory, to be valid\cite{micrela}. 
Small systems often show entirely new behaviour if hard constraints are imposed 
\index{constraints}
leading to confinement in one or more directions. Consider, for example, 
the rich phase behaviour of quasi two-dimensional colloidal 
\index{phase}
solids~\cite{pieranski-q2d,Neser,buckled-1,buckled-2,buckled-3,fortini}
confined between two glass plates showing square, triangular and 
``buckled'' crystalline phases and a, recently observed, re-entrant 
surface melting transition\cite{remelt} of colloidal hard spheres 
not observed in the bulk\cite{al,zo,web,jaster,sura-hdmelt}.

\noindent
Recent studies on various confined Q1D
systems have shown many different
structures depending on the range of interactions and 
commensurability of the natural length scale of the system
with the length scale of confinement~\cite{peeters,doyle-1}. 
These structures play crucial role in determining
the local dynamical properties like asymmetric diffusion, 
viscosity etc. and phase behaviour~\cite{doyle-2}. 
\index{phase}
A study by G. Piacente {\em et. al.}~\cite{peeters}
on confined charged particles interacting via the screened Coulomb
potential and confined in one direction by a parabolic potential 
showed many zero temperature 
layering transitions, i.e. change in the number of layers with 
a change  in  density or range of interaction. 
This also showed regions where these transitions were 
reentrant~\cite{peeters}. Apart from the transition from one to two layers, all
\index{layers}
these layering transitions were shown to be first order~\cite{peeters}.
At high temperature this classical Wigner crystal 
melts and the melting temperature
shows oscillations as a function of density~\cite{peeters}. Such oscillations
are charecteristic of confined systems, arising out of commensurability.
Confined crystals always align one of the lattice planes along the direction
of confinement~\cite{peeters,doyle-1} and confining walls generate 
elongational asymmetry in the local density profile along the walls even 
for the slightest incommensuration. 
A study by R. Haghgooie {\em et. al.} on a system of purely repulsive
dipoles confined in Q1D hard channel showed layering transitions mediated
\index{hard channel}
via an order- disorder transition near the centre of the channel~\cite{doyle-1}.
All the structural properties showed oscillations as the channel width 
increased~\cite{doyle-1}. Wall induced layering was also observed in a 
dusty-plasma study by L.-W. Teng {\em et. al.}~\cite{teng} and  
in shell structures in circular 
confinements~\cite{peeters2,leiderer,lai,bubeck,peeters3}. A similar layering
\index{layering transition}
transition, in which the number of smectic layers in a confined liquid
\index{smectic}
\index{layers}
changes discretely as the wall-to-wall separation is increased, was
noted by P. G. deGennes~\cite{degennes} and 
J. Gao {\em et. al.}~\cite{landman}. J. Gao {\em et. al.}'s 
study also revealed the relations between the local layering structures
and dynamical quantity like diffusion constant~\cite{landman}. deGennes
and Gao {\em et. al.}'s work also calculated the wall to wall force due
to this smectic layering and its commensurability with channel width.
\index{smectic}
For long ranged interactions extreme
localization of wall particles has been observed in studies of
V. M. Bedanov {\em et. al.}~\cite{peeters2} and R. Haghgooie 
{\em et. al.}~\cite{doyle-1}. 
In an earlier experiment on confined steel balls in Q1D vibrated to
\index{experiment}
simulate the effect of temperature, layering transition, phase coexistance
\index{layering transition}
and melting was observed by P. Pieranski {\em et. al.}~\cite{pieranski-1}.
Confinement, in general, can lead to new behaviours in various systems
and processes. Schmidt and Lowen~\cite{buckled-3} studied the phase behaviour
\index{phase}
of a collection of hard spheres confined within a two dimensional slit defined
by two parallel hard plates. Plate separations such that upto two layers
\index{layers}
of solid are accomodated were considered by these authors.
\index{solid}
Recently Fortini and Dijkstra studied the same system for
separations of one to five hard sphere diameters to find a 
rich phase diagram consisting of a dazzling array of upto $26$ distinct 
\index{phase diagram}
\index{phase}
\index{phase diagram}
crystal structures\cite{fortini}.
Similar studies had been carried out long back in 1983
by Pieranski and his group~\cite{pieranski-q2d} to find out many of the 
structures identified by Fortini {\em et. al.}~\cite{fortini}. In biology
specific structures of proteins are required for their specific functioning
which keeps a cell living. Double barreled chaperonins capture protiens
into them to fold them to specific structures. Recently, an array of
genetically engineered chaperonin templates have been used by McMillan 
{\em et. al.}~\cite{mcmillan} to produce ordered nano- particle arrays.
\index{nano}

This chapter is organized as follows. In the next section, we shall introduce
the model confined solid and discuss the geometry and basic definitions of 
\index{confined solid}
\index{solid}
various structural and thermodynamic parameters. We shall then introduce
the various possible structures and phases with their basic characteristics. 
This will be followed by 
the results of computer simulations,
\index{computer}
in the constant NAT (number, area, temperature) ensemble, exploring the 
\index{ensemble}
deformation and failure properties of this system and the relation of the 
\index{failure}
various structures described in the previous section to one another.
In the fifth section we shall try to understand the various transitions 
seen in the computer simulations within simple mean field free volume and 
\index{computer}
density functional approaches. In section {\bf VI} we shall show the 
effect of fluctuations and its role in destruction of long range order in 
this system. We conclude the chapter in section {\bf VII}.

\noindent
\section{The Model System}
\label{system}
The bulk system of hard disks where particles $i$ and $j$, in two dimensions, 
interact with the potential $V_{ij} = 0$ for $|{\bf r}_{ij}| > {d}$ and 
$V_{ij} = \infty$ for $|{\bf r}_{ij}| \leq {d}$, where ${d}$ is 
the hard disk diameter and ${\bf r}_{ij} = {\bf r}_j - {\bf r}_i$ the 
\index{hard disk}
relative position vector 
of the particles, has been extensively\cite{al,zo,web,jaster,sura-hdmelt} 
studied. Apart from being easily 
accessible to theoretical treatment\cite{hansen-macdonald}, experimental systems
with nearly ``hard'' interactions viz. sterically stabilized
colloids\cite{colbook} are 
available. The hard disk free energy is entirely entropic in 
\index{hard disk}
origin and the only thermodynamically relevant variable is the number density   
$\rho = N/V$ or the packing fraction $\eta = (\pi/4) \rho {d}^2$. 
Accurate computer simulations\cite{jaster} of hard 
\index{computer}
disks show that for $\eta > \eta_f = .719$ the system exists as a triangular 
lattice which melts below $\eta_m = .706$. 
Elastic constants of bulk hard disks have been 
calculated in simulations\cite{branka, sura-hdmelt}.
The surface free energy of the hard disk system in contact with a hard wall
\index{hard disk}
has also been obtained\cite{hartmut} taking care that the 
dimensions of the system are compatible with a strain-free 
\index{strain}
triangular lattice. 
\vskip 0.2cm

\begin{figure}[t]
\begin{center}
\includegraphics[width=9.0cm]{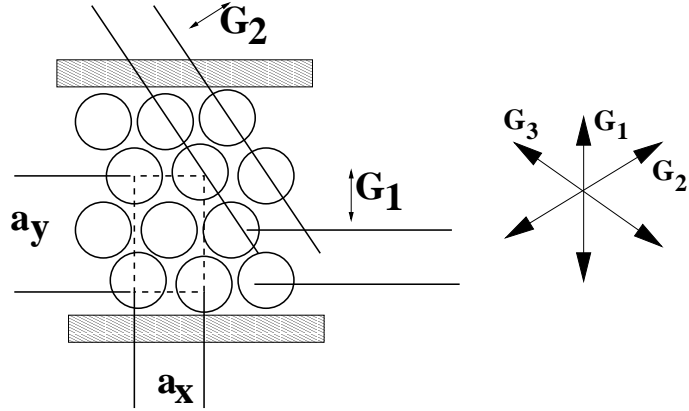}
\end{center}
\caption{The confined solid is shown along with the centered rectangular (CR) unit cell. For 
\index{confined solid}
\index{solid}
an unstrained triangular lattice $a_x = a_0$ and $a_y = \sqrt{3} a_0$. ${\bf G_1}$,  ${\bf G_2}$ and ${\bf G_3}$ denote the directions of the three reciprocal lattice vectors (RLV). The third reciprocal lattice direction ${\bf G_3}$ is equivalent to the direction ${\bf G_2}$, even in presence of the walls.
}
\label{wallpic}
\end{figure}

\noindent
Consider a narrow channel in two dimensions  of width $L_y$ defined by 
hard walls at $y = 0$ and $L_y$ ($V_{\rm wall}(y) = 0$ for 
$ {d/2} < y < L_y - {d/2}$ and $ = \infty$ otherwise) and 
length $L_x$ with $L_x \gg L_y$. Periodic boundary conditions are assumed 
\index{boundary conditions}
in the $x$ direction(Fig.\ref{wallpic}). 
\index{phase}
In order that the channel may accommodate
$n_{l}$ layers of a homogeneous, triangular lattice with lattice parameter
\index{layers}
$a_0$
of hard disks of diameter ${d}$, 
(Fig.\ref{order}) one needs, 
\begin{equation}
L_y = \frac{\sqrt{3}}{2}(n_{l} - 1) a_0 + {d} 
\label{perfect}
\end{equation}
For a system of constant number of particles and $L_y$, $a_0$ is decided
by the packing fraction $\eta$ alone. Note that $L_x=n_x a_0 = N a_0/n_l$, and $a_0$ is given by $\rho = N/L_x L_y$. This gives 
\begin{equation}
a_0 = \frac{\frac{-d}{n_l}+\sqrt{\frac{d^2}{n_l^2}+2\sqrt{3}(1-\frac{1}{n_l})\frac{1}{\rho}}}{\sqrt{3}(1-\frac{1}{n_l})}. 
\label{a0def}
\end{equation}
Defining $\chi(\eta, L_y) = 1 + 2(L_y - {d})/\sqrt{3} a_0$,
the above condition reads $\chi = {\rm integer} = n_{l}$ and violation
of Eqn.(\ref{perfect}) implies a rectangular strain away from the reference
\index{strain}
triangular lattice of $n_l$ layers. The lattice parameters of a centered 
\index{layers}
rectangular (CR) unit cell are $a_x$ and $a_y$ (Fig. \ref{wallpic}). In 
general, for a CR lattice with given  $L_y$ we have, 
$a_y = 2 (L_y - {d})/(n_l-1)$ and, ignoring vacancies,
$a_x = 2/\rho a_y$. 
\vskip 0.2cm

\begin{figure}[t]
\begin{center}
\includegraphics[width=9.0cm]{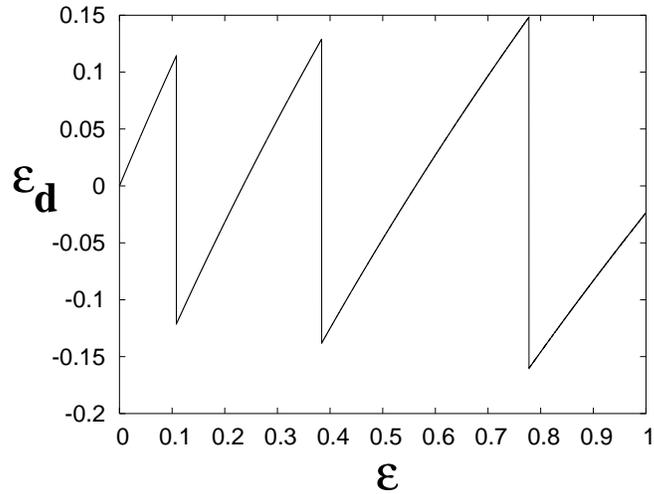}
\end{center}
\caption{ A plot of internal strain $\varepsilon_d$ as a function of external 
strain $\epsilon$. The jumps in $\varepsilon_d$ corresponds to half-integral 
\index{strain}
values of $\chi$.
}
\label{epsd}
\end{figure}

\noindent
Calculation of the deformation strain needs some care
at this stage. Using the initial triangular solid (packing fraction $\eta_0$) 
\index{solid}
as reference, the ``external'' strain associated with changing $L_x$, while 
\index{strain}
keeping $N$ and $L_y$ 
fixed, is $\epsilon =(L_x-L_x^0)/L_x^0
=(\eta_0 - \eta)/\eta $ where $\eta$ is the packing 
fraction of the deformed solid and $\eta = N \pi {d}^2 /4 L_y L_x$. 
Internally, the solid is, however, free to 
\index{solid}
adjust $n_l$ to decrease its energy (strain). Therefore, one needs to calculate
\index{strain}
strains with respect to a reference, distortion-free, triangular lattice 
at $\eta$. Using the definition  $\varepsilon_d = 
\varepsilon_{xx} - \varepsilon_{yy} = (a_x - a_0)/a_0 - (a_y - \sqrt{3}a_0)/\sqrt{3}a_0 = a_x/a_0 - a_y/\sqrt{3}a_0$ and the expressions $a_x = 2/\rho a_y$, $a_y = 2 (L_y - {d})/(n_l-1)$,  $a_0 = 2(L_y-d)/\sqrt{3}(\chi -1) $  we obtain, 
\begin{equation}
\varepsilon_d =  \frac{n_l - 1}{\chi - 1} - \frac{\chi - 1}{n_l - 1},
\label{strain}
\index{strain}
\end{equation}
where the number of layers $n_l$ is the nearest integer to $\chi$ so that 
\index{layers}
$\varepsilon_d$ has a discontinuity at half~-integral values of $\chi$. 
For large $L_y$ this discontinuity and $\varepsilon_d$ itself vanishes as 
$1/L_y$ for all $\eta$. This ``internal'' strain $\varepsilon_d$ is related 
\index{strain}
non-linearly to $\epsilon$ and may remain small even if $\epsilon$ is large
(Fig.\ref{epsd}). 
Note that any pair of variables $\eta$ and $L_y$ (or alternately 
$\epsilon$ and $\chi$) uniquely fixes the state of the system. 

\vskip .2 cm
\section{Structures and Phases}
The different possibilities of structures and phases along with their typical
structure factors are presented in this section. 

\begin{figure}[t]
\begin{center}
\includegraphics[width=6.5cm]{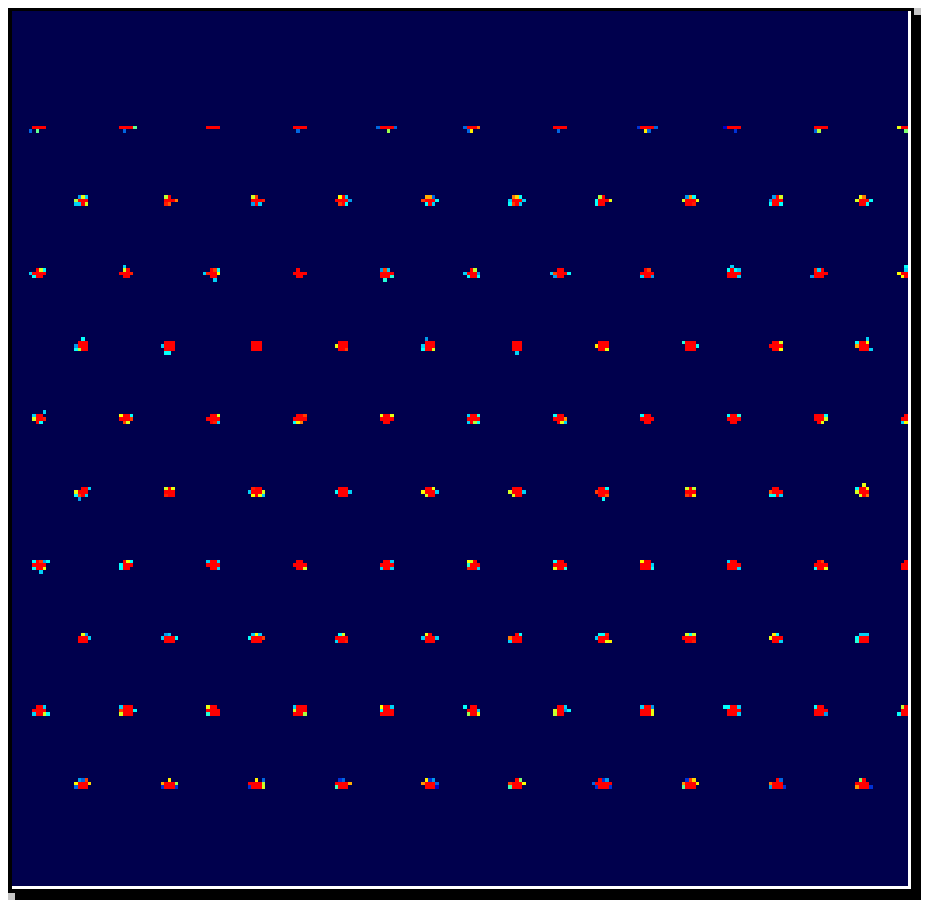}
\includegraphics[width=6.5cm]{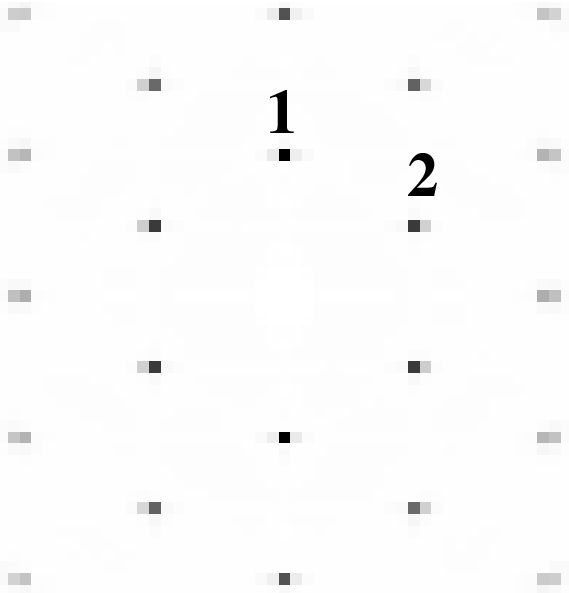}
\end{center}
\caption{Solid: Left panel shows an picture of $10^3$ overlapped configurations
of a high densitiy ($\eta=.85$) solid phase with wall to wall separation 
\index{phase}
commensurate with the density. The color code is such that red means high 
\index{commensurate}
local density and blue means low density. The right panel shows the 
corresponding structure factor which shows the typical pattern for a 
\index{structure factor}
two dimensional hexagonal solid.  
}
\label{sld}
\end{figure}

\noindent
If the separation between the hard walls is kept commensurate such 
\index{commensurate}
that $\chi=n_l$, some 
integer number of layers then the equilibrium phase is a perfect 
\index{layers}
\index{phase}
two dimensional {\em triangular solid} (Fig.\ref{sld}). 
The solid shows a diffraction pattern which is typical of a two dimensional
trangular crystal. We show later that appearances can be deceptive, however.
This triangular ``solid'' is shown to have zero shear modulus which would mean 
\index{shear modulus}
that it can flow without resistance along the length of the channel like a 
liquid. Stretching the solid strip lengthwise, on the other hand, costs energy 
\index{solid}
and is resisted. The strength of the diffraction peaks decreases rapidly with 
the order of the diffraction. In strictly two dimensions this is governed by
a nonuniversal exponent dependent on the elastic constants \cite{kthny1}. 
In Q1D this decay should be faster but larger system sizes 
and averaging over a large number of configurations would be required to 
observe this decay since contraints placed by the hard walls makes the system 
slow to equilibriate at high densities. We return to this question in section 
{\bf VI}.  

\noindent
As discussed in the earlier section, even a small incommensuration due to the 
confining walls in Q1D immediately introduces elongational asymmetry to local 
density profiles along the confining directions. As a result of this, a 
nonzero elongational stress is induced in the system. This causes the two  
\index{stress}
diffraction spots corresponding to planes parallel to the hard walls to 
strengthen at the cost of the other four spots in the smallest reciprocal 
lattice set. This increases the one dimensional character of the system even 
further.

\begin{figure}[t]
\begin{center}
\includegraphics[width=10.0cm]{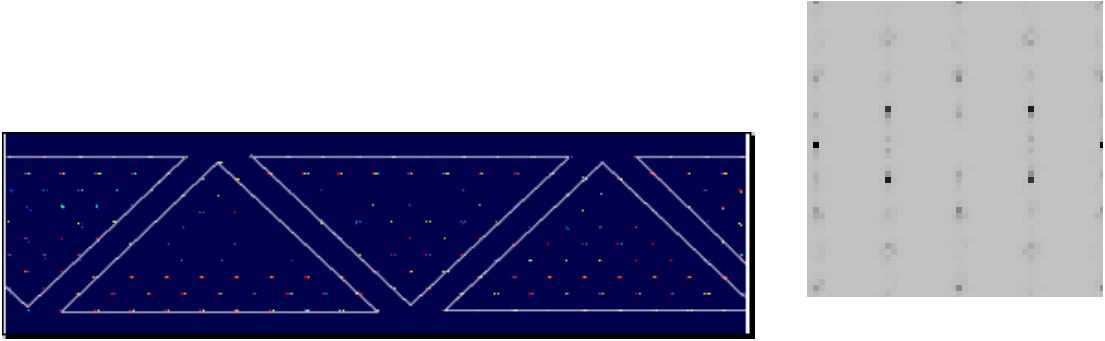}
\includegraphics[width=5.0cm]{buckl-sk-real.eps}
\end{center}
\caption{Buckled phase: A small incommensuration of wall to wall separation
\index{buckled phase}
introduced via a small increase in it from a separation that is commensurate 
\index{commensurate}
with a high density triangular solid at $\eta=0.89$  gives rise to this phase.
\index{solid}
\index{phase}
Increase in channel width reduces the density to $\eta=0.85$.
The left panel shows the picture corresponding to $10^3$ overlapped 
equilibrium configurations. The color code for the local denisties are same as 
before. 
Note different portions of triangular solid separated in $x$- direction are
\index{solid}
displaced along $y$- direction to span the extra space introduced between the
walls. Lines are drawn to identify this shift in triangular regions.
The right panel shows the corresponding structure factor which has
\index{structure factor}
the peak in ${\bf G}_1$ direction diminished. Some extra weak peaks 
corresponding to superlattice reflections appear at lower values of the 
wavenumber.
}
\label{bkld}
\end{figure}

\noindent
A little extra space introduced between the walls starting from a high density
solid phase gives rise to buckling instability in $y$- direction and the system
\index{phase}
breaks into triangular solid regions along the $x$- direction (Fig.\ref{bkld}).
Each of these regions fluctuate with respect to the other in $y$- direction 
giving the impression of a buckling wave travelling along the length of the 
solid. In conformity with the two dimensional analog \cite{buckled-1, 
buckled-2, buckled-3} we call this the {\it buckled solid} and it interpolates
continuously from $\chi = n_l$ to $ n_l\pm1$ layers. This phase can also 
\index{layers}
occur due to the introduction of a compressional strain in $x$-direction 
\index{strain}
keeping $L_y$ fixed. We do not observe the buckled solid at low densities
close to the freezing transition. Extreme incommensuration at such densities 
lead to creation of bands of the smectic phase within a solid eventually 
\index{smectic}
\index{phase}
causing the solid to melt (see next section). The diffraction pattern 
\index{solid}
shows a considerable weakening 
of the spots corresponding to planes parallel to the walls, together with an 
extra spot at smaller wavenumber corresponding to the buckled superlattice. 
The diffraction pattern is therefore almost complementary to that of the 
smectic phase to be discussed below.
\index{smectic}
\index{phase}

\begin{figure}[t]
\begin{center}
\includegraphics[width=8.0cm]{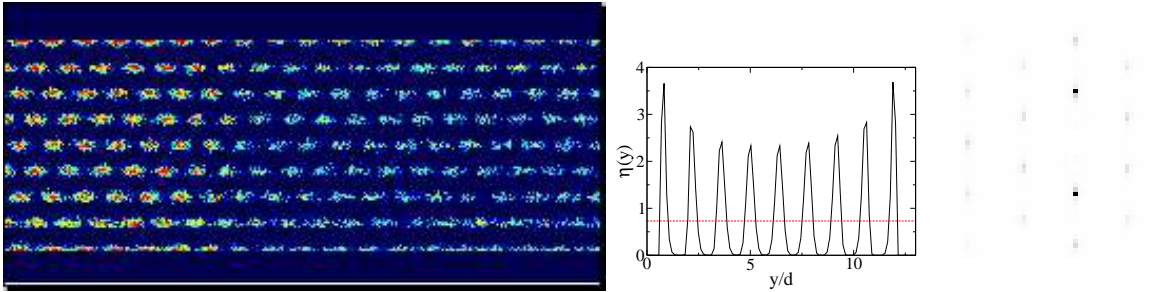}
\includegraphics[width=4.0cm]{smec-73-etay.eps}
\includegraphics[width=4.0cm,angle=90]{Sk-smectic-73.eps}
\end{center}
\caption{Smectic: 
\index{smectic}
The left panel shows the picture of $10^3$ equilibrium 
configurations. The color code for local density is same as before.
The middle panel shows the 
density modulation in $y$- direction in a typical
smectic phase which is solid like in a direction perpendicular to the walls
\index{smectic}
\index{phase}
(${\bf G}_1$ direction) and liquid like in the other direction. A fact 
coroborated by solid- like peak in ${\bf G}_1$ direction in structure factor
\index{solid}
\index{structure factor}
plotted in right hand panel. This is at a packing fraction $\eta=0.73$ obtained
the straining a triangular lattice at $\eta=0.85$ in $x$- direction.
}
\label{smec}
\end{figure}

\noindent
At low enough densities or high enough incommensuration the elongated density 
profiles in the lattice planes
parallel to the walls can overlap to give rise to a {\it smectic} phase
\index{smectic}
\index{phase}
(Fig.\ref{smec})
 in which
local denity peaks are completely smeared out in $x$- direction but are clearly
seperated in $y$-direction giving rise to a solid like order in that direction
\index{solid}
and making the system liquid like in $x$- direction. The diffraction
pattern shows two spots which is typical corresponding to the  
symmetry of a smectic phase.
\index{phase}

\begin{figure}[t]
\begin{center}
\includegraphics[width=10.0cm]{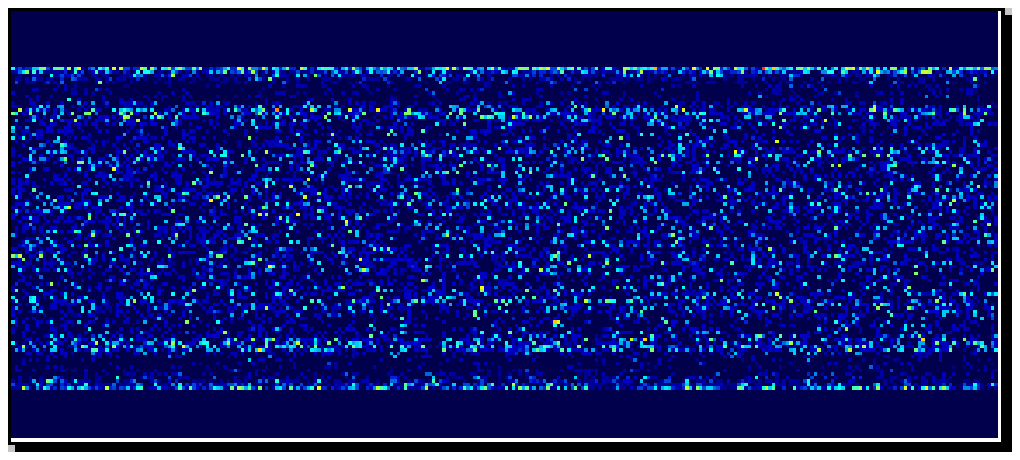}
\vskip .1cm
\includegraphics[width=6.5cm]{modliq-60-etay.eps}
\includegraphics[width=6.5cm]{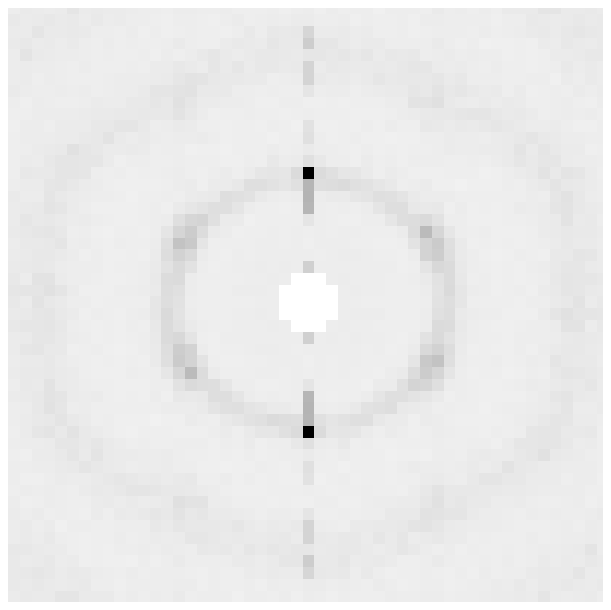}
\end{center}
\caption{Modulated liquid: 
\index{modulated liquid}
The upper panel shows the picture of $10^3$ equilibrium 
configurations. Local denisties are coded in the same color code as 
before.
The lower left hand panel shows the density 
modulation in $y$- direction which is like the smectic phase but the 
\index{phase}
modulation dies out at the centre. The structure factor in right hand 
\index{structure factor}
panel shows a ring structure which is a typical signature of liquid 
superimposed on smectic- like strong peaks in ${\bf G}_1$ direction. 
\index{smectic}
This is the signature of a modulated liquid. This
\index{modulated liquid}
behaviour is found at a packing fraction $\eta=0.6$ obtained
the straining a triangular lattice at $\eta=0.85$ in $x$- direction.
}
\label{lqd}
\end{figure}

At further lower densities the relative Lindeman parameter, a quantity which
measures the relative displacement fluctuations between neighbours (will be
\index{displacement}
defined in following section) diverges and the structure factor shows a ring- 
\index{structure factor}
like feature typical of a liquid appears together with the smectic like peaks 
\index{smectic}
in ${\bf G}_1$ direction. This is a {\it modulated liquid} (Fig.\ref{lqd}). The
\index{modulated liquid}
density modulation decays away from the walls and in channels with larger 
widths, the 
density profile in the middle of the channel becomes uniform.

\section{Mechanical Properties and Failure}
\index{failure}
\label{results}

\noindent
A bulk solid,  
\index{solid}
strained beyond it's critical limit, fails by the nucleation and growth
of cracks\cite{griffith,marder-1,marder-2,langer}.  
The interaction of dislocations  
or zones of plastic deformation\cite{langer,loefsted} with the growing
crack tip determines the failure mechanism viz. either ductile or brittle
\index{crack}
\index{ductile}
fracture. Studies of the fracture of single-walled  
carbon nanotubes\cite{SWCNT-1,SWCNT-2}(SWCNT)  
also show failure driven by bond-breaking  
which produces nano cracks which run along the tube circumference leading to
brittle fracture.  
Thin nano-wires of Ni are known\cite{nano-wire-1,nano-wire-2}, on the
\index{nano}
other hand, to show ductile failure with extensive plastic flow and
\index{failure}
\index{ductile}
amorphization. 

\noindent
In this section we shall present our results for the mechanical behaviour 
and failure of the Q1D solid under tension. As mentioned earlier, we show that 
the Q1D solid behaves anomalously, showing a reversible plastic deformation
and failure in the constant extension ensemble. The failure occurs by the 
\index{failure}
\index{ensemble}
nucleation and growth of smectic regions which occur as distinct bands 
\index{smectic}
spanning the width of the solid strip. 
\index{solid}
\begin{figure}[t]
\begin{center}
\includegraphics[width=9.0cm]{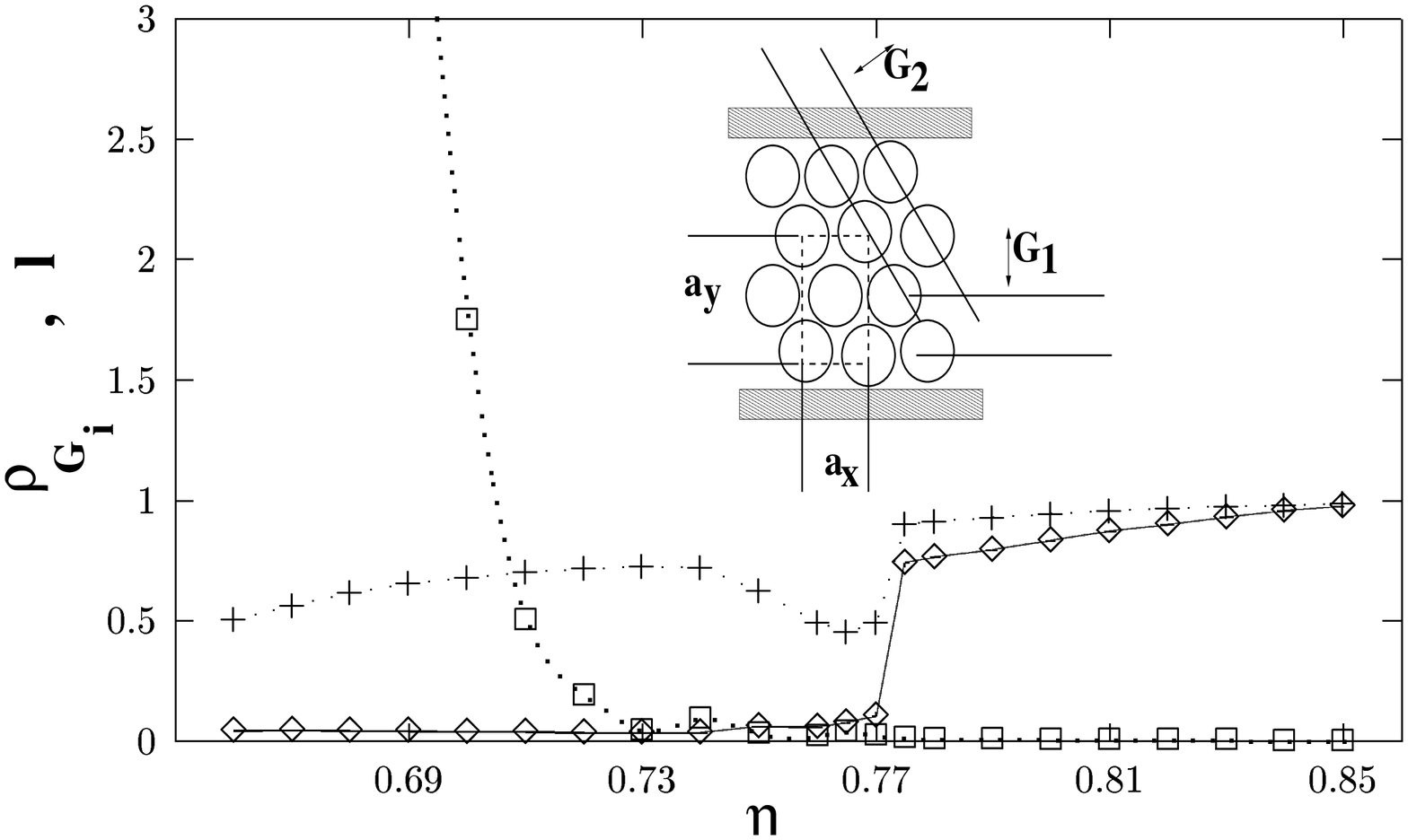}
\end{center}
\caption{Results of NVT ensemble Monte Carlo (MC) simulations of 
\index{ensemble}
$N = n_x \times n_y = 65 \times 10$ hard disks confined between two 
parallel hard walls separated by a distance
$L_y = 9\,{d}$. For each $\eta$, the 
system was equilibrated over $10^6$ MC steps (MCS) and data 
\index{MC}
averaged over a further $10^6$ MCS. 
At $\eta = 0.85$ we have a strain free triangular lattice.
\index{strain}
Plots show the structure factors $\rho_{\bf G_i}, i = 1 (+),2(\diamond)$
for RLVs ${\bf G_i}(\eta)$, averaged over 
symmetry related directions, as a function of $\eta$. 
Also plotted in the same graph is the Lindemann parameter $l(\Box)$.
The lines in the figure are a guide to the eye.
Inset shows the geometry used, 
the reciprocal lattice vectors (RLVs)$\,{\bf G_1}$ and ${\bf G_2}$ and  
the CR unit cell.} 
\label{order}
\end{figure}
\vskip .2cm
\begin{figure}[t]
\begin{center}
\includegraphics[width=9.0cm]{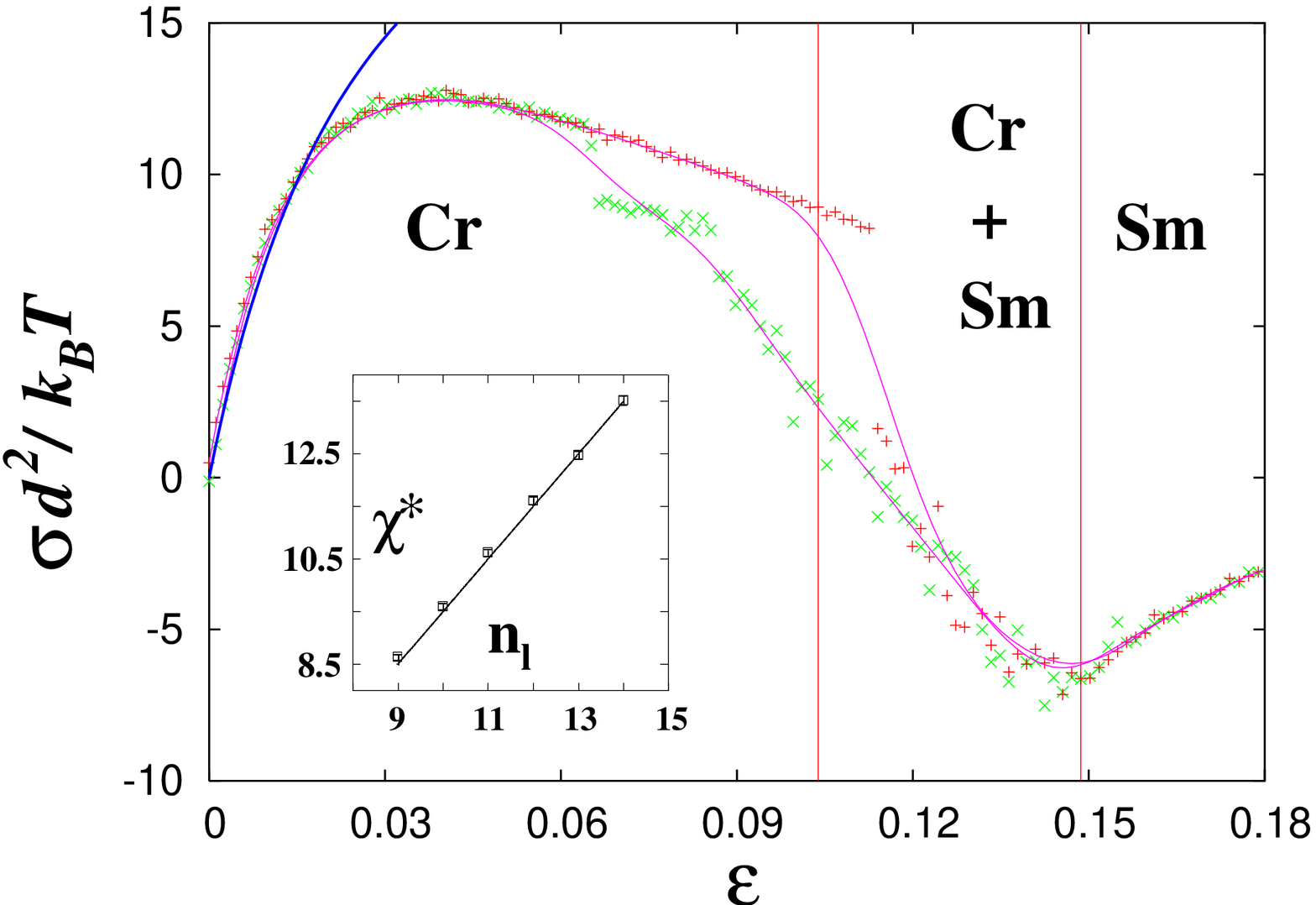}
\end{center}
\caption{ A plot of the conjugate stress $\sigma$ versus external strain 
\index{stress}
$\epsilon$ obtained from our MC simulations of $65 \times 10$
\index{MC}
hard disks initially at $\eta = 0.85$. Data 
is obtained by 
equilibrating at each strain value for $2\times10^4$ MCS and averaging 
\index{strain}
over
a further $3\times10^4$ MCS. 
The stress for the hard disk system 
\index{hard disk}
has been calculated by the 
standard method\cite{elast}. The entire 
\index{stress}
cycle of increasing 
$\epsilon (\diamond)$ and decreasing to zero $(+)$ 
using typical parameters appropriate for an atomic system,
corresponds to a real frequency of $\omega\approx 100 {\rm K\,Hz}$. 
Results do not essentially change for $\omega=10 {\rm K\,Hz}\,- 1{\rm M\,Hz}$.
Inset shows the variation
of the critical $\chi^{\ast}$ with 
\index{layers}
$n_l$, points: simulation data; line: $\chi^{\ast} = n_l-1/2$.}
\index{simulation}
\label{stress}
\index{stress}
\end{figure}
\vskip 0.2cm
We study the effects of strain on the hard disk triangular solid at fixed 
\index{hard disk}
\index{solid}
\index{strain}
$L_y$ large enough to accommodate a small number of layers $n_l \sim 9 - 25$. 
\index{layers}
We monitor the Lindemann parameter 
\bea
l = < ({u^x}_i - {u^x}_j)^2>/a_x^2 + < ({u^y}_i - {u^y}_j)^2>/a_y^2 \nn
\eea
where the angular brackets denote averages over configurations, 
$i$ and $j$ are nearest neighbors and ${u^{\alpha}}_i$ is the $\alpha$-th 
component of the displacement of particle $i$ from it's mean position. 
\index{displacement}
The parameter $l$ diverges at the melting transition \cite{Zahn}.
We also measure the structure factor 
\index{structure factor}
\bea
\rho_{\bf G} =\left| \left< \frac{1}{N^2} \sum_{j,k = 1}^N 
\exp(-i {\bf G}.{\bf r}_{jk})\right> \right|,\nn
\eea
for ${\bf G} = \pm {\bf G_1}(\eta)$, the 
reciprocal lattice vector (RLV) corresponding to the set of close-packed 
lattice planes of the CR lattice perpendicular to the 
wall, and ${\bf \pm G_2}(\eta)$ the 
four equivalent RLVs for close-packed planes at an angle 
($ = \,\pi/3$ and $2\pi/3$ in the triangular lattice) to the wall
(see Fig.\ref{wallpic}). Notice that ${\bf \pm G_2}(\eta)$ and 
${\bf \pm G_2}(\eta)$ as shown Fig.\ref{wallpic} in are equivalent
directions.
\vskip .2cm

\noindent
Throughout, $\rho_{\bf G_2} <  \rho_{\bf G_1} \neq 0$, a 
consequence of the hard wall constraint\cite{hartmut} which manifests 
as an oblate anisotropy of the local density peaks in the solid off from
\index{solid}
commensuration. 
As $\eta$ is decreased both $\rho_{\bf G_1}$ and $\rho_{\bf G_2}$ show 
a jump at $\eta = \eta_{c_1}$ 
where 
$\chi = \chi^{\ast}\approx n_l - 1/2$
(Fig. \ref{stress} (inset)). 
\index{stress}
For $\eta < \eta_{c_1}$ we get  
$\rho_{\bf G_2} = 0$ with $\rho_{\bf G_1} \not= 0$ signifying 
a transition from crystalline to smectic like  order. 
The Lindemann parameter $l$ remains zero and diverges only below 
$\eta = \eta_{c_3}(\approx \eta_m)$ indicating a finite-size-
broadened melting of the smectic to a modulated liquid phase. 
\index{smectic}
\index{modulated liquid}
\index{phase}
The stress, $\sigma = \sigma_{xx} - \sigma_{yy}$, versus strain, $\epsilon$,
\index{strain}
curve is shown in Fig. \ref{stress}. 
For $\eta = \eta_0$ ($\epsilon = 0$) the stress is 
purely hydrostatic with $\sigma_{xx} = \sigma_{yy}$ as expected. 
At this point the system is perfectly commensurate with channel
\index{commensurate}
width and the local density profiles are circularly symmetric.
Initially, the stress increases linearly, flattening out at the  
\index{stress}
onset of plastic behavior at $\eta \stackrel{<}{\sim} \eta_{c_1}$. 
At $\eta_{c_1}$, 
with the nucleation of smectic bands,
$\,\,\sigma$ decreases and eventually becomes negative. 
At $\eta_{c_2}$ the smectic phase spans the entire system and $\sigma$ is minimum.
On further decrease in $\eta$ towards $\eta_{c_3}$ 
\index{smectic}
\index{phase}
,$ \,\,\sigma$ approaches $0$ from below (Fig. \ref{stress}) thus forming a Van der Waals loop. 
If the strain is reversed by increasing $\eta$
back to $\eta_0$ the entire stress-strain curve is traced back 
\index{strain}
with no remnant stress at $\eta = \eta_0$ showing that the 
plastic region is reversible. For the system shown in Figs.\ref{order} and
\ref{stress}, we obtained $\eta_{c_1} \approx .77$,   
\index{stress}
$\eta_{c_2} \approx .74$ and $\eta_{c_3} \approx .7$.
As $L_y$ is increased,
$\eta_{c_1}$ merges with $\eta_{c_3}$ for $n_l \stackrel{>}{\sim} 25$. 
If instead, $L_x$ and $L_y$ are both rescaled to keep $\chi = n_l$ fixed or 
periodic boundary conditions are imposed in both $x$ and $y$ directions, the 
\index{boundary conditions}
transitions in the various quantities occur approximately simultaneously
as expected in the bulk system. Varying $n_x$ in the range $10 - 1000$ produces
no essential change in results.
\vskip .2 cm

\begin{figure}[t]
\begin{center}
\includegraphics[width=10.0cm]{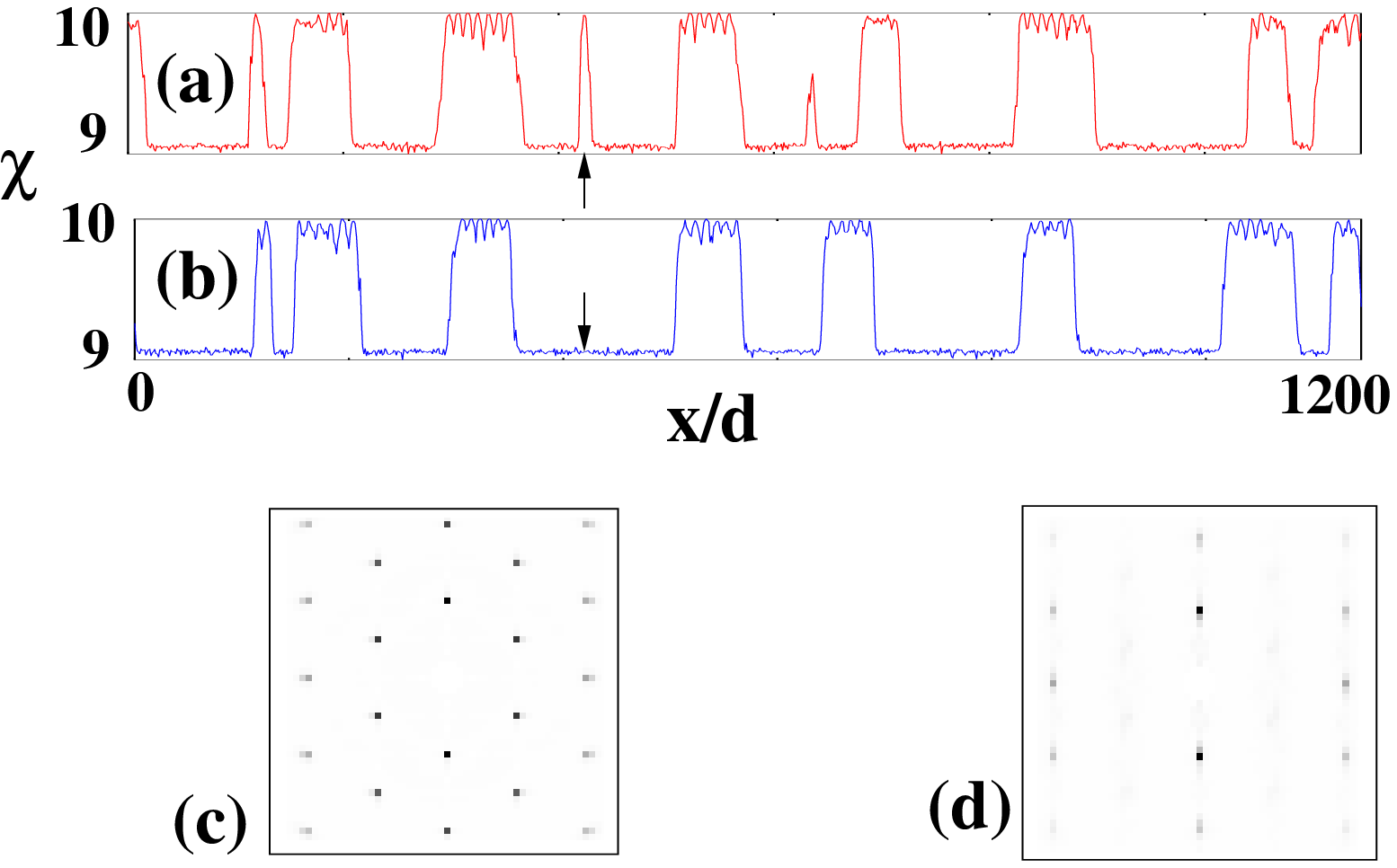}
\vskip 0.2cm
\includegraphics[width=10.0cm]{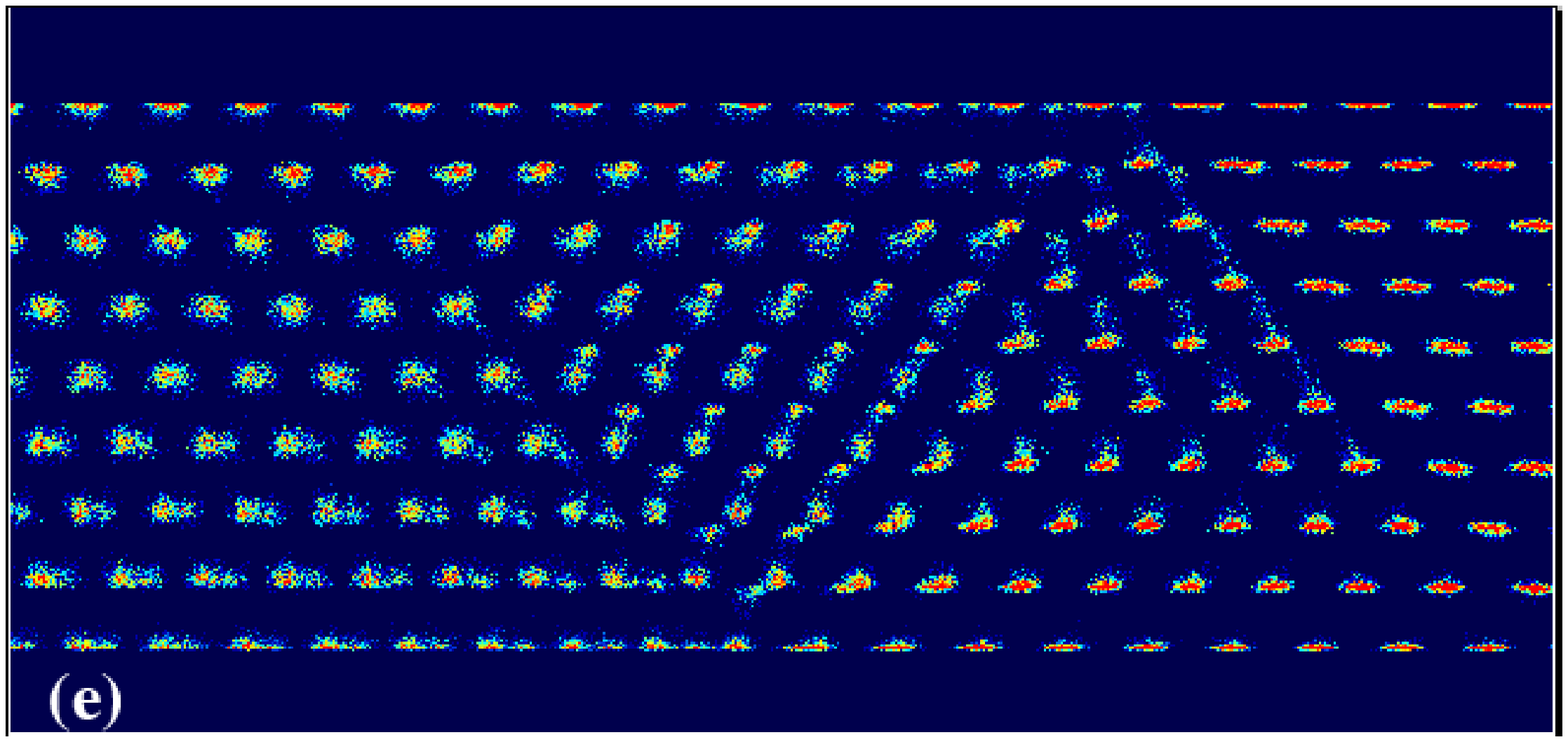}
\end{center}
\caption{
Plot of $\chi(x,t)$ as a function of the 
$x/{d}$ at $\eta = .76$ 
after time $t=$ (a)$5\times10^5$ and (b)$2\times10^6$ MCS for 
$N = 10^3\times 10$. Note that $\chi = 10$ in the solid and $ = 9$ in the 
smectic regions. Arrows show the coalescence of two bands as a function 
of time. Calculated diffraction patterns for 
the (c) solid  and (d) smectic regions.
\index{solid}
(e) Close up view of a crystal-smectic  interface from superimposed 
\index{smectic}
\index{interface}
\index{crystal-smectic interface}
positions of $10^3$ configurations at $\eta = .77$. The colors 
code the local density of points from red/dark (high) to blue/light (low). 
Note the misfit dislocation in the inter-facial region.
\index{dislocation}
\index{dislocation}
}
\label{interface}
\index{interface}
\end{figure}

\noindent
For $\eta_{c_2} < \eta < \eta_{c_1}$ we observe that the smectic  
\index{smectic}
order appears within narrow bands (Fig. \ref{interface}). 
Inside these bands the number of layers is less 
\index{layers}
by one and the system in this range of $\eta$ is in a mixed phase. 
\index{phase}
A plot (Fig.\ref{interface} (a) and (b))
\index{interface}
of $\chi(x,t)$, where we treat $\chi$ as a space and time (MCS) dependent ``order 
parameter'' 
(configuration averaged number of layers over a window in $x$ and $t$),
\index{layers}
shows bands in which $\chi$ is less by one compared to the 
crystalline regions. Once nucleated narrow bands coalesce 
to form wider bands, the dynamics of which is, however, extremely slow. 
\index{dynamics}
The total size of such
bands grow as $\eta$ is decreased. Calculated diffraction 
patterns (Fig. \ref{interface} (c) and (d)) show that, locally, within a 
\index{interface}
smectic band $\rho_{\bf G_1} \gg  \rho_{\bf G_2}$ in contrast to the solid 
\index{solid}
\index{smectic}
region where $\rho_{\bf G_1} \approx \rho_{\bf G_2} \neq 0 $. 
\vskip .2 cm

\begin{figure}[t]
\begin{center}
\includegraphics[width=9.0cm]{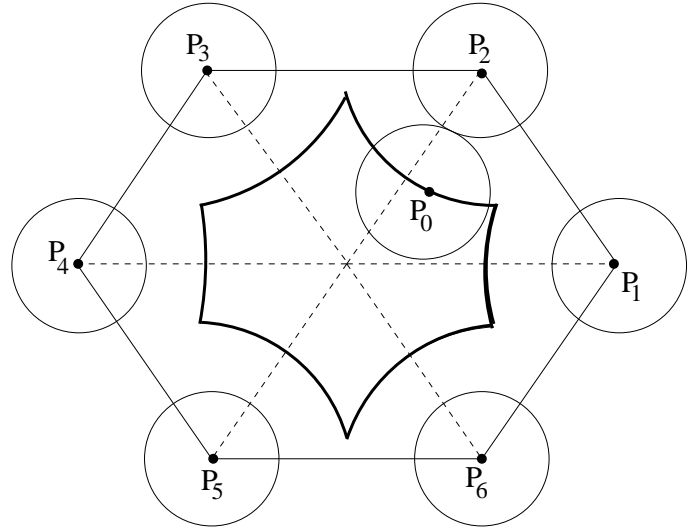}
\end{center}
\caption{In our free-volume theory we assume that the outer six disks
\index{free-volume theory}
  are fixed 
and the central disk moves within this cage of fixed particles. 
The curve in bold line shows the boundary $\cal{B}$ of the free
  volume. A point on this boundary is denoted by $P_0(x,y)$ while the
  centers of the six fixed disks are denoted by $P_i(x_i,y_i)$ with
  $i=1,2...6$.
}
\label{fvol}
\end{figure}

\noindent
The total free energy per unit volume of 
a {\em homogeneous} solid, ${\cal F}^T$, which is in contact with a hard 
\index{solid}
wall and distorted with a (small) strain $\varepsilon_d$ is given by,  
\index{strain}
\bea
{\cal F}^T(\eta,\chi) &=& -\r \ln v_f(\eta,\chi)\nn\\
&\simeq&\frac{1}{2}K^{\Delta}(\eta)\varepsilon_d^2(\chi) + {\cal F}^{\Delta}(\eta)
\label{totf}
\eea
where $K^{\Delta}(\eta)$ is an elastic constant and ${\cal F}^{\Delta}(\eta)$ 
the free energy of the (undistorted) triangular lattice in contact with a 
hard wall\cite{hartmut} at packing fraction $\eta$. The ``fixed neighbor'' free 
volume $v_f(\eta, \chi)$ may be obtained using straight forward, 
though rather tedious, geometrical considerations 
so that ${\cal F}^{\Delta}(\eta) = -k_B T \rho \log v_f(\eta,0)$ and 
$K^{\Delta}(\eta) = \partial^2 {\cal F}^{\Delta}(\eta,\varepsilon_d)/\partial 
\varepsilon_d^2 |_{\varepsilon_d = 0}$ (see Fig.\ref{stress}). $v_f$ is 
\index{stress}
expressed in units of $d^2$.
In fixed neighbour free volume theory (FNFVT), we think
\index{FNFVT}
of a single disk moving in a fixed cage formed by taking the average
positions of its six nearest neighbor disks [~see Fig.~(\ref{fvol})~].
The free volume available to this central particle is given entirely by the
lattice separations $b$ ($=AQ$) and $h$ (See Fig.\ref{fvol}). Note that,
$b=a_0(1+\ex)$ and $h=\sqrt 3 (1+\ey)/2$ where $a_0$ is lattice parameter
of a triangular lattice at any given packing fraction $\eta$ and 
$\ex = (n_l-1)/(\chi-1)$, $\ey=(\chi-1)/(n_l-1)$. As stated in Sec.\ref{system},
$\chi$ is obtained from channel width $L_y$ and packing fraction $\eta$.
$v_f$ is the area enclosed by the boundary $\cal{B}$ in Fig.\ref{fvol}.
 In calculating the free volume we have assumed the geometry
of the free volume to be close to hexagonal (Fig.\ref{fvol}). For small
strains (within $6\%$) around a triangular lattice this assumption is 
valid. However, for large strains the free volume area goes over to a
rhombic shape. At these points our theory fails. 
It is clear that 
${\cal F}^T$ has minima for all $\chi= n_l$. For half integral values 
of $\chi$ the homogeneous crystal is locally unstable. The FNFVT fails
also at these points. This same FNFVT free energy for solid is used in
\index{solid}
\index{FNFVT}
Sec.\ref{theory} to calculate the phase diagram of this confined system.
\index{phase diagram}
\index{phase diagram}

\begin{figure}[h]
\begin{center}
\includegraphics[width=9.0cm]{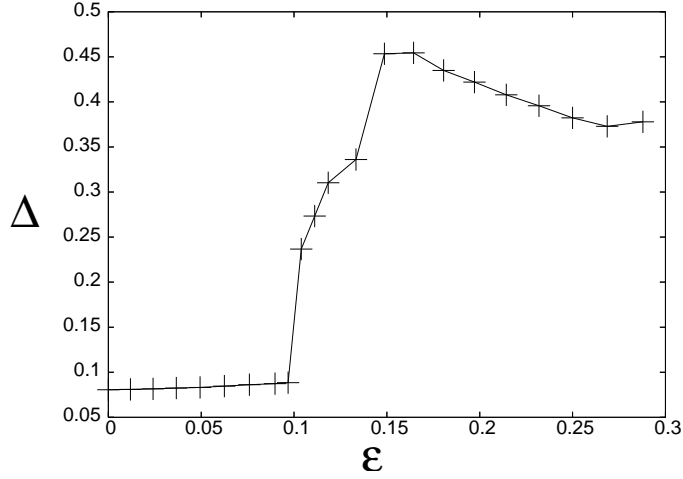}
\end{center}
\caption{Values of $\rho_{G_2}$, $\chi$ and $n_l$ have been calculated from the same simulation which is used in plotting Fig.{\ref{order}}. $C_0$ has been set to one in calculating density profile overlap $\Delta$ at each strain value $\epsilon$.
\index{strain}
\index{simulation}
}
\label{overlap}
\end{figure}

Noting that, 
\index{phase transition}
\index{phase}
$\chi^{\ast} = n_l - 1/2$ (Fig. \ref{stress} inset), it follows 
\index{stress}
from Eqn. \ref{strain}, the critical strain
\index{strain}
$\varepsilon_d^{\ast} = (4 n_l - 5)/(2 n_l - 3)(2 n_l - 2) \sim 1/n_l$ 
which is supported by our simulation data over the range $9 < n_l < 14$.
\index{simulation}
At these strains the solid generates bands 
consisting of regions with one less atomic layer. Within these 
bands adjacent local density peaks of the `atom's 
\index{solid}
overlap in the $x$ 
direction producing a smectic. Indeed, the overlap maybe 
\index{smectic}
calculated approximately using simple density functional arguments\cite{pcmp}
to be $\Delta\equiv\sqrt{<u_x^2>}/a_x = (\chi-1)/4\pi\sqrt{C_0\rho_{{\bf G}_2}}(n_l-1)$ (where $C_0$, direct correlation function for a hard disk uniform liquid, is 
\index{correlation}
\index{hard disk}
a constant of order unity) which, evidently, 
diverges as $\rho_{\bf G_2} \to 0$ (see Sec.\ref{theory}). 
Note that at $\chi = \chi^{\ast}$ the term  
$(\chi-1)/(n_l-1)$ and hence the overlap shows a jump discontinuity even before
$\rho_{\bf G_2} \to 0$. In Fig.{\ref{overlap}} this overlap term has been plotted as a function of strain $\epsilon$ imparted on a ten layered solid.
\index{solid}
\index{strain}
Fig.{\ref{overlap}} shows that the overlap $\Delta$ 
undergoes a jump increase of around $150\%$ at $\epsilon \sim 0.1$ and reaches 
its highest value at $\epsilon \sim 0.15$ as the smectic phase spans the whole 
system. At further higher strain this overlap decreases with melting of the 
smectic into fluid phase. This observation further vindicates the phase 
\index{smectic}
\index{phase}
demarcation in Fig.{\ref{stress}}.
\index{stress}
For large $L_y$, the failure strain $\varepsilon_d^{\ast}$ reduces significantly to wash out the difference in maxima and minima in free energy, therefore, the minima in ${\cal F}^{T}$ merge to 
\index{strain}
produce a smooth free energy surface independent of $\chi$ and more 
conventional modes of failure, viz. cracks, are expected to become operative. 
\index{failure}
\vskip .2 cm

\noindent
For small $L_y$ all regions of the parameter space corresponding to   
non-integral $\chi$ are also {\em globally} unstable as belied by the loop 
in the stress-strain curve (Fig.\ref{stress}). The system should 
\index{strain}
\index{stress}
therefore break up into regions with $n_l$ and $n_l - 1$ layers for 
\index{layers}
infinitesimal $\varepsilon_d$. Such fluctuations are, however, kinetically 
suppressed as we argue below. A superposition of many particle positions 
near such an interface (see Fig. \ref{interface}(e)) shows that: 
$(1)$ The width of the interface is large, spanning about $10 - 15$ atomic 
spacings and the interface is wet by a buckling phase made up of triangular
\index{phase}
groups of nine- layered solid shifted in a direction normal to the walls.
\index{solid}
$(2)$ The interface between $n_l$ layered crystal and $n_l -1$ 
layered smectic contains a {\em dislocation}
\index{dislocation}
\index{dislocation}
\footnote{
A comparison of Fig. \ref{interface} (e) with Fig. 1 of Ref.\cite{Neser} leads
  us to speculate whether a two-dimensional version of the ``buckled'' phase
\index{phase}
  may in-fact wet the solid-smectic interface thereby reducing its energy.
\index{solid}
\index{smectic}
\index{interface}
}
with Burger's 
\index{Burger's vector}
vector in the $y$- direction which makes up for the difference in the number 
of layers.  Each band of width $s$ is therefore held in place by a 
\index{layers}
dislocation-anti-dislocation pair (Fig. \ref{interface}). 
\index{dislocation}
\index{dislocation}
\index{interface}
In analogy with classical nucleation theory\cite{pcmp,cnt}, the 
free energy $F_b$ of a single band can be written as 
\begin{equation}
 F_b = -\delta F s + E_c + \frac{1}{8\pi}b^2 K^\Delta \log \frac{s}{a_0}  
\label{becker-doring}
\end{equation}
where $b = a_y/2$ is the Burger's vector,  
\index{Burger's vector}
$\delta F$ the free energy difference between the crystal 
and the smectic per unit length and $E_c$ the core energy for 
\index{smectic}
a dislocation pair. Bands form when dislocation pairs separated by 
$s > \frac{1}{8\pi}b^2 K^\Delta/\delta F$
arise due to random fluctuations. 
To produce a dislocation pair a large energy barrier of core energy $E_c$
has to be overcome. Though even for very small strains $\varepsilon_d$ the
free energy  ${\cal F}^T$ becomes unstable the random fluctuations can not 
overcome this large energy barrier within finite time scales thereby suppressing the production of $n_l-1$ layered smectic bands up to the point of $\varepsilon_d^{\ast}$. In principle, if one could wait for truely infinite times the fluctuations {\it can} produce such dislocation pairs for any non-zero $\varepsilon_d$
\index{smectic}
though the probability for such productions $\exp(-\beta E_c)$ are indeed very low.
Using a procedure similar to 
that used in Ref.\cite{sura-hdmelt}, we have monitored the dislocation 
\index{dislocation probability}
probability as a function of $\eta$ (Fig.\ref{dislo}). 
\index{probability}
\begin{figure}[t]
\begin{center}
\includegraphics[width=9cm]{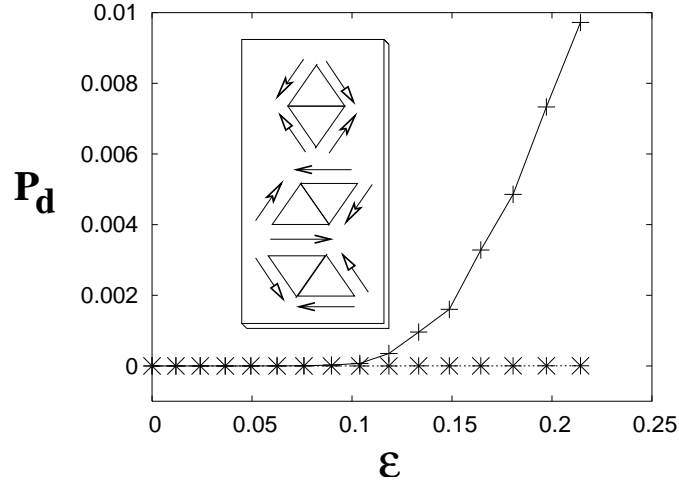}
\end{center}
\caption{Dislocation probabilities of a $65\times 10$ system are plotted as a function of strain starting from a solid commensurate with the interwall spacing $L_y$. The $+$ symbols denote dislocation probabilities for only those Burger's vectors which have components perpendicular to the walls. The corresponding bond-breaking moves are of the type depicted by the uppermost plot in the inset panel. Arrows show the directions of the bond-breaking moves. On the other hand $\ast$ symbols denote other dislocation probabilities corresponding to the other two types of moves shown in the inset.
\index{dislocation}
\index{dislocation}
\index{solid}
\index{strain}
\index{commensurate}
}
\label{dislo}
\end{figure}
Not surprisingly, the probability 
\index{probability}
of obtaining dislocation pairs with the relevant Burger's vector
\index{Burger's vector}
increases dramatically as $\eta \to \eta_{c_1}$ and artificially removing 
configurations with such dislocations suppresses the transition completely. 
Band coalescence occurs by diffusion aided dislocation ``climb'' which at  
\index{dislocation}
\index{dislocation}
high density implies slow kinetics. 
Due to this diffusive nature the size of smectic band $L(t)$ scales as
$L\sim \sqrt{t}$.
Throughout the two-phase region, the crystal is in compression and the 
\index{phase}
smectic in tension along the $y$ direction so that $\sigma$  is 
completely determined by the amount of the co-existing phases;
orientation relationships between the two phases being  
preserved throughout. Again the amount of solid or smectic in the system is entirely governed by the strain value $\epsilon$. 
This means that for a given strain the amount of solid and smectic and 
\index{smectic}
therefore the amount of stress $\sigma$ is entirely determined by the value of strain. This explains the reversible\cite{onions} plastic deformation in Fig. \ref{stress}. 
   
\index{solid}
\index{strain}
\index{stress}

\section{Mean Field Results: The Reversible Failure Transition}
\index{reversible failure}
\label{theory}
The failure of a commensurate solid under 
\index{failure}
\index{commensurate}
tensile strain imposed in the manner discussed in the previous section, 
comes about through the nucleation of smectic bands within the solid. 
\index{solid}
\index{smectic}
Monte-Carlo simulations show, at half-integral $\chi$ where the local 
internal strain $\varepsilon_d$ becomes 
\index{strain}
discontinuous, $\rho({\bf r})$ at nearest neighbour sites 
overlap along the $x$-direction, parallel to the walls, generating 
smectic bands. The stress associated with $\varepsilon_d$ vanishes at 
these points and the solid fails under tension. In this section we shall 
\index{solid}
show, using simple density functional\cite{rama,pcmp} arguments, that 
the phase transition and the consequent tensile failure (a smectic 
cannot support stress parallel
to the smectic layers) is brought about by this overlap in the local density. 
\index{smectic}
\index{layers}
Since mechanical failure in our system is a consequence of a phase transition, 
\index{failure}
\index{phase transition}
\index{phase}
it is reversible --- as the strain is reduced back to zero, the stress also 
\index{strain}
\index{stress}
vanishes and the perfect triangular lattice is recovered\cite{myfail}. 

Within density functional theory\cite{rama}, the excess grand potential of a non-uniform 
\index{density functional theory}
liquid containing a density modulation $\rho({\bf r})$ over the uniform
liquid of density $\rho_l$ is given by,

\begin{eqnarray}
\frac{\Delta \Omega}{k_B T} & = &\int d{\bf r}[\rho({\bf r})\log(\rho({\bf r})/\rho_l)-\delta\rho({\bf r})] \nonumber \\
                           &   & -\frac{1}{2}\int d{\bf r'}C(|{\bf r - r'}|)\delta\rho({\bf r})\delta\rho({\bf r'}). 
\label{rky}
\end{eqnarray}

Here $\delta\rho({\bf r}) =\rho({\bf r})-\rho_l$ and $C(r)$ is the direct 
correlation function of the uniform liquid\cite{hansen-macdonald}. A functional 
\index{correlation}
minimization of the free energy yeilds the following self-consistency 
equation for the density:

\begin{equation}
\frac{\rho({\bf r})}{\rho_l} = \exp[\int d{\bf r'}C(|{\bf r - r'}|)
\delta\rho({\bf r'})]
\end{equation}

In principle one should solve the above equation within the constraints
\index{constraints}
imposed by the walls and obtain the equilibrium $\rho({\bf r})$. Substitution
of this $\rho({\bf r})$ into Eqn.\ref{rky} gives the equilibrium free energy
and phase transitions. While we intend to carry out this procedure in the 
\index{phase}
future, we must point out that for the present problem, this is complicated
by surface terms and anisotropic, external, fields which are difficult to 
incorporate. In this chapter we shall take a much simpler route in exploring
the various conditions for the solid~-smectic transition given the nature 
\index{solid}
\index{smectic}
of the $\rho_{\bf G_i}$ (the order parameters) obtained from our simulations.  

One may expand, therefore, the logarithm of the local density 
profile $\log\rho({\bf r})$ 
in a Fourier series\cite{pcmp} around a lattice point at the origin, 
to get, 
\begin{equation}
\rho({\bf r}) = {\cal N} \exp\left(2 C_0\sum_{i=1}^3 \rho_{\bf G_i}
\cos ({\bf G_i.r} ) \right)
\label{2nd}
\end{equation}
where $C_0$ is a constant, of order unity, denoting the Fourier transform of 
the direct correlation function calculated at a q-vector corresponding to 
\index{correlation}
the smallest RLV set of the solid. We have kept contributions only from 
\index{solid}
this set. 

For a perfect triangular lattice, the RLV's are 
${\bf G_1} = \hat{y} \frac{2\pi}{d_y}$, ${\bf G_2} = \hat{x} \frac{2\pi}{d_y}\cos(\frac{\pi}{6}) + \hat{y} \frac{2\pi}{d_y}\sin(\frac{\pi}{6})$ and ${\bf G_3} = \hat{x} \frac{2\pi}{d_y}\cos(\frac{\pi}{6}) - \hat{y} \frac{2\pi}{d_y}\sin(\frac{\pi}{6})$, 
where $d_y = \frac{\sqrt{3}}{2}a_0$. Using these relations and the fact that 
in the presence of confining walls the Fourier amplitudes denoting solid 
\index{solid}
order are virtually constant upto the transition and 
$\rho_{\bf G_2} = \rho_{\bf G_3} \ne \rho_{\bf G_1} $, 
Eqn.\ref{2nd} gives~\cite{myijp},
\begin{equation}
\rho({\bf r}) = {\cal N} \exp\{C_0 (2 \rho_{\bf G_1} + 4 \rho_{\bf G_2})\}
\exp\left(-\frac{1}{2} C_0 \left(\frac{2\pi}{d_y}\right)^2\left\{ (2 \rho_{\bf G_1} + \rho_{\bf G_2})y^2 + 3 \rho_{\bf G_2} x^2 \right\}\right)
\label{3rd}
\end{equation}
Clearly the density profile is Gaussian, of the form, 
$\rho({\bf r}) \sim exp(-y^2/2\sigma_y^2 - x^2/2\sigma_x^2)$. 
Therefore, the spreads of density profile in $x$ and $y$-directions are 
given by $\sigma_x$ and $\sigma_y$ respectively, with
\begin{eqnarray}
\sigma_x^2 &=& \frac{1}{C_0} \left(\frac{d_y}{2\pi}\right)^2 
\frac{1}{3 \rho_{\bf G_2} }\\
\sigma_y^2 &=& \frac{1}{C_0} \left(\frac{d_y}{2\pi}\right)^2 
\frac{1}{2 \rho_{\bf G_1} + \rho_{\bf G_2}}.
\label{4th}
\end{eqnarray}

In the absence of walls, $\rho_{\bf G_1} = \rho_{\bf G_2}$ making 
$\sigma_x = \sigma_y$, {\em i.e.} the density profile comes out to be 
symmetric in both directions, as expected for the bulk triangular solid. 
\index{solid}
The presence of 
walls make $\sigma_y < \sigma_x$ making the density profile elliptical with 
larger spread in $x$-direction, the direction parallel to the walls. Two 
neighbouring density profiles will overlap to form a smectic if 
$\sigma_x > a_x$. This leads us to the definition of a measure of overlap 
${\cal O}_l = \left(\sigma_x/a_x \right)$.  The condition ${\cal O}_l > 1$ 
is then the Lindemann criterion for nucleation of the smectic phase. 
\index{smectic}
\index{phase}
Remembering $a_x = a_0 (n_l - 1)/(\chi -1)$, we get,
\begin{equation}
{\cal O}_l = \frac{1}{4\pi} \frac{1}{\sqrt{C_0\rho_{\bf G_2}}} \frac{\chi -1}{n_l -1}. 
\end{equation}

Whenever, $\rho_{\bf G_2}\to 0$ {\em i.e.} with the loss of solid order 
${\cal O}_l$ diverges although $\sigma_y$ remains finite, since 
$\rho_{\bf G_1} \neq 0$ in presence of the walls. This indicates a 
solid-smectic transition. However, even before $\rho_{\bf G_2}\to 0$ the 
\index{solid}
\index{smectic}
quantity 
 $\Delta = \frac{\chi -1}{n_l -1}$ and therefore ${\cal O}_l$ shows large 
jumps at those internal strain ($\varepsilon_d$) values where $\chi$ becomes
\index{strain}
 half-integer. It is interesting to note that, at these points 
$\varepsilon_d$ has discontinuities and the system fails\cite{myfail}. This 
shows that the mode of failure predicted by our theory is 
\index{failure}
through a solid-smectic transition. The fact that $\rho_{\bf G_1} $ remains 
\index{solid}
\index{smectic}
non-zero even at very small densities, due to the confinement from the walls, 
gives rise to the density modulations in the confined liquid.

We have shown therefore that the overlap in the density profiles may be used 
as an ``order parameter'' for the solid to smectic transition. We show below
\index{smectic}
that jumps in this order parameter tantamounts to mechanical failure of the 
\index{failure}
solid.
\index{solid}

\begin{figure}[t]
\begin{center}
\includegraphics[width=9.0cm]{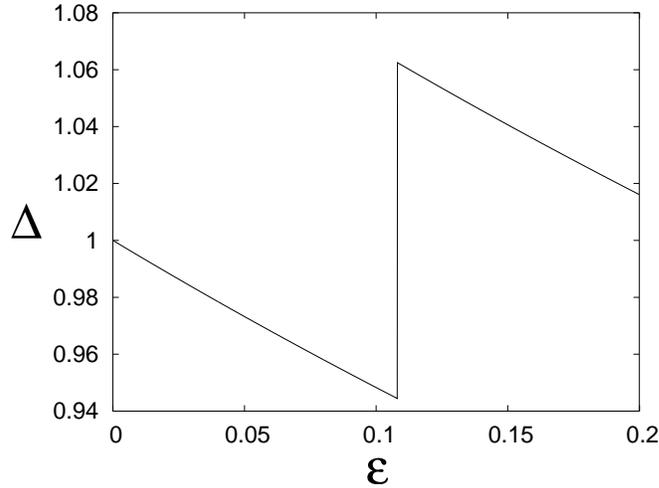}
\end{center}
\caption{For a 10-layered solid with $L_y$ commensurate with the initial strainless triangular structure at $\eta=.85$ overlap term $\Delta$ is plotted as a function of external strain $\epsilon$. Density profile overlap shows a jump increase at strains $\epsilon > .1$, the failure strain value\cite{myfail}. 
\index{solid}
\index{failure}
\index{strain}
\index{commensurate}
}
\label{epsD}
\end{figure}

We begin with studying the overlap $\Delta$ as a function of external 
strain $\epsilon$. For specificity, we start from a triangular solid of 
packing fraction $\eta = .85$ with $L_y$ commensurate with a $n_l = 10$ 
\index{commensurate}
layered solid. With increasing strain initially the overlap $\Delta$ reduces 
\index{solid}
due to increased separation ($a_x$) between neighbouring lattice points. 
But above a strain ($\epsilon$) of about $10\%$, $\chi$ reaches the 
\index{strain}
half-integral mark and $\Delta$ shows a discontinuous increase, indicating 
large 
overlap between neighbouring density profiles along the wall ; indicating a 
solid to smectic transition (Fig.\ref{epsD}) at
\index{solid}
the  failure strain $\epsilon^\ast$. With further 
\index{failure}
increase in strain the overlap reduces, again due to increased separation 
\index{strain}
between neighbouring lattice points. At higher strains the smectic melts 
\index{smectic}
into a modulated liquid due to increased fluctuations connected with the 
\index{modulated liquid}
reduced density\cite{myfail}.

\begin{figure}[t]
\begin{center}
\includegraphics[width=9cm]{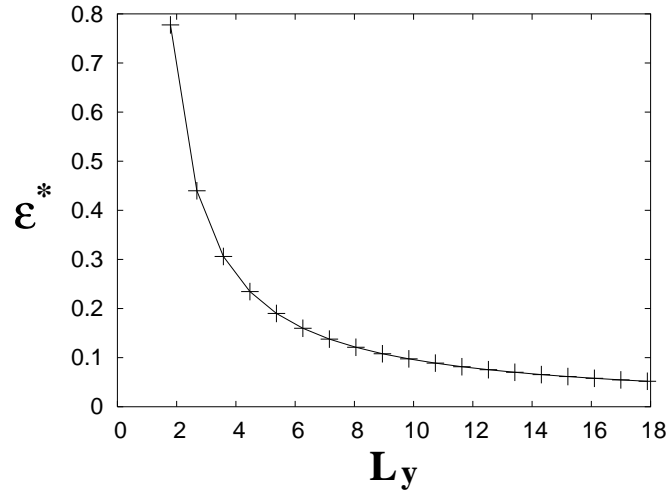}
\end{center}
\caption{Failure strains $\epsilon^\ast$ for various interwall 
separations $L_y$ confining $n_l=2 \to 20$ layered triangular strips
 at $\eta = .85$ is plotted. Failure strain decreases with increase 
\index{failure}
\index{strain}
in $L_y$.
}
\label{fails}
\end{figure}
\begin{figure}[h]
\begin{center}
\includegraphics[width=9cm]{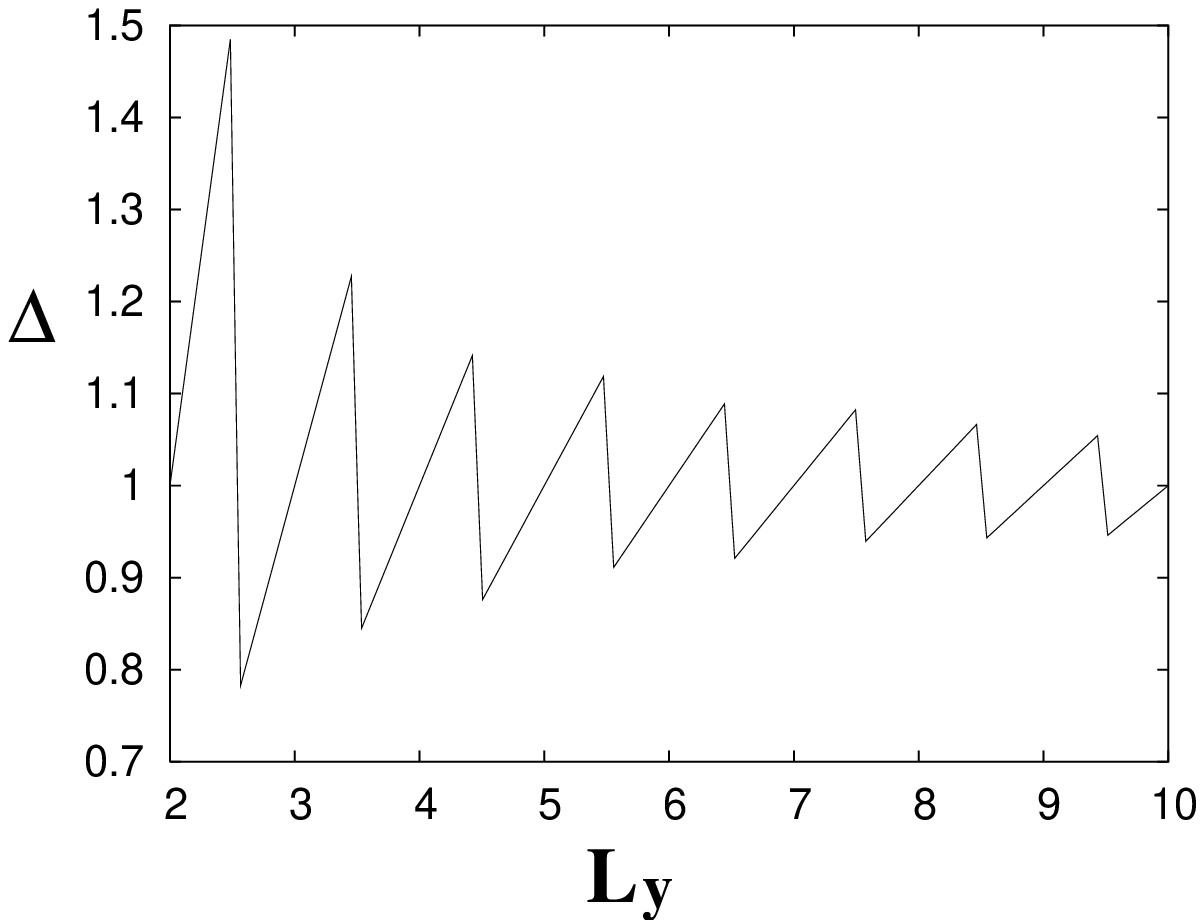}
\end{center}
\caption{Overlap $\Delta$ is plotted for a system at $\eta = .85$ with changing interwall separation $L_y$. Amount of smectic overlaps $\Delta$ at failures reduces with increasing $L_y$.
\index{smectic}
}
\label{ly}
\end{figure}

We have performed this calculation for various $L_y$ values commensurate 
\index{commensurate}
with starting
triangular solids of $n_l = 2, 3 \dots 20$ layers at packing fraction 
\index{layers}
$\eta=.85$. We found out the failure strains $\epsilon^\ast$ at each $L_y$ 
and plotted them in Fig.\ref{fails} as a function of $L_y$. This clearly 
shows that the failure strain reduces with increase in $L_y$. This 
\index{failure}
\index{strain}
demonstrates the fact, derived earlier from Monte-Carlo 
simulations\cite{myfail}, that thinner (smaller $L_y$) strips are 
stronger!
 
In Fig.\ref{ly} we have plotted the overlap term $\Delta$ with increasing 
interwall separation $L_y$ at $\eta = .85$. The jumps, as usual, indicate 
failure strains corresponding to discontinuities in the internal 
strain $\varepsilon_d$
\index{strain}
at half-integral values of $\chi$. The plot shows that the amount of 
overlap at the failure strains $\epsilon^\ast$ reduces with increasing  
$L_y$ indicating that at large interwall separations the system starts 
to behave as a bulk solid and more conventional modes of failure {\em viz.} 
\index{solid}
\index{failure}
through formation and interaction of cracks and twin boundaries starts 
becoming active.

\noindent
\section{Mean Field Results: The Equilibrium Phase Diagram}
\label{theory}
We now calculate the equilibrium phase diagram of the confined two dimensional
system in $\eta - L_y$ plane. For this purpose we utilise the usual common 
tangent construction method to extract the phase diagram in the canonical 
\index{phase diagram}
\index{phase}
\index{phase diagram}
ensemble from the free energies
\index{ensemble}
of the solid and fluid,  the only two unambiguous stable phases of hard disk 
\index{hard disk}
\index{solid}
system in thermodynamic limit. Our computation will, evidently, ignore the 
\index{thermodynamic limit}
other possible phases like smectic and modulating liquid. Physically, 
\index{smectic}
confinement
always induces density modulations and therefore in phase diagram we understand
\index{phase diagram}
\index{phase}
\index{phase diagram}
that the liquid is always a modulated liquid for small channel widths.
\index{modulated liquid}
The phenomenological equation of state due to Santos {\em et. al.}\cite{santos} agrees with bulk fluid simulations of hard disks. The corresponding 
excess free energy per particle is 
\begin{equation}
{\cal F}_{San} = \frac{(2 \eta_c -1)\log\left(1-(2 \eta_c -1)\frac{\eta}{\eta_c}\right)
             -\log(1-\frac{\eta}{\eta_c})}{2(1-\eta_c)},
\end{equation}
$\eta_c$ being the packing fraction of hard disk solid at close-packed limit.
\index{solid}
The total free energy of the fluid will have, moreover, a contribution of free 
particle free energy per particle $(\log\rho -1)$. 
Then the fluid free energy per unit volume becomes ${\cal F}_{Fl} = \rho({\cal F}_{San} + \log\rho -1)$. To incorporate the
effect of structureless walls on the fluid we have incorporated the surface tension of a hard disk fluid obtained from scaled particle theory (SPT) \cite{hartmut},
\index{hard disk}
\begin{equation}
\gamma_{_{Fl}} = -\frac{\eta}{2 (1-\eta)^2} \rho { d} + \frac{1}{2}\beta P {d}
\end{equation}
where the pressure is given by \cite{santos} 
\index{pressure}
\begin{equation}
\frac{\beta P}{\rho} = \left( 1 -2 \eta + \left(\frac{\eta}{\eta_c}\right)^2(2 \eta_c -1)
\right)^{-1}
\end{equation}
Then the liquid free energy per unit volume becomes ${\cal F}_{Fl} = \rho({\cal F}_{San} + \log\rho -1) + \gamma_{_{Fl}}/L_y$.

\begin{figure}[t]
\begin{center}
\includegraphics[width=9cm]{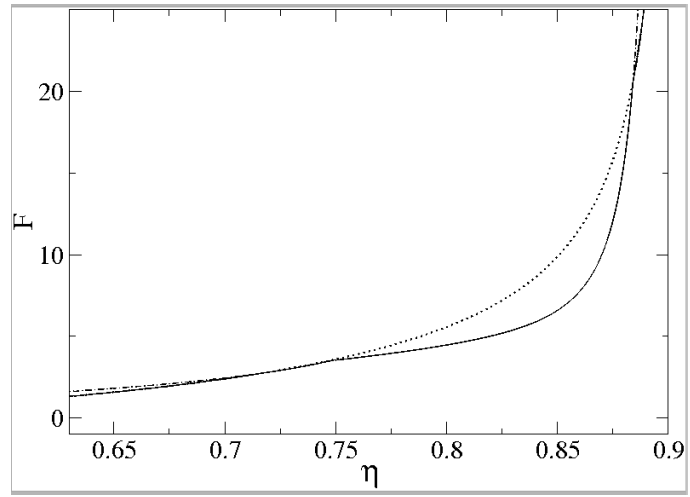}
\end{center}
\caption{Plot of free energy used for constructing the phase diagram. The dotted curve, dashed curve and solid curve denote ${\cal F}_{Fl}$, ${\cal F}_{Cr}$ and $F$ respectively.  
\index{solid}
\index{phase diagram}
\index{phase}
\index{phase diagram}
} 
\label{energy}
\end{figure}

We calculate the solid free energy of hard disk system from free volume theory.
\index{hard disk}
The surface tension of solid in presence of walls is given by\cite{hartmut}
 \begin{equation}
\gamma_{_{Cr}} = \frac{1}{\sqrt 3} \frac{d}{a(a-{d})} \left(1 - \frac{\sqrt 3}{2}\right).
\end{equation}
Hence the solid free energy per unit volume is ${\cal F}_{Cr} = {\cal F}_T(\eta,\chi) + \gamma_{_{Cr}}/L_y$. ${\cal F}_T(\eta,\chi)$ is the FNFVT free energy
\index{solid}
\index{FNFVT}
as described in Sec.\ref{results}.
For every wall to wall separation $L_y$ we vary the
packing fraction $\eta$ and pick up the smaller free energy at each $\eta$ to obtain $F =$ min(${\cal F}_{Cr}$ , ${\cal F}_{Fl}$) in Fig.{\ref{energy}}.

\begin{figure}[t]
\begin{center}
\includegraphics[width=5cm]{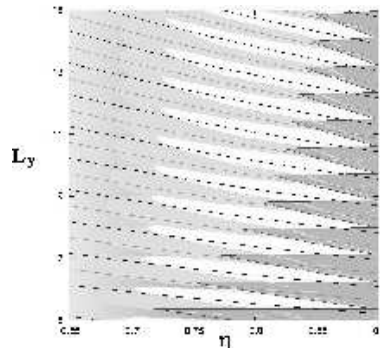}
\end{center}
\caption{Phase diagram. White region denotes solid phase. The darkest region at high densities denote regions of phase space inaccessible to the system due to overlapping hard disks. The darker (aqua green) region at low densities denote fluid phase. All other regions are two phase coexistence regions. The bold (blue) dashed lines denote $\chi = n_l$ and the faint (green) dashed lines denote half integral $\chi$.} 
\index{solid}
\index{phase diagram}
\index{phase}
\index{phase diagram}
\label{phdia}
\end{figure}

Then we use common tangent construction over this free energy $F$ to obtain 
densities of coexisting phases at equal chemical potential. Within this theory,
\index{chemical potential}
 at densities lower than the coexisting fluid density the thermodynamic phase 
is stable fluid and at densities higher than the coexisting solid density it 
is stable solid. At intermediate densities fluid and solid coexist. 
The corresponding phase diagram is given in Fig.{\ref{phdia}}.  

\begin{figure}[h]
\begin{center}
\includegraphics[width=9cm]{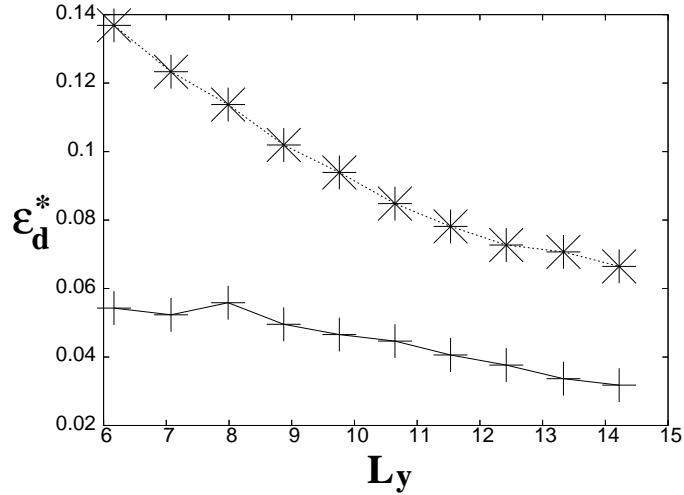}
\end{center}
\caption{The data points denoted by $\ast$ show the deviatoric strain values at which $\chi = n_l - 1/2$ whereas the symbols $+$ denote the arrival of two phase coexistance, {\em i.e.} failure predicted by this theory. Lines are guide to eye. Data have been extracted from Fig.\ref{phdia}.} 
\index{failure}
\index{strain}
\index{phase}
\index{deviatoric strain}
\label{compare}
\end{figure}

From the phase 
\index{phase diagram}
\index{phase diagram}
diagram it is evident that solid phase at lower density gets more and more 
\index{solid}
instable as we increase the wall to wall separation. If we start with solids 
commensurate with $L_y$, {\em i.e.} $\chi = n_l$ at packing fraction 
\index{commensurate}
$\eta = .85$ and start reducing $\eta$ keeping $L_y$ fixed, we can find out 
from the phase diagram the deviatoric strains at which one hits the two-phase 
\index{phase diagram}
\index{phase}
\index{phase diagram}
boundary as well as the strains where one reaches at half-integral values of 
$\chi$. These have been plotted in Fig.{\ref{compare}}.

\noindent
Both the quantities show a decrease in failure strain $\varepsilon_d^\ast$ with
\index{strain}
increase in width of the strip $L_y$ supporting the simulation result, 
\index{simulation}
narrower strips are stronger. Our simple theory therefore predicts a 
first order solid-fluid transition as
a function of $\varepsilon_d$ and failure of the solid as the density 
\index{failure}
enters the region of solid-fluid 
\index{solid}
coexistence at larger critical strain $\varepsilon_d^\ast$ for smaller wall to 
\index{strain}
wall separations. However, details like the 
density modulation, effects of asymmetry in density profile, vanishing 
displacement modes at the walls and most importantly nucleation and dynamics of
\index{displacement}
\index{dynamics}
misfit dislocations crucial to generate the smectic band mediated failure 
\index{smectic}
\index{failure}
observed in simulations are beyond the scope of this theory. Also, the effect 
of kinetic constraints which stabilize the solid phase well inside the 
\index{solid}
\index{constraints}
equilibrium two phase coexistence region is not captured in this 
\index{phase}
approach. We believe, nevertheless that this equilibrium calculation 
may be used as a basis for computations of reaction rates for addressing 
dynamical questions in the future.

\section{Fluctuations and Destruction of Order}
\index{destruction of order}

\begin{figure}[h]
\vskip .5cm
\begin{center}
\includegraphics[width=9.0cm]{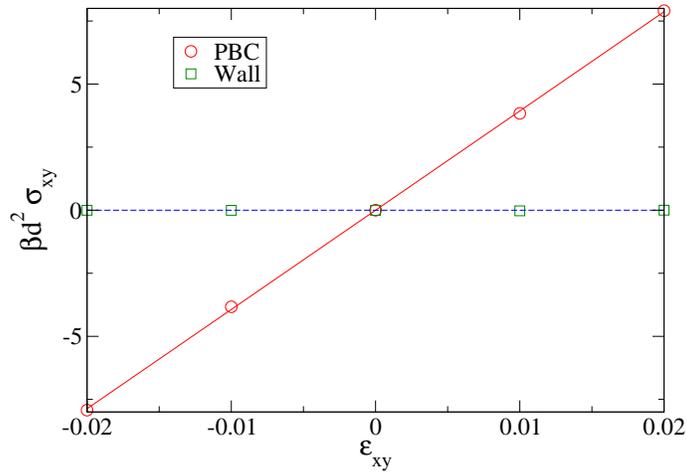}
\end{center}
\caption{Shear stress vs. shear strain at $\eta=0.85$. A system of 
\index{strain}
\index{stress}
\index{shear stress}
$43\times 50$ 2D hard disks simulated with periodic boundary conditions gives
\index{boundary conditions}
a shear modulus $\mu=393.42$ with an error within $1\%$. On the other 
hand when a $40\times 10$ 
triangular lattice of hard disks is confined within a commensurate channel,
\index{commensurate}
that fits $10$-layers of lattice planes, the shear modulus drops drastically 
\index{shear modulus}
\index{layers}
to $\mu=0$!
}
\label{shear}
\end{figure}

\begin{figure}[t]
\begin{center}
\includegraphics[width=9.0cm]{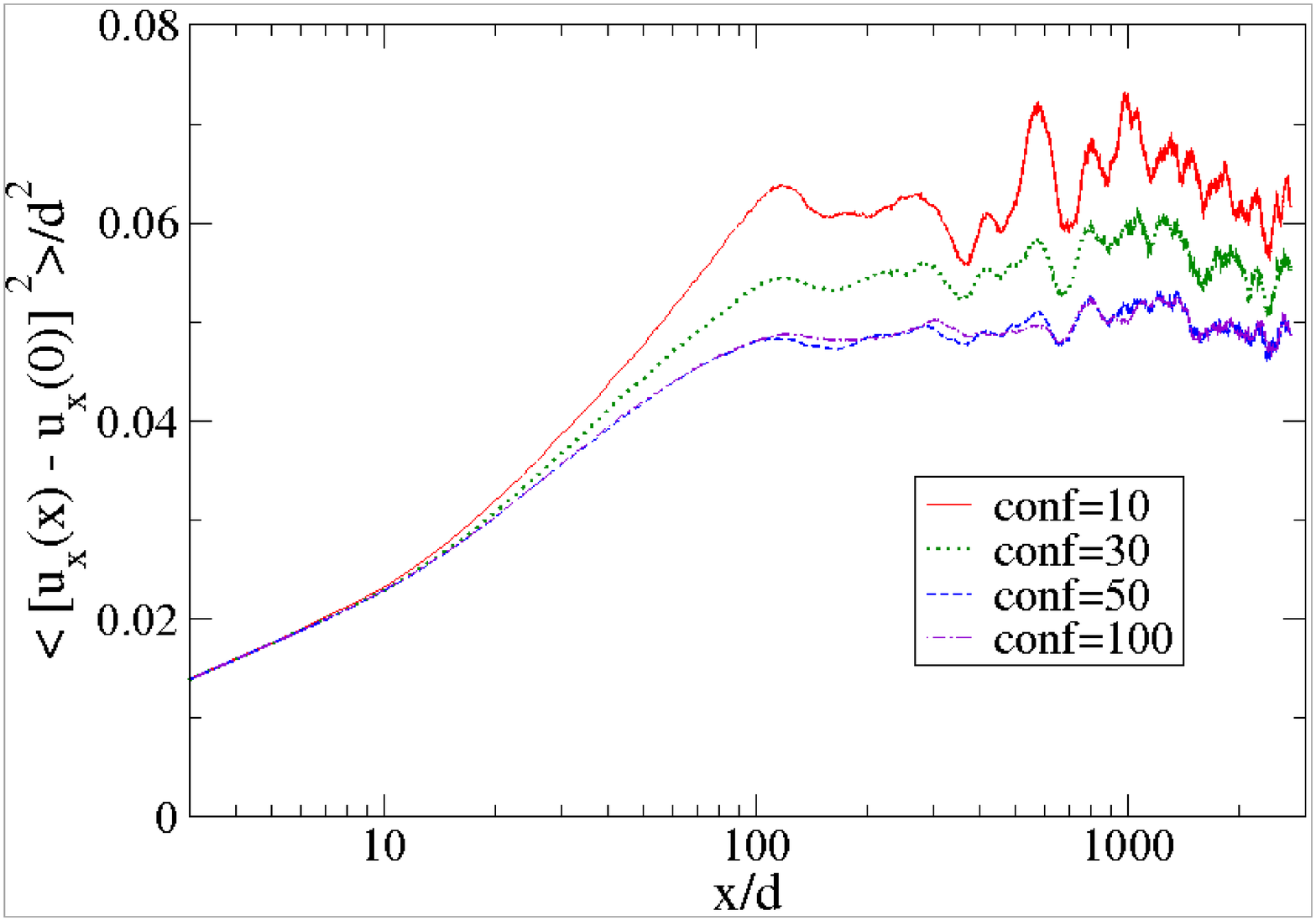}
\vskip 1cm
\includegraphics[width=9.0cm]{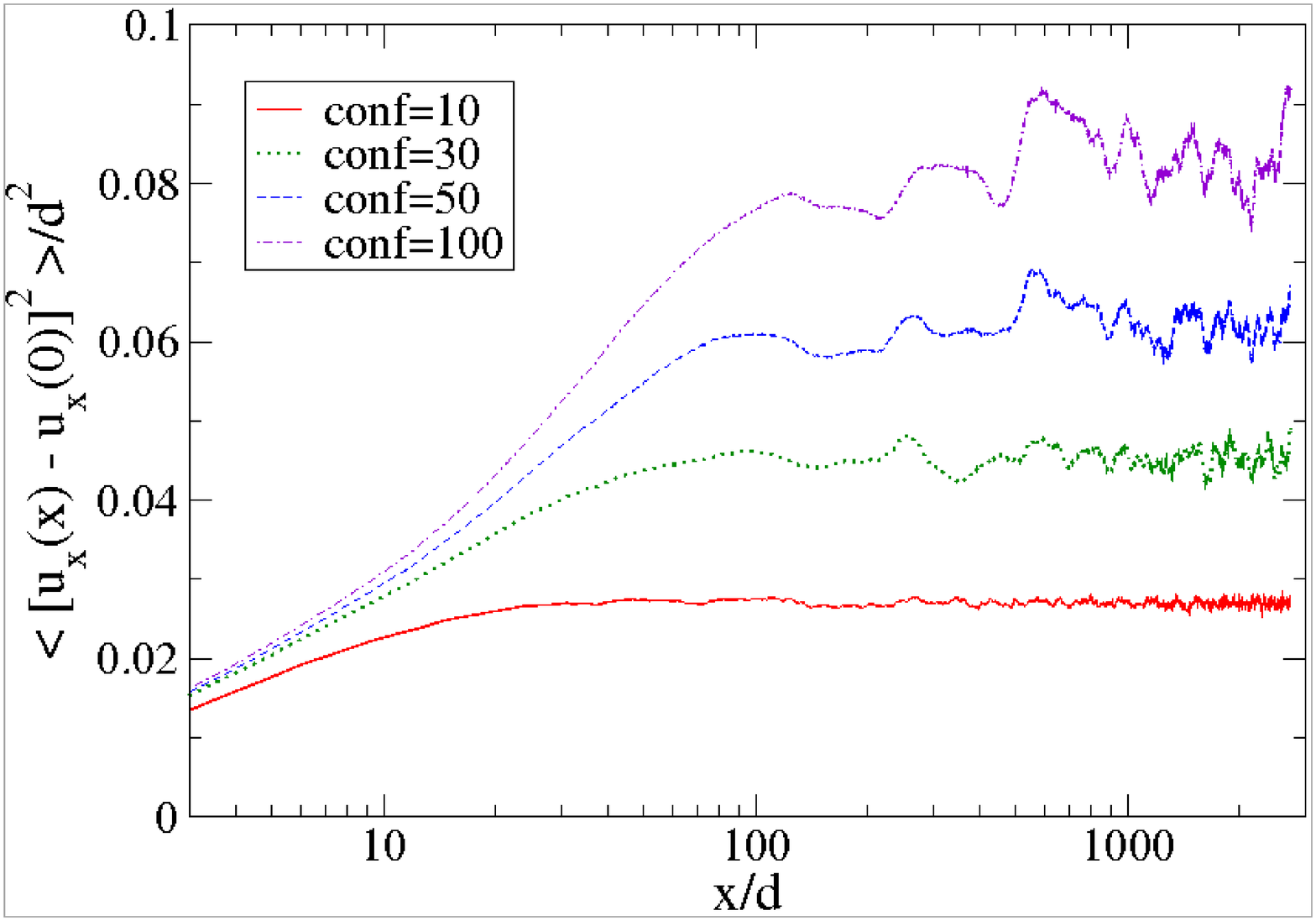}
\end{center}
\caption{
$<({u^x}(x) - {u^x}(0))^2>/d^2$
fluctuations within a single lattice plane (line)
is averaged over equilibrated configurations of $5000\times 10$ system of hard
disks at $\eta=0.75$ and simulation box dimensions which are commensurate with 
\index{simulation}
\index{commensurate}
the lattie parameter of the bulk (unconfined) system. The upper panel shows the
case for a system with periodic boundary condition while the
lower panel shows the same for Q1D confinement using hard walls.
For the confined system, fluctuations increase with increase in the number 
of configurations
over which the displacement correlations are averaged without
\index{displacement}
showing any sign of convergence. Fluctuations in the system with periodic 
boundary conditions, on the other hand, appear to saturate.
\index{boundary conditions}
}
\label{uxux-75}
\end{figure}
\begin{figure}[t]
\begin{center}
\includegraphics[width=9.0cm]{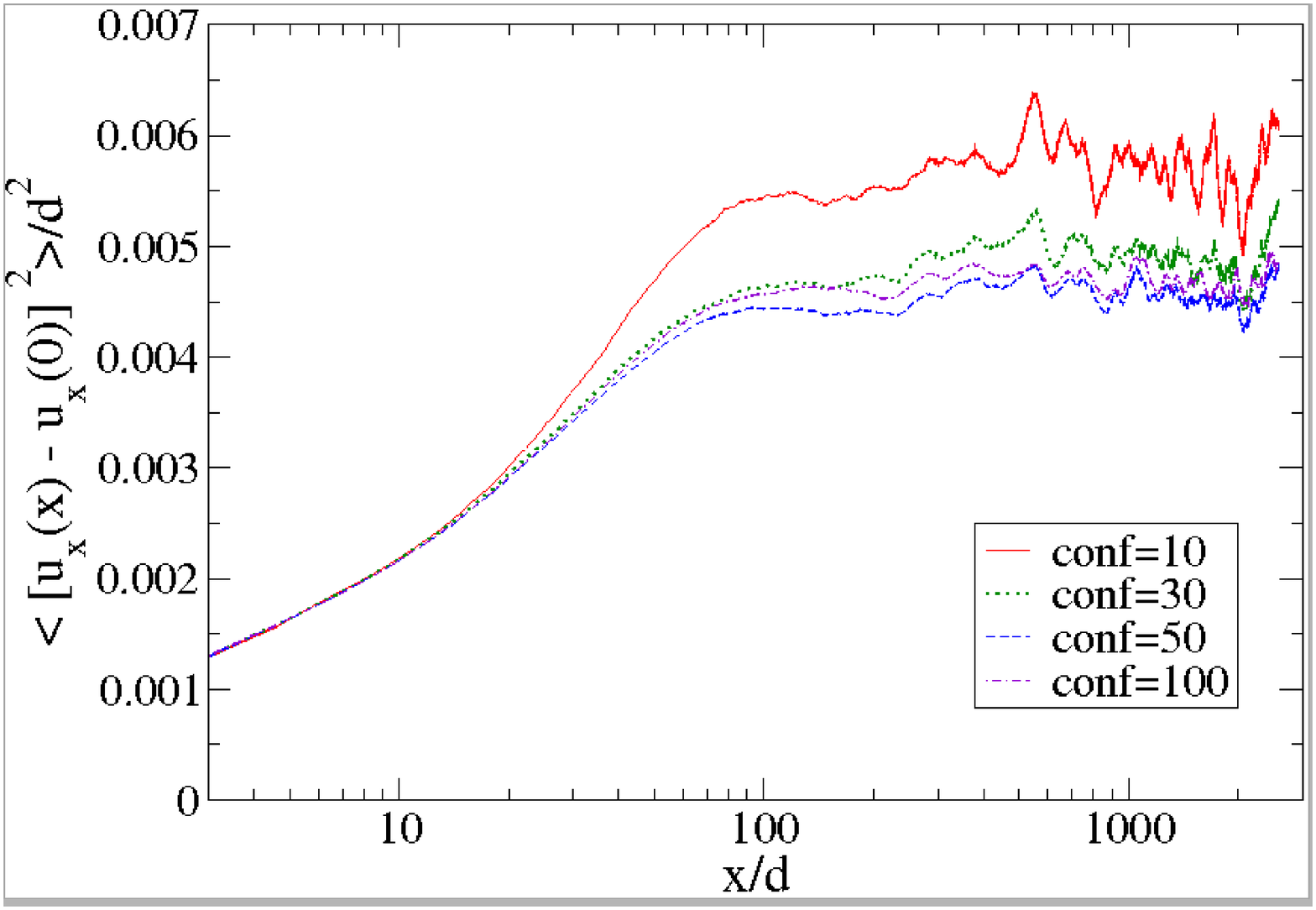}
\vskip 1cm
\includegraphics[width=9.0cm]{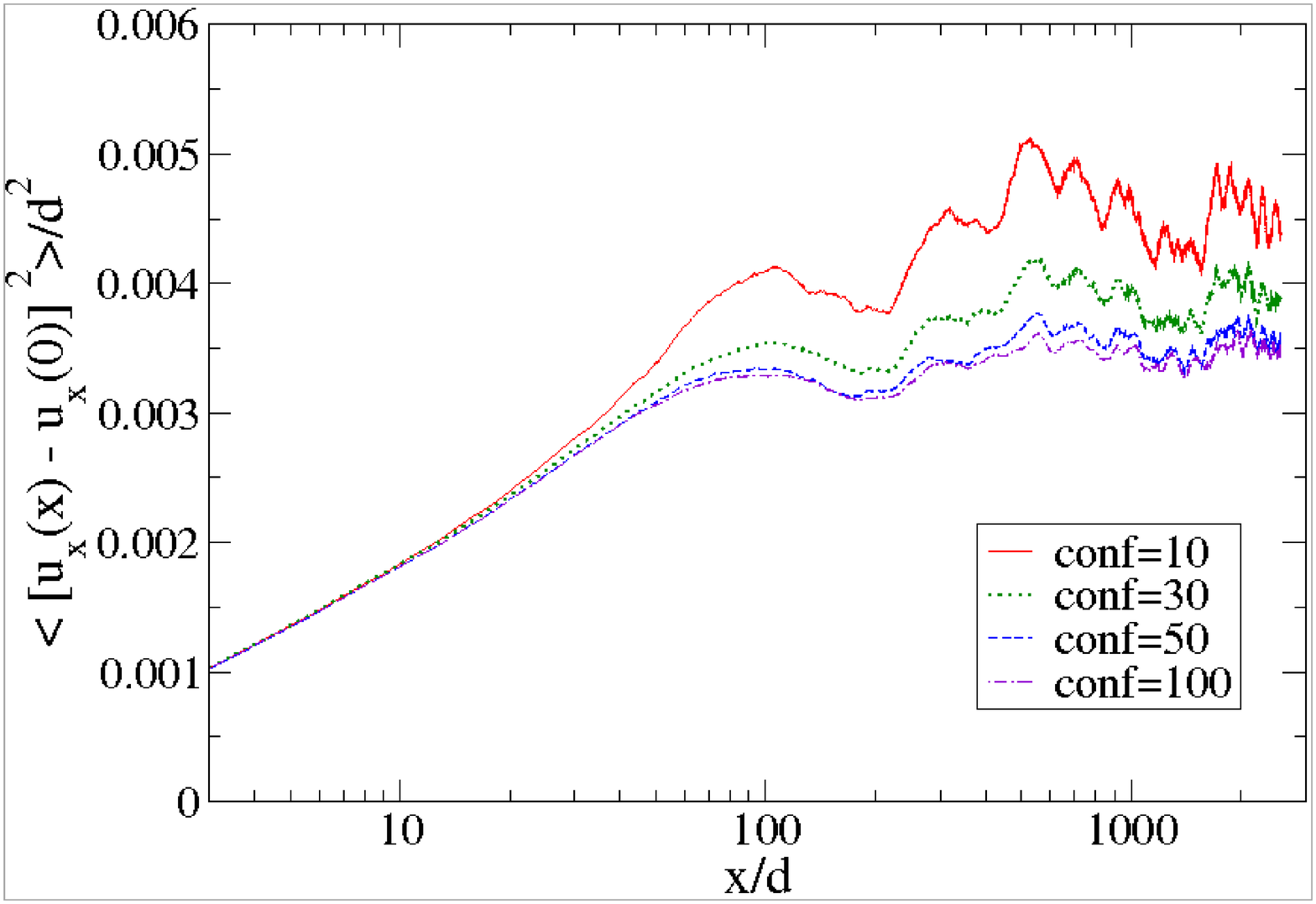}
\end{center}
\caption{The same quantities as in Fig.\ref{uxux-75}
averaged over equilibrated configurations of $5000\times 10$ system of hard
disks at $\eta=0.85$ and a simulation box size that is commensurate with the
\index{simulation}
\index{commensurate}
lattie parameter of the bulk (unconfined) system.
With increase in the number of configurations over which averaging is done, 
fluctuations reduce and converge, both for a system with periodic boundary
condition and for the confined system in channel.
}
\label{uxux-85}
\end{figure}

\noindent
One of the key definitions of a solid states that a solid, as opposed to 
a liquid, is a substance which can retain its shape due to its nonzero 
shear modulus. Going by this definition, a Q1D solid confined within 
planar, structureless walls is not a solid despite its rather striking
triangular crystalline order as well as an apparently solid- like structure 
\index{structure factor}
factor. Indeed, the shear modulus 
of the confined solid at $\eta=0.85$ is zero, though the corresponding system
\index{confined solid}
\index{solid}
with periodic boundary condition show large shear modulus (See Fig.\ref{shear}).
\index{shear modulus}
This is a curious result and is generally true for all values of 
$4 < n_l < 25$ and $\eta$ investigated by us.

\noindent
To understand the nature and amount of fluctuations in the confined Q1D
system we calculate 
correlation between displacement fields along the channel, 
\index{correlation}
$<({u^x}(x) - {u^x}(0))^2>$ for a layer of particles near a boundary. 
The nature of the {\it equilibrium} displacement correlations ultimately 
\index{displacement}
determines the decay of the peak amplitudes of the structure factor 
\index{structure factor}
and the value of the equilibrium elastic moduli \cite{pcmp}. 
In one dimension 
\index{dimension}
$<({u^x}(x) - {u^x}(0))^2> \sim x$ and in two dimensions 
$ <({u^x}(x) - {u^x}(0))^2>\sim \ln (x)$. 
In the Q1D system it is expected that 
for small distances, displacement fluctuations will grow logarithmically 
\index{displacement}
with distance which will crossover to a linear growth at large 
distances~\cite{ricci-0,ricci}.    
We calculate this quantity averaged over
$10,~30,~50,~100$ configurations equlibrated over $10^5$ Monte- Carlo (MC)
\index{Monte- Carlo}
steps and separated by $10^3$ MC steps. 
\index{MC}
We compare the results obtained
from a $5000\times 10$ hard diks solid at $\eta=0.75$ with periodic boundary
\index{boundary conditions}
conditions (PBCs) and with hard channel confinement (Fig.\ref{uxux-75}).
\index{hard channel}
The structure factor for this system is apparently solid like with prominent 
\index{solid}
\index{structure factor}
triangular order. The shear modulus is vanishingly small
\index{shear modulus}
for the confined system and non- zero for a system with PBCs.
We calculate fluctuations averaged over
$10,~30,~50,~100$ configurations equlibrated over $10^5$ Monte- Carlo (MC)
\index{Monte- Carlo}
steps and separated by $10^3$ MC steps.
\index{MC}
The fluctuation of displacement field for the system with
\index{fluctuation}
PBC reduces and converges with the increase in number of configurations
over which averagings are done. However, for the confined solid displacement
\index{displacement}
\index{confined solid}
fluctuations continue to increase with number of configurations over which
the averaging is done.
This clearly shows that as soon as a solid gets confined, even if the
confinement length scale is commensurate with that set by the system density, 
\index{commensurate}
the solid starts to behave like a liquid with zero shear modulus and 
\index{solid}
\index{shear modulus}
linearly increasing displacement correlations. 
\index{displacement}
Density modulations introduced by the hard walls seems to destabilize 
long ranged order in the small channel. This is reminiscent
of large potential strength limit of laser induced
\index{laser}
transitions~\cite{frey-lif-prl,lif-hd,mylif-large} as discussed in the last 
\index{LIF}
chapter.
However, when we compare the results obtained
from a $5000\times 10$ hard disks solid at $\eta=0.85$ with 
\index{solid}
PBCs and hard channel confinement (Fig.\ref{uxux-85}) we obtain a different
\index{hard channel}
result. 
Fig.\ref{uxux-85} clearly shows that with increase in number of configurations 
over which the averagings are done, fluctuation reduce and converge to a 
\index{fluctuation}
small number. This is consistent with the solid like structure factor but
\index{solid}
\index{structure factor}
inconsistent with the vanishing shear modulus of confined system.
\index{shear modulus}
This result is particularly surprising since it can be shown\cite{ricci-0}
that {\it all} Q1D solids whether with PBCs or not and for any 
density $\eta$ would have displacement correlations which increase linearly
\index{displacement}
with system length. The exact value of this crossover length  has 
nonuniversal prefactors which depend on boundary conditions, density, 
\index{boundary conditions}
nature of interactions etc. If the crossover length turns out to be larger
than the system size, then this increase will not be observed and the 
properties of the system will be anomalous and inconsistent. 
The fact that displacement fluctuations saturate at high densities 
\index{displacement}
as seen in our simulations suggests that this may, in fact, be the reason 
behind the puzzling behaviour. An accurate calculation of the crossover length
is required to settle this question conclusively -- a task which is non-trivial
due to the nonuniversal nature of this number.   

\noindent
A separate calculation of Young's moduli (response to elongational
strain) on a commensurate $40\times 10$ hard disk confined solid at $\eta=0.85$
\index{hard disk}
\index{confined solid}
\index{solid}
\index{strain}
\index{commensurate}
shows that $Y_x = 1361$ and $Y_y=1503$ within $3\%$ error. Young modulus
in the longitudinal direction is smaller than that in the 
direction transverse to the 
confinement and both these values are larger than the Young modulus of the
\index{Young modulus}
system under PBC ($Y=1350$). This implies that the non-hydrodynamic component 
of the stress $\s_{xx}-\s_{yy}$ is non-zero for non zero strains as 
shown in Fig.\ref{stress}. Therefore even if we choose to 
regard this Q1D solid as a liquid, it is quite anomalous since it 
\index{solid}
trivially violates Pascal's law which states that the stress tensor in a 
\index{stress}
liquid is always proportional to the identity! Lastly, commensurability seems 
to affect 
strongly the nature and magnitude of the displacement fluctuations which 
\index{displacement}
increase dramatically as the system is made incommensurate.  
Similar behaviour has been  recently noticed by A. Ricci
{\em et. al.}~\cite{ricci-0} for a confined soft disk system. They also noticed
\index{soft disk}
that the 1D structure factor of their system showed liquid like behaviour.
\index{structure factor}
Here we note that in the
work done by A. Ricci {\em et. al.}~\cite{ricci-0}, they used a soft core
interaction potential ($1/r^{12}$) and a soft system- wall potential 
($1/r^{10}$)
and studied the shear modulus and structure factor at density $\r=1.05$,
\index{shear modulus}
\index{structure factor}
very close to bulk phase transition at $\kb T=1$. 
\index{phase transition}
\index{phase}
In that system it is difficult to determine whether the system is 
commensurate or not, since the long ranged soft wall potential would
\index{commensurate}
tend to distort the triangular lattice so that the equilibrium state 
would not be a perfect triangular solid. Our system has the advantage 
\index{solid}
of having much simpler ground states which makes these distinctions 
unambiguous.  
\index{commensurate}
\index{stress}
\index{fluctuation}

\noindent
\section{Conclusion}
\label{conclusion}
In this chapter we explored some of the strange and anomalous properties of 
Q1D solids which are only a few atomic dimensions wide in one direction. 
We have found that these properties persist even if the width is increased 
by a large amount and the approach to bulk behaviour is slow. The bulk 
limit is also approached in an oscillating manner with commensurability 
playing a very important role. 

\noindent
What impact, if any, do our results have for realistic systems?  
Apart from constrained hard sphere colloids\cite{colbook} where our results 
are directly testable, a similar fracture mechanism may be 
observable in experiments on the deformation of mono-layer nano beams 
\index{nano}
or strips of real materials provided the confining channel is made of a 
material which is harder and has a much smaller atomic size than that of
the strip\cite{nanostuff-1,nanostuff-2}. 
The effect of elasticity and corrugations of 
the walls on the fracture process, as well as it's dynamics, are interesting 
\index{dynamics}
directions of future study. The destruction of long ranged solid like order
\index{solid}
should be observable in nano wires and tubes and may lead to fluctuations 
\index{nano}
in transport quantities \cite{akr}. 
\vskip .2 cm

\noindent
{
In the next chapter we shall study the transport of heat accross a Q1D solid
and try to find out the effect of the reversible failure transition in 
the heat transport coefficient.}
\chapter{Heat Conduction in Confined Solid}
\begin{verse}
\begin{flushright}
\it {Jose Arcadio Buendia ventured a murmur: ``It's the largest diamond
in the world". ``No", the gypsy countered, ``It's ice".}
\hfill -- G. G. M{\'a}rquez
\end{flushright}
\end{verse}
\vskip 1cm

In the previous chapter \cite{myfail} we have observed that the properties of
a solid that is confined in a narrow channel can change drastically
\index{solid}
for small changes in applied external strain. This was related to
\index{strain}
structural changes at the microscopic level such as a change in the
number of layers of atoms in the confining direction. These effects
occur basically as a result of the small (few atomic layers in one
direction) dimensions of the system considered and confinement along
some direction. A similar layering
\index{layering transition}
transition, in which the number of smectic layers in a confined liquid
\index{smectic}
\index{layers}
changes discretely as the wall-to-wall separation is increased, was
noted in \cite{degennes,landman}. 
Both \cite{myfail,degennes} look at  equilibrium properties while
\cite{landman} looks at changes in the dynamical properties. 
An interesting question is as to how transport properties, such as
\index{transport}
\index{transport properties}
electrical and thermal conductivity, get affected for these nanoscale
systems under strain. These questions are also important to address in
\index{strain}
view of the  current interest in the properties of nanosystems
both from the point of view of fundamentals and applications 
\cite{datta,nanobook1,nanobook2}. 

In this chapter we consider the effect of strain on the heat current
\index{heat current}
\index{strain}
across a two-dimensional ``solid'' formed by a few layers of interacting
\index{layers}
atoms which are confined in a long narrow channel.  
We note here that, 
in the thermodynamic limit it is expected that there can be no true
\index{thermodynamic limit}
solid phase in this quasi-one-dimensional system. However for a long 
\index{phase}
but finite channel, which is our interest here, and at high
packing fraction the fluctuations are small and the system behaves 
like a solid. We will thus use the word ``solid'' in this sense.  

In previous chapter~\cite{myfail} the anomalous failure, under strain, of
\index{failure}
a narrow strip of  a two dimensional solid formed by hard disks confined within
hard walls [~see Fig.~\ref{cartoon}~] was studied. 
Sharp jumps in the stress-strain were observed. These were related to
\index{strain}
\index{stress}
structural changes in the system which underwent  
transitions from solid-to-smectic-to-modulated liquid phases 
\index{solid}
\index{smectic}
\index{modulated liquid}
\cite{myfail,myijp}. 
In the present chapter we  study changes in the thermal conductance of this 
system as it undergoes elastic deformation and failure
\index{failure}
through a layering transition caused by external elongational
\index{layering transition}
strains applied in different directions.

The calculation of heat conductivity in a many body system is 
a difficult problem. The Kubo 
\index{Kubo}
formula and Boltzmann kinetic theory 
provide formal expressions for the thermal conductivity. In
practice these are usually difficult to evaluate without making
drastic approximations. More importantly a large number of recent
studies \cite{bonet,lepri,lippi,grass} indicate that the heat conductivity of
low-dimensional systems infact diverge. It is then more sensible to
calculate directly the heat current or the conductance of the system rather than
the heat conductivity. In this chapter
we propose a simple-minded calculation of the heat current which can be expected
\index{heat current}
to be good for a hard disk (or hard spheres in the three dimensional
\index{hard disk}
case) system in the solid phase. This reproduces some qualitative
\index{solid}
\index{phase}
features of the simulations and gives  
values for the current which are of the correct order of magnitude.

The organization of this chapter is as follows. In
Sec.~(\ref{simresults}) we explain the model and present the results
from simulations. In Sec.~(\ref{anaresults}) we derive a simple
formula for heat current in a hard-sphere system and evaluate it
\index{heat current}
approximately. We conclude with some discussions in Sec.~(\ref{disc}).  

\section{Results from Simulations}

\label{simresults}
\begin{figure}[t]
\begin{center}
\includegraphics[width=9cm]{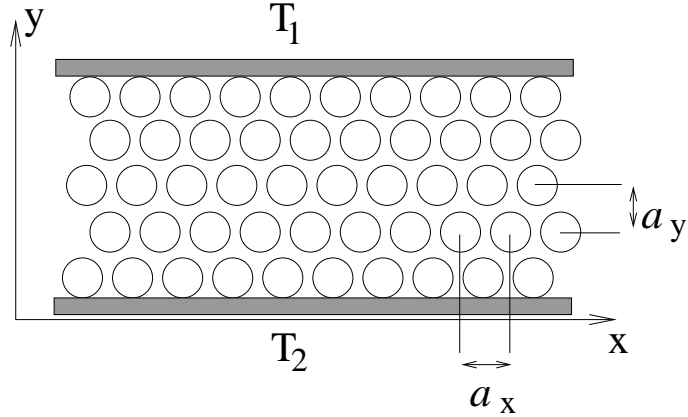}
\end{center}
\caption{ A solid with a triangular lattice structure formed by hard disks confined between two structureless walls at $y=0$ and $y=L_y$.
The walls are maintained at two different temperatures. The lattice
parameters of the unstrained solid are denoted by $a_x^0$ and $a_y^0$. 
\index{solid}
Elongational strains can be imposed by rescaling distances either in
the $x$ or $y$ directions and the lattice parameters change to $a_x$
and $a_y$.
}
\vspace{1cm}
\label{cartoon}
\end{figure}
We consider a two dimensional system of hard disks of diameter $d$ and mass
$m$ which interact with each other through elastic collisions. 
The particles are confined  within a
narrow hard structureless channel [see Fig.~\ref{cartoon}]. 
The hard walls of the channel are located at
$y=0$ and $y=L_y$ and we take periodic boundary conditions in the $x-$direction.
\index{boundary conditions}
The length of the channel along the $x-$direction is $L_x$ and the area is
${\cal A}=L_x\times L_y$.
The confining walls are maintained at  two 
different temperatures ( $T_2$ at $y=0$ and $T_1$ at $y=L_y$ ) so 
that the temperature difference $\D T=T_2-T_1$ gives rise to a heat 
\index{heat current}
current in the $y$-direction.
Initially we start with channel dimensions $L_x^0$ and $L_y^0$ such
that the system is in a phase corresponding to a unstrained solid with
\index{solid}
\index{phase}
a triangular lattice structure.
We then study the heat current in
\index{heat current}
this system when it is strained (a) along the $x-$direction and (b)
along the $y-$direction.

We perform an event-driven collision time dynamics \cite{allen} simulation of 
\index{dynamics}
\index{simulation}
the hard disk system. The upper and lower walls are maintained at temperatures
$T_1 = 1$ and $ T_2 = 2$ (in arbitrary units) respectively by imposing Maxwell 
\index{Maxwell boundary condition}
boundary condition \cite{bonet} at the two confining walls. This means
that whenever 
a hard disk collides with  either the lower or the upper wall it gets reflected
\index{hard disk}
back into the system with a velocity chosen from the distribution
\bea
f(\uu)=\f{1}{\sqrt{2\pi}}\left(\f{m}{\kb T_W}\right)^{3/2}
|u_y|\exp\left(-\f{m\uu^2}{2\kb T_W}\right) 
\eea
where $T_W$ is the temperature ($T_1$ or $T_2$) of the wall on which
the collision occurs.  
During each collision energy is exchanged between the system and the
bath. Thus in our molecular dynamics simulation, the average heat
\index{dynamics}
\index{simulation}
\index{molecular dynamics}
current flowing through the system can be found easily by computing the
net heat loss from the system to the two baths  (say $Q_1$
and $Q_2$ respectively) during a  large time interval $\tau$. 
The steady state heat current from lower to upper bath is given by $ \la I \ra
= \lim_{t \to \infty} Q_1/\tau = -\lim_{t \to \infty} Q_2 /\tau$.    
In the steady state the heat current (the heat flux density integrated over
\index{heat current}
$x$) is independent of $y$. This is a requirement coming from current
conservation. However if the system has inhomogeneities then the flux
density itself can have a spatial dependence and in general we can
have $j=j(x,y)$. In our simulations we have also looked at $j(x,0)$ and $j(x,L_y)$.

Note that the relevant scales in the problem are: $k_B T$ for energy, 
$d$ for length and $\tau_s=\sqrt{md^2/k_B T}$ for time. 
We start from a solid commensurate with its wall to wall separation and
\index{commensurate}
follow two different straining protocols.
In case (a) we strain the solid by rescaling the length in the
\index{solid}
$x-$direction and the imposed external strain is $\e_{xx} = (L_x -
L_x^0)/L_x^0$. In case (b) we rescale the length along the
$y-$direction and the imposed strain is $\e_{yy} = (L_y  
\index{strain}
- L_y^0)/L_y^0$. 

\begin{figure}[t]
\begin{center}
\includegraphics[width=10.0cm]{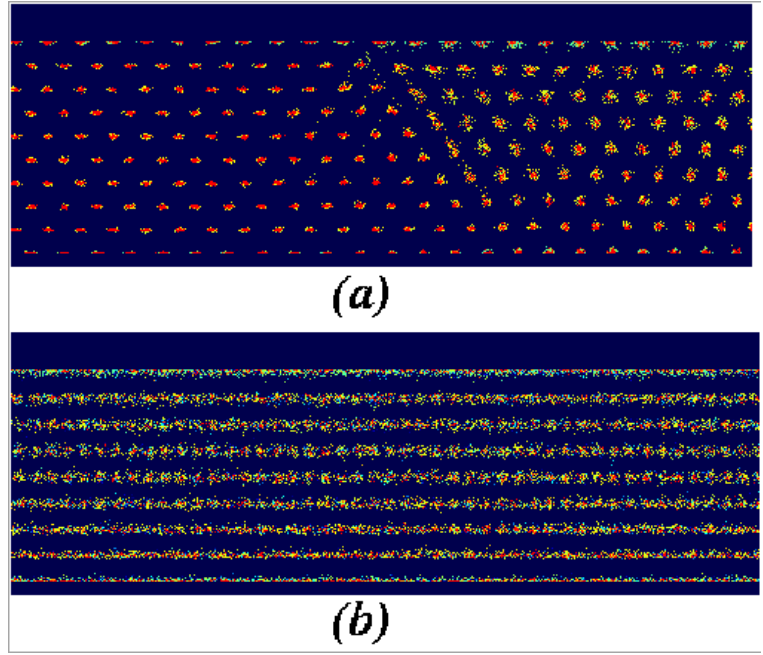}
\end{center}
\caption{
Plots obtained by superposition of $500$ steady state
configurations of a portion of $40\times10$ system taken at equal 
time intervals. Starting from $\eta = 0.85$ imposition of strains 
(a) $\ex=0.1$, (b) $\ex=0.15$ gives rise to these structures.
The colors code local density of points from red/dark (high) to blue/light 
(low). In (a) one can see a $9$-layered structure  nucleated within a 
$10$-layered solid. The corresponding structure factor identifies this
\index{solid}
\index{structure factor}
 to be a smectic~\cite{myfail}. In
(b) the whole system has transformed into a $9$-layered smectic. 
\index{smectic}
}
\label{switch}
\end{figure}

\begin{figure}[t]
\begin{center}
\includegraphics[width=9.0cm]{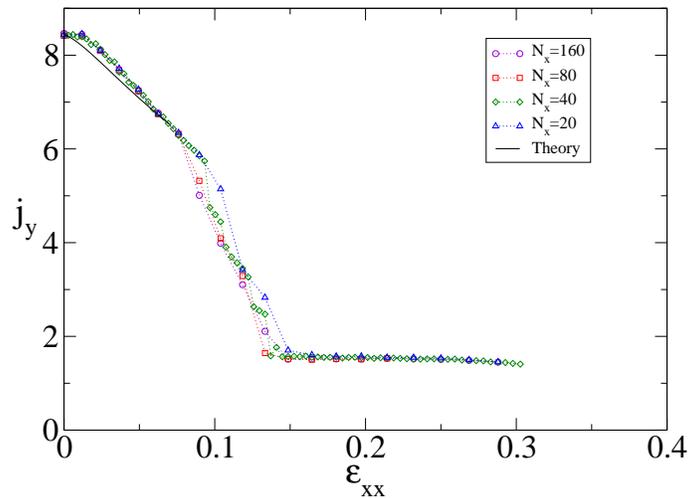}
\end{center}
\caption{ Plot of $j_y$ (in units of $k_BT/\tau_s d$) versus $\ex$ 
   for different lengths of the channel. The width of the channel is
   $N_y=10$ layers. Starting packing fraction is $\eta = 0.85$. 
\index{layers}
The solid line shows the theoretical prediction of
\index{solid}
dependence of the heat current on strain [see Sec.~(\ref{anaresults})]. 
\index{heat current}
\index{strain}
}
\label{jyex}
\end{figure}

The only thermodynamically  relevant variable for a hard disk system is the
\index{hard disk}
packing fraction
$\eta = \pi N  d^2/4 {\cal A}$. For a close packed solid with periodic
boundary condition this value is about $\eta_c=0.9069$. On the other hand
for a confined solid having $N_y$ number of layers 
\index{confined solid}
\index{layers}
$\eta_c = \pi N_y/(2 \sqrt 3(N_y-1)+4)$ and for a $10$- layered solid
$\eta_c=0.893$.
In our simulations we consider initial
values of $\eta$ for the solid to be close to $\eta_c$. The channel
is ``mesoscopic'' in the sense that it has a small width with $N_y=10$
layers of disks in the $y-$direction (in the initially unstrained solid). In
\index{solid}
\index{layers}
the $x-$direction the system can be big and we consider
$N_x=20,~40,~80,~160$ number of disks in the $x-$direction.  
In collision time dynamics
\index{dynamics}
we perform $10^5$ collisions per particle to reach steady state and 
collect data over another $10^5$ collisions per particle.
All the currents calculated in this study are accurate within error bars which
are less than $3\%$ of averaged current. 

Let us briefly recapitulate some of the equilibrium results for
stress-strain behavior obtained in the last chapter.
As the strain $\ex$ is imposed, the perfectly 
triangular solid shows rectangular distortion along with a linear response
in strain versus  stress behavior. Above a critical strain
\index{strain}
\index{stress}
($\ex \approx 0.1$) one finds that smectic bands having a lesser number
of layer nucleate within the solid [this can also be seen in Fig.~(2a),
which is for a nonequilibrium simulation]. 
\index{simulation}
This smectic is liquid-like in $x-$direction (parallel to the walls) and has 
solid-like density modulation order in $y-$direction (perpendicular to the 
\index{solid}
walls). With further increase in strain the size of the smectic region 
\index{strain}
increases and ultimately the whole system goes over to the smectic phase 
\index{phase}
at $\ex \approx 0.15$ [Fig.~(2b)]. At even higher 
strains the smectic melts to a modulated liquid \cite{myfail,myijp}.
The modulated liquid shows typical liquid like ring pattern corresponding to 
\index{modulated liquid}
average inter- particle separation above the smectic like 1D density modulation
\index{smectic}
peaks in the structure factor \cite{myfail-large,myfail}. 
\index{structure factor}
This layering transition is an effect of finite size in the confining
\index{layering transition}
direction. Similar phase behaviors have been observed in experiments on 
\index{phase}
steel balls confined in quasi 1D \cite{pieranski-1}.
We note that to fit a $N_y$ layered triangular solid within 
\index{solid}
a channel of width $L_y$ we require
\bea
L_y = \frac{\sqrt 3}{2}a_x^0 (N_y-1) + d~. 
\eea
This enables us to define a fictitious number of layers 
\index{layers}
\bea
\chi = 2 \frac{L_y- d}{\sqrt 3 a} +1 \nn
\eea
of triangular solid that can span the channel
where $a$ is the lattice parameter at any given density.
The actual number of layers that are present in the strained solid  is
\index{solid}
\index{layers}
$N_y = I(\chi)$ where the function $I(\chi)$ gives the  integer 
part of $\chi$. For confined solids the free energy has minima at
integer values of $\chi$ 
and maxima at half-integral values \cite{myijp,myfail}. The difference
in free-energy between succesive 
maxima and minima gradually decreases with increasing 
$L_y$. Thereby the layering transition washes out for 
\index{layering transition}
$n_l\stackrel{>}{=} 25$ layered 
unstrained solid\cite{myfail}. Up to this number of layers, a triangular solid
\index{solid}
\index{layers}
strip confined between two planar walls fails at a critical deviatoric strain 
$\e^\ast_d\sim1/N_y$, i.e. smaller strips fail at a larger deviatoric
\index{deviatoric strain}
strain ($\e_d = \ex-\ey$).
\index{strain}

We now present the heat conduction simulation results for the two
\index{conduction}
\index{simulation}
cases of straining in $x$ and $y$ directions. 

(a) {\emph Strain in $x$ direction}~-~
In Fig.~\ref{jyex} we plot the heat current density $j_y$ calculated at
different values of the strain $\ex$. Starting from the 
triangular lattice configuration, we find that the heat current
decreases linearly with increase in strain. 
At about the critical strain $\ex \approx 0.1$ we find that the heat
\index{heat current}
current begins to fall at a faster rate.
This is easy to understand physically. At  the onset of critical
strain smectic bands, which have lesser number of particle layers, start
\index{layers}
nucleating (Fig.~\ref{switch}). These regions are much less effective 
in transmitting heat than the solid phase and the  heat current falls 
\index{solid}
\index{phase}
rapidly as the size of the smectic bands grow. 
At about the  strain value $\ex\approx 0.15$ the whole system is spanned by the
smectic.  Beyond this strain there is no appreciable change in the
\index{smectic}
\index{strain}
heat current. 
The solid line in Fig.~\ref{jyex} is an estimate from a simple
\index{solid}
analysis explained in Sec.~(\ref{anaresults}).

\begin{figure}[t]
\begin{center}
\includegraphics[width=9.0cm]{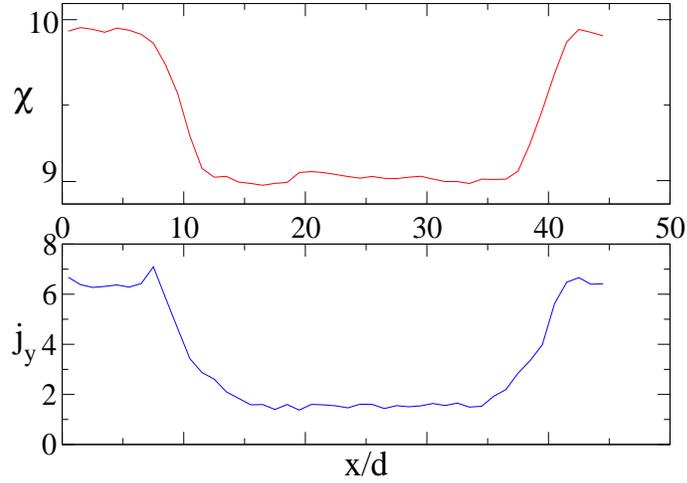}
\end{center}
\caption{ $\chi(x)$ is the local number of layers averaged over $10^3$
steady state configurations for a system of $40\times 10$ hard 
disks. A starting triangular lattice of $\eta=0.85$ is strained to $\ex=0.118$
and the data collected after steady state set in. Also shown is the local 
heat current $j_y(x)$. The regions having lower number of layers conduct 
\index{heat current}
\index{layers}
less effectively.
}
\label{jy-x}
\end{figure}

In Fig.~\ref{jy-x} we plot the local steady state heat current
\index{heat current}
$j_y(x)$ for a system of $40\times 10$ particles  at a 
strain $\ex=0.118$ {\emph{i.e.}} at a strain corresponding to the
solid-smectic phase 
\index{solid}
\index{smectic}
\index{phase}
coexistence. At this same strain the number of layers averaged over $10^3$
\index{strain}
configurations have been plotted. It clearly shows that 
the local heat current is smaller in regions with smaller number of
layers. This is the reason behind getting a sharp drop in average heat
\index{heat current}
\index{layers}
current after the onset of phase coexistence.
\index{phase}

\begin{figure}[t]
\begin{center}
\includegraphics[width=9.0cm]{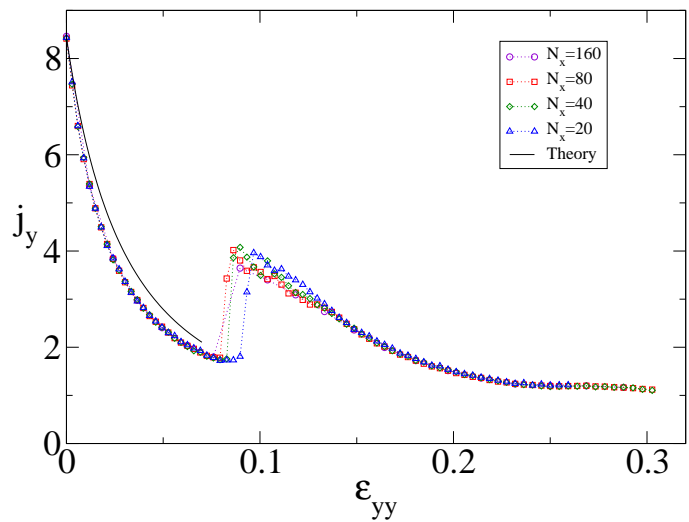}
\end{center} 
\caption{ Plot of $j_y$  (in units of $k_BT/\tau_s d$) versus $\ey$ for different channel
  lengths. The channel width is $N_y=10$ layers. The starting packing
  fraction is $\eta =0.85$. The jump in current occurs at the strain
  value where the number of layers in the $y-$direction increases by
\index{layers}
  one and the system goes to a   smectic phase. 
\index{smectic}
\index{phase}
The solid line shows the theoretical prediction of
\index{solid}
dependence of the heat current on strain [~see
\index{heat current}
\index{strain}
  Sec.~(\ref{anaresults})~].
\vspace{1cm}
}
\label{jyey}
\end{figure}

\begin{figure}[t]
\begin{center}
\includegraphics[width=10.0cm]{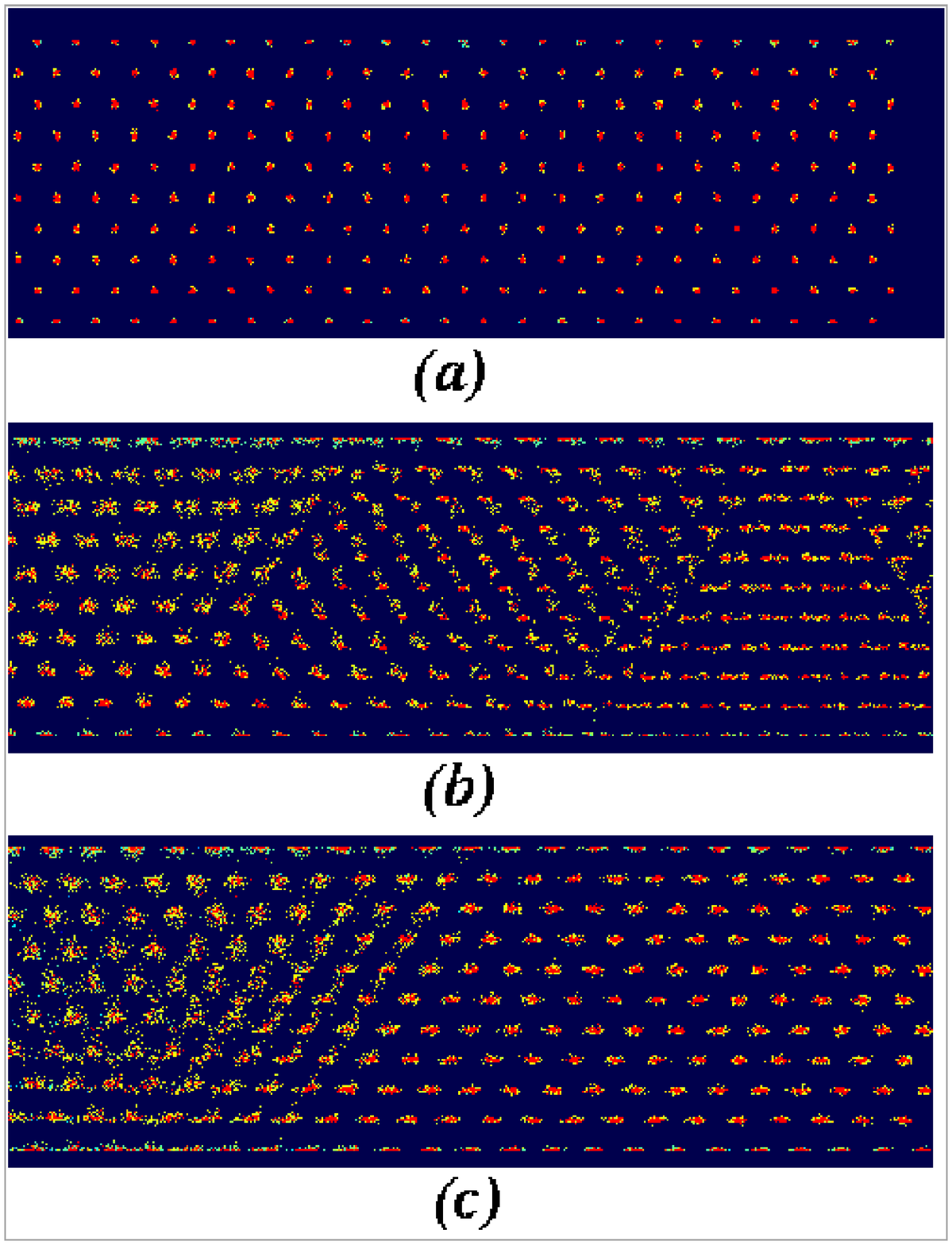}
\end{center}
\caption{
Plots obtained by superposition of $500$ 
steady state configurations of a portion of $40\times10$ system taken at equal 
time intervals. Starting from $\eta = 0.85$ imposition of strains 
(a) $\ey=0.05$, (b) $\ey=0.1$, (c) $\ey=0.12$ gives rise to these structures.
The colors code local density of points from red/dark (high) to blue/light
(low). (a) Solid phase. (b) A mixture of $10$-layered solid and a buckling 
phase. (c) An $11$-layered solid in contact with $10$-layered smectic like
\index{solid}
\index{smectic}
\index{phase}
region. 
}
\label{jump}
\end{figure}

(b){\emph Strain in $y$ direction}~-~ Next we consider the case where, again starting from the density 
$\eta =0.85$, we impose a strain along the $y-$direction. As shown in 
\index{strain}
Fig.~\ref{jyey}, the heat current $j_y$ now has  
\index{heat current}
a completely different nature. The initial fall is much steeper and 
has a form different from the linear drop in Fig.~\ref{jyex}. 
The approximate analytic curve is explained in Sec.~(\ref{anaresults}).  
At about $\ey\approx 0.1$ we see a sharp and presumably discontinuous
jump in the current. At this point the system goes over to a buckled phase
\index{buckled phase}
\index{phase}
(Fig.~\ref{jump}b) in which different parts of solid (along $x$- direction)
are shifted in $y$-direction by small displacement to cover extra space between
\index{displacement}
the walls\cite{buckled-1,buckled-2,buckled-3}. A further small strain induces a layering transition and the system breaks into $N_y=11$ layered solid and 
\index{solid}
\index{strain}
\index{layering transition}
$N_y=10$ layered highly fluctuating smectic like regions. At even higher 
strains ($\ey\sim 0.2$) the whole system eventually melts to a $N_y=11$ 
layered smectic phase. The phase behavior of this system is interesting 
\index{smectic}
\index{phase}
and will be discussed in detail elsewhere\cite{myfail-large}.
Unlike in the case with applied strain in the $x-$direction,
\index{strain}
in the present case the buckling-layering transition is very
\index{layering transition}
sharp. Even though the overall density has {\emph{decreased}}, due to
buckling and increase in number of layers in the conducting direction,
\index{layers}
there is an increase in the energy transferring collisions and hence
the heat current.  
\index{heat current}
The plots in Fig.~\ref{jump} show the structural
changes that occur in the system as one goes through the transition.

\begin{figure}[t]
\begin{center}
\includegraphics[width=9.0cm]{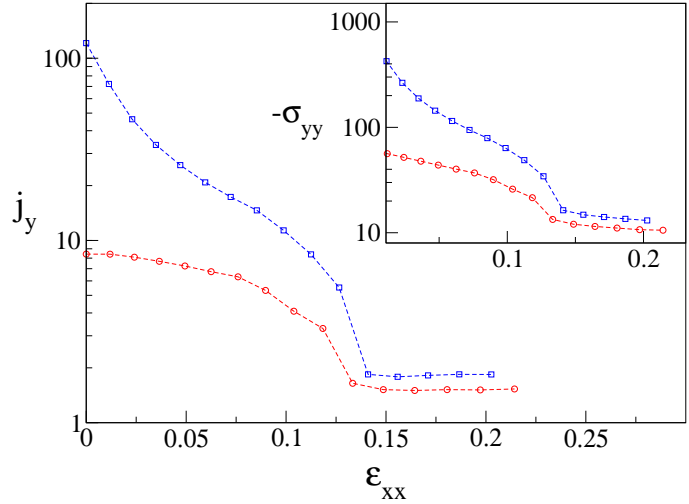}
\end{center}
\caption{Plot of $j_y$ (in units of $k_B T/\tau_s d$ )  versus $\ex$
  for two different starting values of 
  the packing fraction. $\Diamond$ corresponds to a starting value of 
$\eta = 0.89$ while  $+$ is for $\eta = 0.85$. 
In both the cases the initial solid size was $80 \times 10$. The inset shows 
\index{solid}
corresponding plots of $-\sy$ (in units of $k_B T/ d^2$) versus
  $\ex$. Notice that stress-strain curve has the 
\index{strain}
\index{stress}
same qualitative profile as the $j_y$ versus $\ex$ curve.
}
\label{jyex-comp}
\end{figure}

We find in general that the heat current along any direction within the solid
\index{solid}
\index{heat current}
follows the same qualitative features as the stress component along the same
\index{stress}
direction. 
This can be seen in Fig.\ref{jyex-comp} where we have plotted $j_y$
versus $\ex$ for two starting  densities of solids $\eta =
0.85,~0.89$. In the inset we show the corresponding  $-\s_{yy}$ versus $\ex$
curves and see that they follow the same qualitative  
behavior as the heat current curves. 
The reason for this is that  microscopically
they both originate from interparticle collisions.
Infact the microscopic expressions for the total heat current 
\index{heat current}
[~see Eq.~(\ref{totI}) in Sec.~(\ref{anaresults})~] is very 
similar to that for the stress tensor component, with an extra velocity factor. The stress tensor is given by:
\index{stress}
\bea
 {\cal A}\s_{\a \be} &=&     -\sum_{i} \la m u_i^\a u_i^\be \ra 
+\sum_{i< j
  }  \left\la \f{\p \phi(r_{ij})}{\p   r_{ij}} \f{x_{ij}^\a
  x_{ij}^\be}{r_{ij}}  \right\ra ~,
\label{sts}
\eea
where $\{x_i^\a, u_i^\a \}$ refer to the $\a$-th component of 
position and velocity of the $i^{\rm th}$ particle, 
$r_{ij}^2=\sum_\a (x_{ij}^\a)^2$ and $\phi(r_{ij})$ is the interparticle 
potential. For a hard disk system, $\f{\p\phi(r_{ij})}{\p
\index{hard disk}
  r_{ij}}$ 
can be replaced by $-\kb T\d(r_{ij}-{\rm d})$. Also 
in equilibrium we have $\la m u_i^\a u_i^\be \ra = \kb T \d_{\a\be}$
and hence the stress tensor becomes:
\index{stress}
\bea
{\cal A} \s_{\a \be} =  -\kb T \left[  N  \d_{\a\be}+  \left\la \sum_{i<j}  
   \f{x_{ij}^\a x_{ij}^\be}{r_{ij}} \d(r_{ij}(t)- d)\right\ra 
~\right] \nn~.
\eea
Using collision time simulation it is easier to evaluate the 
\index{collision time simulation}
\index{simulation}
stress tensor in the following way. 
\index{stress}
We can rewrite Eq.~(\ref{sts}) as
\bea
{\cal A} \s_{\a \be} &=&     -N \kb T \d_{\a\be} -\sum_{i<j}  
     \left\la  x_{ij}^\a f_{ij}^\be \right\ra \nn~.
\eea 
We use the idea that $\la\dots\ra$ can be
replaced by a time average so that from Eq.~(\ref{sts}) we have
\bea
\ \left\la  x_{ij}^\a f_{ij}^\be \right\ra =-\lim_{\tau \to \infty} \f{1}{\tau} \int_0^\tau dt 
 x_{ij}^\a f_{ij}^\be \nn~.
\eea
Now note that during a collision we have 
$\int dt f_{ij}^\be= \Delta p_{ij}^\be$
where $\Delta \vec p_{ij}$ 
is the  
change in momentum of $i^{\rm th}$ particle due to collision with
$j^{\rm th}$ particle. 
It can be shown that $\Delta \vec p_{ij} = -(\uu_{ij}.\hat r_{ij})\hat
r_{ij}$  where  $ \hat r_{ij} = \rv_{ij}/r_{ij}$ and $\uu_{ij}=
\vec{u}_i-\vec{u}_j$ and $\rv_i,~\uu_i$ are evaluated just before a
collision.  
This change in momentum occurs for a single pair of particle during one
collision event. 
To get the stress tensor we sum over all the 
collision events in the time interval $\tau$ between all pairs of
particles. Therefore in collision time dynamics  
\index{dynamics}
we get the following expression for the stress tensor,
\index{stress}
\bea
{\cal A} \sigma_{\a\be} =  -N \kb T \d_{\a\be}+
\lim_{\tau \to \infty} \f{1}{\tau} \sum_{\tau_c} \sum_{i<j} \Delta
p_{ij}^\a x_{ij}^\be ~,
\eea
where $\sum_{\tau_c}$ denotes a summation over all collisions in time
$\tau$. 

\section{Analysis of Qualitative Features}
\label{anaresults} 
We briefly outline a derivation of the expression for the heat
flux. For the special case of a hard disk system this simplifies
\index{hard disk}
somewhat. We will show  
that starting from this expression and making rather simple minded
approximations  we can explain some of the observed results for  heat flux as a function 
of imposed external strain.
\index{strain}

We consider a system with a general Hamiltonian given by:
\bea
H=\sum_i [\f{m \bu_i^2}{2}+V(\br_i)] +\f{1}{2}\sum_{i,j \neq i} \phi(r_{ij})~, 
\eea
where $V(\br_i)$ is an onsite potential which also includes the wall. 
To define the heat current density we need to write a continuity
\index{heat current}
equation of the form: $\p \e(\br,t)/\p t+\p j_\a(\br,t)/\p x_\a=0$. 
The local energy density is given by: 
\bea
\e(\br,t)&=&\sum_i \d (\br-\br_i) h_i {\rm~~~where} \nn \\ 
h_i &=& \f{m \bu_i^2}{2}+V(\br_i) +\f{1}{2} \sum_{j \neq i} \phi(r_{ij}) \nn
\eea
Taking a derivative with respect to time gives
\bea
\f{\p \e}{\p t} &=& -\f{\p}{\p x_\a} \sum_i \d (\br - \br_i) h_i u_i^\a+
\sum_i \d (\br-\br_i) \dot{h}_i   \\
&=& -\f{\p}{\p x^\a} j_{\a}^K + W^U 
\eea
where ${\bf j}^K=\sum_i \d (\br - \br_i) h_i \bu_i $ is the convective
part of the energy current. We will 
now try to write the remaining part given by $W^U$ as a divergence
term. We have
\bea
W^U &=& \sum_i \d (\br-\br_i) \dot{h}_i \nn \\
&=& \sum_i  \d (\br-\br_i) [m u_i^\a \dot{u}_i^\a + \f{\p V(\br_i)}{\p
    x_i^\a} u_i^\a \nn\\
 &-& \f{1}{2} \sum_{j \neq i} (f_{ij}^\a  u_i^\a + f_{ji}^\a u_j^\a)
]~, \nn
\eea
where $f_{ij}^\a=-\p\phi(r_{ij})/\p x_i^\a$.
Using the equation of motion $m \dot{u}_i^\a = -{\p V}/{\p
  x_i^\a}+\sum_{j\neq i} f_{ij}^\a$ 
we get 
\bea
W^U &=&  \f{1}{2}  \sum_{i,j \neq i} \d (\br-\br_i) (~f_{ij}^\a
u_i^\a - f_{ji}^\a u_j^\a~)~.
\eea
With the identification $W^U=-{\p j_\a^U}/{\p x^\a}$ and using
${\bf{f}}_{ij}=-{\bf{f}}_{ji}$ we finally get:
\bea
j_\a^U(\br) 
= \f{1}{2} \sum_{i,j \neq i} \theta (x_i^\a-x^\a)
\prod_{\nu  \neq \a} \d (x^\nu-x_i^\nu)f_{ij}^\be( u_i^\be + u_j^\be)~   
\eea
where $\theta(x)$ is the Heaviside step function.
This formula has a simple physical
interpretation. First note that we need to sum over only those $i$ for
which $x_i^\a > x^\a$. Then the formula basically gives us the net rate at
which work is done by particles on the left of $x^\a$ on the particles
on the right which is thus the rate at which energy flows from left to
right. The other part, $j_\a^K$, gives the energy flow as a result of
physical motion of particles across $x^\a$.  
Let us look at the total current in the system. Integrating the
current density $j^U_\a$ over all space we get:
\bea
I_\a^U &=& \f{1}{2} \sum_{i,j\neq i} x_i^\a f_{ij}^\be (u_i^\be + u_j^\be) \nn\\
&=& - \f{1}{2} \sum_{i,j\neq i} x_i^\a \f{\p \phi (r_{ij})}{\p r_{ij}}
\f{x^{\be}_{ij}}{r_{ij}} (u_i^\be + u_j^\be) \nn \\
&=& - \f{1}{4} \sum_{i,j\neq i}  \f{\p \phi (r_{ij})}{\p r_{ij}}
\f{x^\a_{ij} x^{\be}_{ij}}{r_{ij}} (u_i^\be + u_j^\be)  ~.
\eea
Including the convective part and taking an average over the steady
state we finally get:
\bea
\la I_\a \ra &=& \la I_\a^K \ra + \la I_\a^U \ra = 
 \sum_i \la~ h_i u_i^\a \ra \nn \\
&-&\f{1}{4} \sum_{i ,j \neq
  i} ~ \la~ \f{\p \phi (r_{ij})}{\p r_{ij}}
\f{x^\a_{ij} x^{\be}_{ij}}{r_{ij}} (u_i^\be + u_j^\be)~ \ra~. 
\label{totI}
\eea
We note that for a general phase space variable $A(\{x_i,u_i\})$ the average $\la
\index{phase}
A \ra$ is the time average $\lim_{\tau \to \infty} (1/\tau)
\int_0^\tau dt A(\{x_i(t),u_i(t)\})$.

{\bf Finding the energy current for hard disk system}: 
\index{hard disk}
The energy current expression
involves the velocities of the colliding particles which change during
a collision so we have to be careful. 
 We use the following expression
for $\la I_\a^U \ra$:
\bea
\la~I_\a^U~\ra &=& \f{1}{4} \sum_{i, j \neq i} \la~ x_{ij}^\a 
({f_{ij}}^{\be} u_i^\be 
-f_{ji}^\be  u_j^\be)~ \ra \nn \\ &=& \lim_{\tau \to \infty} \f{1}{\tau}
\int_0^\tau dt \f{1}{4} \sum_{i, j\neq i}  x_{ij}^\a (f_{ij}^\be
u_i^\be -f_{ji}^\be  u_j^\be)
\eea
Now if we integrate across a collision we see that $\int dt ({\bf
  f}_{ij}.\bu_i)$ gives the change in kinetic energy of the $i^{\rm{th}}$
particle during the collision while $ \int dt ({\bf f}_{ji}.\bu_j)$
gives the change in kinetic energy of the $j^{\rm{th}}$ particle. Hence we get
\bea
\la~I_\a^U~\ra &=& \sum_{i, j \neq i} \lim_{\tau \to \infty} \f{1}{\tau}  \sum_{t_c} \f{1}{4}
x_{ij}^\a (\Delta K_i - \Delta K_j) \nn \\ &=& \sum_{i < j}  \f{\la~ 
x_{ij}^\a \Delta K_i~ \ra_c}{\la~\tau_{ij}~\ra_c}
\label{curr}
\eea
where we have used the fact that for elastic collisions $\Delta K_i =
-\Delta K_j $ and $\sum_{t_c}$ denotes a summation over all
collisions, in the time interval $\tau$, 
between pairs $\{ij\}$. The time interval between successive
collisions between $i^{\rm th}$ and $j^{\rm th}$ particles is
denoted by $\tau_{ij}$ and the average $\la~...~\ra_c$ in the last line  denotes a
{\emph{collisional}} average. Thus $\la \tau_{ij} \ra_c =\lim_{\tau \to
    \infty} \tau/N_{ij}(\tau)$, where $N_{ij}(\tau)$ is the number of
  collisions between $i^{\rm th}$ and $j^{\rm th}$ particles in time $\tau$.  
For hard spheres the convective part of the current involves only the
kinetic energy and is given by
$\la ~I_\a^K~\ra =\sum_i \left\la~ (m\bu_i^2 /2) u_i^\a~ \right\ra$.
Using these expressions we  now try to obtain estimates of the heat
\index{heat current}
current and its dependence on strain in the close packed limit where
\index{strain}
the system looks like a solid with the structure of a strained
\index{solid}
triangular lattice. 

In the close packed limit the convection current can be neglected and
we focus only on the conductive part given by $\la I^U \ra = \la I^U_2
\ra$ (for conduction along the $y-$direction).
\index{conduction}
At this point we assume local thermal equilibrium (LTE) which we prove
from our simulation data at the end of this section.
\index{simulation}
Assuming LTE we write the following approximate
form for the energy change $\Delta K_i$ during a collision:
\bea
\D K_{i} =  \kb (T(y_j)-T(y_i)) = -\kb \f{dT}{dy} y_{ij}=y_{ij}
    \f{\kb \D T}{L_y}~, \nn
\eea
where we have denoted $x^{(\a=2)}_i=y_i$ and $\D T=T_1-T_2$. The
temperature gradient has been assumed  to be 
small and constant. Further we assume that in the close packed limit
that we are considering, only nearest neighbor pairs $\{<ij>\}$ contribute
to the current in Eq.~(\ref{curr}) and that they contribute equally. 
We then get the following approximate form for the total current:
\bea
{\la I_y \ra } \approx \f{3N\kb \D T}{L_y} \f{  y_c^2}{\tau_c}~, 
\label{iu}
\eea
where $\tau_c$ is the average time between successive collisions
between two particles while $y_c^2$ is the mean square separation along 
the $y-$axis of the colliding particles. 
Finally, denoting the density of particles by $\r=N/{\cal A}$ we get
for the current density: 
\beq
j_y = \f{\la I_y \ra}{\cal A} \approx \f{3 \r \kb \D T }{  L_y} \f{  y_c^2}{\tau_c}~. 
\label{ju}
\eeq

For strains $\ex$ and $\ey$ in the $x$ and $y$ directions we have
$\r=\r_0/[(1+\ex)~(1+\ey)]$. We estimate $y_c^2$ and $\tau_c$ from a
simple equilibrium free-volume theory, known as fixed neighbour free
\index{free-volume theory}
volume theory (FNFVT) [~See Sec.5.3 of chapter-5~]. 
\index{FNFVT}
In this picture we think
of a single disk moving in a fixed cage formed by taking the average
positions of its six nearest neighbor disks [see Fig.5.10]. For
different values of the strains we then evaluate the average values 
$[y_c^2]_{fv}$ and $[\tau_c]_{fv}$ for the moving particle from FNFVT.
\index{FNFVT}
\index{free-volume theory}
\begin{figure}[t]
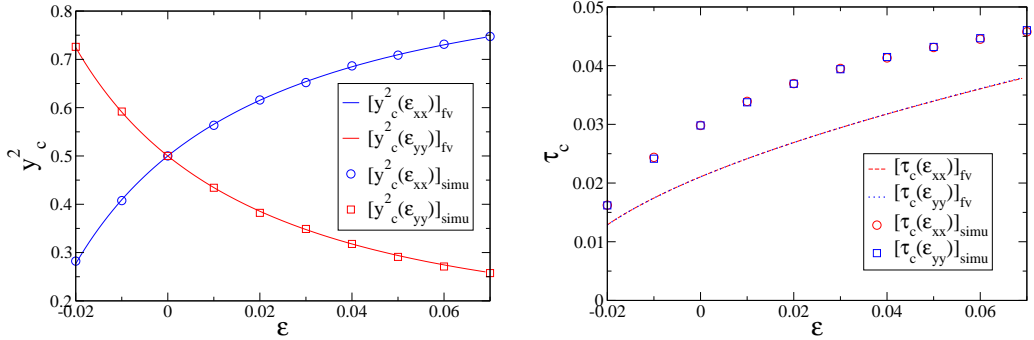

\begin{center}
\includegraphics[width=6.5cm]{y2-eps.eps}
\hskip 0.5cm
\includegraphics[width=6.5cm]{tauc-eps.eps}
\end{center}
\caption{ Plots showing comparision of the analytically calculated
  values of $[y_c^2]_{fv}$ and $[\tau_c]_{fv}$ with those obtained
  from a free volume simulation of a single disk moving within the
\index{simulation}
  free volume cage. The free volume corresponds to a
  starting unstrained triangular lattice at $\eta=0.85$ which is then
  strained along $x$ or $y$ directions.  
}
\label{fv-y2tc}
\end{figure}
We assume that the position of the center of the moving disk
$P_0(x,y)$, at the
time of collision with any one of the six fixed disks, is uniformly
distributed  on the boundary $\cal{B}$ of the free-volume. Hence
$[y_c^2]_{fv}$ is easily calculated using the expression:
\bea
[y_c^2]_{fv}=\f{\sum_i  \int_{{\cal{B}}_i} ds
  (y-y_i)^2}{L_{\cal{B}}}~,\label{ycfv} 
\eea 
where ${\cal{B}}_i$ is the part of the boundary $\cal{B}$ of the free
volume when the middle disk is in contact with the $i^{\rm{th}}$ fixed
disk, $ds$ is the infintesimal length element on $\cal{B}$ while
$L_{\cal{B}}$ is the total length of $\cal{B}$.  
Let the unstrained lattice parameters be
$a_x^0,~a_y^0=\sqrt{3}a_x^0/2$. Under 
strain we have $a_x=a_x^0(1+\e_{xx})$ and $a_y=a_y^0 (1+\e_{yy})$.  
\index{strain}
Using elementary geometry we can then evaluate $[y_c^2]_{fv}$ from
Eq.~(\ref{ycfv}) in terms of $\e_{xx},~\e_{yy}$ and the unstrained
lattice parameter $a_x^0$. An exact calculation of $[\tau_c]_{fv}$ is
nontrivial.  
However we expect $[\tau_c]_{fv}=c~V_{fv}^{1/2}/T^{1/2} $ where
$V_{fv}$ is the ``free 
volume'' [see Fig.~\ref{fvol}] and   $c$ is a constant factor of
$O(1)$ which we will use as a fitting parameter. The calculated values
for $[y_c^2]_{fv}$ and $[\tau_c]_{fv}$ are shown in  Fig.~\ref{fv-y2tc}. 
Also shown are their values obtained from an equilibrium simulation of
\index{simulation}
a single disk moving inside the free volume cage.   
Thus we obtain the following estimate for the heat current:
\index{heat current}
\bea
[j_y]_{fv}= \f{3 \r \kb T^{1/2} \D T }{  L_y} \f{[y_c^2]_{fv} }{c~V_{fv}^{1/2}}~. 
\eea
We plot in Fig.~(\ref{jyex},\ref{jyey}) the above estimate of $[j_y]_{fv}$
along with the results from simulations. We find that the overall features 
of the simulation are reproduced with $c=0.42$. 
\index{simulation}

\begin{figure}[t]
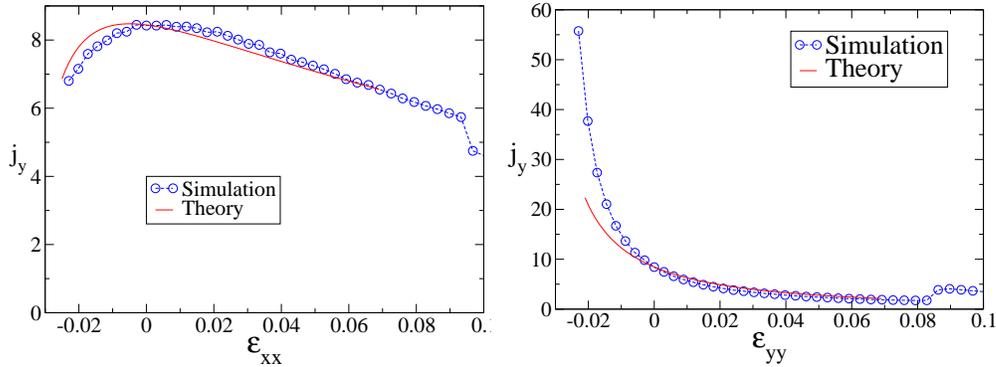

\begin{center}
\includegraphics[width=6.5cm]{jy-ex-negpos.eps}
\includegraphics[width=6.5cm]{jy-ey-negpos.eps}
\end{center}
\caption{ Plot showing effect on $j_y$ of negative strains applied in
  the $x$ and $y$ directions. 
The system is prepared initially in  a triangular lattice at
  $\eta=0.85$. Note that negative $\ex$ reduces $j_y$ whereas negative
  $\ey$ increases $j_y$.
}
\label{comprs}
\end{figure}
\begin{figure}[t]
\begin{center}
\includegraphics[width=9.0cm]{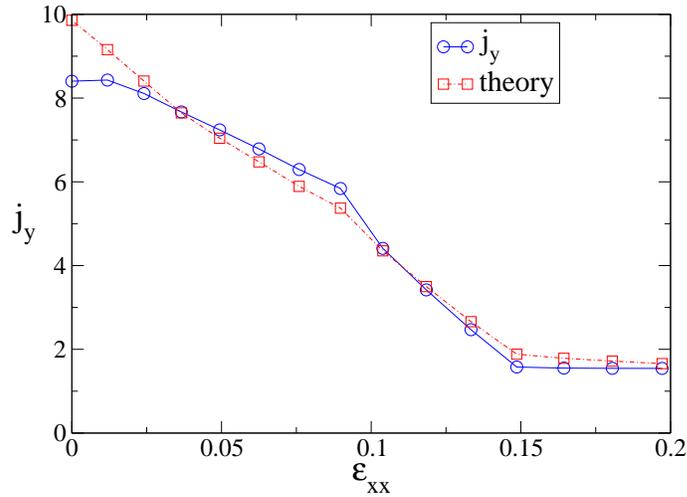}
\end{center}
\caption{Comparison of simulation results for $j_y$ with the
  approximate formula   in Eq.~(\ref{ju}) where   
$\tau_c$ and  $ y_c^2$ are also calculated directly from the same
  simulation. The results are for a $40\times10$ system with starting
\index{simulation}
  value of   $\eta=0.85$ and strained along $x-$direction. 
}
\label{simu-y2tc}
\vspace{1cm} 
\end{figure}
For small strain $[y_c^2(\ex)]_{fv} \sim 0.5+\a \ex - \be_1\ex^2 $,
\index{strain}
$[y_c^2(\ey)]_{fv}\sim0.5-\a \ey+ \be_2\ey^2 $ and
$[\tau_c]_{fv} \sim(\g_1 +\g_2 \e - \g_3 \e^2)$ where $\e$ stands for 
either $\ex$ or $\ey$ and  $\a,~\be_1,~\be_2,~\g_1,\g_2,\g_3$ are all 
positive constants that depend only on
$a_x^0$ (we do not write them explicitly since the expressions are
messy and unilluminating). For 
$\eta=0.85$ we have  $\a=7.62,~\be_1=121.77,~\be_2=124.37$ and 
$\g_1=0.02$, $\g_2=0.33$, $\g_3=1.125$.
From these small strain scaling forms we find that $j_y(\ey)$
\index{scaling}
\index{strain}
always decreases with positive 
$\ey$ and increases with negative or compressive $\ey$ 
(note that we always consider starting configurations of a triangular solid
\index{solid}
of any density). 
On the other hand the sign of the change in $j_y(\ex)$ will
depend on the relative magnitudes of $\a,~ \be_1$ and $\g$ . For
starting density 
$\eta=0.85$, $j_y(\ex)$ decreases both for positive and negative $\ex$. 
In Fig.~\ref{comprs} we show the effect of compressive strains $\ex$ and 
$\ey$ on the heat current $j_y$ and compare the simulation results
\index{heat current}
with the free volume theory. 

It is possible to calculate $y_c^2$ and $\tau_c$ directly  from our  
nonequilibrium collision time dynamics simulation. The mean collision
\index{dynamics}
time $\tau_c$ is obtained by dividing the total simulation time
\index{simulation}
by the total number of collisions per colliding pair.
Similarly $y_c^2$ is evaluated at every collision and we then obtain
its average. Inserting these values of $\tau_c$  and $\la y_c^2\ra$  into
the right hand side of Eq.~(\ref{ju}) we get an estimate of the
current as given by our theory (without making use of free-volume
\index{free-volume theory}
theory). In Fig.~\ref{simu-y2tc} we compare this value of the current
$j_y$, for strain $\e=\ex$, and compare it with the simulation
\index{strain}
\index{simulation}
results. The excellent agreement between the two indicates that our simple
theory is quite accurate.

\begin{figure}[t]
\begin{center}
\includegraphics[width=8.0cm]{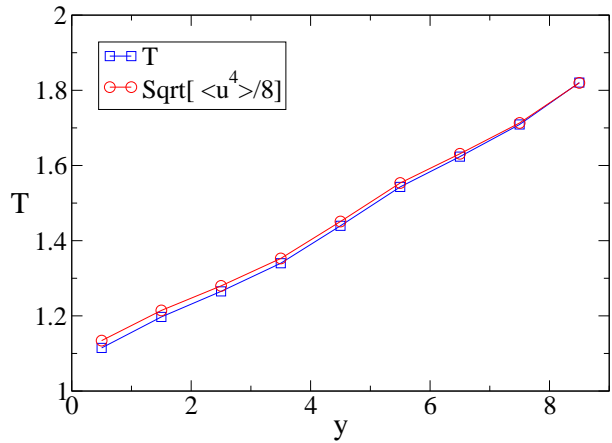}
\end{center}
\caption{ Plot of temperature profile and fourth moment of velocity
  for a strained $40\times 10$ lattice. The unstrained packing
  fraction was $\eta=0.85$ and the system was 
 strained to $\ex=0.0625$.
}
\label{LTE}
\end{figure}

We have also tested the assumptions of a linear temperature
profile and the assumption of local thermal equilibrium (LTE) that we
have used in our theory. In our simulations the local
temperature is defined from the local kinetic energy density,
{\emph{i.e.}} $\kb T = \la m \uu^2 /2  \ra$.  
Local thermal equilibrium requires a close to Gaussian distribution of
the local velocity with a width given by the same temperarture. 
The assumption of LTE can thus be tested by
looking at higher moments of the velocity, evaluated locally. 
Thus we should have $ \la {\uu}^4 \ra = 8(\kb T/m)^2$.
From our simulation we find out $\la {\uu}^4(y)\ra$ and $\kb T(y)$
\index{simulation}
as functions of the distance $y$ from the cold to hot reservoir. The
plot in Fig.~\ref{LTE} shows that the temperature profile is
appoximately linear and LTE is approximately valid.
We use our theory only in the solid phase and in this case there is
\index{solid}
\index{phase}
not much variation in the direction transverse to heat flow
($x-$direction).

\begin{figure}[t]
\begin{center}
\includegraphics[width=8.0cm]{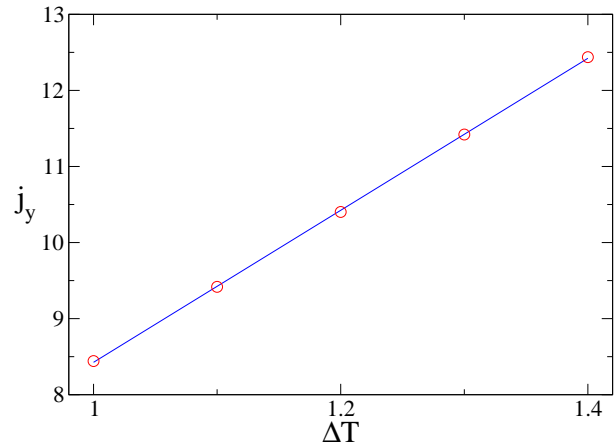}
\end{center}
\caption{
Plot of $j_y$ versus $\D T=T_2-T_1$ for a $40\times 10$ triangular
lattice at $\eta=0.85$. We see that the current increases linearly with
the applied gradient.  
}
\label{linresp}
\end{figure}

Finally we find that the heat conduction in the small confined lattice
\index{conduction}
under small strains shows a linear response behaviour. This can be
seen in Fig.~\ref{linresp} where we plot $j_y$ versus  $\D T =
T_2-T_1$ for  a $40 \times 10$ triangular lattice at $\eta=0.85$.
Note that, as mentioned in the introduction, the bulk thermal
conductivity of a two-dimensional system is expected to be divergent
and the linear response behaviour observed here is only relevant for a
finite system in certain regimes (solid under small strains).  
\index{solid}

\section{Summary and Conclusion}
\label{disc}
In this chapter we have studied heat conduction in a two-dimensional solid 
\index{solid}
\index{conduction}
formed from hard disks confined in a narrow structureless channel. The channel 
has a small width ($\sim 10$
particle layers) and is long ($\sim 100$ particles). Thus our system is in the
\index{layers}
nanoscale regime.  
We have shown that structural changes that occur when this 
solid is strained can lead to sudden jumps in the heat current. 
\index{solid}
\index{heat current}
From the system sizes that we have studied it is not possible to
conclude that these jumps will persist in the limit that the channel
length becomes infinite. However the finite size results are
interesting and relevant since real nano-sized solids {\em are} small.
\index{nano}
We have also proposed a free volume theory type calculation of the heat
\index{heat current}
current. While being heuristic it gives correct order of magnitude
estimates and also reproduces qualitative trends in the current-strain graph.  
\index{strain}
This simple approach should be useful in calculating the heat
conductivity of a hard sphere solid in the high density limit.
\index{solid}

The property of large change of heat current  
could  be utilized to make a system perform as a mechanically controlled
switch of heat current. Similar results are also expected for the
\index{heat current}
electrical conductance and this is shown to be true at least
following one protocol of straining in Ref.\cite{my-econd}. 
From this point of view it seems worthwhile to perform similar
studies on  transport in confined nano-systems in three dimensions and
also with different interparticle interactions.

\appendix



\printindex 

\bibliography{debc}
\end{document}